\documentclass[10pt,letterpaper]{article}
\pdfoutput=1

\usepackage[margin=1in]{geometry}
\usepackage[utf8]{inputenc}
\usepackage[T1]{fontenc}
\usepackage[nopatch=footnote]{microtype}
\usepackage{amsfonts}
\usepackage{amssymb}
\usepackage{amsmath}
\usepackage{amsthm}
\usepackage{booktabs}
\usepackage[inline]{enumitem}
\usepackage[hidelinks,bookmarks,bookmarksopen,bookmarksnumbered]{hyperref}
\usepackage[capitalize]{cleveref}
\usepackage{mathtools}
\usepackage{tikz}
\usepackage[font=small,labelfont=bf]{caption}
\usepackage{titlesec}
\usepackage{thmtools}
\usepackage{wrapfig}
\usepackage{tablefootnote}
\usepackage{thm-restate}
\usepackage{soul}
\usepackage[draft]{fixme}
\usepackage{framed}
\usepackage{tikz}
\usepackage{subcaption}
\usepackage[norefs,nocites,nomsgs]{refcheck}  %
\usepackage{setspace}

\usetikzlibrary{trees}
\usetikzlibrary{arrows}
\usetikzlibrary{arrows.meta,arrows}
\usetikzlibrary{decorations.pathreplacing}
\usetikzlibrary{positioning}
\usetikzlibrary{calc}

\input{settings}

\newcommand{\bigO}{\mathcal{O}}
\newcommand{\Oh}{\bigO}

\DeclareMathOperator*{\argmin}{arg\,min}

\newcommand{\floor}[1]{\left\lfloor #1 \right\rfloor}
\newcommand{\dd}{\mathinner{.\,.}}
\newcommand{\probname}[1]{\text{\sc #1}}

\newcommand{\Pat}{P}
\newcommand{\Text}{T}
\newcommand{\Textinf}{\Text^{\infty}}
\newcommand{\Textlen}{n}
\newcommand{\AlphabetSize}{\sigma}
\newcommand{\IntegerAlphabet}{[0 \dd \AlphabetSize)}
\newcommand{\BinaryAlphabet}{\{{\tt 0}, {\tt 1}\}}
\newcommand{\emptystring}{\varepsilon}
\newcommand{\BitvectorMin}{B_{\rm min}}
\newcommand{\deltatext}{\delta_{\rm text}}
\newcommand{\SSS}{\mathsf{S}}

\newcommand{\Z}{\mathbb{Z}}
\newcommand{\Zz}{\Z_{\ge 0}}
\newcommand{\Zn}{\Zz}
\newcommand{\Zp}{\Z_{>0}}

\newcommand{\SA}[1]{\mathrm{SA}_{#1}}
\newcommand{\ISA}[1]{\mathrm{ISA}_{#1}}

\newcommand{\LCE}[3]{\mathrm{LCE}_{#1}(#2,#3)}

\newcommand{\Int}[3]{\mathrm{int}(#1,#2,#3)}
\newcommand{\Successor}[2]{\mathrm{succ}_{#1}(#2)}
\newcommand{\Substitute}[3]{{\rm sub}(#1,#2,#3)}

\newcommand{\lcp}[2]{\mathrm{lcp}(#1,#2)}
\newcommand{\per}[1]{\mathrm{per}(#1)}
\newcommand{\revstr}[1]{\overline{#1}}
\newcommand{\bin}[2]{{\rm bin}_{#1}(#2)}
\newcommand{\ebin}[2]{{\rm ebin}_{#1}(#2)}

\newcommand{\IndicatorVector}[2]{\mathrm{IndVec}_{#1}(#2)}
\newcommand{\SeqToString}[2]{{\rm SeqToString}_{#1}(#2)}
\newcommand{\BinStrToInt}[1]{{\rm BinStrToInt}(#1)}
\newcommand{\InverseString}[2]{{\rm InvString}_{#1}(#2)}
\newcommand{\PermSeqToString}[3]{{\rm PermSeqToString}_{#1}(#2,#3)}

\newcommand{\OccTwo}[2]{\mathrm{Occ}(#1, #2)}
\newcommand{\RangeBegTwo}[2]{\mathrm{RangeBeg}(#1, #2)}
\newcommand{\RangeEndTwo}[2]{\mathrm{RangeEnd}(#1, #2)}
\newcommand{\LexRange}[3]{\mathrm{LexRange}(#1,#2,#3)}

\newcommand{\DistPrefixPat}[4]{\mathrm{DistPrefix}(#1,#2,#3,#4)}
\newcommand{\DistPrefixPos}[4]{\mathrm{DistPrefix}(#1,#2,#3,#4)}
\newcommand{\DistPrefixes}[3]{\mathcal{D}(#1, #2, #3)}

\newcommand{\RootPos}[3]{\mathrm{root}(#1,#2,#3)}
\newcommand{\RootPat}[2]{\mathrm{root}(#1,#2)}
\newcommand{\HeadPos}[3]{\mathrm{head}(#1,#2,#3)}
\newcommand{\HeadPat}[2]{\mathrm{head}(#1,#2)}
\newcommand{\TailPos}[3]{\mathrm{tail}(#1,#2,#3)}
\newcommand{\TailPat}[2]{\mathrm{tail}(#1,#2)}
\newcommand{\ExpPos}[3]{\mathrm{exp}(#1,#2,#3)}
\newcommand{\ExpPat}[2]{\mathrm{exp}(#1,#2)}
\newcommand{\TypePos}[3]{\mathrm{type}(#1,#2,#3)}
\newcommand{\TypePat}[2]{\mathrm{type}(#1,#2)}
\newcommand{\RunEndFullPos}[3]{e^{\rm full}(#1,#2,#3)}
\newcommand{\RunEndFullPat}[2]{e^{\rm full}(#1,#2)}
\newcommand{\RunEndPos}[3]{e(#1,#2,#3)}
\newcommand{\RunEndPat}[2]{e(#1,#2)}

\newcommand{\RName}{\mathsf{R}}
\newcommand{\RMinusName}{\RName^{-}}
\newcommand{\RPlusName}{\RName^{+}}
\newcommand{\RTwo}[2]{\RName(#1, #2)}
\newcommand{\RThree}[3]{\RName_{#1}(#2, #3)}
\newcommand{\RFour}[4]{\RName_{#1,#2}(#3, #4)}
\newcommand{\RFive}[5]{\RName_{#1,#2,#3}(#4, #5)}
\newcommand{\RMinusTwo}[2]{\RMinusName(#1, #2)}
\newcommand{\RMinusThree}[3]{\RMinusName_{#1}(#2, #3)}
\newcommand{\RMinusFour}[4]{\RMinusName_{#1,#2}(#3, #4)}
\newcommand{\RMinusFive}[5]{\RMinusName_{#1,#2,#3}(#4, #5)}
\newcommand{\RPlusTwo}[2]{\RPlusName(#1, #2)}
\newcommand{\RPlusThree}[3]{\RPlusName_{#1}(#2, #3)}
\newcommand{\RPlusFour}[4]{\RPlusName_{#1,#2}(#3, #4)}
\newcommand{\RPlusFive}[5]{\RPlusName_{#1,#2,#3}(#4, #5)}

\newcommand{\RPrimName}{\RName'}
\newcommand{\RPrimMinusName}{\RPrimName^{-}}
\newcommand{\RPrimPlusName}{\RPrimName^{+}}
\newcommand{\RPrimTwo}[2]{\RPrimName(#1, #2)}
\newcommand{\RPrimMinusTwo}[2]{\RPrimMinusName(#1, #2)}
\newcommand{\RPrimPlusTwo}[2]{\RPrimPlusName(#1, #2)}
\newcommand{\RPrimMinusThree}[3]{\RPrimMinusName_{#1}(#2, #3)}
\newcommand{\RPrimPlusThree}[3]{\RPrimPlusName_{#1}(#2, #3)}
\newcommand{\RMinName}{\RName_{\rm min}}
\newcommand{\RMinMinusTwo}[2]{\RMinName^{-}(#1, #2)}

\newcommand{\Rank}[3]{\mathsf{rank}_{#1}(#2,#3)}
\newcommand{\SpecialRank}[2]{\mathsf{special\mbox{-}rank}_{#1}(#2)}
\newcommand{\Select}[3]{\mathsf{select}_{#1}(#2,#3)}

\newcommand{\PrefixRank}[3]{\mathsf{prefix\mbox{-}rank}_{#1}(#2,#3)}
\newcommand{\PrefixSpecialRank}[3]{\mathsf{prefix\mbox{-}special\mbox{-}rank}_{#1}(#2,#3)}
\newcommand{\PrefixSelect}[3]{\mathsf{prefix\mbox{-}select}_{#1}(#2,#3)}
\newcommand{\PrefixRMQ}[5]{\mathsf{prefix\mbox{-}rmq}_{#1, #2}(#3, #4, #5)}

\newcommand{\TwoSidedRangeCount}[3]{\mathsf{range\mbox{-}count}_{#1}(#2, #3)}
\newcommand{\RangeSelect}[3]{\mathsf{range\mbox{-}select}_{#1}(#2, #3)}
\newcommand{\RMQ}[3]{\mathsf{rmq}_{#1}(#2, #3)}
\newcommand{\ThreeSidedRMQ}[5]{\mathsf{three\mbox{-}sided\mbox{-}rmq}_{#1,#2}(#3, #4, #5)}

\newcommand{\LexSorted}[2]{\mathrm{LexSorted}(#1,#2)}
\newcommand{\Sort}[1]{\mathrm{Sort}(#1)}
\newcommand{\RunsMinusLexSortedTwo}[2]{\mathrm{RunsLexSorted}^{-}(#1,#2)}
\newcommand{\RunsMinusLexSortedThree}[3]{\mathrm{RunsLexSorted}^{-}_{#1}(#2,#3)}
\newcommand{\RunsMinusTextSortedTwo}[2]{\mathrm{RunsTextSorted}^{-}(#1,#2)}
\newcommand{\RunsMinusTextSortedThree}[3]{\mathrm{RunsTextSorted}^{-}_{#1}(#2,#3)}
\newcommand{\RunsMinusLengthSortedTwo}[2]{\mathrm{RunsLengthSorted}^{-}(#1,#2)}

\newcommand{\MinPosBitvectorMinusTwo}[2]{\mathrm{MinPosBitvector}^{-}(#1,#2)}
\newcommand{\MinPosBitvectorMinusFour}[4]{\mathrm{MinPosBitvector}^{-}_{#1,#2}(#3,#4)}
\newcommand{\MinPosBitvectorMinusFive}[5]{\mathrm{MinPosBitvector}^{-}_{#1,#2,#3}(#4,#5)}
\newcommand{\ExpBlockBitvectorMinusTwo}[2]{\mathrm{ExpBlockBitvector}^{-}(#1,#2)}

\newcommand{\NavNonperiodic}[3]{\mathrm{NavNonperiodic}(#1, #2, #3)}
\newcommand{\NavPeriodic}[2]{\mathrm{NavPeriodic}(#1, #2)}

\newcommand{\zero}{{\tt 0}}
\newcommand{\one}{{\tt 1}}
\newcommand{\two}{{\tt 2}}
\newcommand{\three}{{\tt 3}}
\newcommand{\four}{{\tt 4}}

\begin{document}

\title{Explaining the Inherent Tradeoffs for Suffix Array\\ Functionality: Equivalences between String Problems\\ and Prefix Range Queries}

\author{
  \large Dominik Kempa\thanks{Partially funded by the
  NSF CAREER Award 2337891 and the Simons Foundation
  Junior Faculty Fellowship.}\\[-0.3ex]
  \normalsize Department of Computer Science,\\[-0.3ex]
  \normalsize Stony Brook University,\\[-0.3ex]
  \normalsize Stony Brook, NY, USA\\[-0.3ex]
  \normalsize \texttt{kempa@cs.stonybrook.edu}
  \and
  \large Tomasz Kociumaka\\[-0.3ex]
  \normalsize Max Planck Institute for Informatics,\\[-0.3ex]
  \normalsize Saarland Informatics Campus,\\[-0.3ex]
  \normalsize Saarbrücken, Germany\\[-0.3ex]
  \normalsize \texttt{tomasz.kociumaka@mpi-inf.mpg.de}
}

\date{\vspace{-0.5cm}}
\maketitle

\begin{abstract}
  We address a fundamental question in string processing: how much
  time is needed to access suffix array entries when there is
  insufficient space to store it in plain form?  The suffix array
  $\SA{\Text}[1 \dd \Textlen]$ of a text $\Text$ of length $\Textlen$
  is a permutation of $\{1, \dots, \Textlen\}$ that encodes the
  lexicographic ordering of the suffixes of $\Text$.  While its
  canonical application is as a pattern matching index, the suffix
  array’s simplicity, space efficiency, and theoretical guarantees
  have made it a cornerstone of string processing over the past 35
  years, with applications spanning data compression, bioinformatics,
  and information retrieval.

  Prior work has developed unidirectional reductions, showing how
  suffix array queries (given $i \in [1 \dd \Textlen]$, return
  $\SA{\Text}[i]$) can be reduced, among others, to rank queries over
  the Burrows--Wheeler Transform (BWT).  More recently, an alternative
  family of \emph{prefix queries} was introduced, along with a
  reduction that transforms a simple tradeoff for prefix
  queries—requiring little more than a page to develop—into a suffix
  array tradeoff that matches all known space and query time bounds,
  while achieving sublinear construction time. For a binary text
  $\Text \in \BinaryAlphabet^{\Textlen}$, this tradeoff achieves space
  usage $S(\Textlen) = \bigO(\Textlen)$ bits, preprocessing time
  $P_t(\Textlen) = \bigO(\Textlen / \sqrt{\log \Textlen})$,
  preprocessing space $P_s(\Textlen) = \bigO(\Textlen)$ bits, and
  query time $Q(\Textlen) = \bigO(\log^{\epsilon} \Textlen)$ for any constant $\epsilon > 0$. Despite
  this progress, a key question remains: can any of these complexities
  be improved using fundamentally different techniques?

  In this work, we address this question as follows:

  \begin{itemize}
  \item We establish the first \ul{bidirectional} reduction, proving
    that suffix array queries are, up to an additive $\bigO(\log \log
    \Textlen)$ term in query time, \emph{equivalent} to prefix select
    queries in all four aspects. For a sequence $W[1 \dd m]$ of
    $\Oh(\log m)$-bit strings, a prefix select query, given
    $i \in [1 \dd m]$ and a string~$X$, returns the position $j$ of
    the $i$th leftmost string $W[j]$ in $W$ with prefix $X$. Our
    reduction unifies previous approaches and offers a structured
    framework for deriving new tradeoffs without sacrificing
    generality: before this work, prefix select queries constituted
    merely one approach to efficient suffix array queries, whereas we
    prove that, up to the $\bigO(\log \log \Textlen)$ additive overhead in
    query time, they capture all possible suffix array
    representations.
  \item We further show that other core string-processing
    problems—such as inverse suffix array queries, pattern ranking,
    lexicographic range queries, and pattern SA-interval queries—are
    also computationally equivalent to distinct types of prefix
    queries. In total, we identify \emph{six} fundamental problem
    pairs, each consisting of a string problem and a corresponding
    prefix problem, and prove analogous equivalences.
  \end{itemize}
  Taken together, our reductions provide a general method for
  analyzing the efficiency of suffix array and related queries in
  terms of prefix queries, demonstrating that further improvements to
  fundamental string problems can be achieved by focusing solely on
  this recently introduced class of abstract queries.
\end{abstract}

\thispagestyle{empty}
\clearpage
\setcounter{page}{1}

\section{Introduction}\label{sec:intro}

The suffix array (SA)~\cite{ManberM90} is one of the most fundamental
and widely used text indexing structures in computer science. Given a
string $\Text \in \Sigma^{\Textlen}$ of length $\Textlen$ over an
alphabet $\Sigma$, its suffix array $\SA{\Text}[1 \dd \Textlen]$ is a
permutation of $\{1, 2, \dots, \Textlen\}$ that orders the suffixes of
$\Text$ lexicographically, i.e., so that
$\Text[\SA{\Text}[1] \dd \Textlen] \prec \Text[\SA{\Text}[2] \dd \Textlen]
\prec \dots \prec \Text[\SA{\Text}[\Textlen] \dd \Textlen]$.
This simple yet powerful structure enables fast pattern matching:
given a length-$m$ pattern $\Pat \in \Sigma^{m}$, we can determine
whether $\Pat$ appears in $\Text$ via binary search over $\SA{\Text}$
in $\bigO(m \log \Textlen)$ time.

Shortly after it was introduced, researchers recognized that the
suffix array has applications far beyond pattern matching. Due to its
simplicity, space efficiency, strong theoretical guarantees, and
excellent practical performance, the suffix array has, over the past
35 years, become one of the most widely used data
structures~\cite{gusfield,bwtbook,AbouelhodaKO04}, with applications
in string processing, data compression, and information retrieval:

\begin{itemize}

\item Suffix arrays can be augmented~\cite{AbouelhodaKO04} to
  implement the full suffix tree functionality~\cite{Weiner73},
  enabling efficient solutions to classical problems such as
  longest common substring~\cite{gusfield,Charalampopoulos21},
  longest common extension (LCE) queries~\cite{LandauV88,sss},
  longest repeating substring~\cite{gusfield},
  shortest unique substring~\cite{IlieS11}, and
  minimal absent word~\cite{BartonHMP14}.

\item Suffix arrays underlie widely used compression algorithms,
  such as the Burrows--Wheeler Transform (BWT)~\cite{bwt,OkanoharaS09,sss} and
  the Lempel--Ziv (LZ77) factorization~\cite{CrochemoreIS08,OhlebuschG11,KempaP13,GotoB13}.
  This, in turn, determines the complexity of finding
  runs (maximal repeats)~\cite{main1989detecting,KolpakovK99,ChenPS07,CrochemoreI08},
  approximate repetitions~\cite{KolpakovK03},
  local periods~\cite{DuvalKKLL04}, and seeds~\cite{KociumakaKRRW20}.

\item Suffix arrays solve numerous central bioinformatics problems, such as
  sequence alignment~\cite{bowtie2,bwa},
  maximal exact matches (MEMs)~\cite{gusfield},
  maximal unique matches (MUMs)~\cite{kurtz2004versatile},
  tandem repeats~\cite{gusfield2004linear},
  genome assembly~\cite{GonnellaK12,OhlebuschG10},
  sequence clustering~\cite{hazelhurst2011kaboom},
  $k$-mer counting~\cite{kurtz2008new},
  error correction~\cite{ilie2011hitec},
  and many others~\cite{gusfield,ennobook,Grossi11}.

\item Suffix arrays, in addition to
  basic exact pattern matching~\cite{ManberM90},
  solve numerous other variants, including
  approximate pattern matching~\cite{Navarro01,LandauV88},
  gapped pattern matching~\cite{CaceresPZ20},
  parameterized pattern matching~\cite{FujisatoNIBT19},
  pattern matching with ``don't care'' symbols (also known as \emph{wildcards})~\cite{ManberB91},
  regular expression pattern matching~\cite{gusfield,baeza1999modern,ArroyueloBCMN14}, and
  compressed pattern matching~\cite{Gagie2020}.

\end{itemize}

Numerous variants and generalizations of suffix arrays have been developed, including
sparse SAs~\cite{Prezza18,GawrychowskiK17},
multi-string SAs~\cite{shi1996suffix},
dynamic SAs~\cite{dynsa}, on-disk SAs~\cite{GogMCTW14},
distributed SAs~\cite{FlickA19},
enhanced SAs~\cite{AbouelhodaKO04},
gapped SAs~\cite{CrochemoreT10},
compact SAs~\cite{Makinen03},
succinct SAs~\cite{MakinenN05},
entropy-compressed SAs~\cite{GrossiV05,FerraginaM05,sss},
dictionary-compressed SAs~\cite{Gagie2020,collapsing},
SAs for circular strings~\cite{HonLST11},
SAs for weighted strings~\cite{Charalampopoulos20},
SAs for matrices~\cite{KimKP03},
SAs for alignments~\cite{NaPLHLMP13},
SAs for tries~\cite{FerraginaLMM05}, and
SAs for graphs~\cite{BrisaboaCFR18}.

In the majority of the above applications, suffix array construction and query time
determine the complexity of the entire algorithm. Unsurprisingly, 
understanding the tradeoff between the space needed to represent the suffix array,
its query time, and its construction time and working space, is one of the most
studied problems in
stringology~\cite{%
  GrossiV05,FerraginaLMM05,wt,cst,HonSS03,Makinen03,HonLSS03,LamSSY02,%
  MakinenN05,FoschiniGGV06,bwtbook,Grossi11,Belazzougui14,BelazzouguiN14,GonzalezNF14,GogNP15,%
  MunroNN17,NavarroM07,Kempa19,
  Gagie2020,MunroNN20,Gao0N20,MatsudaSST20,CobasGN21,breaking,collapsing%
}.

In this paper, we address the above question in its most fundamental and widely studied variant:
What are the optimal query time, space usage, and construction time/space for a data structure that, for any static text $\Text \in \IntegerAlphabet^{\Textlen}$, can answer suffix array (SA) queries for the text $\Text$ (that is, given any $i \in [1 \dd \Textlen]$, return $\SA{\Text}[i]$)?

The standard suffix array (SA) implementation~\cite{ManberM90}
requires $\Theta(n \log n)$ bits, storing each entry $\SA{\Text}[i]$
with $\lceil \log \Textlen \rceil$ bits. A major breakthrough,
discovered independently by Ferragina and Manzini~\cite{FerraginaM00}
and by Grossi and Vitter~\cite{GrossiV00}, demonstrated SA
representations using $\bigO(n \log \AlphabetSize)$ bits, equivalent
to $\bigO(\Textlen / \log_{\AlphabetSize} \Textlen)$ machine words, with query
support in $\bigO(\log^{\epsilon} \Textlen)$ time for any constant
$\epsilon > 0$. These structures, known as the
\emph{FM-index}~\cite{FerraginaM00} and the
\emph{Compressed Suffix Array (CSA)}~\cite{GrossiV00}, respectively,
match the asymptotic space of storing $\Text$, where each symbol requires
$\Theta(\log \AlphabetSize)$ bits.

Following these results, researchers turned their attention to
optimizing the remaining aspects of the suffix array, with the most
pressing being the time and working space usage of algorithms
constructing data structures with SA functionality. Given that both
the input text and the output SA representation use only $\bigO(n \log
\AlphabetSize)$ bits, an optimal
$\bigO(\Textlen / \log_{\AlphabetSize} \Textlen)$ construction time
is feasible in principle. The first efficient
construction~\cite{HonSS03} achieved
$\bigO(\Textlen \log \log \AlphabetSize)$ time while preserving the
$\bigO(\Textlen \log \AlphabetSize)$-bit working space. Subsequent
improvements reduced this time to randomized
$\bigO(\Textlen)$~\cite{Belazzougui14}, deterministic
$\bigO(\Textlen)$~\cite{MunroNN17}, and finally
$\bigO(n \min(1, \log \AlphabetSize / \sqrt{\log \Textlen}))$~\cite{breaking}.
The latest structure in~\cite{breaking} achieves the optimal size of
$\bigO(\Textlen \log \AlphabetSize)$ bits, answers SA queries on
$\Text$ in $\bigO(\log^{\epsilon} \Textlen)$ time, and, given the text
$\Text$ represented in $\bigO(\Textlen \log \AlphabetSize)$ bits, can
be built in
$\bigO(\Textlen \min(1, \log \AlphabetSize / \sqrt{\log \Textlen}))$
time and within the optimal $\bigO(\Textlen \log \AlphabetSize)$-bit
working space. For example, when $\AlphabetSize = 2$, the structure uses $\bigO(\Textlen)$ bits (that is, $\bigO(\Textlen / \log \Textlen)$ machine words) and the construction time is $\bigO(\Textlen / \sqrt{\log \Textlen})$.

Despite all these advances, the tradeoff between SA space, query
time, and construction time/working space is still not fully understood.
Each of the existing SA structures employs distinct techniques: the
FM-index relies on the Burrows--Wheeler Transform (BWT) and rank
queries~\cite{FerraginaM00}, while the CSA is built on the $\Psi$
function~\cite{GrossiV00}. Recent work introduces yet another
paradigm, reducing SA queries to \emph{prefix rank} and
\emph{select queries}~\cite{breaking}. However, the lack of bidirectional reductions between these approaches leaves open a fundamental question: Are there other, still undiscovered techniques that could further reduce the space, query time, or construction time of space-efficient SAs, or have we reached the theoretical limit?
Answering this question has immediate consequences for
dozens of other problems and is key to understanding more
complex variants of suffix arrays. We thus pose our question as:

\begin{center}
  \emph{Is there a query whose complexity governs the time/space
    tradeoff for space-efficient SA representation?}
\end{center}

\vspace{0.5ex}
\noindent
\textbf{Our Results}$\ \ $
After 35 years since the discovery of the suffix array, we prove the
\ul{first bidirectional reduction}, showing that suffix array queries
are nearly perfectly equivalent to \emph{prefix select}
queries~\cite{breaking} in all four aspects (space usage, query time,
construction time, and construction working space). In other words,
while the reduction in~\cite{breaking} proved that prefix rank/select
queries provide \emph{one} way to achieve efficient SA queries, we
establish here that, up to a very small additive term in the query time,
they in fact capture \emph{every possible} approach to SA
structures and their constructions. Hence, to obtain future tradeoffs,
it essentially suffices to focus on prefix select queries.

Our reduction distills the core difficulty of the SA indexing
problem---which involves handling complex situations such as suffixes
sharing long prefixes---into a nearly perfectly computationally
equivalent, yet much more straightforward, problem on short
bitstrings. Such reductions are important because they reshape our understanding of the problem, allowing researchers to focus on significantly simpler tasks without losing essentially any generality. 
Indeed, as
demonstrated by prior work~\cite{breaking}, the tradeoff for
prefix select queries that leads to the state-of-the-art parameters of
suffix array representations takes only around \emph{one page} to
develop (see~\cite[Proposition~4.3]{breaking}).

To state our results in more detail, let us, for simplicity, consider
the case of a binary alphabet ($\AlphabetSize = 2$).\footnote{We focus
  on strings over a binary alphabet $\Sigma = \BinaryAlphabet$,
  because, as proved later in this paper,
  any larger alphabet can be efficiently reduced to the binary case; see, e.g.,
  \cref{sec:sa-alphabet-reduction,sec:prefix-select-alphabet-reduction}.}
We now formally define the problems studied in this paper. 
As in previous works~\cite{HonSS03,Belazzougui14,MunroNN17,breaking},
we assume that strings in size-$N$ instances of both problems are
represented in the \emph{packed} form, in which a string
$S \in \IntegerAlphabet^{\Textlen}$ is stored using
$\Theta(\lceil\Textlen / \log_{\AlphabetSize} N\rceil)$ machine
words, with each word storing
$\Oh(\log_{\AlphabetSize} N)$ characters from the alphabet
$\IntegerAlphabet$.

\vspace{0.5ex}
\begin{description}[style=sameline,itemsep=1ex,font={\normalfont\bf}]

\item[\probname{Indexing for Suffix Array Queries}:] Given a binary
  text $\Text \in \BinaryAlphabet^{\Textlen}$, represented using
  $\bigO(\Textlen)$ bits, construct a data structure that, given any
  $i \in [1 \dd \Textlen]$, returns $\SA{\Text}[i]$, i.e., the
  starting position of the lexicographically $i$th smallest suffix of
  $\Text$. We define the input size as $N := \Textlen$ bits,
  which matches the length of the text~$T$.

\item[\probname{Indexing for Prefix Select Queries}:] Given a sequence
  $W[1 \dd m]$ of $m \geq 1$ binary strings of length $\ell = 1 +
  \lfloor \log m \rfloor$ (each represented using $\bigO(\ell)$ bits),
  construct a data structure that, given an $\bigO(\ell)$-bit representation
  of any string $X \in \BinaryAlphabet^{\leq \ell}$ and
  any integer $r \in [1 \dd |\{j \in [1 \dd m] : X\text{ is a prefix of }W[j]\}|]$,
  returns the
  $r$th smallest element of the set $\{j \in [1 \dd m] : X\text{ is a
  prefix of }W[j]\}$.
  We define the input size as $N := m \cdot \ell$ bits.
\end{description}

Our main result can be summarized in the following theorem:

\begin{theorem}[Main result of this paper]\label{th:main}
  The problems of \probname{Indexing} \probname{for} \probname{Suffix}
  \probname{Array} \probname{Queries} and \probname{Indexing}
  \probname{for} \probname{Prefix} \probname{Select}
  \probname{Queries} are equivalent in the following sense. Consider
  any data structure that, for an input of at most
  $N$ bits to one of the problems, achieves the following
  complexities:
  \begin{itemize}
  \item space usage of $S(N)$ bits,
  \item preprocessing time $P_t(N)$,
  \item preprocessing space of $P_s(N)$ bits (working space),
  \item query time $Q(N)$.
  \end{itemize}
  Then, for every integer $N'$, there exists $N=\Oh(N')$ such that, given an
  input instance of at most $N'$ bits to the other problem, we can
  construct, in $\bigO(P_t(N))$ time and using $\bigO(P_s(N))$ bits of
  working space, a data structure of size $\bigO(S(N))$ bits that
  answers queries in $\bigO(\log \log N + Q(N))$ time.
\end{theorem}

\begin{table}[t!]
  \centering
  \begin{tabular}{l l r}
    \toprule
    String Problem & Prefix Problem & Reference\\
    \midrule
    \probname{Suffix Array Queries} & \probname{Prefix Select Queries} & \cref{sec:prefix-select}\\
    \probname{Inverse Suffix Array Queries} & \probname{Prefix Special Rank Queries} & \cref{sec:prefix-special-rank}\\
    \probname{Pattern Ranking Queries} & \probname{Prefix Rank Queries} & \cref{sec:prefix-rank}\\
    \probname{Lex-Range Reporting Queries} & \probname{Prefix Range Reporting Queries} & \cref{sec:prefix-range-reporting}\\
    \probname{Lex-Range Emptiness Queries} & \probname{Prefix Range Emptiness Queries} & \cref{sec:prefix-range-emptiness}\\
    \probname{Lex-Range Minimum Queries} & \probname{Prefix Range Minimum Queries} & \cref{sec:prefix-rmq}\\
    \bottomrule
  \end{tabular}
  \caption{Near-perfect equivalences established in this paper between central string-processing problems (left) and fundamental prefix problems (right).
    Additionally, in \cref{sec:equiv-pattern-ranking-and-pattern-sa-interval}, we prove
    that the problem of indexing for
    \probname{Pattern Ranking Queries} is perfectly equivalent to the problem of
    \probname{Pattern SA-Interval Queries}.}\label{tab:equiv-problems}
\end{table}

\vspace{3ex}
\noindent
\textbf{Beyond Suffix Array Queries}$\ \ $
The above reduction raises natural questions: Are other fundamental string queries nearly equivalent to certain prefix queries? 
Conversely, are basic prefix queries nearly equivalent to natural string queries? 
For example, one may ask these questions for \emph{Inverse Suffix Array} (ISA) queries (given a position $j\in[1\dd \Textlen]$, return the index $i\in[1\dd \Textlen]$ such that $\SA{\Text}[i]=j$) and for \emph{prefix rank} queries (given $i\in[1\dd m]$ and $X\in\BinaryAlphabet^{\le \ell}$, compute $\PrefixRank{W}{i}{X}\coloneqq |\{j\in[1\dd i]:X\text{ is a prefix of }W[j]\}|$).

Remarkably, the answer to both questions is affirmative. 
In total, we establish \emph{six} equivalences between fundamental string queries, such as SA and ISA, and basic prefix queries, including prefix select and prefix rank. 
These six problems and their corresponding prefix formulations are summarized in \cref{tab:equiv-problems}.

Perhaps surprisingly, although ISA queries were previously answered via a reduction to prefix rank queries~\cite{breaking}, we do not obtain the converse reduction. 
Instead, we show that ISA queries correspond to what we call \emph{Prefix Special Rank} queries.
A prefix special rank query is defined on the same instance as prefix rank and select queries; given an index $i\in[1\dd m]$ and a position $p\in[0\dd \ell]$, it returns $\PrefixRank{W}{i}{W[i][1\dd p]}$ (see \cref{def:prefix-rank-and-select}). 
This equivalence is analogous to \cref{th:main}, incurring only an additional $\bigO(\log\log N)$ factor in query time.

By contrast, the fully general prefix rank queries are equivalent to another fundamental problem, \emph{Pattern Ranking}: computing the lexicographic rank of a pattern $\Pat\in\Sigma^{m}$ among all text suffixes (i.e., counting suffixes lexicographically smaller than $\Pat$).
This equivalence is more subtle because the input pattern may occupy $\omega(1)$ machine words (equivalently, $\omega(\log N)$ bits) under the standard word-RAM model with word size $\Theta(\log N)$. We therefore state both problems more formally below to clarify how this input-size issue affects the reductions.

\vspace{0.5ex}
\begin{description}[style=sameline,itemsep=1ex,font={\normalfont\bf}]

\item[\probname{Indexing for Pattern Ranking Queries}:] Given a binary
  text $\Text \in \BinaryAlphabet^{\Textlen}$, represented using
  $\bigO(\Textlen)$ bits, construct a data structure that, given any
  pattern $\Pat \in \BinaryAlphabet^{m}$ represented in $\bigO(m)$
  bits, returns $|\{i \in [1 \dd \Textlen] :
  \Text[i \dd \Textlen] \prec \Pat\}|$. 
  We define the input size as $N := \Textlen$ bits and the pattern size as $M:=m$~bits.

\item[\probname{Indexing for Prefix Rank Queries}:] Given a sequence
  $W[1 \dd m]$ of $m \geq 1$ binary strings of length $\ell = 1 +
  \lfloor \log m \rfloor$ (each represented using $\bigO(\ell)$ bits),
  construct a data structure that, given any $i \in [0 \dd m]$ and an
  $\bigO(\ell)$-bit representation of any
  $X \in \BinaryAlphabet^{\leq \ell}$,
  returns $\PrefixRank{W}{i}{X} = |\{j \in [1 \dd i] :
  X\text{ is a prefix of }W[j]\}|$.  
  We define the input size as $N := m \cdot \ell$ bits.
\end{description}

\noindent
Unlike the earlier two-step reductions, our reduction sequence involves \emph{three} problems, with a special variant of pattern ranking as the intermediate problem. In more detail:
\begin{enumerate}

\item We prove that,
  given any data structure for prefix rank queries with
  complexities $S(N)$, $P_t(N)$, $P_s(N)$, and $Q(N)$ (defined as in
  \cref{th:main}), we can derive a structure for pattern ranking with complexities $S'(N) = \Oh(S(N'))$,
  $P'_t(N) = \Oh(P_t(N'))$, $P'_s(N) = \Oh(P_s(N'))$, and
  $Q'(N,M) = \bigO(\log \log N + M / \log N + Q(N'))$ for some $N'=\Oh(N)$, where $Q'(N,M)$ is the time
  complexity of answering ranking queries on $M$-bit patterns.

\item We then prove that,
  given any structure for pattern ranking queries \emph{restricted for patterns
  of length $\bigO(\log N)$} that achieves complexities $S(N)$,
  $P_t(N)$, $P_s(N)$, and $Q(N)$, we can achieve $S'(N) = \Oh(S(N'))$,
  $P'_t(N) = \Oh(P_t(N'))$, $P'_s(N) = \Oh(P_s(N'))$, and $Q'(N) = \Oh(Q(N'))$ for prefix rank for some $N'=\Oh(N)$.

\item To complete the cycle, we observe that, by definition, if we can
  answer pattern ranking queries in $Q(N,M)=\bigO(M / \log N + Q(N))$ time
  for patterns of any length $M$, then we can answer them for patterns of
  length $M = \bigO(\log N)$ in $\bigO(Q(N))$ time.

\end{enumerate}

This cycle of reductions thus shows that:
(1) the pattern ranking problem achieves maximal hardness (that is,
  the overhead beyond the time necessary to read the query input) already
  for patterns of length $\bigO(\log N)$, and
(2) this restricted pattern ranking problem is nearly perfectly
  equivalent to prefix rank queries.

\vspace{3ex}
\noindent
\textbf{More Complex Prefix Queries}$\ \ $
Having established the string problems corresponding to the prefix
rank, prefix select, and prefix special rank problems, we turn our
attention to more complex prefix \emph{range} queries. Let again
$W[1 \dd m]$ be a sequence of $m$ binary strings of length
$\ell := 1 + \lfloor \log m \rfloor$, and let $b, e \in [0 \dd m]$ and
$X \in \BinaryAlphabet^{\leq \ell}$.  The next two types of queries we
study ask about the set
$\mathcal{J} := \{j \in (b \dd e] : X\text{ is a prefix of }W[j]\}$ described using the positions $b,e$ and the string $X$.
Specifically, in \emph{Prefix Range Reporting}, we ask to report all
elements of the set $\mathcal{J}$.  In \emph{Prefix Range Emptiness}, we simply ask whether
$\mathcal{J} \neq \emptyset$. In addition to these two prefix range
queries, recent work on sublinear-time LZ77
factorization~\cite{sublinearlz} defined yet another, more complex,
variant of prefix queries:
\emph{Prefix Range Minimum Queries (Prefix RMQ)}. In the prefix RMQ
problem, in addition to the sequence $W[1 \dd m]$ of $m$ strings, we are
also given an array $A[1 \dd m]$ of $m$ integers. At query time, given
any $b, e \in [0 \dd m]$ and $X \in \BinaryAlphabet^{\leq \ell}$,
we ask to return $\argmin\{A[j] : j\in \mathcal{J}\}$.
In other words, we seek the position of the smallest value in the array $A$ restricted to indices $j\in (b\dd e]$ for which the corresponding string $W[j]$ has $X$ as a prefix.

In this paper, we prove that each of these three types of prefix range
queries is nearly perfectly equivalent (again, up to an additional
$\bigO(\log \log N)$ term in the query time) to a distinct fundamental string
problem in which, given two patterns
$\Pat_1, \Pat_2 \in \BinaryAlphabet^{*}$, we ask about the set
$\LexRange{\Pat_1}{\Pat_2}{\Text} :=
\{j \in [1 \dd \Textlen] : \Pat_1 \preceq \Text[j \dd \Textlen] \prec \Pat_2\}$
(\cref{def:lex-range}), that is, the contiguous block in the suffix
array of $\Text$ containing all suffixes that are lexicographically
between $\Pat_1$ and $\Pat_2$ (see \cref{rm:lex-range}).
These types of problems arise, for example, in algorithms for
suffix sorting and in the construction of the
Burrows--Wheeler Transform (BWT)~\cite{srm,FerraginaGM12,pSAscan}.
We show that:

\begin{itemize}

\item Prefix range reporting is nearly perfectly equivalent to enumerating
  $\LexRange{\Pat_1}{\Pat_2}{\Text}$ (\cref{sec:prefix-range-reporting}).

\item Prefix range emptiness is nearly perfectly equivalent to checking if
  $\LexRange{\Pat_1}{\Pat_2}{\Text} \neq \emptyset$
  (\cref{sec:prefix-range-emptiness}).

\item Prefix range minimum is nearly perfectly equivalent to computing
  $\min \LexRange{\Pat_1}{\Pat_2}{\Text}$ (\cref{sec:prefix-rmq}).

\end{itemize}

These six near-perfect equivalences summarized in
\cref{tab:equiv-problems} unveil a remarkably clean structure
underlying many central string problems, which, at their core, all
reduce to much simpler rank/select and range queries over short
bitstrings.

\vspace{3ex}
\noindent
\textbf{Alphabet Reductions}$\ \ $
As an auxiliary result of the above reductions, we establish a very
useful fact: All twelve of the studied problems, which naturally
generalize to strings over an integer alphabet $\IntegerAlphabet$,
already reach maximal computational hardness for the binary alphabet
($\AlphabetSize = 2$). Moreover, any instance with a larger alphabet
can be efficiently reduced to the binary case.

To illustrate this reduction principle, let us consider the most
fundamental of the twelve problems: Indexing a string
$\Text \in \IntegerAlphabet^{\Textlen}$ for suffix array
queries. In~\cite{breaking}, it was shown that, given the
$\bigO(\Textlen \log \AlphabetSize)$-bit representation of a text
$\Text \in \IntegerAlphabet^{\Textlen}$ and any constant $\epsilon > 0$,
we can construct a data structure of size
$\bigO(\Textlen \log \AlphabetSize)$ bits in
$\bigO((\Textlen \log \AlphabetSize) / \sqrt{\log \Textlen})$ time and
using $\bigO(\Textlen \log \AlphabetSize)$ bits of working space. This
structure allows us to retrieve $\SA{\Text}[i]$ for any
$i \in [1 \dd \Textlen]$ in $\bigO(\log^{\epsilon} \Textlen)$ time.
For $\AlphabetSize = 2$, we thus obtain an $\Oh(\Textlen)$-bit structure which can be constructed in
$\bigO(\Textlen / \sqrt{\log \Textlen})$ time using $\bigO(\Textlen)$
bits of working space. In this paper, we show that to
achieve the same tradeoff for any alphabet size $\AlphabetSize \geq 2$,
it would suffice to develop the above result only for
$\AlphabetSize = 2$. More precisely, in
\cref{sec:sa-alphabet-reduction} we prove that, given the packed
representation of $\Text \in \IntegerAlphabet^{\Textlen}$, we can compute, in
$\bigO(\Textlen / \log_{\AlphabetSize} \Textlen)$ time,
integers $\alpha$, $\beta$, $\gamma$, $\mu$, and the packed
representation of a binary string $\Text_{\rm bin}$ of length $|\Text_{\rm bin}| = \Theta(\Textlen \log \AlphabetSize)$ such that, for every $i \in [1 \dd \Textlen]$, it holds
\[
  \SA{\Text}[i] = \frac{\SA{\Text_{\rm bin}}[\alpha + i] - \beta}{\gamma} + \mu.
\]
Thus, even if the construction from~\cite{breaking} were available
only for the binary alphabet, our reduction would allow us to extend it to
larger alphabets in a black-box manner, significantly simplifying the
original algorithm. In
\cref{sec:prefix-select-alphabet-reduction,%
    sec:isa-alphabet-reduction,%
    sec:prefix-special-rank-alphabet-reduction,%
    sec:pattern-ranking-alphabet-reduction,%
    sec:prefix-rank-alphabet-reduction,%
    sec:lex-range-reporting-alphabet-reduction,%
    sec:prefix-range-reporting-alphabet-reduction,%
    sec:lex-range-emptiness-alphabet-reduction,%
    sec:prefix-range-emptiness-alphabet-reduction,%
    sec:lex-range-minimum-alphabet-reduction,%
    sec:prefix-rmq-alphabet-reduction},
we also establish analogous results for the remaining problems analyzed in
this paper.

\vspace{3ex}
\noindent
\textbf{Related Work}$\ \ $
Many of the queries studied in this paper have also been examined in
the compressed setting, where the input string is highly repetitive,
and the goal is to answer queries using space close to the size of the text in
compressed form. Parallel work~\cite{dichotomy} significantly advances
the understanding of string queries in this setting, showing that
in $\bigO(\delta(\Text) \log^{\bigO(1)} \Textlen)$ space (a well-studied and
widely accepted space bound for compressed data structures),
central string queries fall into two categories: those with complexity
$\Theta(\tfrac{\log \Textlen}{\log \log \Textlen})$ (including
suffix array (SA), inverse suffix array (ISA), longest common
prefix (LCP) array, and longest common extension (LCE) queries) and
those with complexity $\Theta(\log \log \Textlen)$ (including
the Burrows--Wheeler transform (BWT), permuted longest common prefix
(PLCP) array, last-to-first (LF) mapping, inverse last-to-first
(LF$^{-1}$) mapping, lexicographical predecessor ($\Phi$), and inverse
lexicographical predecessor ($\Phi^{-1}$) queries).

Another fundamental data structure in string processing that has been
extensively studied in terms of trade-offs between space usage, query
time, and construction efficiency is the suffix tree---a predecessor
of the suffix array. Over the years, numerous (unidirectional)
reductions have been demonstrated. For example, Sadakane~\cite{cst}
expresses the runtime of his \emph{Compressed Suffix Tree (CST)}
operations primarily in terms of the so-called $\Phi$ and $\Psi$
functions. Russo, Navarro, and Oliveira~\cite{RussoNO11} achieve
different trade-offs using the same functions. Fischer, M\"akinen, and
Navarro~\cite{FischerMN09} use a different set of operations for their
CST, expressing all runtimes in terms of suffix array queries, range
minimum queries (RMQ), next smaller value (NSV), and previous smaller
value (PSV) queries. Gagie, Navarro, and Prezza~\cite{Gagie2020} use
a similar set of operations, additionally augmented with the
LF-mapping query and access to the
\mbox{longest common prefix (LCP) array.}

Similar reductions for \emph{algorithms} are less common. A recent
step in this direction is~\cite{hierarchy}, which develops a comprehensive
hierarchy among many string and combinatorial problems (such as LZ77
factorization, BWT construction, batched ISA/LPF queries, longest common
factor, and inversion counting) whose state-of-the-art algorithms run
in $\bigO(\Textlen / \sqrt{\log \Textlen})$ time for inputs of $\Textlen$ bits
(i.e., $\Theta(\Textlen / \log \Textlen)$ machine words).

\vspace{3ex}
\noindent
\textbf{Organization of the Paper}$\ \ $
First, in \cref{sec:prelim}, we formally introduce the notation
and definitions used throughout the paper. In
\cref{sec:overview}, we provide an overview of our
techniques. Next, in
\cref{%
    sec:prefix-select,%
    sec:prefix-special-rank,%
    sec:prefix-rank,%
    sec:prefix-range-reporting,%
    sec:prefix-range-emptiness,%
    sec:prefix-rmq},
we present our main equivalences (see \cref{tab:equiv-problems}).
Finally, in \cref{sec:equiv-pattern-ranking-and-pattern-sa-interval},
we present an additional equivalence between
\probname{Pattern Ranking Queries} and
\probname{Pattern SA-Interval Queries}.

\section{Preliminaries}\label{sec:prelim}

\subsection{Basic Definitions}

\begin{figure}
  \centering
  \begin{tikzpicture}[yscale=0.35]
    \foreach \x [count=\i] in {a, aababa, aababababaababa,
      aba, abaababa, abaababababaababa, ababa, ababaababa,
      abababaababa, ababababaababa, ba, baababa,
      baababababaababa, baba, babaababa, babaababababaababa,
      bababaababa, babababaababa, bbabaababababaababa}
        \draw (1.9, -\i) node[right]
          {$\texttt{\x}$};
    \draw(1.9,0) node[right] {\scriptsize $\Text[\SA{\Text}[i]\dd \Textlen]$};
    \foreach \x [count=\i] in {b, b, b, b, b, b, a, b, b,
                               a, a, a, a, a, a, b, a, a, a}{
      \draw (0.5, -\i) node {\footnotesize $\i$};
    }
    \draw (0.5,0) node{\scriptsize $i$};
    \foreach \x [count=\i] in {19,14,5,17,12,3,15,10,
                               8,6,18,13,4,16,11,2,9,7,1}
      \draw (1.3, -\i) node {$\x\vphantom{\textbf{\underline{7}}}$};
    \draw(1.3,0) node{\scriptsize $\SA{\Text}$};
  \end{tikzpicture}

  \vspace{-1.3ex}
  \caption{A list of all sorted suffixes of $\Text =
    \texttt{bbabaababababaababa}$ along with
    the suffix array.}\label{fig:sa-example}
\end{figure}

A \emph{string} is a finite sequence of characters from a given
\emph{alphabet} $\Sigma$.  The length of a string $S$ is denoted $|S|$. For $i
\in [1\dd |S|]$,\footnote{For $i,j\in \mathbb{Z}$, we let
  $[i \dd j] = \{k \in \Z : i \leq k \leq j\}$,
  $[i \dd j) = \{k \in \Z : i \leq k < j\}$, and
  $(i \dd j] = \{k \in \Z: i < k \leq j\}$.}
the $i$th character of $S$ is denoted $S[i]$.
A~\emph{substring} or a \emph{factor} of $S$ is a string of the form $S[i \dd j) =
S[i]S[{i+1}]\cdots S[{j-1}]$ for some $1\le i \le j \le |S|+1$.
Substrings of the form $S[1\dd j)$ and $S[i\dd |S|{+}1)$ are called
\emph{prefixes} and \emph{suffixes}, respectively.
We use $\revstr{S}$ to denote the \emph{reverse} of $S$, i.e.,
$S[|S|]\cdots S[2]S[1]$.
We denote the \emph{concatenation} of two strings $U$ and $V$, that
is, the string $U[1]\cdots U[|U|]V[1]\cdots V[|V|]$, by $UV$ or $U\cdot V$.
Furthermore, $S^k = \bigodot_{i=1}^k S$ is the concatenation of $k \in
\Zz$ copies of $S$; note that $S^0 = \emptystring$ is the \emph{empty
string}. A nonempty string $S$ is said to be \emph{primitive} if it
cannot be written as $S = U^k$, where $k \geq 2$.  An integer $p\in
[1\dd |S|]$ is a \emph{period} of $S$ if $S[i] = S[i + p]$ holds for
every $i \in [1 \dd |S|-p]$. We denote the smallest period of $S$ as
$\per{S}$.  For every $S \in \Sigma^{+}$, we define the
infinite power $S^{\infty}$ so that $S^{\infty}[i] = S[1 + (i-1) \bmod |S|]$
for $i \in \Z$.  In particular, $S = S^{\infty}[1 \dd |S|]$.
By $\lcp{U}{V}$
we denote the length of
the longest common prefix
of $U$ and $V$. For any
string $S \in \Sigma^{*}$ and any $j_1, j_2 \in [1 \dd |S|]$, we
denote $\LCE{S}{j_1}{j_2} = \lcp{S[j_1 \dd |S|]}{S[j_2 \dd |S|]}$.
We use $\preceq$ to denote the order on $\Sigma$, extended to the
\emph{lexicographic} order on $\Sigma^*$ so that $U,V\in \Sigma^*$
satisfy $U \preceq V$ if and only if either
\begin{enumerate*}[label=(\alph*)]
  \item $U$ is a prefix of $V$, or
  \item $U[1 \dd i) = V[1 \dd i)$ and
    $U[i]\prec V[i]$ holds for some $i\in [1\dd \min(|U|,|V|)]$.
\end{enumerate*}

\begin{definition}[Pattern occurrences and SA-interval]\label{def:occ}
  For any $\Pat \in \Sigma^{*}$ and $\Text \in \Sigma^*$, we define
  \begin{align*}
    \OccTwo{\Pat}{\Text}
      &= \{j \in [1 \dd |\Text|] : j + |\Pat| \leq |\Text| + 1\text{ and }\Text[j \dd j + |\Pat|) = \Pat\},\\
    \RangeBegTwo{\Pat}{\Text}
      &= |\{j \in [1 \dd |\Text|] : \Text[j \dd |\Text|] \prec \Pat\}|,\\
    \RangeEndTwo{\Pat}{\Text}
      &= \RangeBegTwo{\Pat}{\Text} + |\OccTwo{\Pat}{\Text}|.
  \end{align*}
\end{definition}

\subsection{Suffix Array}

\begin{definition}[Suffix array and its inverse~\cite{sa}]\label{def:sa}
  For any string $\Text \in \Sigma^{\Textlen}$ of length $\Textlen \geq 1$, the \emph{suffix
  array} $\SA{\Text}[1 \dd \Textlen]$ of $\Text$ is a permutation of $[1 \dd \Textlen]$
  such that $\Text[\SA{\Text}[1] \dd \Textlen] \prec \Text[\SA{\Text}[2] \dd \Textlen] \prec \cdots
  \prec \Text[\SA{\Text}[\Textlen] \dd \Textlen]$, i.e., $\SA{\Text}[i]$ is the starting position
  of the lexicographically $i$th suffix of $\Text$; see \cref{fig:sa-example}
  for an example. The \emph{inverse suffix array} $\ISA{\Text}[1 \dd \Textlen]$
  is the inverse permutation of $\SA{\Text}$, i.e., $\ISA{\Text}[j] = i$ holds if
  and only if $\SA{\Text}[i] = j$. Intuitively, $\ISA{\Text}[j]$ stores the
  lexicographic \emph{rank} of $\Text[j \dd \Textlen]$ among the suffixes of~$\Text$.
\end{definition}

\begin{remark}\label{rm:occ}
  Note that the two values $\RangeBegTwo{\Pat}{\Text}$
  and $\RangeEndTwo{\Pat}{\Text}$ (\cref{def:occ}) are the endpoints of the so-called
  \emph{SA-interval} representing the occurrences of $\Pat$ in $\Text$, i.e.,
  \[
    \OccTwo{\Pat}{\Text} =
      \{\SA{\Text}[i] : i \in (\RangeBegTwo{\Pat}{\Text} \dd \RangeEndTwo{\Pat}{\Text}]\}
  \]
  holds for every $\Text \in \Sigma^{+}$ and $\Pat \in \Sigma^{*}$, including when
  $\Pat = \emptystring$ and when $\OccTwo{\Pat}{\Text} = \emptyset$.
\end{remark}

\begin{definition}[Lexicographical range]\label{def:lex-range}
  For any $\Pat_1, \Pat_2 \in \Sigma^{*}$ and $\Text \in \Sigma^{*}$, we define
  \begin{align*}
    \LexRange{\Pat_1}{\Pat_2}{\Text}
      &= \{j \in [1 \dd |\Text|] : \Pat_1 \preceq \Text[j \dd |\Text|] \prec \Pat_2\}.
  \end{align*}
\end{definition}

\begin{remark}\label{rm:lex-range}
  Observe that for any $\Pat_1, \Pat_2 \in \Sigma^{*}$ and any nonempty $\Text \in \Sigma^{+}$,
  it holds
  \[
    \LexRange{\Pat_1}{\Pat_2}{\Text} =
      \{\SA{\Text}[i] : i \in (\RangeBegTwo{\Pat_1}{\Text} \dd \RangeBegTwo{\Pat_2}{\Text}]\}.
  \]
\end{remark}

\subsection{Rank and Selection Queries}\label{sec:rank-select}

\begin{definition}[Rank and selection queries]\label{def:rank-select}
  For a string $S\in \Sigma^n$, we define:
  \begin{description}[style=sameline,itemsep=1ex]
  \item[$\Rank{S}{j}{a}$:] Given $a\in \Sigma$ and $j\in
    [0\dd n]$, compute $|\{i\in [1\dd j]: S[i]=a\}|$.
  \item[$\SpecialRank{S}{j}$:] Given $j\in
    [1\dd n]$, compute $|\{i\in [1\dd j]: S[i]=S[j]\}|$.
  \item[$\Select{S}{r}{a}$:] Given $a\in \Sigma$ and
    $r\in [1\dd \Rank{S}{n}{a}]$, find the $r$th smallest element of
    $\{i\in [1\dd n] : S[i]=a\}$.
  \end{description}
\end{definition}

\begin{theorem}[Rank and selection queries in
    bitvectors~\cite{WaveletSuffixTree,Clark98,Jac89,MunroNV16}]\label{th:bin-rank-select}
  For every string $S\in \{{\tt 0},{\tt 1}\}^*$, there exists a data structure of
  $\bigO(|S|)$ bits answering rank and selection queries in $\bigO(1)$
  time. Moreover, given the packed representations of $m$ binary
  strings of total length $n$, the data structures for all these
  strings can be constructed in $\bigO(m + n/\log n)$ time.
\end{theorem}

\subsection{Prefix Queries}\label{sec:prelim-prefix-queries}

\begin{definition}[Prefix rank and selection queries~\cite{breaking}]\label{def:prefix-rank-and-select}
  For a sequence $S \in (\Sigma^{*})^m$ of strings over alphabet $\Sigma$, we define:
  \begin{description}[style=sameline,itemsep=1ex]
  \item[$\PrefixRank{S}{j}{X}$:]
    Given $X \in \Sigma^{*}$ and $j \in [0 \dd m]$, compute
    $|\{i \in [1 \dd j]: X\text{ is a prefix of }S[i]\}|$.
  \item[$\PrefixSpecialRank{S}{j}{p}$:]
    Given $j \in [1 \dd m]$ and $p \in [0 \dd |S[j]|]$, compute
    $\PrefixRank{S}{j}{S[j][1 \dd p]}$.
  \item[$\PrefixSelect{S}{r}{X}$:]
    Given $X \in \Sigma^{*}$ and $r \in [1\dd \PrefixRank{S}{m}{X}]$,
    find the $r$th smallest element of
    $\{i \in [1 \dd m] : X\text{ is a prefix of }S[i]\}$.
  \end{description}
\end{definition}

\begin{definition}[Prefix range minimum queries (prefix RMQ)~\cite{sublinearlz}]\label{def:prefix-rmq}
  Let $A \in \Zn^m$ be a sequence of $m$ nonnegative integers and $S \in (\Sigma^{*})^m$ be
  a sequence of $m$ strings over alphabet $\Sigma$.
  For every $b, e \in [0 \dd m]$ and $X \in \Sigma^{*}$
  we define\footnote{We assume that $\argmin_{x \in S} f(x)$
    returns the \emph{smallest} $y \in S$ such that
    $f(y) = \min\{ f(x) : x\in S\}$.}
  \[
    \PrefixRMQ{A}{S}{b}{e}{X} :=
    \begin{cases}
      \argmin_{i \in \mathcal{I}} A[i] & \text{ if }\mathcal{I} \neq \emptyset,\\
      \infty & \text{ otherwise},
    \end{cases}
  \]
  where $\mathcal{I} = \{i \in (b \dd e] : X\text{ is a prefix of }S[i]\}$.
\end{definition}

\subsection{Range Queries}\label{sec:prelim-range-queries}

\begin{definition}[Range counting and selection]\label{def:two-sided-range-count-and-select}
  Let $A[1 \dd m]$ be an array of $m \geq 0$ nonnegative integers.
  For every $j \in [0 \dd m]$ and $v \geq 0$, we define
  \[
    \TwoSidedRangeCount{A}{j}{v} := |\{j \in (0 \dd j] : A[i] \geq v\}|.
  \]
  For every $v \geq 0$ and every $r \in [1 \dd \TwoSidedRangeCount{A}{m}{v}]$,
  by $\RangeSelect{A}{r}{v}$ we define the $r$th smallest element of
  $\{i \in [1 \dd m] : A[i] \geq v\}$.
\end{definition}

\begin{definition}[Range minimum queries (RMQ)]\label{def:rmq}
  Let $A[1 \dd m]$ be a sequence of $m$ integers.
  For every $b, e \in [0 \dd m]$,
  we define
  \[
    \RMQ{A}{b}{e} :=
    \begin{cases}
      \argmin_{i \in (b \dd e]}A[i] & \text{ if }b < e,\\
      \infty & \text{ otherwise}.
    \end{cases}
  \]
\end{definition}

\begin{definition}[Three-sided RMQ]\label{def:three-sided-rmq}
  Let $A[1 \dd m]$ and $B[1 \dd m]$ be two arrays of $m$
  integers. For every $b, e \in [0 \dd m]$ and $v \in \Z$, we define
  \[
    \ThreeSidedRMQ{A}{B}{b}{e}{v} :=
    \begin{cases}
      \argmin_{i \in \mathcal{I}} A[i] & \text{ if }\mathcal{I} \neq \emptyset,\\
      \infty & \text{ otherwise},
    \end{cases}
  \]
  where $\mathcal{I} = \{i \in (b \dd e] : B[i] \geq v\}$.
\end{definition}

\begin{theorem}[{\cite{FischerH11}}]\label{th:rmq}
  For every array $A[1\dd m]$ of $m$ integers, there is a data structure
  of $\bigO(m)$ bits that answers range minimum queries over $A$ in
  $\bigO(1)$ time and can be constructed in $\bigO(m)$ time.
\end{theorem}

\begin{corollary}\label{cr:two-sided-range-reporting}
  An array $A[1 \dd m]$ of $m$ integers
  can be preprocessed in $\bigO(m)$ time, so that
  given any $b,e \in [0 \dd m]$ and $v \geq 0$, we can compute
  the set $\mathcal{I} = \{i \in (b \dd e] : A[i] \geq v\}$ in
  $\bigO(1 + |\mathcal{I}|)$ time.
\end{corollary}
\begin{proof}

  The structure consists of a single component: the data structure from \cref{th:rmq} using $O(m)$ bits.

  The query algorithm is a recursive function that, given $b, e \in [0 \dd m]$ and $v \geq 0$,
  returns $\{i \in (b \dd e] : A[i] \geq v\}$. If $b \geq e$, we immediately exit.
  Otherwise, compute $i_{\min} = \RMQ{A}{b}{e}$. If $A[i_{\min}] \geq v$,
  report $i_{\min}$ and recurse on $(b \dd i-1]$ and $(i \dd e]$. If $A[i_{\min}] < v$, we again exit.
  In total, this takes $\bigO(1 + |\mathcal{I}|)$ time.

  The construction of the data structure takes $\bigO(m)$ time (see \cref{th:rmq}).
\end{proof}

\begin{theorem}[{\cite{breaking}}]\label{th:two-sided-range-count-and-range-select}
  An array $A[1 \dd m']$ of $m' \leq m$ nonnegative integers
  such that
  $\sum_{i=1}^{m'} A[i] = \bigO(m \log m)$
  can be preprocessed in $\bigO(m)$ time, so that
  two-sided range counting and range selection queries (\cref{def:two-sided-range-count-and-select})
  on $A$ can be answered in $\bigO(\log \log m)$ time and $\bigO(1)$ time, respectively.
\end{theorem}

\begin{theorem}[{\cite{sublinearlz}}]\label{th:three-sided-rmq}
  Any two arrays $A[1 \dd m']$ and $B[1 \dd m']$ of $m' \leq m$ nonnegative
  satisfying
  $\max_{i=1}^{m'} A[i] = \bigO(m \log m)$ and $\sum_{i=1}^{m'} B[i] = \bigO(m \log m)$
  can be preprocessed in $\bigO(m)$ time, so that
  three-sided RMQ queries (\cref{def:three-sided-rmq})
  on $A$ and $B$ can be answered in $\bigO(\log \log m)$ time.\footnote{The
  structure in~\cite{sublinearlz} assumes that the arguments $b,e,v$ satisfy
  $\{i \in (b \dd e] : B[i] \geq v\} \neq \emptyset$. To adapt the structure from~\cite{sublinearlz}
  to the more general definition of prefix RMQ queries (\cref{def:prefix-rmq}), we simply include
  the structure from \cref{th:two-sided-range-count-and-range-select} as a component of the structure
  from \cref{th:three-sided-rmq}. At the beginning of the query, we check if
  $\{i \in (b \dd e] : B[i] \geq v\} \neq \emptyset$ in $\bigO(\log \log m)$
  time using two range counting queries.}
\end{theorem}

\subsection{String Synchronizing Sets}\label{sec:prelim-sss}

String synchronizing sets~\cite{sss} allow for a locally-consistent selection of positions in a given text $\Text$.
The underlying parameter $\tau$ governs the context size (with respect to which the selection is consistent) and the achievable size of the synchronizing set. 

\begin{definition}[$\tau$-synchronizing set~\cite{sss}]\label{def:sss}
  Let $\Text\in \Sigma^{\Textlen}$ be a string, and let $\tau \in
  [1 \dd \lfloor\frac{\Textlen}{2}\rfloor]$ be a parameter. A set $\SSS
  \subseteq [1 \dd \Textlen - 2\tau + 1]$ is called a
  \emph{$\tau$-synchronizing set} of $\Text$ if it satisfies the
  following \emph{consistency} and \emph{density} conditions:
  \begin{enumerate}
  \item\label{def:sss-consistency}
    If $\Text[i \dd i + 2\tau) = \Text[j\dd j + 2\tau)$, then $i \in \SSS$
    holds if and only if $j \in \SSS$
    (for $i, j \in [1 \dd \Textlen - 2\tau + 1]$),
  \item\label{def:sss-density}
    $\SSS\cap[i \dd i + \tau) = \emptyset$ if and only if
    $i \in \RTwo{\tau}{\Text}$ (for $i \in [1 \dd \Textlen - 3\tau + 2]$),
    where
    \[
      \RTwo{\tau}{\Text} := \{i \in [1 \dd \Textlen - 3\tau + 2] :
      \per{\Text[i \dd i + 3\tau - 2]} \leq \tfrac{1}{3}\tau\}.
    \]
  \end{enumerate}
\end{definition}

\begin{remark}\label{rm:sss-size}
  In most applications, we want to minimize $|\SSS|$. Note, however, that
  the density condition imposes a lower bound
  $|\SSS| = \Omega(\frac{\Textlen}{\tau})$ for strings of length
  $\Textlen \ge 3\tau-1$ that do not contain substrings of length $3\tau - 1$
  with period at most $\frac{1}{3}\tau$.  Thus, we cannot hope to achieve an
  upper bound improving in the worst case upon the following ones.
\end{remark}

\begin{theorem}[{\cite[Proposition~8.10]
      {sss}}]\label{th:sss-existence-and-construction}
  For every string $\Text$ of length $\Textlen$ and parameter $\tau \in
  [1 \dd \lfloor\frac{\Textlen}{2}\rfloor]$, there exists a $\tau$-synchronizing
  set $\SSS$ of size $|\SSS| =
  \bigO\left(\frac{\Textlen}{\tau}\right)$.
  Moreover, if $\Text \in \IntegerAlphabet^{\Textlen}$, where
  $\AlphabetSize = \Textlen^{\bigO(1)}$, such $\SSS$ can be deterministically
  constructed in $\bigO(\Textlen)$ time.
\end{theorem}

\begin{theorem}[{\cite[Theorem~8.11]{sss}}]\label{th:sss-packed-construction}
  For every constant $\mu < \tfrac{1}{5}$, given the packed
  representation of a text $\Text \in \IntegerAlphabet^{\Textlen}$
  and a positive integer $\tau \leq \mu\log_\AlphabetSize \Textlen$,
  one can deterministically construct in $\bigO(\frac{\Textlen}{\tau})$
  time a $\tau$-synchronizing set of $\Text$ of size $\bigO(\frac{\Textlen}{\tau})$.
\end{theorem}

\subsection{Model of Computation}

We use the standard word RAM model of computation~\cite{Hagerup98}
with $w$-bit \emph{machine words}, where $w \ge \log \Textlen$, and all
standard bitwise and arithmetic operations take $\bigO(1)$ time.
Unless explicitly stated otherwise, we measure space complexity in machine words.

In the RAM model, strings are usually represented as arrays, with each
character occupying one memory cell. A single character, however, only needs
$\lceil \log \AlphabetSize \rceil$ bits, which might be much less
than $w$. We can therefore store (the \emph{packed representation} of)
a text $\Text \in \IntegerAlphabet^{\Textlen}$ using
$\bigO(\lceil \tfrac{\Textlen \log \AlphabetSize}{w} \rceil)$ words.

\section{Technical Overview}\label{sec:overview}

\subsection{Equivalence of Suffix Array and
  Prefix Select Queries}\label{sec:overview-sa-and-prefix-select}

In this section, we outline the equivalence between indexing for suffix array (SA) queries
and indexing for prefix select queries. Our focus is
on the reduction showing how prefix select queries can be answered
using suffix array queries. The reduction in the opposite direction (from
suffix array to prefix select queries), which motivated our research,
was proved in~\cite{breaking}. In \cref{sec:sa-to-prefix-select}, we
augment this result with the alphabet reduction for prefix select
queries (\cref{sec:prefix-select-alphabet-reduction}), obtaining a
clean reduction between the two problems on a binary alphabet (in
particular, without requiring an extra alphabet symbol to serve as a
sentinel at the end of the text).

\vspace{3ex}
\noindent
\textbf{Reducing Prefix Select to Suffix Array Queries}$\ \ $
Let $W[1 \dd m]$ be an array of $m \geq 1$ bitstrings of length
$\ell = 1 + \lfloor \log m \rfloor$, each represented in $\bigO(1)$ machine
words; that is, the input to the prefix select indexing problem
(see \cref{sec:prefix-select-and-sa-problem-def}).
Consider the string
$\Text = \bigodot_{i=1}^{m} \revstr{W[i]} \cdot \four \cdot (i+4)$
over the alphabet $\{\zero, \one, \dots, m+4\}$, where $\revstr{Y}$
denotes the reversal of $Y$.

Consider any $X \in \BinaryAlphabet^{\leq \ell}$, and let
$q = \PrefixRank{W}{m}{X}$ be the number of indices $i \in [1 \dd m]$
such that $X$ is a prefix of $W[i]$. Moreover, let
$(p_1, p_2, \ldots, p_q)$ be the sequence of these indices in increasing order;
formally, $p_i = \PrefixSelect{W}{i}{X}$. Our
reduction is based on the following observation:

\begin{description}[style=sameline,itemsep=1ex,font={\normalfont\itshape}]
\item[Observation: The lexicographically sorted sequence of suffixes starting in
  $\OccTwo{\revstr{X} \cdot \four}{\Text}$ lets us identify the sequence
  $(p_1, p_2, \dots, p_q)$.]
  There are exactly $m$ occurrences of the symbol $\four$ in $\Text$.
  Thus, every occurrence of $\revstr{X} \cdot \four$ aligns with the end
  of $\revstr{W[i]} \cdot \four$ for some $i \in [1 \dd m]$.
  The existence of such an occurrence is equivalent to $X$ being a prefix
  of $W[i]$. Hence, $\OccTwo{\revstr{X} \cdot \four}{\Text} =
  \{p_i \cdot \delta - (1 + |X|) : i \in [1 \dd q] \}$, where
  $\delta = \ell + 2$. Since $p_1 < p_2 < \dots < p_q$, the elements of
  $\OccTwo{\revstr{X} \cdot \four}{\Text}$ appear in the same order in
  $\Text$ and in its suffix array. Denoting
  $\OccTwo{\revstr{X} \cdot \four}{\Text} =
  \{a_1, a_2, \dots, a_q\}$ with $a_1 < a_2 < \dots < a_q$, we thus have
  $a_i = p_i \cdot \delta - (1 + |X|)$ for all $i \in [1 \dd q]$, and
  also
  $\SA{\Text}(b \dd b+q] = (a_1, \ldots, a_q)$, where
  $b = \RangeBegTwo{\revstr{X} \cdot \four}{\Text}$.
  Consequently, 
  $\SA{\Text}[b+i] = a_i = p_i \cdot \delta - (1+|X|)$ holds for every
  $i \in [1 \dd q]$. Since
  $\PrefixSelect{W}{i}{X} = p_i$ by definition, it follows that
  \[
    \PrefixSelect{W}{i}{X}
      = p_i
      = \frac{\SA{\Text}[b+i] + (1+|X|)}{\delta}
      = \left\lceil \frac{\SA{\Text}[b+i]}{\delta} \right\rceil,
  \]
  where the last equality follows from $1 \leq 1 + |X| < \delta$.
\end{description}

The first challenge in this reduction is that the resulting string
$\Text$ is over a large alphabet $\Sigma$ of size $|\Sigma| = \Theta(m)$. Any
approach that maps each symbol to $\Theta(\log |\Sigma|)$ bits would
produce a string of length $\Theta(|\Text| \cdot \log m) =
\Theta(m \log^2 m)$, which is too large. This issue is resolved by
slightly redefining~$\Text$. For $i \in [0 \dd m]$, define $s(i)$ as
a string of length $k := 1 + \lfloor \log m \rfloor$, obtained by
replacing every occurrence of $\zero$ with $\two$ and every occurrence
of $\one$ with $\three$ in the length-$k$ binary representation of the
integer~$i$. (We include leading zeros to pad the string to the
required length $k$.) Then, define
\[
  \Text = \bigodot_{i=1}^{m} \revstr{W[i]} \cdot \four \cdot s(i-1).
\]
The resulting string has an alphabet of size only five and retains all
the properties required in the above analysis, except that the value
of $\delta$ in the formula for $\PrefixSelect{W}{i}{X}$ becomes
$\delta = \ell + k + 1$.

The next challenge in the reduction is that, to answer a prefix select
query via a suffix array query (using the formula above), we need to
quickly obtain the value
$b = \RangeBegTwo{\revstr{X} \cdot \four}{\Text}$. Since the number of
possible strings $X \in \BinaryAlphabet^{\leq \ell}$ is
$\Theta(2^{\ell}) = \Theta(m)$, we can store this information for all
$X$. In general, it is not known how to compute all such values for an
arbitrary text~$\Text$. However, the symbol $\four$ in
$\Text$ provides separation that reduces the problem to: (1) computing
the frequencies of all $X \in \BinaryAlphabet^{\leq \ell}$ as prefixes of
strings in $W[1 \dd m]$, and then (2) computing a prefix sum over these
frequencies during a postorder traversal of a complete binary tree of
height $\ell$. These computations can be performed in $\bigO(m)$
time; see \cref{pr:range-beg}.

The final challenge is reducing the string $\Text$ to a binary
alphabet. We now sketch a more general reduction proving that, given
the packed representation of $\Text \in \IntegerAlphabet^{\Textlen}$,
we can compute the packed representation of a binary string
$\Text_{\rm bin}$ of length $|\Text_{\rm bin}| =
\Theta(\Textlen \log \AlphabetSize)$ and integers $\alpha$, $\beta$,
$\gamma$, and $\mu$ such that, for every $i \in [1 \dd \Textlen]$,
\vspace{1ex}
\[
  \SA{\Text}[i] = \frac{\SA{\Text_{\rm bin}}[\alpha + i] - \beta}{\gamma} + \mu.
\]

For any $k \in \Zp$ and $x \in [0 \dd 2^k)$, let
$\bin{k}{x} \in \BinaryAlphabet^{k}$ denote the length-$k$ string
containing the binary representation of $x$ (with leading zeros). For
a string $S$ over $\Sigma = [0 \dd 2^k)$, we then let
$\bin{k}{S} := \textstyle\bigodot_{i=1}^{|S|} \bin{k}{S[i]}$. For the
same $k$, $x$, and $S$, we define
\[
  \ebin{k}{x} :=
    \one^{k+1} \cdot
    \zero \cdot
    \bin{k}{x} \cdot
    \zero,
  \qquad
  \text{and}
  \qquad
  \ebin{k}{S} :=
    \textstyle\bigodot_{i=1}^{|S|} \ebin{k}{S[i]}.
\]

With these definitions, we can state our alphabet reduction. For any
$\Text \in \IntegerAlphabet^{\Textlen}$, we define
$\Text_{\rm bin} = \ebin{k}{\Text} = \bigodot_{i=1}^{\Textlen} \ebin{k}{\Text[i]}$,
where $k = \lceil \log \AlphabetSize \rceil$. To emulate suffix array
queries on $\Text$ using $\Text_{\rm bin}$, note that, for any
$S_1, S_2 \in \IntegerAlphabet^{*}$, $S_1 \prec S_2$ holds if and only
if $\ebin{k}{S_1} \prec \ebin{k}{S_2}$. This implies that the suffixes of
$\Text_{\rm bin}$ starting at positions in the set
$\mathcal{P} = \{1 + (i-1)\gamma : i \in [1 \dd \Textlen]\}$ (where
$\gamma = 2\lceil \log \AlphabetSize \rceil + 3$) occur in the same
relative order as the corresponding suffixes in $\Text$. Moreover,
these are precisely the suffixes prefixed with $\one^{k+1}$, so they
occupy a contiguous block of $\SA{\Text_{\rm bin}}$. It is easy to
see that this block is located at the very end of the suffix
array. Putting everything together, letting
$\alpha = |\Text_{\rm bin}| - |\Text|$, we have for every
$i \in [1 \dd \Textlen]$ that
$\SA{\Text_{\rm bin}}[\alpha + i] = (\SA{\Text}[i] - 1) \gamma + 1$. By
rewriting this, we obtain
$\SA{\Text}[i] =
(\SA{\Text_{\rm bin}}[\alpha + i] - 1) / \gamma + 1$. Thus, we obtain
our formula with $\beta = \mu = 1$, and $\alpha$ and $\gamma$ defined
as above. Finally, note that the string $\Text_{\rm bin}$ can be
constructed from $\Text$ using lookup tables in
$\bigO(\Textlen / \log_{\AlphabetSize} \Textlen)$ time (see
\cref{pr:sa-alphabet-reduction-poly-to-binary}).

\subsection{Equivalence of Lex-Range Reporting and
  Prefix Range Reporting Queries}\label{sec:overview-lex-range-reporting-and-prefix-range-reporting}

In this section, we provide an overview of a novel equivalence between
lex-range reporting and prefix range reporting queries.
This reduction differs from
similar reductions in earlier works (e.g., the reduction from
pattern ranking to prefix rank presented in~\cite{breaking}), where
the query algorithm proceeds differently depending on whether the
input pattern $\Pat \in \Sigma^{m}$ is highly periodic or not. In the
lex-range reporting queries considered here, the input consists of
\emph{two} patterns, $\Pat_1$ and $\Pat_2$, which may differ in their
periodicity (e.g., only one may be periodic),
leading to more complex query algorithms.

\vspace{3ex}
\noindent
\textbf{Reducing Lex-Range Reporting to Prefix Range Reporting Queries}$\ \ $
Consider a data structure answering prefix range reporting queries
that, for any sequence $W$ of $m \geq 1$ binary strings of length
$\ell = 1 + \lfloor \log m \rfloor$, uses $S(m)$ space, takes $P_t(m)$
time and $P_s(m)$ working space to build, and, given any
$b,e \in [0 \dd m]$ and the packed representation of
$X \in \BinaryAlphabet^{\leq \ell}$, returns $\mathcal{I} :=
\{i \in (b \dd e] : X\text{ is a prefix of }W[i]\}$ in
$Q_{\rm base}(m) + |\mathcal{I}| \cdot Q_{\rm elem}(m)$ time. In this
section, we outline the proof that, for every $\Text \in
\BinaryAlphabet^{\Textlen}$, there exists
$m = \Theta(\Textlen / \log \Textlen)$ such that, given the packed
representation of $\Text$, we can in
$\bigO(\Textlen / \log \Textlen + P_t(m))$ time and using
$\bigO(\Textlen / \log \Textlen + P_s(m))$ working space construct a
data structure of size $\bigO(\Textlen / \log \Textlen + S(m))$ that,
given the packed representations of any
$\Pat_1, \Pat_2 \in \BinaryAlphabet^{*}$, computes
$\mathcal{J} := \LexRange{\Pat_1}{\Pat_2}{\Text}$ (see
\cref{def:lex-range}) in
$\bigO(\log \log \Textlen + Q_{\rm base}(m) +
  |\mathcal{J}| \cdot Q_{\rm elem}(m) +
  |\Pat_1| / \log \Textlen +
  |\Pat_2| / \log \Textlen)$ time.

To simplify the reduction, we assume that $\Text[\Textlen] = \two$,
i.e., that there is a sentinel symbol at the end of $\Text$. This
increases the alphabet size, but since at the very end we reduce the
alphabet back to binary, adding the sentinel does not affect the
parameters of the reduction. We set $\AlphabetSize = 3$. We also
assume that $\Textlen > \AlphabetSize^{13}$ and denote
$\tau = \mu\log_{\AlphabetSize} \Textlen \geq 1$, where
$\mu < \tfrac{1}{12}$ is a constant (such $\mu$ and $\tau$ exist by
our assumption about $\Textlen$).

We say that a pattern $\Pat \in \IntegerAlphabet^{m}$ is
\emph{$\tau$-periodic} if $m \geq 3\tau-1$ and $\per{\Pat[1 \dd
3\tau-1]} \leq \tfrac{1}{3}\tau$.
Otherwise, $\Pat$ is \emph{$\tau$-nonperiodic}.  For simplicity, below
we outline how to reduce the computation of
$\LexRange{\Pat_1}{\Pat_2}{\Text}$ to prefix range reporting when
one of the following three cases holds:
\begin{itemize}
\item $|\Pat_1| \leq 3\tau - 1$ and $|\Pat_2| \leq 3\tau - 1$,
\item $\Pat_1$ and $\Pat_2$ are $\tau$-nonperiodic and
  $\lcp{\Pat_1}{\Pat_2} \geq 3\tau - 1$,
\item $\Pat_1$ and $\Pat_2$ are $\tau$-periodic and
  $\lcp{\Pat_1}{\Pat_2} \geq 3\tau - 1$.
\end{itemize}
Our reduction also works for all other combinations. In those
cases, we decompose the set
$\LexRange{\Pat_1}{\Pat_2}{\Text}$ into at most three subsets, each of
which is computed using a variant of one of the above three cases; see
the proof of
\cref{pr:lex-range-reporting-to-prefix-range-reporting-nonbinary} for
details.

\vspace{3ex}
\noindent
\textit{Short Patterns}$\ \ $
Consider $\Pat_1, \Pat_2 \in \IntegerAlphabet^{*}$ satisfying
$|\Pat_1| \leq 3\tau-1$ and $|\Pat_2| \leq 3\tau-1$. Denote
$\mathcal{S}_{\rm short} = \{\Text[i \dd \min(\Textlen + 1, i + 3\tau
- 1)) : i \in [1 \dd \Textlen]\}$.  Note that $|\mathcal{S}_{\rm
short}| = \bigO(\AlphabetSize^{3\tau-1}) =
\bigO(\AlphabetSize^{3\mu\log_{\AlphabetSize} \Textlen}) =
\bigO(\Textlen^{3\mu}) = \bigO(\sqrt{\Textlen})$.  Let $k =
|\mathcal{S}_{\rm short}|$ and let $A_{\rm short}[1 \dd k]$ be an array
containing all elements of $\mathcal{S}_{\rm short}$ in
lexicographic order. Finally, let $B_{3\tau-1}[1 \dd \Textlen]$
be a bitvector marking every index $i \in [1 \dd \Textlen]$ such that
either $i = \Textlen$, or $i < \Textlen$ and
$\LCE{\Text}{\SA{\Text}[i]}{\SA{\Text}[i+1]} < 3\tau-1$. In other words,
$B_{3\tau-1}$ marks boundaries between blocks of suffixes of $\Text$
sharing the same string from $\mathcal{S}_{\rm short}$.  Assume that
$B_{3\tau-1}$ has been augmented with support for $\bigO(1)$-time rank
and select queries using \cref{th:bin-rank-select}. Note that, thanks
to the sentinel symbol at the end of $\Text$, the set
$\mathcal{S}_{\rm short}$ is prefix-free (i.e., no string in
$\mathcal{S}_{\rm short}$ is a proper prefix of another string in
$\mathcal{S}_{\rm short}$).

We enumerate the set
$\LexRange{\Pat_1}{\Pat_2}{\Text}$ as follows.  First, using lookup
tables, we compute $b_i = \RangeBegTwo{\Pat_i}{\Text}$ for $i \in
\{1,2\}$ (see \cref{pr:nav-index-short}). Observe that for every $i
\in \{1,2\}$, $b_i > 0$ implies that $B_{3\tau-1}[b_i] = \one$. Thus,
using rank/select queries on $B_{3\tau-1}$, we can compute $i_1, i_2
\in [0 \dd k]$ such that
\[
  \LexRange{\Pat_1}{\Pat_2}{\Text} =
    \bigcup_{i \in (i_1 \dd i_2]} \OccTwo{A_{\rm short}[i]}{\Text}.
\]
We have thus reduced computing
$\LexRange{\Pat_1}{\Pat_2}{\Text}$ to enumerating $\OccTwo{X}{\Text}$
for $X \in \IntegerAlphabet^{\leq 3\tau-1}$; see
\cref{pr:lex-range-reporting-short}.

To solve the latter problem, we split the text $\Text$ into
$\bigO(\tfrac{\Textlen}{\tau})$ blocks of size $2\ell-1$, where $\ell
= 3\tau-1$, so that each adjacent block overlaps the next by exactly
$\ell-1$ symbols. We then sort all the blocks (represented by their
starting positions) lexicographically and store the resulting sequence
in an array $A_{\rm blk}[1 \dd m]$. We also prepare a lookup table
$L_{\rm blocks}$ that, for every $X \in \IntegerAlphabet^{\leq
  3\tau-1}$, stores a list ${\rm blocks}'(X)$ of pointers to
\emph{distinct} blocks in the above list that contain at least one
occurrence of~$X$ among the first $\ell$ positions.
Along with each pointer, we also keep a bitmask
marking the occurrences of $X$ within the first $\ell$ positions of
that block. At query time, we
first obtain the list of distinct blocks containing $X$ from
$L_{\rm blocks}$. For each distinct block in ${\rm blocks}'(X)$, we then
obtain from $A_{\rm blk}$ the list of all occurrences of that block
in the text. Altogether, this computation provides enough information to
retrieve all occurrences of $X$ in $\Text$ in time linear in their
number. See \cref{pr:short-occ-reporting} for details.

\vspace{3ex}
\noindent
\textit{Nonperiodic Patterns}$\ \ $
Let $\Pat_1, \Pat_2 \in \IntegerAlphabet^{*}$ be $\tau$-nonperiodic
patterns (\cref{def:periodic-pattern}) satisfying
$\lcp{\Pat_1}{\Pat_2} \geq 3\tau - 1$. Denote $\Pat' = \Pat_1[1 \dd
3\tau-1] = \Pat_2[1 \dd 3\tau-1]$.  To compute
$\LexRange{\Pat_1}{\Pat_2}{\Text}$, we first check using lookup tables
(\cref{pr:nav-index-short}) whether $\OccTwo{\Pat'}{\Text} =
\emptyset$. If so, we conclude that $\LexRange{\Pat_1}{\Pat_2}{\Text}
= \emptyset$. Let us therefore assume that $\OccTwo{\Pat'}{\Text} \neq
\emptyset$. Let $\SSS$ be a $\tau$-synchronizing set of $\Text$ computed
in $\bigO(\Textlen / \log_{\AlphabetSize} \Textlen)$ time using
\cref{th:sss-packed-construction}.  Denote $\Textlen' = |\SSS| =
\bigO(\tfrac{\Textlen}{\tau})$.

\begin{description}[style=sameline,itemsep=1ex,font={\normalfont\itshape}]
\item[Observation: The computation of
  $\LexRange{\Pat_1}{\Pat_2}{\Text}$ can be reduced to prefix range
  reporting.]  Since $\Pat$ is $\tau$-nonperiodic (i.e.,
  $\per{\Pat'} > \tfrac{1}{3}\tau$), it follows from the consistency and
  density conditions of $\SSS$ (\cref{def:sss}) that
  $\Successor{\SSS}{i} - i = \Successor{\SSS}{i'} - i' < \tau$ holds
  for every $i,i' \in \OccTwo{\Pat'}{\Text}$. Let $\deltatext \geq 0$ denote
  this common offset, i.e., such that for every
  $i \in \OccTwo{\Pat'}{\Text}$, it holds $\Successor{\SSS}{i} - i =
  \deltatext$.  Let $(s_i)_{i \in [1 \dd \Textlen']}$ be the sequence
  containing positions in $\SSS$, sorted according to the
  lexicographic order of the corresponding suffixes in $\Text$.
  Let $\Pat'_k = \Pat_k(\deltatext \dd |\Pat_k|]$, where $k \in \{1,2\}$.
  By the above observation, to identify $\LexRange{\Pat_1}{\Pat_2}{\Text}$,
  we need to find the range of positions in the sequence
  $(s_i)_{i \in [1 \dd \Textlen']}$ containing suffixes of the text that
  are lexicographically between $\Pat'_1$ and $\Pat'_2$. Such computation
  can be done in $\bigO(\log \log \Textlen +
    |\Pat_1| / \log_{\AlphabetSize} \Textlen +
    |\Pat_2| / \log_{\AlphabetSize} \Textlen)$ time
  (see \cref{pr:nav-index-nonperiodic}). Denote
  $b_k = |\{i \in [1 \dd \Textlen'] : \Text[s_i \dd \Textlen] \prec \Pat'_k\}|$
  for $k \in \{1,2\}$. After computing $b_1$ and $b_2$, it remains to select
  the positions from $(s_i)_{i \in (b_1 \dd b_2]}$ that correspond to
  suffixes preceded by $\Pat'[1 \dd \deltatext]$ in $\Text$.
  To implement this, define an array 
  $A_{\rm str}[1 \dd \Textlen']$ such that $A_{\rm str}[i] =
  \revstr{\Textinf[s_i - \tau \dd s_i + 2\tau)}$. Then, letting
  $D = \Pat'[1 \dd \deltatext + 2\tau]$, it holds
  $\LexRange{\Pat_1}{\Pat_2}{\Text} = \{s_i - \deltatext : i \in (b_1
  \dd b_2]\text{ and }\revstr{D} \text{ is a prefix of }A_{\rm str}[i]\}$; see
  \cref{lm:lex-range-nonperiodic,lm:lex-range-nonperiodic-array}.
  In other words, the remaining step in the computation of
  $\LexRange{\Pat_1}{\Pat_2}{\Text}$ is a prefix range reporting query.
  See \cref{pr:lex-range-reporting-nonperiodic} for the complete algorithm.
\end{description}

\vspace{3ex}
\noindent
\textit{Periodic Patterns}$\ \ $
Let $\Pat_1, \Pat_2 \in \IntegerAlphabet^{*}$ be $\tau$-periodic
patterns (\cref{def:periodic-pattern}) such that $\lcp{\Pat_1}{\Pat_2}
\geq 3\tau - 1$.  Denote $\Pat' = \Pat_1[1 \dd 3\tau-1] = \Pat_2[1 \dd
3\tau-1]$. Since, by \cref{def:periodic-pattern}, we have
$\per{\Pat'} \leq \tfrac{1}{3}\tau$, and since every $j \in
\LexRange{\Pat_1}{\Pat_2}{\Text}$ satisfies $j \in
\OccTwo{\Pat'}{\Text}$, we obtain
$\LexRange{\Pat_1}{\Pat_2}{\Text} \subseteq \RTwo{\tau}{\Text} = \{j
\in [1 \dd \Textlen - 3\tau + 2] : \per{\Text[j \dd j + 3\tau - 1)}
\leq \tfrac{1}{3}\tau\}$ (\cref{def:sss}).

\begin{description}[style=sameline,itemsep=1ex,font={\normalfont\itshape}]

\item[Observation: $\LexRange{\Pat_1}{\Pat_2}{\Text}$ can be expressed
  using maximal periodic runs in $\Text$.]  To efficiently compute
  $\LexRange{\Pat_1}{\Pat_2}{\Text}$ in this case, we group positions
  in $\RTwo{\tau}{\Text}$ into maximal blocks corresponding to
  \emph{$\tau$-runs}, i.e., maximal substrings $\Text[i \dd j)$
  satisfying $j-i \geq 3\tau - 1$ and $\per{\Text[i \dd j)} \leq
  \tfrac{1}{3}\tau$.  The gap between $|Y|$ and $\per{Y}$ ensures that
  $\tau$-runs overlap by fewer than $\tfrac{2}{3}\tau$ symbols, so the
  number of $\tau$-runs is $\bigO(\tfrac{\Textlen}{\tau}) =
  \bigO(\Textlen / \log_{\AlphabetSize} \Textlen)$ (see
  \cref{lm:runs}). To efficiently process $\tau$-runs, we
  introduce the following definitions.
  Let $\Text[x \dd y)$ be a $\tau$-run.
  Since $y-x \geq 3\tau - 1$ and $p := \per{\Text[x \dd y)} \leq
  \tfrac{1}{3}\tau$, we can uniquely write $\Text[x \dd y) =
  H'H^{k}H''$, where $H = \min\{\Text[x + \delta \dd x + \delta + p) :
  \delta \in [0 \dd p)\}$ is the \emph{Lyndon root} of the
  run, and $H'$ (resp.\ $H''$) is a proper suffix (resp.\ prefix) of
  $H$. We then denote
  $\HeadPos{x}{\tau}{\Text} = |H'|$,
  $\TailPos{x}{\tau}{\Text} = |H''|$,
  $\RootPos{x}{\tau}{\Text} = H$, and
  $\ExpPos{x}{\tau}{\Text} = k$.
  Using these properties, we partition $\RTwo{\tau}{\Text}$ as well as
  all $\tau$-runs into classes that allow us to reduce the search space
  during query time (see \cref{def:R-subsets,def:R-prim}).
  Specifically, using the above notation, we can express
  certain ranges in the suffix array in terms of $\tau$-runs
  (\cref{lm:exp-subblock-characterization}). This characterization
  is then gradually extended into a complete characterization (as
  well as an efficient enumeration) of the set $\LexRange{\Pat_1}{\Pat_2}{\Text}$;
  see \cref{pr:RskH-minus-report,%
   pr:periodic-lex-range-minus-reporting-subblock,%
   pr:periodic-lex-range-minus-reporting,%
   pr:periodic-lex-range-reporting,%
   pr:lex-range-reporting-periodic}.
\end{description}

\vspace{3ex}
The above three cases are combined in
\cref{pr:lex-range-reporting-to-prefix-range-reporting-nonbinary}.
Recall, however, that we added a sentinel symbol at the end of $\Text$
to ensure that every suffix of $\Text$ has exactly one occurrence in~$\Text$.
To reduce the alphabet back to binary, we therefore need to
perform an alphabet reduction for prefix range reporting.
To this end, observe that if,
given any sequence $W[1 \dd m]$ of $m \geq 1$
strings of length $\ell \geq 1$ over the alphabet $\IntegerAlphabet$, we
set $k = \lceil \log \AlphabetSize \rceil$ and let $W_{\rm bin}[1 \dd
m]$ be any sequence of binary strings such that, for every $j \in [1
\dd m]$, $\bin{k}{W[j]}$ (\cref{def:bin}) is a prefix of $W_{\rm
bin}[j]$, then for every $X \in \IntegerAlphabet^{\leq \ell}$ and
$b,e \in [0 \dd m]$, it holds that
\[
  \{i \in (b \dd e] : X\text{ is a prefix of }W[i]\} =
  \{i \in (b \dd e] : X_{\rm bin}\text{ is a prefix of }W_{\rm bin}[i]\},
\]
where $X_{\rm bin} = \bin{k}{X}$;
see \cref{lm:prefix-queries-alphabet-reduction}. Moreover, the reduction
can be implemented in time optimal in the bit-encoding of
$W$; see \cref{pr:prefix-range-reporting-alphabet-reduction}.

\vspace{3ex}
\noindent
\textbf{Reducing Prefix Range Reporting to Lex-Range Reporting Queries}$\ \ $
Let $W[1 \dd m]$ be an array of $m \geq 1$ bitstrings of length $\ell
= 1 + \lfloor \log m \rfloor$, each represented in $\bigO(1)$ machine
words, i.e., the input to the problem of indexing for prefix range
reporting queries (see
\cref{sec:prefix-range-reporting-and-lex-range-reporting-problem-def}).
The basic idea of the reduction is similar to that in
\cref{sec:overview-sa-and-prefix-select}; namely, the text $\Text$ is
defined as $\Text = \bigodot_{i=1}^{m} \revstr{W[i]} \cdot \four \cdot
s(i-1)$, where $s(i)$ is a length-$k$ binary encoding of $i \in [0 \dd
  m]$, with each $\zero$ (resp.\ $\one$) replaced by $\two$
(resp.\ $\three$), where $k := 1 + \lfloor \log m \rfloor$.  We then
prove (see \cref{lm:seq-to-string}) that, for every $b,e \in [0 \dd m]$
and $X \in \BinaryAlphabet^{\leq \ell}$, it holds
\[
  \{i \in (b \dd e] : X\text{ is a prefix of }W[i]\} =
  \{\lceil j/\beta \rceil : j \in \LexRange{\Pat_1}{\Pat_2}{\Text}\},
\]
where $\Pat_1 = \revstr{X} \cdot \four \cdot s(b)$, $\Pat_2 =
\revstr{X} \cdot \four \cdot s(e)$, and $\beta = \ell + k + 1$.
Thus, an algorithm for lex-range
reporting can be used for prefix range reporting.

As in other problems, the above reduction increases the size
of the alphabet. To reduce the alphabet back to binary, in
\cref{sec:lex-range-reporting-alphabet-reduction}, we prove a general
alphabet reduction for lex-range reporting. Specifically, we show that,
given the packed representation of $\Text \in
\IntegerAlphabet^{\Textlen}$, we can, in $\bigO(\Textlen /
\log_{\AlphabetSize} \Textlen)$ time, construct the packed
representation of a string $\Text_{\rm bin} \in \BinaryAlphabet^{*}$
satisfying $|\Text_{\rm bin}| = \Theta(\Textlen \log \AlphabetSize)$,
as well as integers $\alpha$, $\gamma$, and $\delta$, and a data structure that,
given the packed representation of any $\Pat_1, \Pat_2 \in
\IntegerAlphabet^{\leq m}$, in $\bigO(1 + m / \log_{\AlphabetSize}
\Textlen)$ time, returns the packed representations of $\Pat'_1, \Pat'_2
\in \BinaryAlphabet^{*}$ such that $|\Pat'_i| = |\Pat_i| \cdot
(2\lceil \log \AlphabetSize \rceil + 3)$ (for $i \in \{1,2\}$) and
\[
  \LexRange{\Pat_1}{\Pat_2}{\Text} =
  \left\{\frac{j-\alpha}{\gamma} + \delta :
    j \in \LexRange{\Pat'_1}{\Pat'_2}{\Text_{\rm bin}}\right\};
\]
see \cref{pr:lex-range-reporting-alphabet-reduction}. Together,
these two reductions yield
\cref{th:prefix-range-reporting-to-lex-range-reporting}.

\section{Equivalence of Suffix Array and Prefix Select Queries}\label{sec:prefix-select}

\subsection{Problem Definitions}\label{sec:prefix-select-and-sa-problem-def}
\vspace{-1.5ex}

\begin{framed}
  \noindent
  \probname{Indexing for Suffix Array Queries}
  \begin{bfdescription}
  \item[Input:]
    The packed representation of a string
    $\Text \in \BinaryAlphabet^{\Textlen}$.
  \item[Output:]
    A data structure that, given
    any $i \in [1 \dd \Textlen]$, returns $\SA{\Text}[i]$, i.e., the
    starting position of the lexicographically $i$th smallest suffix
    of $\Text$.
  \end{bfdescription}
\end{framed}

\begin{framed}
  \noindent
  \probname{Indexing for Prefix Select Queries}
  \begin{bfdescription}
  \item[Input:]
    A sequence $W[1 \dd m]$ of $m \geq 1$ binary strings of length $\ell = 1 + \lfloor \log m \rfloor$, with all
    strings represented in the packed form.
  \item[Output:]
    A data structure that, given
    the packed representation of any $X \in \BinaryAlphabet^{\leq \ell}$ and any
    $i \in [1 \dd \PrefixRank{W}{m}{X}]$ (see \cref{def:prefix-rank-and-select}),
    returns $\PrefixSelect{W}{i}{X}$, i.e.,
    the $i$th smallest element of the set $\{j \in [1 \dd m] : X\text{ is a prefix of }W[j]\}$.
  \end{bfdescription}
\end{framed}
\vspace{2ex}

\subsection{Reducing Prefix Select to Suffix Array Queries}\label{sec:prefix-select-to-sa}

\subsubsection{Preliminaries}\label{sec:prefix-select-to-sa-prelim}

\begin{definition}[Binary mapping]\label{def:bin}
  For any $k \in \Zp$ and $x \in [0 \dd 2^k)$, by $\bin{k}{x} \in \BinaryAlphabet^{k}$
  we denote the length-$k$ string containing the binary representation of $x$ (with leading zeros).
  For any string $S$ over
  alphabet $\Sigma = [0 \dd 2^k)$, we then let
  $\bin{k}{S} := \textstyle\bigodot_{i=1}^{|S|} \bin{k}{S[i]}$.
\end{definition}

\begin{proposition}\label{pr:bin}
  Let $u \geq 1$ and $\AlphabetSize$ be such that $2 \leq \AlphabetSize < u^{\bigO(1)}$.
  Denote $k = \lceil \log \AlphabetSize \rceil$.
  In $\bigO(u/\log_{\AlphabetSize} u)$ time we can construct a data structure that,
  given the packed representation of any $S \in \IntegerAlphabet^{*}$,
  computes the packed representation of $\bin{k}{S}$ (\cref{def:bin}) in
  $\bigO(1 + |S| / \log_{\AlphabetSize} u)$ time.
\end{proposition}
\begin{proof}
  The proof proceeds similarly as the proof of~\cite[Proposition~5.47]{hierarchy}.
  If $\AlphabetSize > u^{1/4}$, then we do not need any preprocessing. Otherwise, in
  $\bigO(u / \log_{\AlphabetSize} u)$ time we construct a lookup table
  that implements the mapping for substring of length $\leq \ell$ for some
  $\ell = \Theta(\log_{\AlphabetSize} u)$ in $\bigO(1)$ time, yielding
  an overall query time of $\bigO(1 + |S| / \log_{\AlphabetSize} u)$.
\end{proof}

\begin{definition}[Extended binary mapping]\label{def:ebin}
  Let $k \in \Zp$. For every $x \in [0 \dd 2^k)$, define
  \[
    \ebin{k}{a} := \one^{k+1} \cdot \zero \cdot \bin{k}{x} \cdot \zero,
  \]
  where $\bin{k}{x}$ is as in \cref{def:bin}. For any string $S$ over
  alphabet $\Sigma = [0 \dd 2^k)$, we then let
  \[
    \ebin{k}{S} := \textstyle\bigodot_{i=1}^{|S|} \ebin{k}{S[i]}.
  \]
\end{definition}

\begin{lemma}[{\cite{hierarchy}}]\label{lm:ebin-order}
  Let $k \in \Zp$ and let $\AlphabetSize = 2^k$. For every
  $S_1, S_2 \in \IntegerAlphabet^{*}$, $S_1 \prec S_2$ holds
  if and only if $\ebin{k}{S_1} \prec \ebin{k}{S_2}$.
\end{lemma}

\begin{proposition}[{\cite{hierarchy}}]\label{pr:ebin}
  Let $u \geq 1$ and $\AlphabetSize$ be such that $2 \leq \AlphabetSize < u^{\bigO(1)}$.
  Denote $k = \lceil \log \AlphabetSize \rceil$.
  In $\bigO(u/\log_{\AlphabetSize} u)$ time we can construct a data structure that,
  given the packed representation of any $S \in \IntegerAlphabet^{*}$,
  computes the packed representation of $\ebin{k}{S}$ (\cref{def:ebin}) in
  $\bigO(1 + |S| / \log_{\AlphabetSize} u)$ time.
\end{proposition}

\begin{definition}[Symbol-substitute function]\label{def:sub}
  Let $u, v \in \Sigma^{+}$ and $c \in \Sigma$. We define
  $\Substitute{u}{c}{v}$ as a string obtained by replacing all occurrences
  of $c$ in $u$ with $v$. Formally, $\Substitute{u}{c}{v} =
  \bigodot_{i=1,\dots,|u|} f(u[i])$, where $f: \Sigma \rightarrow
  \Sigma^{*}$ is such that for every $a \in \Sigma$:
  \[
    f(a) =
      \begin{cases}
        a & \text{if }a \neq c, \\
        v & \text{otherwise}.
      \end{cases}
  \]
\end{definition}

\begin{definition}[Sequence-to-string mapping]\label{def:seq-to-string}
  Let $W[1 \dd m]$ be a sequence of $m \geq 1$ strings of length $\ell \geq 1$ over alphabet $\BinaryAlphabet$.
  Denote $k = 1 + \lfloor \log m \rfloor$. For any $i \in [0 \dd m]$, let
  $s(i) = \Substitute{\Substitute{\bin{k}{i}}{\zero}{\two}}{\one}{\three}$ (see \cref{def:bin,def:sub}).
  We then define
  \[
    \SeqToString{\ell}{W} := \textstyle\bigodot_{i=1}^{m} \revstr{W[i]} \cdot \four \cdot s(i-1).
  \]
\end{definition}

\begin{definition}[Integer mapping for binary strings]\label{def:bin-str-to-int}
  For every $X \in \BinaryAlphabet^{*}$, we let
  $\BinStrToInt{X} = 1$ (if $X = \emptystring$) and
  $\BinStrToInt{X} = 2^k + x$,
  where $k = |X|$ and $x \in [0 \dd 2^k)$ is such that $\bin{k}{x} = X$ (if $X \neq \emptystring$); see \cref{def:bin}.
\end{definition}

\begin{remark}\label{rm:bin-str-to-int}
  Note that the mapping in \cref{def:bin-str-to-int}
  is injective, since given $y = \BinStrToInt{X}$ we can decode $X$ by first computing
  the length of $|X|$ (using the position of the most significant bit in $y$), and
  then interpreting the least significant $|X|$ bits as symbols of $X$.
  Alternatively, this
  follows by noting that $\BinStrToInt{X}$ is the identifier of a node with a root-to-node label $X$
  in a trie of the set $\BinaryAlphabet^{*}$, in which nodes are numbered as in
  a heap (with 1 as the identifier of the root).
\end{remark}

\begin{lemma}\label{lm:seq-to-string-occ}
  Let $W[1 \dd m]$ be a sequence of $m \geq 1$ strings of length $\ell \geq 1$ over alphabet $\BinaryAlphabet$.
  Let $\Text = \SeqToString{\ell}{W}$ (\cref{def:seq-to-string}), $k = 1 + \lfloor \log m \rfloor$,
  and $\beta = \ell + k + 1$.
  Finally, let $X \in \BinaryAlphabet^{\leq \ell}$ and
  $\mathcal{P} = \{i \in [1 \dd m] : X\text{ is a prefix of }W[i]\}$.
  Then, it holds
  \[
    \OccTwo{\revstr{X} \cdot \four}{\Text} = \{i \cdot \beta - (k+|X|) : i \in \mathcal{P}\}.
  \]
\end{lemma}
\begin{proof}

  Denote $A = \{i \cdot \beta - (k+|X|) : i \in \mathcal{P}\}$.

  First, we show the inclusion $\OccTwo{\revstr{X} \cdot \four}{\Text} \subseteq A$.
  Let $j \in \OccTwo{\revstr{X} \cdot \four}{\Text}$. Note that $\OccTwo{\four}{\Text} = \{i \cdot \beta - k : i \in [1 \dd m]\}$.
  Thus, $j \in \OccTwo{\revstr{X} \cdot \four}{\Text}$ implies that for some $i \in [1 \dd m]$, it holds
  $j = i \cdot \beta - (k + |X|)$. Since $\Text[i \cdot \beta - k \dd |\Text|]$ is preceded
  in $\Text$ with $\revstr{W[i]}$, it follows that $\revstr{X}$ is a suffix of $\revstr{W[i]}$, or
  equivalently, $X$ is a prefix of $W[i]$. This implies that $i \in \mathcal{P}$, and hence, by definition
  of $A$, $i \cdot \beta - (k+|X|) \in A$.
  It remains to note that $i \cdot \beta - (k+|X|) = j$.

  We now show the opposite inclusion. Let $j \in A$. Then, there exists $i \in \mathcal{P}$ such that
  $j = i \cdot \beta - (k+|X|)$. By $i \in \mathcal{P}$, it holds $i \in [1 \dd m]$ and $X$ is a prefix of $W[i]$.
  Note that, by \cref{def:seq-to-string}, $\Text(i \cdot \beta \dd |\Text|]$
  is preceded in $\Text$ with $\revstr{W[i]} \cdot \four \cdot s(i-1)$.
  Since $X$ is a prefix of $W[i]$, or equivalently, $\revstr{X}$ is a suffix
  of $\revstr{W[i]}$, we thus obtain that $\Text(i \cdot \beta \dd |\Text|]$ is preceded in $\Text$
  with $\revstr{X} \cdot \four \cdot s(i-1)$. Thus, $i \cdot \beta - (k+|X|) \in \OccTwo{\revstr{X} \cdot \four}{\Text}$.
  By $j = i \cdot \beta - (k+|X|)$, we thus obtain $j \in \OccTwo{\revstr{X} \cdot \four}{\Text}$.
\end{proof}

\begin{proposition}\label{pr:range-beg}
  Let $W[1 \dd m]$ be an array of $m \geq 1$ binary string of length $\ell = 1 + \lfloor \log m \rfloor$
  and let $\Text = \SeqToString{\ell}{W}$ (\cref{def:seq-to-string}).
  Given the array $W$, with all strings represented in the packed form, we can in $\bigO(m)$
  time construct a data structure that, given the packed representation of any $X \in \BinaryAlphabet^{\leq \ell}$ (i.e.,
  a pair $(k,x)$, where $k = |X|$ and $x \in [0 \dd 2^k)$ is such that $\bin{k}{x} = X$),
  returns $\RangeBegTwo{\revstr{X} \cdot \four}{\Text}$ in $\bigO(1)$ time.
\end{proposition}
\begin{proof}

  \DSComponents
  The data structure consists of a single component: an array
  $R[1 \dd 2^{\ell+1})$ defined such that for every
  $X \in \BinaryAlphabet^{\leq \ell}$,
  $R[\BinStrToInt{X}] = \RangeBegTwo{\revstr{X} \cdot \four}{\Text}$ (see \cref{def:bin-str-to-int}).
  The array $R$, and hence the whole data structure, needs
  $\bigO(2^{\ell}) = \bigO(m)$ space.

  \DSQueries
  The queries are implemented as follows.
  Suppose that we are given the packed representation of
  $X$, i.e., a pair $(k,x)$, where $k = |X|$ and $x \in [0 \dd 2^k)$ is
  such that $\bin{k}{x} = X$. First, in $\bigO(1)$ time
  we compute $y := \BinStrToInt{X} = 2^k + x$. We then return $R[y]$.

  \DSConstruction
  The data structure is constructed as follows:
  \begin{enumerate}

  \item First, we compute an auxiliary array
    $C[1 \dd 2^{\ell+1})$ defined such that, for every
    $X \in \BinaryAlphabet^{\leq \ell}$, it holds $C[\BinStrToInt{X}] = |\OccTwo{X \cdot \four}{\Text}|$, i.e.,
    $C$ stores the frequency in $\Text$ of every string of the form $X \cdot \four$. We begin by
    initializing all entries of $C$ to zero. We then compute $C[\BinStrToInt{X}]$
    for all $X \in \BinaryAlphabet^{\ell}$. To this
    end, it suffices to scan the array $W$. We then iterate $k = \ell-1, \ell-2, \ldots, 1$, and for each
    $k$ we compute $C[\BinStrToInt{X}]$ for all $X \in \BinaryAlphabet^{k}$ using the equality
    $C[\BinStrToInt{X}] = C[\BinStrToInt{\zero \cdot X}] + C[\BinStrToInt{\one \cdot X}]$.
    More precisely, given $k \in (\ell \dd 1]$ and some $x \in [0 \dd 2^k)$
    such that $\bin{k}{x} = X$, we have $\BinStrToInt{X} = 2^k + x$,
    $\BinStrToInt{\zero \cdot X} = 2^{k+1} + x$, and $\BinStrToInt{\one \cdot X} = 2^{k+1} + 2^k + x$,
    and hence we set $C[2^k + x] = C[2^{k+1} + x] + C[2^{k+1} + 2^{k} + x]$.
    Finally, we set $C[\BinStrToInt{\emptystring}] = C[1] = m$.
    In total, the computation of $C$ takes $\bigO(2^{\ell}) = \bigO(m)$ time.

  \item Next, we compute an auxiliary array $R'[1 \dd 2^{\ell+1})$ defined such that
    for every $X \in \BinaryAlphabet^{\leq \ell}$, it holds $R'[\BinStrToInt{X}] = \RangeBegTwo{X \cdot \four}{\Text}$.
    The case $X = \emptystring$ is easy to handle, and we set $R'[\BinStrToInt{\emptystring}] = R'[1] = |\Text| - m$.
    To compute the remaining values, observe that since the
    set $\{X \cdot \four : X \in \BinaryAlphabet^{*}\}$ is prefix-free (i.e., no string in the set is a prefix of another),
    it follows that for every $X \in \BinaryAlphabet^{\leq \ell} \setminus \{\emptystring\}$, it holds
    \[
      \RangeBegTwo{X \cdot \four}{\Text} =
        \sum_{X' \in \BinaryAlphabet^{\leq \ell} \setminus \{\emptystring\} : X' \cdot \four \prec X \cdot \four}
          |\OccTwo{X' \cdot \four}{\Text}|.
    \]
    Consequently, to compute $R'$, it suffices to enumerate all strings in the set
    $\{X \cdot \four : X \in \BinaryAlphabet^{\leq \ell} \setminus \{\emptystring\}\}$ in lexicographical order,
    while keeping the cumulative frequency of smaller strings. To perform this enumeration, we observe that such
    order corresponds to a post-order traversal of a trie of $\BinaryAlphabet^{\ell}$. Such traversal
    is easily implemented as a recursive function. To navigate during this traversal, we maintain
    the packed representation of the root-to-node label for the currently visited node, i.e., for a node labeled $X$,
    we keep $k = |X|$ and an integer $x \in [0 \dd 2^k)$ such that $\bin{k}{x} = X$. The corresponding representation
    for the left (resp.\ right) node is then $(k+1,2x)$ (resp.\ $(k+1,2x+1)$). Using this representation,
    we can always compute $|\OccTwo{X \cdot \four}{\Text}| = C[\BinStrToInt{X}] = C[2^k + x]$ to update the cumulative
    frequency, and write the current cumulative frequency to $R'[\BinStrToInt{X}]$.
    In total, we spend $\bigO(2^{\ell}) = \bigO(m)$ time.

  \item Let $\ell' = \lceil \ell / 2 \rceil$. In this step, we compute an auxiliary lookup table
    $L_{\rm rev}[1 \dd 2^{\ell'+1})$ defined such that for every
    $X \in \BinaryAlphabet^{\leq \ell'} \setminus \{\emptystring\}$,
    letting $k = |X|$ and $y \in [0 \dd 2^k)$ be such that $\bin{k}{y} = \revstr{X}$, it holds
    $L_{\rm rev}[\BinStrToInt{X}] = y$. The computation of $y$ for any
    $X \in \BinaryAlphabet^{\leq \ell'} \setminus \{\emptystring\}$
    takes $\bigO(|X|) = \bigO(\ell')$ time, and hence in total we spend $\bigO(2^{\ell'} \ell') = \bigO(\sqrt{m} \log m)
    = \bigO(m)$ time. Given $L_{\rm rev}$ and the packed representation of any $X \in \BinaryAlphabet^{\leq \ell}$
    (i.e., the length $k = |X|$ and an integer $x \in [0 \dd 2^k)$ satisfying $\bin{k}{x} = X$),
    we can in $\bigO(1)$ time compute $y \in [0 \dd 2^k)$ such that $\bin{k}{y} = \revstr{X}$. Consequently, with
    such input we can compute $\BinStrToInt{\revstr{X}}$.

  \item In the last step, we compute the array $R$. To this end, we iterate over all $X \in \BinaryAlphabet^{\leq \ell}$.
    Given the packed representation of $X$, in $\bigO(1)$ time we first compute $\BinStrToInt{X}$ and
    $\BinStrToInt{\revstr{X}}$, and then write $R[\BinStrToInt{X}] := R'[\BinStrToInt{\revstr{X}}]$.
    In total, this takes $\bigO(2^{\ell}) = \bigO(m)$ time.
  \end{enumerate}
  In total, the construction takes $\bigO(m)$ time.
\end{proof}

\begin{proposition}\label{pr:seq-to-string}
  Let $W[1 \dd m]$ be an array of $m \geq 1$ binary strings of
  length $\ell = 1 + \lfloor \log m \rfloor$. Given array $W$, with all
  strings represented in the packed form, we can in $\bigO(m)$ time compute
  the packed representation of the string $\SeqToString{\ell}{W}$ (\cref{def:seq-to-string}).
\end{proposition}
\begin{proof}

  The algorithm proceeds as follows:
  \begin{enumerate}

  \item Let $\ell' = \lceil \ell / 2 \rceil$. We compute an auxiliary lookup table
    $L_{\rm rev}$ such that for every $X \in \BinaryAlphabet^{\leq \ell'}$, $L_{\rm rev}$ maps
    $X$ to the packed representation of $\revstr{X}$ (over alphabet $\{\zero, \ldots, \four\}$).
    When accessing $L_{\rm rev}$, we map $X$ to $\BinStrToInt{X}$ (\cref{def:bin-str-to-int}).
    The computation takes $\bigO(\sqrt{m} \log m) = \bigO(m)$ time.
    Given the packed representation of any $X \in \BinaryAlphabet^{\leq \ell}$ (over alphabet $\BinaryAlphabet$),
    we can then in $\bigO(1)$ time compute the packed representation of $\revstr{X}$ (over $\{\zero, \ldots, \four\}$).

  \item Let $\ell'$ be defined as above. In this step, we compute an auxiliary lookup table
    $L_{\rm map}$ defined such that for every $X \in \BinaryAlphabet^{\leq \ell'}$,
    $L_{\rm map}$ maps $X$ to the packed representation
    of $\Substitute{\Substitute{X}{\zero}{\two}}{\one}{\three}$ (over alphabet $\{\zero, \ldots, \four\}$).
    Similarly as above, the construction takes $\bigO(\sqrt{m} \log m) = \bigO(m)$ time.
    Given $i \in [0 \dd m]$, we can then compute the packed representation of $s(i)$
    (defined as in \cref{lm:prefix-select-to-sa})
    over alphabet $\{\zero, \dots, \four\}$ in $\bigO(1)$ time.

  \item With the above lookup tables, the construction of the packed
    representation of $\SeqToString{\ell}{W}$ takes $\bigO(m)$ time.
  \end{enumerate}
  In total, we spend $\bigO(m)$ time.
\end{proof}

\subsubsection{Problem Reduction}

\begin{lemma}\label{lm:prefix-select-to-sa}
  Let $W[1 \dd m]$ be an array of $m \geq 1$ strings of length $\ell \geq 1$ over alphabet $\BinaryAlphabet$.
  Denote $\Text = \SeqToString{\ell}{W}$ (\cref{def:seq-to-string}). 
  For every $X \in \BinaryAlphabet^{\leq \ell}$ and $i \in [1 \dd \PrefixRank{W}{m}{X}]$, it holds
  \[
    \PrefixSelect{W}{i}{X} = \left\lceil \frac{\SA{\Text}[\alpha + i]}{\beta} \right\rceil,
  \]
  where $\alpha = \RangeBegTwo{\revstr{X}\cdot\four}{\Text}$ and $\beta = \ell + \lfloor \log m \rfloor + 2$.
\end{lemma}
\begin{proof}
  Let $k$ and $s(i)$ (where $i \in [0 \dd m]$) be as in \cref{def:seq-to-string}.
  Denote $q = \PrefixRank{W}{m}{X}$. Let $(a_i)_{i \in [1 \dd q]}$ and $(b_i)_{i \in [1 \dd q]}$ be sequences
  defined by $a_i = \PrefixSelect{W}{i}{X}$ and $b_i = a_i \cdot \beta - (k+|X|)$. The proof proceeds
  in three steps:
  \begin{enumerate}
  \item First, observe that by \cref{lm:seq-to-string-occ}, it holds
    $\OccTwo{\revstr{X} \cdot \four}{\Text} = \{b_1, \ldots, b_q\}$.
  \item Next, we prove that $\Text[b_1 \dd |\Text|] \prec \Text[b_2 \dd |\Text|] \prec \cdots \prec
    \Text[b_q \dd |\Text|]$. To this end, it suffices to note that for every $i \in [1 \dd q]$,
    the suffix $\Text[b_i \dd |\Text|]$ starts with the string $\revstr{X} \cdot \four \cdot s(a_i-1)$.
    Since by definition we have $a_1 < a_2 < \cdots < a_q$ and
    $s(0) \prec s(1) \prec \cdots \prec s(m-1)$, it follows
    that $\revstr{X} \cdot \four \cdot s(a_i) \prec \revstr{X} \cdot \four \cdot s(a_{i+1})$ holds for
    every $i \in [1 \dd q)$. This immediately yields $\Text[b_i \dd |\Text|] \prec \Text[b_{i+1} \dd |\Text|]$.
  \item By the above two steps, and the definition of $\alpha$, it follows that for every
    $i \in [1 \dd q]$, it holds $\SA{\Text}[\alpha + i] = b_i$. Noting that for every
    $i \in [1 \dd q]$, it holds $\lceil b_i / \beta \rceil = a_i$, we thus obtain
    \[
      \PrefixSelect{W}{i}{X}
        = a_i
        = \left\lceil \frac{b_i}{\beta} \right\rceil
        = \left\lceil \frac{\SA{\Text}[\alpha + i]}{\beta} \right\rceil.
        \qedhere
    \]
  \end{enumerate}
\end{proof}

\subsubsection{Alphabet Reduction for Suffix Array Queries}\label{sec:sa-alphabet-reduction}

\begin{lemma}\label{lm:sa-alphabet-reduction}
  Let $\AlphabetSize \geq 2$ and $\Text \in \IntegerAlphabet^{+}$.
  For every $i \in [1 \dd |\Text|]$, it holds
  \[
    \SA{\Text}[i] = \frac{\SA{\Text'}[\Delta + i]-1}{\delta} + 1,
  \]
  where
  $k = \lceil \log \AlphabetSize \rceil$,
  $\Text' = \ebin{k}{\Text}$ (\cref{def:ebin}),
  $\Delta = |\Text'| - |\Text|$, and 
  $\delta = 2k + 3$.
\end{lemma}
\begin{proof}
  By \cref{lm:ebin-order}, for
  every $j_1, j_2 \in [1 \dd n]$, $\Text[j_1 \dd |\Text|] \prec \Text[j_2 \dd |\Text|$ holds if
  and only if $\Text'[j'_1 \dd |\Text'|] \prec \Text'[j'_2 \dd |\Text'|]$, where $j'_k = (j_k-1)\delta + 1$ ($k \in \{1,2\}$).
  On the other hand, since $\one^{k+2}$ does not occur in $\Text'$, it follows that
  $\{\SA{\Text'}[\Delta+i] : i \in [1 \dd |\Text|]\} = \{(j-1)\delta + 1 : j \in [1 \dd |\Text|]\}$. Putting these
  together implies
  that for every $i \in [1 \dd |\Text|]$, it holds $\SA{\Text'}[\Delta + i] = (\SA{\Text}[i]-1)\delta + 1$. Rewriting this
  formula yields the claim.
\end{proof}

\begin{proposition}\label{pr:sa-alphabet-reduction-poly-to-binary}
  Let $\Text \in \IntegerAlphabet^{\Textlen}$ be a nonempty string, where $2 \leq \AlphabetSize < \Textlen^{\bigO(1)}$. Given
  the packed representation of $\Text$, we can in $\bigO(\Textlen / \log_{\AlphabetSize} \Textlen)$ time compute
  integers $\alpha$, $\beta$, $\gamma$, $\mu$, and the packed representation of a string
  $\Text_{\rm bin} \in \BinaryAlphabet^{+}$ such that $|\Text_{\rm bin}| = \Theta(\Textlen \log \AlphabetSize)$,
  and for every $i \in [1 \dd \Textlen]$, it holds
  \[
    \SA{\Text}[i] = \frac{\SA{\Text_{\rm bin}}[\alpha + i] - \beta}{\gamma} + \mu.
  \]
\end{proposition}
\begin{proof}
  Let $k = \lceil \log \AlphabetSize \rceil$. The algorithm proceeds as follows:
  \begin{enumerate}
  \item In $\bigO(\Textlen / \log_{\AlphabetSize} \Textlen)$ time we construct the data structure from
    \cref{pr:ebin} with $u = \Textlen$. Given the packed representation of any $S \in \IntegerAlphabet^{*}$, we can then
    compute the packed representation of $\ebin{k}{S}$ (\cref{def:ebin}) in
    $\bigO(1 + |S| / \log_{\AlphabetSize} \Textlen)$ time.
  \item Using the above structure, we compute the packed representation of the string
    $\Text_{\rm bin} = \ebin{k}{\Text}$. This takes
    $\bigO(\Textlen / \log_{\AlphabetSize} \Textlen)$ time.
  \item In $\bigO(1)$ time we set $\alpha = |\Text_{\rm bin}| - |\Text|$, $\beta = 1$, $\gamma = 2k + 3$, and $\mu = 1$.
  \end{enumerate}
  In total, the above algorithm takes $\bigO(\Textlen / \log_{\AlphabetSize} \Textlen)$ time.

  The correctness of the construction follows by \cref{lm:sa-alphabet-reduction}.
\end{proof}

\subsubsection{Summary}\label{sec:prefix-select-to-sa-summary}

\begin{theorem}
  Consider a data structure answering suffix array queries
  (see \cref{sec:prefix-select-and-sa-problem-def})
  that,
  for any text $\Text \in \BinaryAlphabet^{\Textlen}$, achieves the following
  complexities (where in the preprocessing we assume that we are given as input the packed representation of $\Text$):
  \begin{itemize}
  \item space usage $S(\Textlen)$,
  \item preprocessing time $P_t(\Textlen)$,
  \item preprocessing space $P_s(\Textlen)$,
  \item query time $Q(\Textlen)$.
  \end{itemize}
  For every sequence $W[1 \dd m]$ of $m \geq 1$ binary strings of length
  $\ell = 1 + \lfloor \log m \rfloor$, there exists $\Textlen = \bigO(m \log m)$ such that,
  given the sequence $W$ with all strings represented in the packed form,
  we can in $\bigO(m + P_t(\Textlen))$ time and
  $\bigO(m + P_s(\Textlen))$ working space build a data structure of size
  $\bigO(m + S(\Textlen))$ that, given the packed representation of any
  $X \in \BinaryAlphabet^{\leq \ell}$ and any $i \in [1 \dd \PrefixRank{W}{m}{X}]$,
  in $\bigO(Q(\Textlen))$ time computes $\PrefixSelect{W}{i}{X}$.
\end{theorem}
\begin{proof}

  We use the following definitions.
  Let $\Text_{\rm aux} = \SeqToString{\ell}{W}$ (\cref{def:seq-to-string}).
  Denote $\Textlen_{\rm aux} = |\Text_{\rm aux}| = m \cdot (2\ell + 1) = \Theta(m \log m)$.
  Let $\alpha$, $\beta$, $\gamma$, $\mu$, and $\Text_{\rm bin}$ denote the four integers
  and the text resulting from applying \cref{pr:sa-alphabet-reduction-poly-to-binary}
  to text $\Text_{\rm aux}$. Denote $\Textlen = |\Text_{\rm bin}|$. By \cref{pr:sa-alphabet-reduction-poly-to-binary},
  we have $\Textlen = \Theta(\Textlen_{\rm aux}) = \Theta(m \log m)$,
  and for every $i \in [1 \dd \Textlen_{\rm aux}]$,
  it holds $\SA{\Text_{\rm aux}}[i] = (\SA{\Text_{\rm bin}}[\alpha + i] - \beta) / \gamma + \mu$.

  \DSComponents
  The data structure consists of the following components:
  \begin{enumerate}
  \item The integers $\alpha$, $\beta$, $\gamma$, and $\mu$ defined as above.
  \item The data structure from the claim for the text $\Text_{\rm bin}$.
    It needs $\bigO(S(\Textlen))$ space.
  \item The structure resulting from applying \cref{pr:range-beg} to the array $W$.
    It needs $\bigO(m)$ space.
  \end{enumerate}
  In total, the structure needs $\bigO(m + S(\Textlen))$ space.

  \DSQueries
  Given the packed representation of any string
  $X \in \BinaryAlphabet^{\leq \ell}$ and any index $i \in [1 \dd \PrefixRank{W}{m}{X}]$,
  we compute $\PrefixSelect{W}{i}{X}$ as follows:
  \begin{enumerate}
  \item Using \cref{pr:range-beg}, in $\bigO(1)$ time we compute
    $\alpha_{\rm aux} := \RangeBegTwo{\revstr{X} \cdot \four}{\Text_{\rm aux}}$.
  \item Using the structure from the claim
    constructed for text $\Text_{\rm bin}$, in $\bigO(Q(\Textlen))$ time we compute
    $j := \SA{\Text_{\rm bin}}[\alpha + \alpha_{\rm aux} + i]$.
    In $\bigO(1)$ time we set
    $j_{\rm aux} = (j - \beta) / \gamma + \mu$. By \cref{pr:sa-alphabet-reduction-poly-to-binary} (and the discussion above),
    it holds $j_{\rm aux} = \SA{\Text_{\rm aux}}[\alpha_{\rm aux} + i]$.
  \item In $\bigO(1)$ time we compute $s := \lceil j_{\rm aux} / \beta_{\rm aux} \rceil$, where
    $\beta_{\rm aux} = 2\ell + 1$.
    By \cref{lm:prefix-select-to-sa}, it holds $s = \PrefixSelect{W}{i}{X}$. Thus, we return $s$ as the answer.
  \end{enumerate}
  In total, the query takes $\bigO(Q(\Textlen))$ time.

  \DSConstruction
  The components of the data structure are constructed as follows:
  \begin{enumerate}
  \item To compute $\alpha$, $\beta$, $\gamma$, and $\mu$ (defined above), we proceed as follows.
    First, in $\bigO(m)$ time we apply \cref{pr:seq-to-string}
    to the sequence $W$ to compute the packed representation of the
    string $\Text_{\rm aux}$ (defined above). Then, we apply \cref{pr:sa-alphabet-reduction-poly-to-binary} for
    $\Text_{\rm aux}$ to
    compute the packed representation of $\Text_{\rm bin}$ (as defined above),
    together with values $\alpha$, $\beta$, $\gamma$, and $\mu$. Since $\Text_{\rm aux}$ is over
    alphabet $\{\zero, \ldots, \four\}$, the application of \cref{pr:sa-alphabet-reduction-poly-to-binary}
    takes $\bigO(|\Text_{\rm aux}| / \log |\Text_{\rm aux}|) = \bigO(m)$ time.
  \item In the second step, we apply the preprocessing from the claim to string $\Text_{\rm bin}$.
    This takes $\bigO(P_t(\Textlen))$ time and uses $\bigO(P_s(\Textlen))$ working space.
  \item Finally, we apply \cref{pr:range-beg} to the sequence $W$. This takes
    $\bigO(m)$ time.
  \end{enumerate}
  In total, the construction takes $\bigO(m + P_t(\Textlen))$ time and uses $\bigO(m + P_s(\Textlen))$ working space.
\end{proof}

\subsection{Reducing Suffix Array to Prefix Select Queries}\label{sec:sa-to-prefix-select}

\subsubsection{Preliminaries}\label{sec:sa-to-prefix-select-prelim}

\begin{observation}\label{ob:prefix-rank-and-select-padding}
  Let $W[1 \dd m]$ be a sequence of $m \geq 1$ strings of length $\ell \geq 1$ over alphabet $\Sigma$ and let
  $W'[1 \dd m']$ be any sequence of $m' \geq m$ strings of length $\ell' \geq \ell$ over alphabet $\Sigma$
  such that for every $j \in [1 \dd m]$, $W[j]$ is a prefix of $W'[j]$.
  Consider any string $X \in \Sigma^{\leq \ell}$. Then,
  \begin{itemize}
  \item For every $i \in [0 \dd m]$, it holds \[\PrefixRank{W}{i}{X} = \PrefixRank{W'}{i}{X}.\]
  \item For every $i \in [1 \dd \PrefixRank{W}{m}{X}]$, it holds
    \[\PrefixSelect{W}{i}{X} = \PrefixSelect{W'}{i}{X}.\]
  \end{itemize}
  Furthermore, for every $i \in [1 \dd m]$ and every $p \in [0 \dd \ell]$, it holds
  \[\PrefixSpecialRank{W}{i}{p} = \PrefixSpecialRank{W'}{i}{p}.\]
\end{observation}

\subsubsection{Problem Reduction}\label{sec:sa-to-prefix-select-problem-reduction}

\begin{theorem}[{\cite{breaking}}]\label{th:sa-to-prefix-select-nonbinary}
  Consider a data structure that, given any sequence $W$ of $k$ strings of length $\ell$ over alphabet $\IntegerAlphabet$
  achieves the following complexities (where the input strings
  during construction are given in the packed representation, and at query time we are given
  the packed representation of any $X \in \IntegerAlphabet^{\leq \ell}$ and any $i \in [1 \dd \PrefixRank{W}{k}{X}]$, and
  we return $\PrefixSelect{W}{i}{X}$; see \cref{sec:prelim-prefix-queries}):
  \begin{itemize}
  \item space usage $S(k,\ell,\AlphabetSize)$,
  \item preprocessing time $P_t(k,\ell,\AlphabetSize)$,
  \item preprocessing space $P_s(k,\ell,\AlphabetSize)$,
  \item query time $Q(k,\ell,\AlphabetSize)$.
  \end{itemize}
  For every $\Text \in \IntegerAlphabet^{\Textlen}$ such that
  $2 \leq \AlphabetSize < \Textlen^{1/13}$ and $\Text[\Textlen]$ does not occur in $\Text[1 \dd \Textlen)$,
  there exist positive integers $k = \Theta(\Textlen / \log_{\AlphabetSize} \Textlen)$ and
  $\ell \leq (1 + \lfloor \log k \rfloor) / \lceil \log \AlphabetSize \rceil$
  such that, given the packed representation
  of $\Text$, we can in
  $\bigO(\Textlen / \log_{\AlphabetSize} \Textlen + P_t(k,\ell,\AlphabetSize))$ time and using
  $\bigO(\Textlen / \log_{\AlphabetSize} \Textlen + P_s(k,\ell,\AlphabetSize))$ working space build a data structure of size
  $\bigO(\Textlen / \log_{\AlphabetSize} \Textlen + S(k,\ell,\AlphabetSize))$ that answers suffix
  array queries on $\Text$ (see \cref{sec:prelim})
  in $\bigO(\log \log \Textlen + Q(k,\ell,\AlphabetSize))$ time.
\end{theorem}
\begin{proof}
  The above result, with cosmetically different parameters, was proved in~\cite[Theorem~5.2]{breaking}.
  To obtain the above claim, we make the following minor changes:
  \begin{itemize}
  \item First, we note that the original statement mentions both
    prefix rank and select queries. To obtain the above result it suffices to observe that during SA queries
    only prefix select queries are used (see~\cite[Proposition~5.5]{breaking}).
  \item In the original
    claim it holds $k = \bigO(\Textlen / \log_{\AlphabetSize} \Textlen)$. To obtain the claim with
    $k = \Theta(\Textlen / \log_{\AlphabetSize} \Textlen)$, we simply pad the input sequence of strings
    into length $k = \max(k', k'') = \Theta(\Textlen / \log_{\AlphabetSize} \Textlen)$, where
    $k' = \bigO(\Textlen / \log_{\AlphabetSize} \Textlen)$ is the original
    parameter, and $k'' = \lceil \Textlen / \log_{\AlphabetSize} \Textlen \rceil$.
    By \cref{ob:prefix-rank-and-select-padding}, the queries on the padded sequence are
    the same as on the original sequence.
  \item In the above claim, we have $\ell \leq (1 + \lfloor \log k \rfloor) / \lceil \log \AlphabetSize \rceil$
    and $2 \leq \AlphabetSize < \Textlen^{1/13}$.
    However,
    the constraints in the original claim~\cite{breaking} are $\ell = \bigO(\log_{\AlphabetSize} \Textlen)$ and
    $2 \leq \AlphabetSize < \Textlen^{1/7}$.
    To obtain the claim with modified parameters, we recall that
    in the proof of the original claim,
    we have $\ell = 3\tau$, where $\tau = \lfloor \mu\log_{\AlphabetSize} \Textlen \rfloor$ and $\mu$ is
    any positive constant smaller than $\tfrac{1}{6}$ such that $\tau \geq 1$ (such $\mu$ exists by the assumption
    $\AlphabetSize < \Textlen^{1/7}$).
    Here we modify the value of $\tau$ and instead use
    $\tau = \lfloor \mu' \log_{\AlphabetSize} \Textlen \rfloor$, where $\mu'$ is a positive constant smaller than $\tfrac{1}{12}$
    such that $\tau \geq 1$ (such $\mu'$ exists by the assumption $\AlphabetSize < \Textlen^{1/13}$). This change does
    not affect the correctness of the data structure, since it makes $\tau$ smaller, and the only constraints on $\tau$
    in~\cite{breaking} are upper bounds.
    To show that
    $\ell \leq (1 + \lfloor \log k \rfloor) / \lceil \log \AlphabetSize \rceil$, we proceed as follows:
    \begin{itemize}
    \item First, observe that by the assumption $2 \leq \AlphabetSize < \Textlen^{1/13}$
      implies that $\Textlen$ is sufficiently large so that
      $k^2 \geq k''^2 \geq (\Textlen / \log \Textlen)^2 \geq \Textlen$. Consequently,
      $\log \Textlen \leq 2\log k$.
    \item On the other hand, note that for every $x \geq 1$, it holds $\lceil x \rceil \leq 2x$.
    \end{itemize}
    Putting the above together, we obtain:
    \begin{align*}
      \ell &= 3 \lfloor \mu' \log_{\AlphabetSize} \Textlen \rfloor
           \leq \lfloor 3\mu' \log_{\AlphabetSize} \Textlen \rfloor
           \leq \Big\lfloor \tfrac{1}{4} \tfrac{\log \Textlen}{\log \AlphabetSize} \Big\rfloor
           \leq \Big\lfloor \tfrac{1}{2} \tfrac{\log k}{\log \AlphabetSize} \Big\rfloor
           \leq \Big\lfloor \tfrac{\log k}{\lceil \log \AlphabetSize\rceil } \Big\rfloor
           \leq \Big\lfloor \tfrac{1 + \lfloor \log k \rfloor}{\lceil \log \AlphabetSize \rceil} \Big\rfloor
           \leq \tfrac{1 + \lfloor \log k \rfloor}{\lceil \log \AlphabetSize \rceil}.
      \qedhere
    \end{align*}
  \end{itemize}
\end{proof}

\subsubsection{Alphabet Reduction for Prefix Select Queries}\label{sec:prefix-select-alphabet-reduction}

\begin{lemma}\label{lm:prefix-queries-alphabet-reduction}
  Let $\AlphabetSize \geq 2$ and let $W[1 \dd m]$ be an array of $m \geq 1$ strings of length $\ell \geq 1$ over
  alphabet $\IntegerAlphabet$. Let $k = \lceil \log \AlphabetSize \rceil$,
  and $W_{\rm bin}[1 \dd m]$ be any sequence of binary strings such that
  for every $j \in [1 \dd m]$, $\bin{k}{W[j]}$ (\cref{def:bin}) is a prefix of $W_{\rm bin}[j]$.
  \begin{enumerate}
  \item\label{lm:prefix-queries-alphabet-reduction-it-1}
    Let $X \in \IntegerAlphabet^{\leq \ell}$ and $X_{\rm bin} = \bin{k}{X}$. Then:
    \begin{enumerate}
    \item\label{lm:prefix-queries-alphabet-reduction-it-1a}
      For every $i \in [0 \dd m]$, it holds
      \[
        \PrefixRank{W}{i}{X} = \PrefixRank{W_{\rm bin}}{i}{X_{\rm bin}}.
      \]
    \item\label{lm:prefix-queries-alphabet-reduction-it-1b}
      For every $i \in [1 \dd \PrefixRank{W}{m}{X}]$, it holds
      \[
        \PrefixSelect{W}{i}{X} = \PrefixSelect{W_{\rm bin}}{i}{X_{\rm bin}}.
      \]
    \item\label{lm:prefix-queries-alphabet-reduction-it-1c}
      For every $b,e \in [0 \dd m]$, it holds
      \[
        \{i \in (b \dd e] : X\text{ is a prefix of }W[i]\} =
        \{i \in (b \dd e] : X_{\rm bin}\text{ is a prefix of }W_{\rm bin}[i]\}.
      \]
    \end{enumerate}
  \item\label{lm:prefix-queries-alphabet-reduction-it-2}
    For every $i \in [1 \dd m]$ and every $p \in [0 \dd \ell]$, it holds
    \[
      \PrefixSpecialRank{W}{i}{p} = \PrefixSpecialRank{W_{\rm bin}}{i}{p \cdot k}.
    \]
  \end{enumerate}
\end{lemma}
\begin{proof}
  The proof follows by observing that for every $A,B \in \IntegerAlphabet^{*}$, $A$ is a prefix of $B$ if and only if
  $\bin{k}{A}$ is a prefix of $\bin{k}{B}$.
\end{proof}

\begin{proposition}\label{pr:prefix-select-alphabet-reduction}
  Consider a data structure answering prefix select queries
  (see \cref{sec:prefix-select-and-sa-problem-def}) that, for any
  sequence of $m \geq 1$ binary strings of length
  $1 + \lfloor \log m \rfloor$, achieves the following
  complexities (where both the string at query time as well as
  input strings during construction are given in the packed representation):
  \begin{itemize}
  \item space usage $S(m)$,
  \item preprocessing time $P_t(m)$,
  \item preprocessing space $P_s(m)$,
  \item query time $Q(m)$.
  \end{itemize}
  Let $W[1 \dd m']$ be a sequence of $m' \geq 1$ equal-length strings over alphabet
  $\IntegerAlphabet$ (where $\AlphabetSize \geq 2$) of length
  $\ell \leq (1 + \lfloor \log m' \rfloor) / \lceil \log \AlphabetSize \rceil$,
  Given the
  sequence $W$, with all strings represented in the packed form, we can in
  $\bigO(m' + P_t(m'))$ time and using $\bigO(m' + P_s(m'))$ working space
  construct a data structure of size $\bigO(m' + S(m'))$ that, given the packed
  representation of any $X \in \IntegerAlphabet^{\leq \ell}$
  and any $i \in [1 \dd \PrefixRank{W}{m'}{X}]$, returns
  $\PrefixSelect{W}{i}{X}$ in $\bigO(Q(m'))$ time.
\end{proposition}
\begin{proof}

  Let $k = \lceil \log \AlphabetSize \rceil$.
  Let $W_{\rm bin}[1 \dd m']$ denote a sequence of binary strings of length
  $\ell_{\rm bin} = 1 + \lfloor \log m' \rfloor$ defined so that for every $j \in [1 \dd m']$,
  $W_{\rm bin}[j]$ is a prefix of length $\ell_{\rm bin}$ of the string $\bin{k}{W[j]} \cdot \zero^{\infty}$.
  Note that since we assumed $\ell \leq (1 + \lfloor \log m' \rfloor) / \lceil \log \AlphabetSize \rceil$,
  it follows (see \cref{def:bin})
  that $\ell \cdot k \leq 1 + \lfloor \log m' \rfloor = \ell_{\rm bin}$, i.e., 
  for every $j \in [1 \dd m']$, $\bin{k}{W[j]}$ is a prefix of $W_{\rm bin}[j]$.

  \DSComponents
  The data structure consists of the following components:
  \begin{enumerate}
  \item The data structure from \cref{pr:bin} constructed for $u = m'$.
    Given the packed representation of any $S \in \IntegerAlphabet^{*}$,
    we can then compute the packed representation of $\bin{k}{S}$ (\cref{def:bin})
    in $\bigO(1 + |S|/\log_{\AlphabetSize} m')$ time. It needs
    $\bigO(m' / \log_{\AlphabetSize} m') = \bigO(m')$ space.
  \item The data structure from the claim constructed for
    $W_{\rm bin}[1 \dd m']$. It needs $\bigO(S(m'))$ space.
  \end{enumerate}
  In total, the data structure needs $\bigO(m' + S(m'))$ space.

  \DSQueries
  The queries are answered as follows. Assume that we are given the
  packed representation of some $X \in \IntegerAlphabet^{\leq \ell}$
  and some position $i \in [1 \dd \PrefixRank{W}{m'}{X}]$. The value
  $\PrefixSelect{W}{i}{X}$ is computed as follows:
  \begin{enumerate}
  \item Using \cref{pr:bin}, in
    $\bigO(1 + |X| / \log_{\AlphabetSize} m') =
    \bigO(1 + \ell / \log_{\AlphabetSize} m') = \bigO(1)$ time
    we compute the packed representation of the string $X_{\rm bin} = \bin{k}{X}$.
  \item Using the structure from the claim constructed for $W_{\rm bin}$,
    we compute $j = \PrefixSelect{W_{\rm bin}}{i}{X_{\rm bin}}$ in $\bigO(Q(m'))$ time.
    Note that $|X_{\rm bin}| \leq |X| \cdot k \leq \ell \cdot k \leq \ell_{\rm bin}$.
    Thus, we can indeed ask such query for $W_{\rm bin}$. Moreover, note that by
    \cref{lm:prefix-queries-alphabet-reduction}\eqref{lm:prefix-queries-alphabet-reduction-it-1a}, it holds
    $\PrefixRank{W}{m'}{X} = \PrefixRank{W_{\rm bin}}{m'}{X_{\rm bin}}$.
    Thus, $i \in [1 \dd \PrefixRank{W_{\rm bin}}{m'}{X_{\rm bin}}]$, i.e.,
    $\PrefixSelect{W_{\rm bin}}{i}{X_{\rm bin}}$ is well-defined. Finally, note that by
    \cref{lm:prefix-queries-alphabet-reduction}\eqref{lm:prefix-queries-alphabet-reduction-it-1b}, it also holds
    $\PrefixSelect{W}{i}{X} = \PrefixSelect{W_{\rm bin}}{i}{X_{\rm bin}}$.
    We thus return $j$ as the answer.
  \end{enumerate}
  In total, the query takes $\bigO(Q(m'))$ time.

  \DSConstruction
  The components of the data structure are constructed as follows:
  \begin{enumerate}
  \item In $\bigO(m' / \log_{\AlphabetSize} m') = \bigO(m')$ time we
    construct the structure from \cref{pr:bin} for $u = m'$.
  \item To construct the second component, we first
    construct the sequence $W_{\rm bin}[1 \dd m']$ using \cref{pr:bin}.
    Computing $\ebin{k}{W[j]}$ for a single position $j \in [1 \dd m']$ takes
    $\bigO(1 + |W[j]| / \log_{\AlphabetSize} m') = \bigO(1 + \ell / \log_{\AlphabetSize} m') = \bigO(1)$ time.
    Padding with zeros to the length $\ell_{\rm bin}$ also takes $\bigO(1)$ time.
    Thus, over all $j \in [1 \dd m']$, we spend $\bigO(m')$ time.
    We then apply the preprocessing from the claim to the sequence $W_{\rm bin}[1 \dd m']$.
    It takes $\bigO(m' + P_t(m'))$ time and uses $\bigO(m' + P_s(m'))$ working space.
  \end{enumerate}
  In total, the construction takes
  $\bigO(m' + P_t(m'))$ time and uses
  $\bigO(m' + P_s(m'))$ working space.
\end{proof}

\subsubsection{Summary}\label{sec:sa-to-pref-select-summary}

\begin{theorem}\label{th:sa-to-prefix-select}
  Consider a data structure answering prefix select queries
  (see \cref{sec:prefix-select-and-sa-problem-def}) that, for any
  sequence of $m \geq 1$ binary strings of length
  $1 + \lfloor \log m \rfloor$, achieves the following
  complexities
  (where both the string at query time as well as
  input strings during construction are given in the packed representation):
  \begin{itemize}
  \item space usage $S(m)$,
  \item preprocessing time $P_t(m)$,
  \item preprocessing space $P_s(m)$,
  \item query time $Q(m)$.
  \end{itemize}
  For every $\Text \in \BinaryAlphabet^{\Textlen}$, there exists
  $m = \Theta(\Textlen / \log \Textlen)$ such that, given the packed
  representation of $\Text$, we can in
  $\bigO(\Textlen / \log \Textlen + P_t(m))$ time and using
  $\bigO(\Textlen / \log \Textlen + P_s(m))$ working space construct
  a data structure of size $\bigO(\Textlen / \log \Textlen + S(m))$ that
  answers suffix array queries on $\Text$
  (see \cref{sec:prefix-select-and-sa-problem-def})
  in $\bigO(\log \log \Textlen + Q(m))$ time.
\end{theorem}
\begin{proof}

  Assume that $\Textlen > 3^{13}-1$ (otherwise, the claim holds trivially).

  We use the following definitions. Let $\AlphabetSize = 3$ and
  let $\Text' \in \IntegerAlphabet^{\Textlen + 1}$ be a string defined so that $\Text'[1 \dd \Textlen] = \Text$ and
  $\Text'[\Textlen + 1] = \two$. Denote $\Textlen' = |\Text'|$. Note that $\Text'[\Textlen']$ does not occur
  in $\Text'[1 \dd \Textlen')$ and it holds $2 \leq \AlphabetSize < (\Textlen')^{1/13}$.
  Let $D$ denote the data structure resulting from applying \cref{pr:prefix-select-alphabet-reduction} with alphabet size
  $\AlphabetSize$
  to the structure (answering prefix select queries on binary strings) form the claim. Given any sequence $W[1 \dd m']$
  of $m' \geq 1$ equal-length strings over alphabet $\IntegerAlphabet$ of length
  $\ell' \leq (1 + \lfloor \log m' \rfloor) / \lceil \log \AlphabetSize \rceil$ as input (with all strings represented in the packed
  form), the structure $D$ achieves the following complexities (where at query time it is given
  the packed representation of any $X \in \IntegerAlphabet^{\leq \ell'}$ and any
  $i \in [1 \dd \PrefixRank{W}{m'}{X}]$, and returns $\PrefixSelect{W}{i}{X}$):
  \begin{itemize}
  \item space usage $S'(m',\ell',\AlphabetSize) = \bigO(m' + S(m'))$,
  \item preprocessing time $P'_t(m',\ell',\AlphabetSize) = \bigO(m' + P_t(m'))$,
  \item preprocessing space $P'_s(m',\ell',\AlphabetSize) = \bigO(m' + P_s(m'))$,
  \item query time $Q'(m',\ell',\AlphabetSize) = \bigO(Q(m'))$,
  \end{itemize}
  where the complexities $S(m')$, $P_t(m')$, $P_s(m')$, and $Q(m')$ refer to the structure from the claim (answering
  prefix select for binary strings).

  \DSComponents
  The data structure answering suffix array queries
  consists of a single component: the data structure from \cref{th:sa-to-prefix-select-nonbinary}
  applied for text $\Text'$ and with the structure $D$ as the structure answering prefix select queries.
  Note that we can use $D$, since it supports prefix select queries for the combination of parameters required
  in \cref{th:sa-to-prefix-select-nonbinary}, i.e., for sequences over alphabet $\IntegerAlphabet$, with the
  length $\ell$ of all strings satisfying
  $\ell \leq (1 + \lfloor \log k \rfloor) / \lceil \log \AlphabetSize \rceil$ (where $k$ is the
  length of the input sequence).
  To bound the space usage of this component, note that by
  \cref{th:sa-to-prefix-select-nonbinary} and the above discussion, there exists
  $m = \Theta(\Textlen' / \log_{\AlphabetSize} \Textlen') = \Theta(\Textlen / \log \Textlen)$ (recall that $\AlphabetSize = 3$)
  such that the structure needs
  $\bigO(\Textlen' / \log_{\AlphabetSize} \Textlen' + S'(m,\ell,\AlphabetSize)) =
  \bigO(\Textlen' / \log_{\AlphabetSize} \Textlen' + m + S(m)) = \bigO(\Textlen / \log \Textlen + S(m))$ space.

  \DSQueries
  The suffix array queries are answered as follows. Let $i \in [1 \dd \Textlen]$. Note that by definition of
  $\Text'$, it follows that $\SA{\Text}[i] = \SA{\Text'}[i]$. By \cref{th:sa-to-prefix-select-nonbinary} and the above
  discussion, computing $\SA{\Text'}[i]$ takes $\bigO(\log \log \Textlen' + Q'(m,\ell,\AlphabetSize)) =
  \bigO(\log \log \Textlen + Q(m))$ time.

  \DSConstruction
  By \cref{th:sa-to-prefix-select-nonbinary} and the above discussion, construction of the above data structure
  answering suffix array queries on $\Text'$ (and hence also on $\Text$) takes
  $\bigO(\Textlen' / \log_{\AlphabetSize} \Textlen' + P'_t(m,\ell,\AlphabetSize)) =
  \bigO(\Textlen' / \log_{\AlphabetSize} \Textlen' + m + P_t(m)) =
  \bigO(\Textlen / \log \Textlen + P_t(m))$
  time and uses
  $\bigO(\Textlen' / \log_{\AlphabetSize} \Textlen' + P'_s(m,\ell,\AlphabetSize)) =
  \bigO(\Textlen' / \log_{\AlphabetSize} \Textlen' + m + P_s(m)) =
  \bigO(\Textlen / \log \Textlen + P_s(m))$ working space.
\end{proof}

\section{Equivalence of Inverse Suffix Array and Prefix Special Rank Queries}\label{sec:prefix-special-rank}

\subsection{Problem Definitions}\label{sec:prefix-special-rank-and-isa-problem-def}
\vspace{-1.5ex}

\begin{framed}
  \noindent
  \probname{Indexing for Inverse Suffix Array Queries}
  \begin{bfdescription}
  \item[Input:]
    The packed representation of a string
    $\Text \in \BinaryAlphabet^{\Textlen}$.
  \item[Output:]
    A data structure that, given
    any $i \in [1 \dd \Textlen]$, returns $\ISA{\Text}[i]$, i.e.,
    the value
    \[
      |\{j \in [1 \dd \Textlen] : \Text[j \dd \Textlen] \preceq \Text[i \dd \Textlen]\}|.
    \]
  \end{bfdescription}
\end{framed}

\begin{framed}
  \noindent
  \probname{Indexing for Prefix Special Rank Queries}
  \begin{bfdescription}
  \item[Input:]
    A sequence $W[1 \dd m]$ of $m \geq 1$ binary strings of length $\ell = 1 + \lfloor \log m \rfloor$, with all
    strings represented in the packed form.
  \item[Output:]
    A data structure that, given
    any index $i \in [1 \dd m]$ and any length $p \in [0 \dd \ell]$,
    returns $\PrefixSpecialRank{W}{i}{p}$ (\cref{def:prefix-rank-and-select}), i.e., the value
    \[
      |\{j \in [1 \dd i] : X\text{ is a prefix of }W[j]\}|,
    \]
    where
    $X = W[i][1 \dd p]$.
  \end{bfdescription}
\end{framed}
\vspace{2ex}

\subsection{Reducing Prefix Special Rank to Inverse Suffix Array Queries}\label{sec:prefix-special-rank-to-isa}

\subsubsection{Problem Reduction}

\begin{lemma}\label{lm:prefix-special-rank-to-isa}
  Let $W[1 \dd m]$ be an array of $m \geq 1$ strings of length $\ell \geq 1$ over alphabet $\BinaryAlphabet$.
  Denote $\Text = \SeqToString{\ell}{W}$ (\cref{def:seq-to-string}).
  Then, for every $i \in [1 \dd m]$ and every $p \in [0 \dd \ell]$, it holds
  \[
    \PrefixSpecialRank{W}{i}{p} = \ISA{\Text}[(i\beta - (k + p)] - \alpha,
  \]
  where $X$ is a prefix of $W[i]$ of length $p$,
  $\alpha = \RangeBegTwo{\revstr{X}\cdot\four}{\Text}$, and $\beta = \ell + \lfloor \log m \rfloor + 2$.
\end{lemma}
\begin{proof}
  Let $k$ and $s(i)$ (where $i \in [0 \dd m]$) be as in \cref{def:seq-to-string}.
  By definition of the prefix special rank query,
  $\PrefixSpecialRank{W}{i}{p} = j$ holds if and only if
  $j \in [1 \dd \PrefixRank{W}{m}{X}]$ and $\PrefixSelect{W}{j}{X} = i$. We will prove
  the claim by showing that the latter two conditions are satisfied
  for $j = \ISA{\Text}[i\beta - (k+p)] - \alpha$.
  \begin{itemize}
  \item First, we prove that $j \in [1 \dd \PrefixRank{W}{m}{X}]$.
    To show $j \geq 1$, observe that
    the suffix $\Text(i \beta \dd |\Text|]$ is preceded with a string $\revstr{W[i]} \cdot \four \cdot s(i-1)$.
    Since $X$ is a prefix of $W[i]$ of length $p$ (which is
    equivalent to $\revstr{X}$ being a suffix of length $p$ of $\revstr{W[i]}$), it follows that
    $\Text[i\beta - (k+p) \dd |\Text|]$ has the substring $\revstr{X} \cdot \four$ as a prefix.
    Consequently, letting $\alpha' = \RangeEndTwo{\revstr{X} \cdot \four}{\Text}$,
    we have $\ISA{\Text}[i\beta - (k+p)] \in (\alpha \dd \alpha']$.
    In particular, $j = \ISA{\Text}[i\beta - (k+p)] - \alpha \geq 1$.
    To show $j \leq \PrefixRank{W}{m}{X}$, recall that in the proof of \cref{lm:prefix-select-to-sa}, we showed
    that $|\OccTwo{\revstr{X} \cdot \four}{\Text}| = \PrefixRank{W}{m}{X}$.
    Thus, $j = \ISA{\Text}[i\beta - (k+p) - \alpha \leq \alpha' - \alpha =
    |\OccTwo{\revstr{X} \cdot \four}{\Text}| = \PrefixRank{W}{m}{X}$.
  \item To prove that $\PrefixSelect{W}{j}{X} = i$, we apply \cref{lm:prefix-select-to-sa}. More precisely,
    it holds
    \begin{align*}
      \PrefixSelect{W}{j}{X}
        &= \PrefixSelect{W}{\ISA{\Text}[i\beta - (k+p)] - \alpha}{X} \\
        &= \left\lceil \frac{\SA{\Text}[\alpha + \ISA{\Text}[i\beta - (k+p)] - \alpha]}{\beta} \right\rceil \\
        &= \left\lceil \frac{\SA{\Text}[\ISA{\Text}[i\beta - (k+p)]]}{\beta} \right\rceil
        = \left\lceil \frac{i\beta - (k+p)}{\beta} \right\rceil
        = i.
        \qedhere
    \end{align*}
  \end{itemize}
\end{proof}

\subsubsection{Alphabet Reduction for Inverse Suffix Array Queries}\label{sec:isa-alphabet-reduction}

\begin{proposition}[{\cite{hierarchy}}]\label{pr:isa-alphabet-reduction-poly-to-binary}
  Let $\Text \in \IntegerAlphabet^{\Textlen}$ be a nonempty string, where $\AlphabetSize < \Textlen^{\bigO(1)}$. Given
  the packed representation of $\Text$, we can in $\bigO(\Textlen / \log_{\AlphabetSize} \Textlen)$ time compute
  integers $\Delta$ and $\delta$, and the packed representation of a string
  $\Text' \in \BinaryAlphabet^{+}$ such that $|\Text'| = \Theta(\Textlen \log \AlphabetSize)$,
  and for every $j \in [1 \dd \Textlen]$, it holds
  \[
    \ISA{\Text}[j] = \ISA{\Text'}[(j-1)\delta + 1] - \Delta.
  \]
\end{proposition}

\subsubsection{Summary}\label{sec:prefix-special-rank-to-isa-summary}

\begin{theorem}
  Consider a data structure answering inverse suffix array queries
  (see \cref{sec:prefix-special-rank-and-isa-problem-def})
  that, for any text $\Text \in \BinaryAlphabet^{\Textlen}$,
  achieves the following complexities (where in the preprocessing we assume that we are given as input the packed
  representation of $\Text$):
  \begin{itemize}
  \item space usage $S(\Textlen)$,
  \item preprocessing time $P_t(\Textlen)$,
  \item preprocessing space $P_s(\Textlen)$,
  \item query time $Q(\Textlen)$.
  \end{itemize}
  For every sequence $W[1 \dd m]$ of $m \geq 1$ binary strings of length $\ell = 1 + \lfloor \log m \rfloor$, there
  exists $\Textlen = \bigO(m \log m)$ such that, given the sequence $W$ with all strings represented in the packed form,
  we can in $\bigO(m + P_t(\Textlen))$ time and using $\bigO(m + P_s(\Textlen))$ working space construct a data
  structure of size $\bigO(m + S(\Textlen))$ that, given any $i \in [1 \dd m]$ and $p \in [0 \dd \ell]$, computes
  $\PrefixSpecialRank{W}{i}{p}$ in $\bigO(Q(\Textlen))$ time.
\end{theorem}
\begin{proof}

  We use the following definitions.
  Let $\Text_{\rm aux} = \SeqToString{\ell}{W}$ (\cref{def:seq-to-string}).
  Denote $\Textlen_{\rm aux} = |\Text_{\rm aux}| = m \cdot (2\ell + 1) = \Theta(m \log m)$.
  Let $\Delta$, $\delta$, and $\Text_{\rm bin}$ denote the two integers
  and the text resulting from applying \cref{pr:isa-alphabet-reduction-poly-to-binary}
  to text $\Text_{\rm aux}$. Denote $\Textlen = |\Text_{\rm bin}|$. By \cref{pr:isa-alphabet-reduction-poly-to-binary},
  we have $\Textlen = \Theta(\Textlen_{\rm aux}) = \Theta(m \log m)$,
  and for every $j \in [1 \dd \Textlen_{\rm aux}]$,
  it holds $\ISA{\Text_{\rm aux}}[j] = \ISA{\Text_{\rm bin}}[(j-1)\delta + 1] - \Delta$.

  \DSComponents
  The data structure consists of the following components:
  \begin{enumerate}
  \item The integers $\Delta$ and $\delta$ defined as above.
  \item The data structure from the claim for the text $\Text_{\rm bin}$.
    It needs $\bigO(S(\Textlen))$ space.
  \item The structure resulting from applying \cref{pr:range-beg} to the array $W$.
    It needs $\bigO(m)$ space.
  \end{enumerate}
  In total, the structure needs $\bigO(m + S(\Textlen))$ space.

  \DSQueries
  Given any $i \in [1 \dd m]$ and $p \in [0 \dd \ell]$, we compute
  $\PrefixSpecialRank{W}{i}{p}$ as follows:
  \begin{enumerate}
  \item In $\bigO(1)$ time compute the packed representation of $X$, defined
    as a length-$p$ prefix of $W[i]$.
  \item Using \cref{pr:range-beg}, in $\bigO(1)$ time compute
    $\alpha_{\rm aux} := \RangeBegTwo{\revstr{X} \cdot \four}{\Text_{\rm aux}}$.
  \item Set
    $j_{\rm aux} := i\beta - (\ell+p)$, where $\beta = 2\ell + 1$.
    Using the structure from the claim
    constructed for text $\Text_{\rm bin}$, in $\bigO(Q(\Textlen))$ time compute
    $i_{\rm aux} := \ISA{\Text_{\rm bin}}[(j_{\rm aux} - 1)\delta + 1] - \Delta$.
    By \cref{pr:isa-alphabet-reduction-poly-to-binary} (and the discussion above),
    $i_{\rm aux} = \ISA{\Text_{\rm aux}}[j_{\rm aux}] = \ISA{\Text_{\rm aux}}[i\beta - (\ell+p)]$.
  \item In $\bigO(1)$ time compute $r := i_{\rm aux} - \alpha_{\rm aux}$.
    By \cref{lm:prefix-special-rank-to-isa}, it holds $r = \PrefixSpecialRank{W}{i}{p}$.
    Thus, we return $r$ as the answer.
  \end{enumerate}
  In total, the query takes $\bigO(Q(\Textlen))$ time.

  \DSConstruction
  The components of the data structure are constructed as follows:
  \begin{enumerate}
  \item To compute $\Delta$ and $\delta$ (defined above), we proceed as follows.
    First, in $\bigO(m)$ time we apply \cref{pr:seq-to-string}
    to the sequence $W$ to compute the packed representation of the
    string $\Text_{\rm aux}$ (defined above). Then, we apply \cref{pr:isa-alphabet-reduction-poly-to-binary} for
    $\Text_{\rm aux}$ to
    compute the packed representation of $\Text_{\rm bin}$ (as defined above),
    together with values $\Delta$ and $\delta$. Since $\Text_{\rm aux}$ is over
    alphabet $\{\zero, \ldots, \four\}$, the application of \cref{pr:isa-alphabet-reduction-poly-to-binary}
    takes $\bigO(|\Text_{\rm aux}| / \log |\Text_{\rm aux}|) = \bigO(m)$ time.
  \item In the second step, we apply the preprocessing from the claim to string $\Text_{\rm bin}$.
    This takes $\bigO(P_t(\Textlen))$ time and uses $\bigO(P_s(\Textlen))$ working space.
  \item Finally, we apply \cref{pr:range-beg} to the sequence $W$. This takes
    $\bigO(m)$ time.
  \end{enumerate}
  In total, the construction takes $\bigO(m + P_t(\Textlen))$ time and uses $\bigO(m + P_s(\Textlen))$ working space.
\end{proof}

\subsection{Reducing Inverse Suffix Array to Prefix Special Rank Queries}\label{sec:isa-to-prefix-special-rank}

\subsubsection{Problem Reduction}\label{sec:isa-to-prefix-special-rank-problem-reduction}

\begin{theorem}[{\cite{breaking}}]\label{th:isa-to-prefix-special-rank-nonbinary}
  Consider a data structure that, given any sequence $W$ of $k$ strings of length $\ell$ over alphabet $\IntegerAlphabet$
  achieves the following complexities (where the input strings
  during construction are given in the packed representation, and at query time we are given any
  $i \in [1 \dd k]$ and any $p \in [0 \dd \ell]$, and we return $\PrefixSpecialRank{W}{i}{p}$; see \cref{sec:prelim-prefix-queries}):
  \begin{itemize}
  \item space usage $S(k,\ell,\AlphabetSize)$,
  \item preprocessing time $P_t(k,\ell,\AlphabetSize)$,
  \item preprocessing space $P_s(k,\ell,\AlphabetSize)$,
  \item query time $Q(k,\ell,\AlphabetSize)$.
  \end{itemize}
  For every $\Text \in \IntegerAlphabet^{\Textlen}$ such that
  $2 \leq \AlphabetSize < \Textlen^{1/13}$ and $\Text[\Textlen]$ does not occur in $\Text[1 \dd \Textlen)$,
  there exist positive integers $k = \Theta(\Textlen / \log_{\AlphabetSize} \Textlen)$
  and $\ell \leq (1 + \lfloor \log k \rfloor) / \lceil \log \AlphabetSize \rceil$
  such that, given the packed representation
  of $\Text$, we can in $\bigO(\Textlen / \log_{\AlphabetSize} \Textlen + P_t(k,\ell,\AlphabetSize))$ time
  and using $\bigO(\Textlen / \log_{\AlphabetSize} \Textlen + P_s(k,\ell,\AlphabetSize))$ working space
  build a data structure of size
  $\bigO(\Textlen / \log_{\AlphabetSize} \Textlen + S(k,\ell,\AlphabetSize))$ that answers inverse suffix
  array queries on $\Text$ (see \cref{sec:prelim})
  in $\bigO(\log \log \Textlen + Q(k,\ell,\AlphabetSize))$ time.
\end{theorem}
\begin{proof}
  The proof is almost identical to the proof of \cref{th:sa-to-prefix-select-nonbinary},
  and reduces to performing cosmetic fixes to~\cite[Theorem~5.2]{breaking}. The main
  difference here is to observe that during inverse suffix array queries, the structure
  from~\cite{breaking} performs only prefix special rank queries;
  see~\cite[Proposition~5.4]{breaking}. Observe also that performing the padding of
  the input for prefix special rank queries does not change the answers by
  \cref{ob:prefix-rank-and-select-padding}.
\end{proof}

\subsubsection{Alphabet Reduction for Prefix Special Rank Queries}\label{sec:prefix-special-rank-alphabet-reduction}

\begin{proposition}\label{pr:prefix-special-rank-alphabet-reduction}
  Consider a data structure answering prefix special rank queries
  (see \cref{sec:prefix-special-rank-and-isa-problem-def}) that, for any
  sequence of $m \geq 1$ binary strings of length
  $1 + \lfloor \log m \rfloor$, achieves the following
  complexities (where the input strings during construction are
  given in the packed representation):
  \begin{itemize}
  \item space usage $S(m)$,
  \item preprocessing time $P_t(m)$,
  \item preprocessing space $P_s(m)$,
  \item query time $Q(m)$.
  \end{itemize}
  Let $W[1 \dd m']$ be a sequence of $m' \geq 1$ equal-length strings over alphabet
  $\IntegerAlphabet$ (where $\AlphabetSize \geq 2$) of length
  $\ell \leq (1 + \lfloor \log m' \rfloor) / \lceil \log \AlphabetSize \rceil$,
  Given the
  sequence $W$, with all strings represented in the packed form, we can in
  $\bigO(m' + P_t(m'))$ time and using $\bigO(m' + P_s(m'))$ working space
  construct a data structure of size $\bigO(m' + S(m'))$ that, given any
  $i \in [1 \dd m']$ and $p \in [0 \dd \ell]$, returns
  $\PrefixSpecialRank{W}{i}{p}$ in $\bigO(Q(m'))$ time.
\end{proposition}
\begin{proof}

  Let $k = \lceil \log \AlphabetSize \rceil$.
  Let $W_{\rm bin}[1 \dd m']$ denote a sequence of binary strings of length
  $\ell_{\rm bin} = 1 + \lfloor \log m' \rfloor$ defined so that for every $j \in [1 \dd m']$,
  $W_{\rm bin}[j]$ is a prefix of length $\ell_{\rm bin}$ of the string $\bin{k}{W[j]} \cdot \zero^{\infty}$.
  Note that since we assumed $\ell \leq (1 + \lfloor \log m' \rfloor) / \lceil \log \AlphabetSize \rceil$,
  it follows (see \cref{def:bin})
  that $\ell \cdot k \leq 1 + \lfloor \log m' \rfloor = \ell_{\rm bin}$, i.e.,
  for every $j \in [1 \dd m']$, $\bin{k}{W[j]}$ is a prefix of $W_{\rm bin}[j]$.

  \DSComponents
  The data structure answering prefix special rank queries on $W[1 \dd m']$ consist of single component:
  the data structure from the claim constructed for $W_{\rm bin}[1 \dd m']$. It needs $\bigO(S(m'))$ space.

  \DSQueries
  The queries are answered as follows. Assume that we are given some
  $i \in [1 \dd m']$ and $p \in [0 \dd \ell]$.
  To compute the value of $\PrefixSpecialRank{W}{i}{p}$, we simply return the value
  $\PrefixSpecialRank{W_{\rm bin}}{i}{p \cdot k}$ in $\bigO(Q(m'))$ time.
  This is correct by \cref{lm:prefix-queries-alphabet-reduction}\eqref{lm:prefix-queries-alphabet-reduction-it-2}.

  \DSConstruction
  To construct the above data structure, we proceed as follows.
  First, in $\bigO(m' / \log_{\AlphabetSize} m') = \bigO(m')$ time we
  construct the structure from \cref{pr:bin} for $u = m'$.
  We then construct the sequence $W_{\rm bin}[1 \dd m']$ using \cref{pr:bin}.
  Computing $\ebin{k}{W[j]}$ for a single position $j \in [1 \dd m']$ takes
  $\bigO(1 + |W[j]| / \log_{\AlphabetSize} m') = \bigO(1 + \ell / \log_{\AlphabetSize} m') = \bigO(1)$ time.
  Padding with zeros to the length $\ell_{\rm bin}$ also takes $\bigO(1)$ time.
  Thus, over all $j \in [1 \dd m']$, we spend $\bigO(m')$ time.
  We then apply the preprocessing from the claim to the sequence $W_{\rm bin}[1 \dd m']$.
  It takes $\bigO(m' + P_t(m'))$ time and uses $\bigO(m' + P_s(m'))$ working space.
  In total, the construction takes
  $\bigO(m' + P_t(m'))$ time and uses
  $\bigO(m' + P_s(m'))$ working space.
\end{proof}

\subsubsection{Summary}\label{sec:isa-to-prefix-special-rank-summary}

\begin{theorem}\label{th:isa-to-prefix-special-rank}
  Consider a data structure answering prefix special rank queries
  (see \cref{sec:prefix-special-rank-and-isa-problem-def}) that, for any
  sequence of $m \geq 1$ binary strings of length
  $1 + \lfloor \log m \rfloor$, achieves the following
  complexities
  (where the input strings during construction
  are given in the packed representation):
  \begin{itemize}
  \item space usage $S(m)$,
  \item preprocessing time $P_t(m)$,
  \item preprocessing space $P_s(m)$,
  \item query time $Q(m)$.
  \end{itemize}
  For every $\Text \in \BinaryAlphabet^{\Textlen}$, there exists
  $m = \Theta(\Textlen / \log \Textlen)$ such that, given the packed
  representation of $\Text$, we can in
  $\bigO(\Textlen / \log \Textlen + P_t(m))$ time and using
  $\bigO(\Textlen / \log \Textlen + P_s(m))$ working space construct
  a data structure of size $\bigO(\Textlen / \log \Textlen + S(m))$ that
  answers inverse suffix array queries on $\Text$
  (see \cref{sec:prefix-special-rank-and-isa-problem-def})
  in $\bigO(\log \log \Textlen + Q(m))$ time.
\end{theorem}
\begin{proof}

  Assume that $\Textlen > 3^{13}-1$ (otherwise, the claim holds trivially).

  We use the following definitions. Let $\AlphabetSize = 3$ and
  let $\Text' \in \IntegerAlphabet^{\Textlen + 1}$ be a string defined so that $\Text'[1 \dd \Textlen] = \Text$ and
  $\Text'[\Textlen + 1] = \two$. Denote $\Textlen' = |\Text'|$. Note that $\Text'[\Textlen']$ does not occur
  in $\Text'[1 \dd \Textlen')$ and it holds $2 \leq \AlphabetSize < (\Textlen')^{1/13}$.
  Let $D$ denote the data structure resulting from applying \cref{pr:prefix-special-rank-alphabet-reduction} with alphabet size
  $\AlphabetSize$
  to the structure (answering prefix special rank queries on binary strings) form the claim. Given any sequence $W[1 \dd m']$
  of $m' \geq 1$ equal-length strings over alphabet $\IntegerAlphabet$ of length
  $\ell' \leq (1 + \lfloor \log m' \rfloor) / \lceil \log \AlphabetSize \rceil$ as input (with all strings represented in the packed
  form), the structure $D$ achieves the following complexities (where at query time it is given
  some $i \in [1 \dd m']$ and $p \in [0 \dd \ell]$, and returns the value $\PrefixSpecialRank{W}{i}{p}$):
  \begin{itemize}
  \item space usage $S'(m',\ell',\AlphabetSize) = \bigO(m' + S(m'))$,
  \item preprocessing time $P'_t(m',\ell',\AlphabetSize) = \bigO(m' + P_t(m'))$,
  \item preprocessing space $P'_s(m',\ell',\AlphabetSize) = \bigO(m' + P_s(m'))$,
  \item query time $Q'(m',\ell',\AlphabetSize) = \bigO(Q(m'))$,
  \end{itemize}
  where the complexities $S(m')$, $P_t(m')$, $P_s(m')$, and $Q(m')$ refer to the structure from the claim (answering
  prefix special rank for binary strings).

  \DSComponents
  The data structure answering inverse suffix array queries
  consists of a single component: the data structure from \cref{th:isa-to-prefix-special-rank-nonbinary}
  applied for text $\Text'$ and with the structure $D$ as the structure answering prefix special rank queries.
  Note that we can use $D$, since it supports prefix special rank queries for the combination of parameters required
  in \cref{th:isa-to-prefix-special-rank-nonbinary}, i.e., for sequences over alphabet $\IntegerAlphabet$, with the
  length $\ell$ of all strings satisfying
  $\ell \leq (1 + \lfloor \log k \rfloor) / \lceil \log \AlphabetSize \rceil$ (where $k$ is the
  length of the input sequence).
  To bound the space usage of this component, note that by
  \cref{th:isa-to-prefix-special-rank-nonbinary} and the above discussion, there exists
  $m = \Theta(\Textlen' / \log_{\AlphabetSize} \Textlen') = \Theta(\Textlen / \log \Textlen)$ (recall that $\AlphabetSize = 3$)
  such that the structure needs
  $\bigO(\Textlen' / \log_{\AlphabetSize} \Textlen' + S'(m,\ell,\AlphabetSize)) =
  \bigO(\Textlen' / \log_{\AlphabetSize} \Textlen' + m + S(m)) = \bigO(\Textlen / \log \Textlen + S(m))$ space.

  \DSQueries
  The inverse suffix array queries are answered as follows. Let $i \in [1 \dd \Textlen]$. Note that by definition of
  $\Text'$, it follows that $\ISA{\Text}[i] = \ISA{\Text'}[i]$. By \cref{th:isa-to-prefix-special-rank-nonbinary} and the above
  discussion, computing $\ISA{\Text'}[i]$ takes $\bigO(\log \log \Textlen' + Q'(m,\ell,\AlphabetSize)) =
  \bigO(\log \log \Textlen + Q(m))$ time.

  \DSConstruction
  By \cref{th:isa-to-prefix-special-rank-nonbinary} and the above discussion,
  construction of the above data structure
  answering inverse suffix array queries on $\Text'$ (and hence also on $\Text$) takes
  $\bigO(\Textlen' / \log_{\AlphabetSize} \Textlen' + P'_t(m,\ell,\AlphabetSize)) =
  \bigO(\Textlen' / \log_{\AlphabetSize} \Textlen' + m + P_t(m)) =
  \bigO(\Textlen / \log \Textlen + P_t(m))$
  time and uses
  $\bigO(\Textlen' / \log_{\AlphabetSize} \Textlen' + P'_s(m,\ell,\AlphabetSize)) =
  \bigO(\Textlen' / \log_{\AlphabetSize} \Textlen' + m + P_s(m)) =
  \bigO(\Textlen / \log \Textlen + P_s(m))$ working space.
\end{proof}

\section{Equivalence of Pattern Ranking and Prefix Rank Queries}\label{sec:prefix-rank}

\subsection{Problem Definitions}\label{sec:prefix-rank-and-pattern-ranking-problem-def}
\vspace{-1.5ex}

\begin{framed}
  \noindent
  \probname{Indexing for Pattern Ranking Queries}
  \begin{bfdescription}
  \item[Input:]
    The packed representation of a string
    $\Text \in \BinaryAlphabet^{\Textlen}$.
  \item[Output:]
    A data structure that, given the packed representation of any
    $\Pat \in \BinaryAlphabet^{*}$, returns
    \[
      \RangeBegTwo{\Pat}{\Text} = |\{j \in [1 \dd \Textlen] : \Text[j \dd \Textlen] \prec \Pat\}|.
    \]
  \end{bfdescription}
\end{framed}

\begin{framed}
  \noindent
  \probname{Indexing for Prefix Rank Queries}
  \begin{bfdescription}
  \item[Input:]
    A sequence $W[1 \dd m]$ of $m \geq 1$ binary strings of length $\ell = 1 + \lfloor \log m \rfloor$, with all
    strings represented in the packed form.
  \item[Output:]
    A data structure that, given
    the packed representation of any $X \in \BinaryAlphabet^{\leq \ell}$ and any
    $i \in [0 \dd m]$, returns $\PrefixRank{W}{i}{X}$, i.e.,
    the value $|\{j \in [1 \dd i] : X\text{ is a prefix of }W[j]\}|$.
  \end{bfdescription}
\end{framed}
\vspace{2ex}

\subsection{Reducing Prefix Rank to Pattern Ranking Queries}\label{sec:prefix-rank-to-pattern-ranking}

\subsubsection{Problem Reduction}

\begin{lemma}\label{lm:prefix-rank-to-pattern-rank}
  Let $W[1 \dd m]$ be an array of $m \geq 1$ strings of length $\ell \geq 1$
  over alphabet $\BinaryAlphabet$.
  Denote $\Text = \SeqToString{\ell}{W}$ (\cref{def:seq-to-string}).
  Let $s(i)$ (where $i \in [0 \dd m]$) be as in \cref{def:seq-to-string}.
  For every $X \in \BinaryAlphabet^{\leq \ell}$ and $i \in [0 \dd m]$,
  it holds
  \[
    \PrefixRank{W}{i}{X} = \RangeBegTwo{X'}{\Text} - \alpha,
  \]
  where $X' = \revstr{X} \cdot \four \cdot s(i)$ and
  $\alpha = \RangeBegTwo{\revstr{X}\cdot\four}{\Text}$.
\end{lemma}
\begin{proof}
  Denote
  \[
    A = \{j \in \OccTwo{\revstr{X}\cdot\four}{\Text} : \Text[j \dd \Textlen] \prec X'\}.
  \]
  Observe that because $\revstr{X}\cdot \four$ is a prefix of $X'$, it follows that
  $\RangeBegTwo{X'}{\Text}
      = \RangeBegTwo{\revstr{X}\cdot\four}{\Text} + |A|
      = \alpha + |A|$.
  This implies that we only need to show that $\PrefixRank{W}{i}{X} = |A|$.

  Let $k = 1 + \lfloor \log m \rfloor$ and $\beta = \ell + k + 1$.
  Denote $q = \PrefixRank{W}{m}{X}$.
  Let $(a_i)_{i \in [1 \dd q]}$ and $(b_i)_{i \in [1 \dd q]}$ be sequences
  defined by $a_i = \PrefixSelect{W}{i}{X}$ and $b_i = a_i \cdot \beta - (k+|X|)$. Recall that in
  the proof of \cref{lm:prefix-select-to-sa}, we showed that
  $\OccTwo{\revstr{X} \cdot \four}{\Text} = \{b_1, \ldots, b_q\}$ and
  $\Text[b_1 \dd |\Text|] \prec \Text[b_2 \dd |\Text|] \prec \cdots \prec \Text[b_q \dd |\Text|]$.
  Denote $r = \PrefixRank{W}{i}{X}$, and
  consider three cases:
  \begin{itemize}
  \item First, assume that $0 < r < q$. By definition of the prefix rank query, this implies
    that $a_{r} \leq i < a_{r+1}$. Thus, we also have $s(a_{r}-1) \prec s(i) \preceq s(a_{r+1}-1)$.
    Observe now (see also the proof of \cref{lm:prefix-select-to-sa}), that
    for every $i \in [1 \dd q]$, $\revstr{X} \cdot \four \cdot s(a_i-1)$ is a prefix of $\Text[b_i \dd |\Text|]$.
    In particular $\revstr{X} \cdot \four \cdot s(a_r-1)$ is a prefix of $\Text[b_r \dd |\Text|]$
    and $\revstr{X}\cdot\four\cdot s(a_{r+1}-1)$ is a prefix of $\Text[b_{r+1} \dd |\Text|]$. By
    $X' = \revstr{X} \cdot \four \cdot s(i)$ and
    $s(a_{r}-1) \prec s(i) \preceq s(a_{r+1}-1)$ (and the fact that all strings $s(a_{r}-1)$, $s(i)$, and $s(a_{r+1}-1)$ are of the
    same length), we thus obtain that
    $\Text[b_r \dd |\Text|] \prec X' \preceq \Text[b_{r+1} \dd |\Text|]$.
    By $\Text[b_1 \dd |\Text|] \prec \Text[b_2 \dd |\Text|] \prec \cdots \prec \Text[b_q \dd |\Text|]$,
    this implies
    that $A = \{b_1, \ldots, b_r\}$, i.e., $|A| = r$.
  \item Let us now assume that $r = 0$. This implies that $i < a_{0}$. Hence, $s(i) \preceq s(a_{0}-1)$, and thus
    $X' \preceq \Text[b_1 \dd |\Text|]$. We thus have $A = \emptyset$, and hence $|A| = r$.
  \item Let us now assume that $r = q$. We then have $a_{q} \leq i$. Thus, $s(a_{q}-1) \prec s(i)$, and hence
    $\Text[b_q \dd |\Text|] \prec X'$. We thus have $A = \{b_1, \ldots, b_q\}$, and hence
    $|A| = r$.
  \end{itemize}
  We have thus proved that in all cases it holds $|A| = r = \PrefixRank{W}{i}{X}$. By the above discussion, this
  concludes the proof.
\end{proof}

\subsubsection{Alphabet Reduction for Pattern Ranking Queries}\label{sec:pattern-ranking-alphabet-reduction}

\begin{lemma}\label{lm:pattern-rank-alphabet-reduction}
  Let $\AlphabetSize \geq 2$, $\Text \in \IntegerAlphabet^{*}$, and $\Pat \in \IntegerAlphabet^{*}$.
  Then, it holds
  \[
    \RangeBegTwo{\Pat}{\Text} = \RangeBegTwo{\Pat'}{\Text'} - \Delta,
  \]
  where $k = \lceil \log \AlphabetSize \rceil$,
  $\Pat' = \ebin{k}{\Pat}$ (\cref{def:ebin}),
  $\Text' = \ebin{k}{\Text}$, and
  $\Delta = |\Text'| - |\Text|$.
\end{lemma}
\begin{proof}
  Denote
  \[
    A = \{j \in \OccTwo{\one^{k+1}}{\Text'} : \Text'[j \dd |\Text'|] \prec \Pat'\}.
  \]
  Observe that since $\one^{k+1}$ is a prefix of $\Pat'$, it follows that
  $\RangeBegTwo{\Pat'}{\Text'} = \RangeBegTwo{\one^{k+1}}{\Text'} + |A| = \Delta + |A|$.
  Thus, it suffices to prove that $\RangeBegTwo{\Pat}{\Text} = |A|$.

  First, note that $\OccTwo{\one^{k+1}}{\Text'} = \{1 + \delta \cdot (j-1) : j \in [1 \dd |\Text|]\}$, where $\delta = 2k + 3$.
  This implies that, letting $\mathcal{S} = \{\ebin{k}{\Text[j \dd |\Text|]} : j \in [1 \dd |\Text|]\}$,
  we can equivalently write $\mathcal{S} = \{\Text'[j \dd |\Text'|] : j \in \OccTwo{\one^{k+1}}{\Text'}\}$.
  By \cref{lm:ebin-order} and the definition of $\RangeBegTwo{\Pat}{\Text}$, we thus have
  \begin{align*}
    \RangeBegTwo{\Pat}{\Text}
      &= |\{j \in [1 \dd |\Text|] : \Text[j \dd |\Text|] \prec \Pat\}|\\
      &= |\{S \in \mathcal{S} : S \prec \Pat'\}|\\
      &= |\{j \in \OccTwo{\one^{k+1}}{\Text} : \Text'[j \dd |\Text'|] \prec \Pat'\}|
      = |A|.
      \qedhere
  \end{align*}
\end{proof}

\begin{proposition}\label{pr:pattern-ranking-alphabet-reduction}
  Let $\Text \in \IntegerAlphabet^{\Textlen}$ be a nonempty string, where
  $2 \leq \AlphabetSize < \Textlen^{\bigO(1)}$. Given the packed representation of
  $\Text$, we can in $\bigO(\Textlen / \log_{\AlphabetSize} \Textlen)$ time
  construct the packed representation of a
  text $\Text_{\rm bin} \in \BinaryAlphabet^{*}$ satisfying
  $|\Text_{\rm bin}| = \Theta(\Textlen \log \AlphabetSize)$, an integer
  $\alpha$, and a data structure that, given the packed representation of
  any $\Pat \in \IntegerAlphabet^{m}$, in
  $\bigO(1 + m / \log_{\AlphabetSize} \Textlen)$ time returns the
  packed representation of a string $\Pat_{\rm bin} \in \BinaryAlphabet^{*}$
  satisfying $|\Pat_{\rm bin}| =
  m \cdot (2\lceil \log \AlphabetSize \rceil + 3)$ and
  \[
    \RangeBegTwo{\Pat}{\Text} =
      \RangeBegTwo{\Pat_{\rm bin}}{\Text_{\rm bin}} - \alpha
  \]
\end{proposition}
\begin{proof}

  Let $k = \lceil \log \AlphabetSize \rceil$.
  To compute $\alpha$ and the packed representation of $\Text_{\rm bin}$,
  we proceed as follows:
  \begin{enumerate}
  \item In $\bigO(\Textlen / \log_{\AlphabetSize} \Textlen)$ time we construct the data structure from
    \cref{pr:ebin} for $u = \Textlen$.
    Given the packed representation of any $S \in \IntegerAlphabet^{*}$, we can then
    compute the packed representation of $\ebin{k}{S}$ (\cref{def:ebin}) in
    $\bigO(1 + |S| / \log_{\AlphabetSize} \Textlen)$ time.
  \item Using the above structure, we compute the packed representation of
    $\Text_{\rm bin} = \ebin{k}{\Text}$ in $\bigO(\Textlen / \log_{\AlphabetSize} \Textlen)$ time.
  \item In $\bigO(1)$ time we set $\alpha = |\Text_{\rm bin}| - |\Text|$.
  \end{enumerate}
  In total, the construction of the packed representation of $\Text_{\rm bin}$ and $\alpha$
  takes $\bigO(\Textlen / \log_{\AlphabetSize} \Textlen)$. By \cref{def:ebin}, it
  holds $|\Text_{\rm bin}| = \Textlen \cdot (2k + 3) = \Theta(\Textlen \log \AlphabetSize)$.

  Next, we address the construction of the data structure from the claim.
  It consists of a single component: the data structure from \cref{pr:ebin} for $u = \Textlen$.
  It uses $\bigO(\Textlen / \log_{\AlphabetSize} \Textlen)$ space.

  The queries are implemented as follows. Given the packed representation of
  $\Pat \in \IntegerAlphabet^{m}$, we compute the packed representation
  of $\Pat_{\rm bin}$ defined as $\Pat_{\rm bin} = \ebin{k}{\Pat}$.
  Using the structure from \cref{pr:ebin}, this takes $\bigO(1 + m / \log_{\AlphabetSize} \Textlen)$ time.
  By \cref{def:ebin}, it holds $|\Pat_{\rm bin}| = m \cdot (2k + 3)$,
  and by \cref{lm:pattern-rank-alphabet-reduction},
  we have $\RangeBegTwo{\Pat}{\Text} = \RangeBegTwo{\Pat_{\rm bin}}{\Text_{\rm bin}} - \alpha$.

  By \cref{pr:ebin}, construction of the data structure takes
  $\bigO(\Textlen / \log_{\AlphabetSize} \Textlen)$ time.
\end{proof}

\subsubsection{Summary}\label{sec:prefix-rank-to-pattern-ranking-summary}

\begin{theorem}\label{th:prefix-rank-to-pattern-ranking}
  Consider a data structure answering pattern ranking queries
  (see \cref{sec:prefix-rank-and-pattern-ranking-problem-def})
  that, for any
  text $\Text \in \BinaryAlphabet^{\Textlen}$, achieves the following complexities
  (where in the preprocessing we assume that we are given as input the packed
  representation of $\Text$, and at query time we are given a packed
  representation of $\Pat \in \BinaryAlphabet^{\leq k}$):
  \begin{itemize}
  \item space usage $S(\Textlen)$,
  \item preprocessing time $P_t(\Textlen)$,
  \item preprocessing space $P_s(\Textlen)$,
  \item query time $Q(\Textlen, k)$.
  \end{itemize}
  For every sequence $W[1 \dd m]$ of $m \geq 1$ binary strings of length
  $\ell = 1 + \lfloor \log m \rfloor$, there exists
  $\Textlen = \bigO(m \log m)$ and $k = \bigO(\log m)$ such that, given the sequence $W$ with
  all strings represented in the packed form, we can in
  $\bigO(m + P_t(\Textlen))$ time and using $\bigO(m + P_s(\Textlen))$
  working space construct a data structure of size $\bigO(m + S(\Textlen))$
  that, given any $i \in [0 \dd m]$ and the packed representation of any
  $X \in \BinaryAlphabet^{\leq \ell}$, computes
  $\PrefixRank{W}{i}{X}$ in $\bigO(Q(\Textlen, k))$ time.
\end{theorem}
\begin{proof}

  We use the following definitions.
  Let $\Text_{\rm aux} = \SeqToString{\ell}{W}$ (\cref{def:seq-to-string}).
  Denote $\Textlen_{\rm aux} =
  |\Text_{\rm aux}| = m \cdot (2\ell+1) = \Theta(m \log m)$. Let
  $\Text_{\rm bin}$ and $\alpha_{\rm bin}$ denote the text and an integer obtained
  by applying \cref{pr:pattern-ranking-alphabet-reduction} to text
  $\Text_{\rm aux}$. Denote $\Textlen = |\Text_{\rm bin}|$.
  Since $\Text_{\rm aux}$ is over alphabet $\{\zero, \dots, \four\}$,
  by \cref{pr:pattern-ranking-alphabet-reduction} we have
  $\Textlen = \Theta(|\Text_{\rm aux}|) = \Theta(m \log m)$.
  Denote $\ell' = \lceil \ell / 2 \rceil$ and
  $k = 18 \lfloor \log m \rfloor + 27 = \bigO(\log m)$.

  \DSComponents
  The data structure consists of the following components:
  \begin{enumerate}
  \item The structure from \cref{pr:pattern-ranking-alphabet-reduction} applied
    to text $\Text_{\rm aux}$. Since $\Text_{\rm aux}$ is over alphabet
    $\{\zero, \dots, \four\}$, the structure needs
    $\bigO(|\Text_{\rm aux}| / \log |\Text_{\rm aux}|) =
    \bigO(m)$ space.
  \item The data structure from the claim (i.e., answering pattern ranking
    queries) for the text $\Text_{\rm bin}$. The structure needs
    $\bigO(S(\Textlen))$ space.
  \item The structure resulting from applying \cref{pr:range-beg} to the array $W$.
    It needs $\bigO(m)$ space.
  \item A lookup table $L_{\rm rev}$ such that for every $X \in \BinaryAlphabet^{\leq \ell'}$,
    $L_{\rm rev}$ maps $X$ to the packed representation of $\revstr{X}$ (over alphabet $\{\zero, \dots, \four\}$).
    When accessing $L_{\rm rev}$, we map $X$ to $\BinStrToInt{X}$ (\cref{def:bin-str-to-int}). Thus, $L_{\rm rev}$ needs
    $\bigO(2^{\ell'}) = \bigO(\sqrt{\Textlen})$ space.
    Given the packed representation of any $X \in \BinaryAlphabet^{\leq \ell}$ (over alphabet $\BinaryAlphabet$),
    we can then in $\bigO(1)$ time compute the packed representation of $\revstr{X}$ (over $\{\zero, \ldots, \four\}$).
  \item A lookup table $L_{\rm map}$ defined such that for every $X \in \BinaryAlphabet^{\leq \ell'}$,
    $L_{\rm map}$ maps $X$ to the packed representation of
    $\Substitute{\Substitute{X}{\zero}{\two}}{\one}{\three}$ (over alphabet $\{\zero, \ldots, \four\}$).
    Similarly as above, $L_{\rm map}$ needs $\bigO(\sqrt{\Textlen})$ space.
    Given $i \in [0 \dd m]$, we can then compute the packed representation
    (over alphabet $\{\zero, \dots, \four\}$) of the string $s(i)$ 
    (\cref{def:seq-to-string}) in $\bigO(1)$ time.
  \item The integer $\alpha_{\rm bin}$ (defined  above).
  \end{enumerate}
  In total, the structure needs $\bigO(m + S(\Textlen))$ space.

  \DSQueries
  The queries are answered as follows. Assume that we are given some $i \in [0 \dd m]$
  and the packed representation of some $X \in \BinaryAlphabet^{\leq \ell}$. The query
  algorithm proceeds as follows:
  \begin{enumerate}
  \item Using $L_{\rm rev}$ and $L_{\rm map}$, in $\bigO(1)$ time we compute the packed
    representation of the string $X_{\rm aux} = \revstr{X} \cdot \four \cdot s(i)$
    (over alphabet $\{\zero, \dots, \four\}$).
    Note that $|X_{\rm aux}| = |X| + 1 + \ell \leq 2\ell + 1 \leq 2\lfloor \log m \rfloor + 3$.
  \item Using \cref{pr:range-beg}, in $\bigO(1)$ time we compute
    $\alpha_{\rm aux} := \RangeBegTwo{\revstr{X} \cdot \four}{\Text_{\rm aux}}$.
  \item Using \cref{pr:pattern-ranking-alphabet-reduction}, in
    $\bigO(1 + |X_{\rm aux}| / \log \Textlen_{\rm aux}) =
    \bigO(1 + \ell / \log m) = \bigO(1)$ time, we compute
    the packed representation of a string $X_{\rm bin} \in \BinaryAlphabet^{*}$
    of length $|X_{\rm bin}| = 9 \cdot |X_{\rm aux}| \leq 18 \lfloor \log m \rfloor + 27 \leq k$
    such that
    \[
      \RangeBegTwo{X_{\rm aux}}{\Text_{\rm aux}} = \RangeBegTwo{X_{\rm bin}}{\Text_{\rm bin}} - \alpha_{\rm bin}.
    \]
  \item Using the structure from the claim, we compute $b_{\rm bin} = \RangeBegTwo{X_{\rm bin}}{\Text_{\rm bin}}$.
    By $|X_{\rm bin}| \leq k$, this takes $\bigO(Q(\Textlen, k))$ time.
  \item In $\bigO(1)$ time, we compute $b_{\rm aux} = b_{\rm bin} - \alpha_{\rm bin}$.
    By the above, $b_{\rm aux} = \RangeBegTwo{X_{\rm aux}}{\Text_{\rm aux}}$.
  \item In $\bigO(1)$ time we compute and return $r = b_{\rm aux} - \alpha_{\rm aux}$.
    By \cref{lm:prefix-rank-to-pattern-rank}, $\PrefixRank{W}{i}{X} = r$.
  \end{enumerate}
  In total, the query takes $\bigO(Q(\Textlen, k))$ time.

  \DSConstruction
  The components of the data structure are constructed as follows:
  \begin{enumerate}
  \item In $\bigO(m)$ time we apply \cref{pr:seq-to-string} to the
    sequence $W$ to compute the packed representation of the string
    $\Text_{\rm aux}$. We then apply \cref{pr:pattern-ranking-alphabet-reduction}
    to $\Text_{\rm aux}$. Since $\Text_{\rm aux}$ is over alphabet
    $\{\zero, \dots, \four\}$, this takes
    $\bigO(|\Text_{\rm aux}| / \log |\Text_{\rm aux}|) = \bigO(m)$
    time. Note that \cref{pr:pattern-ranking-alphabet-reduction}, in addition
    to the data structure, also returns the integer $\alpha_{\rm bin}$ and
    the packed representation of the string $\Text_{\rm bin}$ (defined above).
  \item We apply the preprocessing from the claim to the string $\Text_{\rm bin}$.
    This takes $\bigO(P_t(|\Text_{\rm bin}|)) = \bigO(P_t(\Textlen))$ time
    and uses $\bigO(P_s(|\Text_{\rm bin}|)) = \bigO(P_s(\Textlen))$ working space.
  \item We apply \cref{pr:range-beg} to the array $W$. This takes
    $\bigO(m)$ time.
  \item The lookup table $L_{\rm rev}$ is computed in $\bigO(m)$ time similarly
    as in the proof of \cref{pr:seq-to-string}.
  \item Next, we compute the lookup table $L_{\rm map}$. Similarly as above,
    we proceed as in \cref{pr:seq-to-string}, and spend $\bigO(m)$ time.
  \item Lastly, we store the integer $\alpha_{\rm bin}$, which was
    already computed above.
  \end{enumerate}
  In total, the construction takes
  $\bigO(m + P_t(\Textlen))$ time and uses
  $\bigO(m + P_s(\Textlen))$ working space.
\end{proof}

\subsection{Reducing Pattern Ranking to Prefix Rank Queries}\label{sec:pattern-ranking-to-prefix-rank}

\subsubsection{Problem Reduction}\label{sec:pattern-ranking-to-prefix-rank-problem-reduction}

\begin{theorem}[{\cite{breaking}}]\label{th:pattern-ranking-to-prefix-rank-nonbinary}
  Consider a data structure that, given any sequence $W$ of $k$ strings of length $\ell$ over alphabet $\IntegerAlphabet$
  achieves the following complexities (where the input strings
  during construction are given in the packed representation, and at query time we are given
  the packed representation of any $X \in \IntegerAlphabet^{\leq \ell}$ and any $i \in [0 \dd k]$, and
  we return $\PrefixRank{W}{i}{X}$; see \cref{sec:prelim-prefix-queries}):
  \begin{itemize}
  \item space usage $S(k,\ell,\AlphabetSize)$,
  \item preprocessing time $P_t(k,\ell,\AlphabetSize)$,
  \item preprocessing space $P_s(k,\ell,\AlphabetSize)$,
  \item query time $Q(k,\ell,\AlphabetSize)$.
  \end{itemize}
  For every $\Text \in \IntegerAlphabet^{\Textlen}$ such that
  $\AlphabetSize < \Textlen^{1/13}$ and $\Text[\Textlen]$ does not occur in $\Text[1 \dd \Textlen)$,
  there exist positive integers $k = \Theta(\Textlen / \log_{\AlphabetSize} \Textlen)$ and
  $\ell \leq (1 + \lfloor \log k \rfloor) / \lceil \log \AlphabetSize \rceil$
  such that, given the packed representation
  of $\Text$, we can in $\bigO(\Textlen / \log_{\AlphabetSize} \Textlen + P_t(k,\ell,\AlphabetSize))$ time
  and using $\bigO(\Textlen / \log_{\AlphabetSize} \Textlen + P_s(k,\ell,\AlphabetSize))$ working space
  build a data structure of size
  $\bigO(\Textlen / \log_{\AlphabetSize} \Textlen + S(k,\ell,\AlphabetSize))$ that, given the packed
  representation of any $\Pat \in \IntegerAlphabet^{*}$, returns
  $\RangeBegTwo{\Pat}{\Text}$ (see \cref{sec:prelim}) in
  $\bigO(\log \log \Textlen + Q(k,\ell,\AlphabetSize) + |\Pat| / \log_{\AlphabetSize} \Textlen)$ time.
\end{theorem}
\begin{proof}
  The proof is almost identical to the proof of \cref{th:sa-to-prefix-select-nonbinary},
  and reduces to performing cosmetic fixes to~\cite[Theorem~6.3]{breaking}.
  Note that performing the padding of
  the input for prefix rank queries does not change the answers by
  \cref{ob:prefix-rank-and-select-padding}.
\end{proof}

\subsubsection{Alphabet Reduction for Prefix Rank Queries}\label{sec:prefix-rank-alphabet-reduction}

\begin{proposition}\label{pr:prefix-rank-alphabet-reduction}
  Consider a data structure answering prefix rank queries
  (see \cref{sec:prefix-rank-and-pattern-ranking-problem-def}) that, for any
  sequence of $m \geq 1$ binary strings of length
  $1 + \lfloor \log m \rfloor$, achieves the following
  complexities (where both the string at query time as well as
  input strings during construction are given in the packed representation):
  \begin{itemize}
  \item space usage $S(m)$,
  \item preprocessing time $P_t(m)$,
  \item preprocessing space $P_s(m)$,
  \item query time $Q(m)$.
  \end{itemize}
  Let $W[1 \dd m']$ be a sequence of $m' \geq 1$ equal-length strings over alphabet
  $\IntegerAlphabet$ (where $\AlphabetSize \geq 2$) of length
  $\ell \leq (1 + \lfloor \log m' \rfloor) / \lceil \log \AlphabetSize \rceil$,
  Given the
  sequence $W$, with all strings represented in the packed form, we can in
  $\bigO(m' + P_t(m'))$ time and using $\bigO(m' + P_s(m'))$ working space
  construct a data structure of size $\bigO(m' + S(m'))$ that, given the packed
  representation of any $X \in \IntegerAlphabet^{\leq \ell}$
  and any $i \in [0 \dd m']$, returns
  $\PrefixRank{W}{i}{X}$ in $\bigO(Q(m'))$ time.
\end{proposition}
\begin{proof}
  The above reduction is essentially the same as the one presented in
  \cref{pr:prefix-select-alphabet-reduction}, except we only need
  \cref{lm:prefix-queries-alphabet-reduction}\eqref{lm:prefix-queries-alphabet-reduction-it-1a}.
\end{proof}

\subsubsection{Summary}\label{sec:pattern-ranking-to-pref-rank-summary}

\begin{theorem}\label{th:pattern-ranking-to-prefix-rank}
  Consider a data structure answering prefix rank queries
  (see \cref{sec:prefix-rank-and-pattern-ranking-problem-def}) that, for any
  sequence of $m \geq 1$ binary strings of length
  $1 + \lfloor \log m \rfloor$, achieves the following
  complexities
  (where both the string at query time as well as
  input strings during construction are given in the packed representation):
  \begin{itemize}
  \item space usage $S(m)$,
  \item preprocessing time $P_t(m)$,
  \item preprocessing space $P_s(m)$,
  \item query time $Q(m)$.
  \end{itemize}
  For every $\Text \in \BinaryAlphabet^{\Textlen}$, there exists
  $m = \Theta(\Textlen / \log \Textlen)$ such that, given the packed
  representation of $\Text$, we can in
  $\bigO(\Textlen / \log \Textlen + P_t(m))$ time and using
  $\bigO(\Textlen / \log \Textlen + P_s(m))$ working space construct
  a data structure of size $\bigO(\Textlen / \log \Textlen + S(m))$ that
  answers pattern ranking queries, i.e.,
  given the packed representation of any $\Pat \in \BinaryAlphabet^{*}$, returns
  $\RangeBegTwo{\Pat}{\Text}$ (see \cref{sec:prefix-rank-and-pattern-ranking-problem-def})
  in $\bigO(\log \log \Textlen + Q(m) + |\Pat| / \log \Textlen)$ time.
\end{theorem}
\begin{proof}

  Assume that $\Textlen > 3^{13}-1$ (otherwise, the claim holds trivially).

  We use the following definitions. Let $\AlphabetSize = 3$ and
  let $\Text' \in \IntegerAlphabet^{\Textlen + 1}$ be a string defined so that $\Text'[1 \dd \Textlen] = \Text$ and
  $\Text'[\Textlen + 1] = \two$. Denote $\Textlen' = |\Text'|$. Note that $\Text'[\Textlen']$ does not occur
  in $\Text'[1 \dd \Textlen')$ and it holds $2 \leq \AlphabetSize < (\Textlen')^{1/13}$.
  Let $D$ denote the data structure resulting from applying \cref{pr:prefix-rank-alphabet-reduction} with alphabet size
  $\AlphabetSize$
  to the structure (answering prefix rank queries on binary strings) form the claim. Given any sequence $W[1 \dd m']$
  of $m' \geq 1$ equal-length strings over alphabet $\IntegerAlphabet$ of length
  $\ell' \leq (1 + \lfloor \log m' \rfloor) / \lceil \log \AlphabetSize \rceil$ as input (with all strings represented in the packed
  form), the structure $D$ achieves the following complexities (where at query time it is given
  the packed representation of any $X \in \IntegerAlphabet^{\leq \ell'}$ and any
  $i \in [0 \dd m']$, and returns $\PrefixRank{W}{i}{X}$):
  \begin{itemize}
  \item space usage $S'(m',\ell',\AlphabetSize) = \bigO(m' + S(m'))$,
  \item preprocessing time $P'_t(m',\ell',\AlphabetSize) = \bigO(m' + P_t(m'))$,
  \item preprocessing space $P'_s(m',\ell',\AlphabetSize) = \bigO(m' + P_s(m'))$,
  \item query time $Q'(m',\ell',\AlphabetSize) = \bigO(Q(m'))$,
  \end{itemize}
  where the complexities $S(m')$, $P_t(m')$, $P_s(m')$, and $Q(m')$ refer to the structure from the claim (answering
  prefix rank for binary strings).

  \DSComponents
  The data structure answering pattern ranking queries
  consists of a single component: the data structure from \cref{th:pattern-ranking-to-prefix-rank-nonbinary}
  applied for text $\Text'$ and with the structure $D$ as the structure answering prefix rank queries.
  Note that we can use $D$, since it supports prefix rank queries for the combination of parameters required
  in \cref{th:pattern-ranking-to-prefix-rank-nonbinary}, i.e., for sequences over alphabet $\IntegerAlphabet$, with the
  length $\ell$ of all strings satisfying
  $\ell \leq (1 + \lfloor \log k \rfloor) / \lceil \log \AlphabetSize \rceil$ (where $k$ is the
  length of the input sequence).
  To bound the space usage of this component, note that by
  \cref{th:pattern-ranking-to-prefix-rank-nonbinary} and the above discussion, there exists
  $m = \Theta(\Textlen' / \log_{\AlphabetSize} \Textlen') = \Theta(\Textlen / \log \Textlen)$ (recall that $\AlphabetSize = 3$)
  such that the structure needs
  $\bigO(\Textlen' / \log_{\AlphabetSize} \Textlen' + S'(m,\ell,\AlphabetSize)) =
  \bigO(\Textlen' / \log_{\AlphabetSize} \Textlen' + m + S(m)) = \bigO(\Textlen / \log \Textlen + S(m))$ space.

  \DSQueries
  The pattern ranking queries are answered as follows. Let $\Pat \in \BinaryAlphabet^{*}$.
  Note that by definition of
  $\Text'$, it follows that $\RangeBegTwo{\Pat}{\Text} = \RangeBegTwo{\Pat}{\Text'}$.
  By \cref{th:pattern-ranking-to-prefix-rank-nonbinary} and the above
  discussion, computing $\RangeBegTwo{\Pat}{\Text'}$ takes
  $\bigO(\log \log \Textlen' + Q'(m,\ell,\AlphabetSize) + |\Pat| / \log_{\AlphabetSize} \Textlen') =
  \bigO(\log \log \Textlen + Q(m) + |\Pat| / \log \Textlen)$ time.

  \DSConstruction
  By \cref{th:pattern-ranking-to-prefix-rank-nonbinary}
  and the above discussion, construction of the above data structure
  answering pattern ranking queries on $\Text'$ (and hence also on $\Text$) takes
  $\bigO(\Textlen' / \log_{\AlphabetSize} \Textlen' + P'_t(m,\ell,\AlphabetSize)) =
  \bigO(\Textlen' / \log_{\AlphabetSize} \Textlen' + m + P_t(m)) =
  \bigO(\Textlen / \log \Textlen + P_t(m))$
  time and uses
  $\bigO(\Textlen' / \log_{\AlphabetSize} \Textlen' + P'_s(m,\ell,\AlphabetSize)) =
  \bigO(\Textlen' / \log_{\AlphabetSize} \Textlen' + m + P_s(m)) =
  \bigO(\Textlen / \log \Textlen + P_s(m))$ working space.
\end{proof}

\section{Equivalence of Lex-Range Reporting and Prefix Range Reporting Queries}\label{sec:prefix-range-reporting}

\subsection{Problem Definitions}\label{sec:prefix-range-reporting-and-lex-range-reporting-problem-def}
\vspace{-1.5ex}

\begin{framed}
  \noindent
  \probname{Indexing for Lex-Range Reporting Queries}
  \begin{bfdescription}
  \item[Input:]
    The packed representation of a string
    $\Text \in \BinaryAlphabet^{\Textlen}$.
  \item[Output:]
    A data structure that, given the packed representation of any
    $\Pat_1, \Pat_2 \in \BinaryAlphabet^{*}$, returns the set
    $\LexRange{\Pat_1}{\Pat_2}{\Text}$ (\cref{def:lex-range}).
  \end{bfdescription}
\end{framed}

\begin{framed}
  \noindent
  \probname{Indexing for Prefix Range Reporting Queries}
  \begin{bfdescription}
  \item[Input:]
    A sequence $W[1 \dd m]$ of $m \geq 1$ binary strings of length $\ell = 1 + \lfloor \log m \rfloor$, with all
    strings represented in the packed form.
  \item[Output:]
    A data structure that, given
    the packed representation of any $X \in \BinaryAlphabet^{\leq \ell}$ and any
    $b,e \in [0 \dd m]$, returns the set
    \[
      \{i \in (b \dd e] : X\text{ is a prefix of }W[i]\}.
    \]
  \end{bfdescription}
\end{framed}
\vspace{2ex}

\subsection{Reducing Prefix Range Reporting to Lex-Range Reporting Queries}\label{sec:prefix-range-reporting-to-lex-range-reporting}

\subsubsection{Problem Reduction}\label{sec:prefix-range-reporting-to-lex-range-reporting-problem-reduction}

\begin{lemma}\label{lm:seq-to-string}
  Let $W[1 \dd m]$ be a sequence of $m \geq 1$ strings of length $\ell \geq 1$ over alphabet $\BinaryAlphabet$.
  Let $s(i)$ (where $i \in [0 \dd m]$) be as in \cref{def:seq-to-string}.
  Denote
  $\Text = \SeqToString{\ell}{W}$ (\cref{def:seq-to-string}).
  Consider a string $X \in \BinaryAlphabet^{\leq \ell}$ and let $b, e \in [0 \dd m]$.
  Denote
  $\Pat_1 = \revstr{X} \cdot \four \cdot s(b)$,
  $\Pat_2 = \revstr{X} \cdot \four \cdot s(e)$,
  and $\beta = \ell + \lfloor \log m \rfloor + 2$.
  Let also
  $\mathcal{P} = \{i \in (b \dd e] : X\text{ is a prefix of }W[i]\}$
  and $\mathcal{Q} = \LexRange{\Pat_1}{\Pat_2}{\Text}$.
  Then, it holds:
  \begin{enumerate}
  \item $\mathcal{P} = \{\lceil j/\beta \rceil : j \in \mathcal{Q}\}$,
  \item $|\mathcal{P}| = |\mathcal{Q}|$.
  \end{enumerate}
\end{lemma}
\begin{proof}

  Denote $k = 1 + \lfloor \log m \rfloor$ and $\mathcal{Q}' = \{\lceil j/\beta \rceil : j \in \mathcal{Q}\}$.

  First, we prove the inclusion $\mathcal{P} \subseteq \mathcal{Q}'$. Let $i \in \mathcal{P}$. Then,
  $i \in (b \dd e]$ and $X$ is a prefix of $W[i]$.
  Denote $j = i \cdot \beta - (k + |X|)$. By \cref{def:seq-to-string}, the
  string $\Text[i \cdot \beta + 1 \dd |\Text|]$ is preceded in $\Text$ with $\revstr{W[i]} \cdot \four \cdot s(i-1)$.
  Since $X$ is a prefix of $W[i]$, we thus obtain that
  $\Text[j \dd |\Text|]$ has $\revstr{X} \cdot \four \cdot s(i-1)$ as a prefix. The assumption $b < i \leq e$, or equivalently,
  $b \leq i-1 < e$, implies that $s(b) \preceq s(i-1) \prec s(e)$. We thus obtain
  that $\Pat_1 \preceq \revstr{X} \cdot \four \cdot s(i-1) \prec \Pat_2$, and hence
  $\Pat_1 \preceq \Text[j \dd |\Text|] \prec \Pat_2$. We have thus proved that $j \in \LexRange{\Pat_1}{\Pat_2}{\Text}$.
  Since $\lceil j/\beta \rceil = i$, we thus obtain $i \in \mathcal{Q}'$.

  We now prove the second inclusion. Let $q \in \mathcal{Q}'$.
  Then, there exists $j \in \LexRange{\Pat_1}{\Pat_2}{\Text}$
  such that $\lceil j/\beta \rceil = q$. The assumption $j \in \LexRange{\Pat_1}{\Pat_2}{\Text}$ implies
  that $\Pat_1 \preceq \Text[j \dd |\Text|] \prec \Pat_2$. By definition of $\Pat_1$ and $\Pat_2$, this implies
  that $\Text[j \dd |\Text|]$ has the string $\revstr{X} \cdot \four$ as a prefix. Consequently, by
  \cref{lm:seq-to-string-occ}, there exists $i \in [1 \dd m]$ such that $X$ is a prefix of $W[i]$ and
  $j = i \cdot \beta - (k + |X|)$. By \cref{def:seq-to-string}, the suffix
  $\Text(i \cdot \beta - k \dd |\Text|]$ has the string $s(i-1)$ as a prefix.
  Putting together with the previous observation, we thus obtain that $\Text[j \dd |\Text|]$ has
  $\revstr{X} \cdot \four \cdot s(i-1)$ as a prefix. From the definition of $\Pat_1$ and $\Pat_2$, and the assumption
  $\Pat_1 \preceq \Text[j \dd |\Text|] \prec \Pat_2$, this implies that $s(b) \preceq s(i-1) \prec s(e)$, and hence
  $b \leq i-1 < e$, or equivalently, $b < i \leq e$. Combining this with $X$ being a prefix of $W[i]$ (see above),
  we obtain $i \in \mathcal{P}$. It remains to note that $j = i \cdot \beta - (k+|X|)$ implies that
  $q = \lceil j/\beta \rceil = i$. We have thus proved that $q \in \mathcal{P}$.

  It remains to show that $|\mathcal{P}| = |\mathcal{Q}|$. Observe that since above we proved that
  $|\mathcal{P}| = |\mathcal{Q}'|$, it suffices to show that $|\mathcal{Q}| = |\mathcal{Q}'|$. To show this, first note
  that, by definition of $\mathcal{Q}'$, it immediately follows that $|\mathcal{Q}'| \leq |\mathcal{Q}|$. To show
  $|\mathcal{Q}| \leq |\mathcal{Q}'|$, consider any $j_1, j_2 \in Q$ such that $j_1 \neq j_2$. Above we observed that
  this implies that there exist $i_1,i_2$ such that $j_1 = i_1 \cdot \beta - (k + |X|)$ and $j_2 = i_2 \cdot \beta - (k + |X|)$.
  Since we assumed that $j_1 \neq j_2$, we thus must have $i_1 \neq i_2$. It remains to note that for $x \in \{1, 2\}$, it holds
  $\lceil j_x /\beta \rceil = \lceil (i_x \cdot \beta - (k+|X|)) / \beta \rceil = i_x$.
  We have thus proved that the mapping $f : \mathcal{Q} \rightarrow \mathcal{Q}'$
  defined by $f(j) = \lceil j/\beta \rceil$ is injective. Thus, $|\mathcal{Q}| \leq |\mathcal{Q}'|$, and hence
  we obtain $|\mathcal{Q}| = |\mathcal{Q'}| = |\mathcal{P}|$, i.e., the last claim.
\end{proof}

\subsubsection{Alphabet Reduction for Lex-Range Reporting Queries}\label{sec:lex-range-reporting-alphabet-reduction}

\begin{lemma}\label{lm:lex-range-alphabet-reduction}
  Let $\AlphabetSize \geq 2$ and $\Text, \Pat_1, \Pat_2 \in \IntegerAlphabet^{*}$.
  Then, it holds (see \cref{def:lex-range})
  \[
    \LexRange{\Pat'_1}{\Pat'_2}{\Text'} =
      \{(j-1)\delta + 1 : j \in \LexRange{\Pat_1}{\Pat_2}{\Text}\},
  \]
  where
  $k = \lceil \log \AlphabetSize \rceil$,
  $\delta = 2k + 3$,
  $\Pat'_1 = \ebin{k}{\Pat_1}$,
  $\Pat'_2 = \ebin{k}{\Pat_2}$, and
  $\Text' = \ebin{k}{\Text}$ (\cref{def:ebin}).
\end{lemma}
\begin{proof}

  Denote $A = \{(j-1)\delta + 1 : j \in \LexRange{\Pat_1}{\Pat_2}{\Text}\}$.

  First, we prove the inclusion $A \subseteq \LexRange{\Pat'_1}{\Pat'_2}{\Text'}$.
  Let $j' \in A$. Then, there exists a position $j \in \LexRange{\Pat_1}{\Pat_2}{\Text}$
  such that $j' = (j-1)\delta + 1$. By \cref{def:lex-range}, we have
  $\Pat_1 \preceq \Text[j \dd |\Text|] \prec \Pat_2$.
  By \cref{def:ebin}, it holds $\Text'[j' \dd |\Text'|] =
  \bigodot_{t=j}^{|\Text|} \ebin{k}{\Text[t]} = \ebin{k}{\Text[j \dd |\Text|]}$.
  By \cref{lm:ebin-order}, $\Pat_1 \preceq \Text[j \dd |\Text|]$ holds if and only if
  $\ebin{k}{\Pat_1} \preceq \ebin{k}{\Text[j \dd |\Text|]}$. Thus, we have
  $\Pat'_1 \preceq \Text'[j' \dd |\Text'|]$. Analogously, $\Text'[j' \dd |\Text'|] \prec \Pat'_2$.
  Thus, we have $j' \in \LexRange{\Pat'_1}{\Pat'_2}{\Text'}$.

  We now prove the opposite inclusion. Let $j' \in \LexRange{\Pat'_1}{\Pat'_2}{\Text'}$.
  This implies that
  $\Pat'_1 \preceq \Text'[j' \dd |\Text'|] \prec \Pat'_2$. Note that both $\Pat'_1$ and $\Pat'_2$ have the string
  $\one^{k+1}$ as a prefix. Thus, $\one^{k+1}$ is also a prefix of $\Text'[j' \dd |\Text'|]$. By
  $\OccTwo{\one^{k+1}}{\Text'} = \{(t-1) \delta + 1 : t \in [1 \dd |\Text|]\}$, it follows that
  there exists $j \in [1 \dd |\Text|]$ such that $j' = (j-1) \delta + 1$.
  By definition of $\Text'$, it follows
  that $\Text'[j' \dd |\Text'|] = \bigodot_{t=j}^{|\Text|} \ebin{k}{\Text[t]} = \ebin{k}{\Text[j \dd |\Text|]}$.
  Observe that by \cref{lm:ebin-order}, $\Pat_1 \preceq \Text[j \dd |\Text|]$ holds if and only if
  $\Pat'_1 \preceq \ebin{k}{\Text[j \dd |\Text|]}$. By $\ebin{k}{\Text[j \dd |\Text|]} = \Text'[j' \dd |\Text'|]$
  and the assumption $\Pat'_1 \preceq \Text'[j' \dd |\Text'|]$, we thus obtain $\Pat_1 \preceq \Text[j \dd |\Text|]$.
  Analogously, it holds $\Text[j \dd |\Text|] \prec \Pat_2$. Thus, $j \in \LexRange{\Pat_1}{\Pat_2}{\Text}$, and hence
  we obtain $j' \in A$.
\end{proof}

\begin{proposition}\label{pr:lex-range-reporting-alphabet-reduction}
  Let $\Text \in \IntegerAlphabet^{\Textlen}$ be a nonempty string, where
  $2 \leq \AlphabetSize < \Textlen^{\bigO(1)}$. Given the packed representation of
  $\Text$, we can in $\bigO(\Textlen / \log_{\AlphabetSize} \Textlen)$ time
  construct the packed representation of a
  text $\Text_{\rm bin} \in \BinaryAlphabet^{*}$ satisfying
  $|\Text_{\rm bin}| = \Theta(\Textlen \log \AlphabetSize)$, integers
  $\alpha$, $\gamma$, and $\delta$, and a data structure that,
  given the packed representation of any
  $\Pat_1, \Pat_2 \in \IntegerAlphabet^{\leq m}$, in
  $\bigO(1 + m / \log_{\AlphabetSize} \Textlen)$ time returns the
  packed representation of $\Pat'_1, \Pat'_2 \in \BinaryAlphabet^{*}$
  such that
  $|\Pat'_i| = |\Pat_i| \cdot (2\lceil \log \AlphabetSize \rceil + 3)$ (where $i \in \{1,2\}$)
  and
  \[
    \LexRange{\Pat_1}{\Pat_2}{\Text} =
    \left\{\frac{j-\alpha}{\gamma} + \delta : j \in \LexRange{\Pat'_1}{\Pat'_2}{\Text_{\rm bin}}\right\}.
  \]
\end{proposition}
\begin{proof}

  Let $k = \lceil \log \AlphabetSize \rceil$.
  To compute $\alpha$, $\gamma$, $\delta$, and $\Text_{\rm bin}$,
  we proceed as follows:
  \begin{enumerate}
  \item In $\bigO(\Textlen / \log_{\AlphabetSize} \Textlen)$ time we construct the data structure from
    \cref{pr:ebin} for $u = \Textlen$.
    Given the packed representation of any $S \in \IntegerAlphabet^{*}$, we can then
    compute the packed representation of $\ebin{k}{S}$ (\cref{def:ebin}) in
    $\bigO(1 + |S| / \log_{\AlphabetSize} \Textlen)$ time.
  \item Compute the packed representation of
    $\Text_{\rm bin} = \ebin{k}{\Text}$ in $\bigO(\Textlen / \log_{\AlphabetSize} \Textlen)$ time.
  \item In $\bigO(1)$ time set $\alpha = 1$, $\gamma = 2k + 3$, and $\delta = 1$.
  \end{enumerate}
  In total, the construction of the packed representation of $\Text_{\rm bin}$ and integers $\alpha$, $\gamma$, and $\delta$
  takes $\bigO(\Textlen / \log_{\AlphabetSize} \Textlen)$. By \cref{def:ebin},
  $|\Text_{\rm bin}| = \Textlen \cdot (2k + 3) = \Theta(\Textlen \log \AlphabetSize)$.

  Next, we address the construction of the data structure from the claim.
  It consists of a single component: the data structure from \cref{pr:ebin} for $u = \Textlen$.
  It uses $\bigO(\Textlen / \log_{\AlphabetSize} \Textlen)$ space.

  The queries are implemented as follows. Given the packed representation of
  $\Pat_1,\Pat_2 \in \IntegerAlphabet^{\leq m}$, we compute the packed representation
  of $\Pat'_1$ and $\Pat'_2$ defined as $\Pat'_i = \ebin{k}{\Pat_i}$.
  Using the structure from \cref{pr:ebin}, this takes $\bigO(1 + m / \log_{\AlphabetSize} \Textlen)$ time.
  By \cref{def:ebin}, it holds $|\Pat'_i| = |\Pat_i| \cdot (2k + 3)$ for $i \in \{1,2\}$,
  and by \cref{lm:lex-range-alphabet-reduction},
  $\LexRange{\Pat_1}{\Pat_2}{\Text} = \{(j-\alpha)/\gamma+\delta : j \in \LexRange{\Pat'_1}{\Pat'_2}{\Text_{\rm bin}}\}$.

  By \cref{pr:ebin}, construction of the data structure takes
  $\bigO(\Textlen / \log_{\AlphabetSize} \Textlen)$ time.
\end{proof}

\subsubsection{Summary}\label{sec:prefix-range-reporting-to-lex-range-reporting-summary}

\begin{theorem}\label{th:prefix-range-reporting-to-lex-range-reporting}
  Consider a data structure answering lex-range reporting queries
  (see \cref{sec:prefix-range-reporting-and-lex-range-reporting-problem-def})
  that, for any
  text $\Text \in \BinaryAlphabet^{\Textlen}$, achieves the following complexities
  (where in the preprocessing we assume that we are given as input the packed
  representation of $\Text$, and at query time we are given a packed
  representation of $\Pat_1,\Pat_2 \in \BinaryAlphabet^{\leq k}$, and we
  denote $p = |\LexRange{\Pat_1}{\Pat_2}{\Text}|$):
  \begin{itemize}
  \item space usage $S(\Textlen)$,
  \item preprocessing time $P_t(\Textlen)$,
  \item preprocessing space $P_s(\Textlen)$,
  \item query time $Q_{\rm base}(\Textlen, k) + p \cdot Q_{\rm occ}(\Textlen, k)$.
  \end{itemize}
  For every sequence $W[1 \dd m]$ of $m \geq 1$ binary strings of length
  $\ell = 1 + \lfloor \log m \rfloor$, there exists
  $\Textlen = \bigO(m \log m)$ and $k = \bigO(\log m)$ such that, given the sequence $W$, with
  all strings represented in the packed form, we can in
  $\bigO(m + P_t(\Textlen))$ time and using $\bigO(m + P_s(\Textlen))$
  working space construct a data structure of size $\bigO(m + S(\Textlen))$
  that, given the packed representation of any $X \in \BinaryAlphabet^{\leq \ell}$
  and any pair of integers $b,e \in [0 \dd m]$, computes the
  set $\mathcal{P} = \{i \in (b \dd e] : X\text{ is a prefix of }W[i]\}$ in
  $\bigO(Q_{\rm base}(\Textlen, k) + p \cdot Q_{\rm occ}(\Textlen, k))$ time,
  where $p = |\mathcal{P}|$.
\end{theorem}
\begin{proof}

  We use the following definitions.
  Let $\Text_{\rm aux} = \SeqToString{\ell}{W}$ (\cref{def:seq-to-string}).
  Denote $\Textlen_{\rm aux} =
  |\Text_{\rm aux}| = m \cdot (2\ell+1) = \Theta(m \log m)$. Let
  $\Text_{\rm bin}$ and $\alpha_{\rm bin}$, $\gamma_{\rm bin}$, and $\delta_{\rm bin}$
  denote the text and the three integers obtained
  by applying \cref{pr:lex-range-reporting-alphabet-reduction} to text
  $\Text_{\rm aux}$. Denote $\Textlen = |\Text_{\rm bin}|$.
  Since $\Text_{\rm aux}$ is over alphabet $\{\zero, \dots, \four\}$,
  by \cref{pr:lex-range-reporting-alphabet-reduction} we have
  $\Textlen = \Theta(|\Text_{\rm aux}|) = \Theta(m \log m)$.
  Lastly, denote $k = 18 \lfloor \log m \rfloor + 27 = \bigO(\log m)$.

  \DSComponents
  The data structure consists of the following components:
  \begin{enumerate}
  \item The structure from \cref{pr:lex-range-reporting-alphabet-reduction} applied
    to text $\Text_{\rm aux}$. Since $\Text_{\rm aux}$ is over alphabet
    $\{\zero, \dots, \four\}$, the structure needs
    $\bigO(|\Text_{\rm aux}| / \log |\Text_{\rm aux}|) =
    \bigO(m)$ space.
  \item The data structure from the claim (i.e., answering lex-range reporting
    queries) for the text $\Text_{\rm bin}$. The structure needs
    $\bigO(S(\Textlen))$ space.
  \item A lookup table $L_{\rm rev}$ defined as in
    \cref{th:prefix-rank-to-pattern-ranking}. It needs
    $\bigO(\sqrt{\Textlen}) = \bigO(m)$ space.
  \item A lookup table $L_{\rm map}$ defined as in
    \cref{th:prefix-rank-to-pattern-ranking}. It also needs
    $\bigO(\sqrt{\Textlen}) = \bigO(m)$ space.
  \item The integers $\alpha_{\rm bin}$, $\gamma_{\rm bin}$, and $\delta_{\rm bin}$.
  \end{enumerate}
  In total, the structure needs $\bigO(m + S(\Textlen))$ space.

  \DSQueries
  The queries are answered as follows. Assume that we are given
  the packed representation of some $X \in \BinaryAlphabet^{\leq \ell}$ and
  two integers $b,e \in [0 \dd m]$.
  The query algorithm proceeds as follows:
  \begin{enumerate}
  \item Using $L_{\rm rev}$ and $L_{\rm map}$, in $\bigO(1)$ time we compute the packed
    representation of strings $\Pat_1 = \revstr{X} \cdot \four \cdot s(b)$
    and $\Pat_2 = \revstr{X} \cdot \four \cdot s(e)$
    (over alphabet $\{\zero, \dots, \four\}$).
    Note that $|\Pat_1| = |\Pat_2| = |X| + 1 + \ell \leq 2\ell + 1 \leq 2\lfloor \log m \rfloor + 3$.
  \item Using \cref{pr:lex-range-reporting-alphabet-reduction}, in
    $\bigO(1 + |\Pat_1| / \log \Textlen_{\rm aux} + |\Pat_2| / \log \Textlen_{\rm aux}) =
    \bigO(1 + \ell / \log m) = \bigO(1)$ time, we compute
    the packed representation of strings $\Pat'_1,\Pat'_2 \in \BinaryAlphabet^{*}$
    of length $|\Pat'_1| = |\Pat'_2| = 9 \cdot |\Pat_1| \leq 18 \lfloor \log m \rfloor + 27 \leq k$
    such that
    \[
      \LexRange{\Pat_1}{\Pat_2}{\Text_{\rm aux}} = 
      \{(j-\alpha_{\rm bin})/\gamma_{\rm bin} + \delta_{\rm bin} : j \in \LexRange{\Pat'_1}{\Pat'_2}{\Text_{\rm bin}}\}.
    \]
  \item Using the structure from the claim, we compute the set
    $\mathcal{Q}_{\rm bin} = \LexRange{\Pat'_1}{\Pat'_2}{\Text_{\rm bin}}$.
    Letting $p = |\mathcal{Q}_{\rm bin}|$, by $|\Pat'_1|, |\Pat'_2| \leq k$, this takes
    $\bigO(Q_{\rm base}(\Textlen, k) + p \cdot Q_{\rm occ}(\Textlen, k))$ time.
  \item In $\bigO(p)$ time we compute the set
    $\mathcal{Q}_{\rm aux} = \{(j-\alpha_{\rm bin})/\gamma_{\rm bin} + \delta_{\rm bin} :
    j \in \mathcal{Q}_{\rm bin}\}$. Note that since each element of $\mathcal{Q}_{\rm bin}$ yields a different element
    of $\mathcal{Q}_{\rm aux}$, we have $p = |\mathcal{Q}_{\rm bin}| = |\mathcal{Q}_{\rm aux}|$.
    By the above equation, it holds $\mathcal{Q}_{\rm aux} = \LexRange{\Pat_1}{\Pat_2}{\Text_{\rm aux}}$.
  \item In $\bigO(p)$ time we compute the set
    $\mathcal{P} = \{\lceil j / \beta \rceil : j \in \mathcal{Q}_{\rm aux}\}$, where
    $\beta = \ell + \lfloor \log m \rfloor + 2 = 2\ell + 1$.
    By \cref{lm:seq-to-string}, it holds
    $\mathcal{P} = \{i \in (b \dd e] : X\text{ is a prefix of }W[i]\}$. Moreover, it holds
    $|\mathcal{P}| = |\mathcal{Q}_{\rm aux}|$, which implies $|\mathcal{P}| = p$.
    We thus return $\mathcal{P}$ as the answer.
  \end{enumerate}
  In total, the query takes $\bigO(Q_{\rm base}(\Textlen, k) + p \cdot Q_{\rm occ}(\Textlen, k))$ time.

  \DSConstruction
  The components of the data structure are constructed as follows:
  \begin{enumerate}
  \item The first component of the structure is computed as follows:
    \begin{enumerate}
    \item In $\bigO(m)$ time we apply \cref{pr:seq-to-string} to
      the sequence $W$ to compute the packed representation of the string
      $\Text_{\rm aux} = \SeqToString{\ell}{W}$ (\cref{def:seq-to-string}).
    \item We apply \cref{pr:lex-range-reporting-alphabet-reduction}
      to $\Text_{\rm aux}$. Since the string $\Text_{\rm aux}$ is over alphabet
      $\{\zero, \dots, \four\}$, this takes
      $\bigO(|\Text_{\rm aux}| / \log |\Text_{\rm aux}|) = \bigO(m)$
      time. Note that \cref{pr:lex-range-reporting-alphabet-reduction}, in addition
      to the data structure, also returns integers $\alpha_{\rm bin}$, $\gamma_{\rm bin}$, and
      $\delta_{\rm bin}$, and the packed representation of the string $\Text_{\rm bin}$ (defined above).
    \end{enumerate}
  \item We apply the preprocessing from the claim to the string $\Text_{\rm bin}$.
    This takes $\bigO(P_t(|\Text_{\rm bin}|)) = \bigO(P_t(\Textlen))$ time
    and uses $\bigO(P_s(|\Text_{\rm bin}|)) = \bigO(P_s(\Textlen))$ working space.
  \item The lookup table $L_{\rm rev}$ is computed in $\bigO(m)$ time similarly
    as in the proof of \cref{pr:seq-to-string}.
  \item Next, we compute the lookup table $L_{\rm map}$. Similarly as above,
    we proceed as in \cref{pr:seq-to-string}, and spend $\bigO(m)$ time.
  \item Lastly, we store integers $\alpha_{\rm bin}$, $\gamma_{\rm bin}$,
    and $\delta_{\rm bin}$, which were already computed above.
  \end{enumerate}
  In total, the construction takes
  $\bigO(m + P_t(\Textlen))$ time and uses
  $\bigO(m + P_s(\Textlen))$ working space.
\end{proof}

\subsection{Reducing Lex-Range Reporting to Prefix
  Range Reporting Queries}\label{sec:lex-range-reporting-to-prefix-range-reporting}

\subsubsection{Problem Reduction}

\paragraph{Preliminaries}\label{sec:lex-range-reporting-to-pref-range-reporting-prelim}

\begin{definition}[$\tau$-periodic and $\tau$-nonperiodic patterns]\label{def:periodic-pattern}
  Let $\Pat \in \Sigma^{m}$ and $\tau \geq 1$. We say that $\Pat$ is
  \emph{$\tau$-periodic} if it holds $m \geq 3\tau - 1$ and
  $\per{\Pat[1 \dd 3\tau - 1]} \leq \tfrac{1}{3}\tau$. Otherwise, it
  is called \emph{$\tau$-nonperiodic}.
\end{definition}

\begin{definition}[String-to-integer mapping]\label{def:int}
  Let $m > 0$, $\AlphabetSize > 1$. For every $X \in \IntegerAlphabet^{\leq m}$,
  by $\Int{m}{\AlphabetSize}{X}$ we denote an integer
  constructed by appending $2m - 2|X|$ zeros to $X$ and $|X|$ $c$s (where $c = \AlphabetSize - 1$)
  to $X$, and then interpreting the resulting string as a base-$\AlphabetSize$
  representation of a number in $[0 \dd \AlphabetSize^{2m})$.
\end{definition}

\begin{lemma}[\cite{sublinearlz}]\label{lm:int}
  Let $m > 0$ and $\AlphabetSize > 1$.
  For all $X, X' \in \IntegerAlphabet^{\leq m}$,
  $X \prec X'$ implies $\Int{m}{\AlphabetSize}{X} <
  \Int{m}{\AlphabetSize}{X'}$. In particular, $X \neq X'$ implies
  $\Int{m}{\AlphabetSize}{X} \neq \Int{m}{\AlphabetSize}{X'}$.
\end{lemma}

\begin{proposition}\label{pr:per}
  Let $\AlphabetSize \geq 2$ and $\tau \geq 1$. Given the values of $\AlphabetSize$ and $\tau$,
  we can in $\bigO(\AlphabetSize^{3\tau} \cdot \tau^2)$ time construct a data
  structure that, given the packed representation of any $\Pat \in \IntegerAlphabet^{3\tau-1}$, in
  $\bigO(1)$ time determines if $\Pat$ is $\tau$-periodic (\cref{def:periodic-pattern}).
\end{proposition}
\begin{proof}
  The data structure consists of a lookup table $L_{\rm per}$.
  Its definition, space requirement, usage, and construction is as
  described in Section~5.1.1 of~\cite{breaking}, except to reduce its size,
  we index the lookup table using simply the packed representation of
  $\Pat$ (and hence reduce its size to $\bigO(\AlphabetSize^{3\tau})$).
\end{proof}

\paragraph{The Short Patterns}

\subparagraph{Preliminaries}

\begin{definition}[Indicator vector of a set]\label{def:indicator-vector}
  For any $\ell \in \Zp$ and any set $A \subseteq [1 \dd \ell]$, by
  $\IndicatorVector{\ell}{A}$ we denote a bitvector $V[1 \dd \ell]$ defined so
  that for $i \in [1 \dd \ell]$, $V[i] = \one$ holds if and only if
  $i \in A$.
\end{definition}

\subparagraph{Basic Navigational Primitives}

\begin{proposition}\label{pr:nav-index-short}
  Let $\Text \in \IntegerAlphabet^{\Textlen}$ be such that $2 \leq \AlphabetSize < \Textlen^{1/13}$ and $\Text[\Textlen]$ does not
  occur in $\Text[1 \dd \Textlen)$. Let $\tau = \lfloor \mu\log_{\AlphabetSize} \Textlen \rfloor$, where $\mu$ is a positive
  constant smaller than $\tfrac{1}{12}$ such that $\tau \geq 1$.
  Given the packed representation of $\Text$, we can in $\bigO(\Textlen / \log_{\AlphabetSize} \Textlen)$ time
  construct a data structure that, given the packed representation of any $X \in \IntegerAlphabet^{\leq 3\tau-1}$,
  returns the pair $(\RangeBegTwo{X}{\Text},\allowbreak \RangeEndTwo{X}{\Text})$ (see \cref{def:occ})
  in $\bigO(1)$ time.
\end{proposition}
\begin{proof}
  The data structure consists of the lookup table $L_{\rm range}$.
  Its definition, space requirement, usage, and construction is as
  described in Section~5.1 of~\cite{breaking}.
\end{proof}

\subparagraph{Algorithms}

\begin{proposition}\label{pr:short-occ-reporting}
  Let $\Text \in \IntegerAlphabet^{\Textlen}$ be such that $2 \leq \AlphabetSize < \Textlen^{1/13}$ and $\Text[\Textlen]$ does not
  occur in $\Text[1 \dd \Textlen)$. Let $\tau = \lfloor \mu\log_{\AlphabetSize} \Textlen \rfloor$, where $\mu$ is a positive
  constant smaller than $\tfrac{1}{12}$ such that $\tau \geq 1$.
  Given the packed representation of $\Text$, we can in $\bigO(\Textlen / \log_{\AlphabetSize} \Textlen)$ time
  construct a data structure that given the packed representation of any $X \in \IntegerAlphabet^{\leq 3\tau-1}$,
  returns $\mathcal{J} := \OccTwo{X}{\Text}$ (\cref{def:occ}) in $\bigO(1 + |\mathcal{J}|)$ time.
\end{proposition}
\begin{proof}

  We use the following definitions. Let $\ell = 3\tau - 1$.
  For any $i \in [1 \dd \Textlen]$, denote $B(i) := \Text[i \dd \min(\Textlen + 1, i + 2\ell - 1))$.
  Denote $m = \lceil \tfrac{\Textlen}{\ell} \rceil$, and let $(c_j)_{j \in [1 \dd m]}$ be a sequence
  containing a permutation of the set $\{1 + (i-1)\ell : i \in [1 \dd m]\}$ such that, for every $j, j' \in [1 \dd m]$,
  $j < j'$ implies that it holds $B(c_j) \prec B(c_{j'})$, or $B(c_j) = B(c_{j'})$ and $c_j < c_{j'}$.
  Let $A_{\rm blk}[1 \dd m]$ be an array defined by $A_{\rm blk}[i] = c_i$.
  Let
  \[
    \mathcal{F} := \{1\} \cup \{i \in [2 \dd m] : B(c_i) \neq B(c_{i-1})\}.
  \]
  For every $X \in \IntegerAlphabet^{\leq 3\tau - 1}$, we define (see \cref{def:indicator-vector})
  \begin{align*}
    {\rm blocks}(X)
      &:= \{i \in \mathcal{F} : \OccTwo{X}{B(c_i)} \cap [1 \dd \ell] \neq \emptyset\},\\
    {\rm blocks}'(X)
      &:= \{(i,\IndicatorVector{\ell}{\OccTwo{X}{B(c_i)} \cap [1 \dd \ell]} : i \in {\rm blocks}(X)\}.
  \end{align*}
  Let $L_{\rm blocks}$ be a mapping such that, for every $X \in \IntegerAlphabet^{\leq 3\tau - 1}$, $L_{\rm blocks}$
  maps $X$ to the set ${\rm blocks}'(X)$ defined above.
  Finally, let $L_{\rm bit}$ be a mapping such that, for every $x \in [1 \dd 2^{\ell})$, $L_{\rm bit}$ maps $x$
  to the position of the least significant $\one$-bit in the binary encoding of $x$.

  \DSComponents
  The data structure consists of the following components:
  \begin{enumerate}
  \item The array $A_{\rm blk}[1 \dd m]$ in plain form using $\bigO(m) = \bigO(\tfrac{\Textlen}{\ell}) =
    \bigO(\Textlen / \log_{\AlphabetSize} \Textlen)$ space.
  \item The lookup table $L_{\rm blocks}$, where for every $X \in \IntegerAlphabet^{\leq 3\tau-1}$, $L_{\rm blocks}[X]$
    stores the pointer to the head of a linked list containing all elements of the set ${\rm blocks}'(X)$. Each element
    $(i,V) \in {\rm blocks}'(X)$ is stored as a pair of integers, with the bitvector $V$ encoded as a number
    in the range $[0 \dd 2^{\ell})$. When accessing $L_{\rm blocks}[X]$, the string $X \in \IntegerAlphabet^{\leq 3\tau-1}$
    is represented as
    a number $\Int{3\tau}{\AlphabetSize}{X} \in [0 \dd \AlphabetSize^{6\tau})$ (\cref{def:int}).
    Thus, the range of indexes for $L_{\rm blocks}$ is bounded by
    $\bigO(\AlphabetSize^{6\tau}) = \bigO(\sqrt{\Textlen}) = \bigO(\Textlen / \log_{\AlphabetSize} \Textlen)$.
    To bound the total length of all lists, i.e.,
    $\sum_{X \in \IntegerAlphabet^{\leq 3\tau-1}} |{\rm blocks}(X)|$, we proceed as follows:
    \begin{itemize}
    \item First, observe that by definition of the sequence $(c_j)_{j \in [1 \dd m]}$ and the set $\mathcal{F}$,
      for every $i,i' \in \mathcal{F}$, $i \neq i'$ implies that
      $B(c_i) \neq B(c_{i'})$. Since for every $j \in [1 \dd \Textlen]$, $B(j) \in \IntegerAlphabet^{\leq 2\ell-1}$,
      it follows that $|\mathcal{F}| = \bigO(\AlphabetSize^{2\ell}) = \bigO(\AlphabetSize^{6\tau}) =
      \bigO(\sqrt{\Textlen})$.
    \item Second, note that $|{\rm blocks}(\emptystring)| = |\mathcal{F}| = \bigO(\sqrt{\Textlen})$.
    \item Finally, observe that if for some $i \in \mathcal{F}$ and $X \in \IntegerAlphabet^{\leq \ell} \setminus \{\emptystring\}$,
      it holds $i \in {\rm blocks}(X)$, then for some
      $\delta \in [1 \dd \ell]$ satisfying $\delta + |X| \leq |B(c_i)| + 1$, it holds $B(c_i)[\delta \dd \delta + |X|) = X$.
      This implies that $i$ can occur in ${\rm blocks}(X)$ for at most $\ell^2$ distinct nonempty strings $X$.
      Consequently,
      $\sum_{X \in \IntegerAlphabet^{\leq 3\tau-1} \setminus \{\emptystring\}} |{\rm blocks}(X)| \leq |\mathcal{F}| \cdot \ell^2$.
    \end{itemize}
    Putting everything together, we thus obtain that
    \[
      \sum_{X \in \IntegerAlphabet^{\leq 3\tau-1}} |{\rm blocks}(X)| =
      \bigO(|\mathcal{F}| \cdot \ell^2) =
      \bigO(\sqrt{\Textlen} \log^2 \Textlen) =
      \bigO(\Textlen / \log_{\AlphabetSize} \Textlen).
    \]
  \item The lookup table $L_{\rm bit}$ using $\bigO(2^{\ell}) = \bigO(\AlphabetSize^{\ell}) = \bigO(\sqrt{\Textlen})
    = \bigO(\Textlen / \log_{\AlphabetSize} \Textlen)$ space.
  \end{enumerate}
  In total, the structure needs $\bigO(\Textlen / \log_{\AlphabetSize} \Textlen)$ space.

  \DSQueries
  Given the packed representation of any $X \in \IntegerAlphabet^{\leq 3\tau-1}$, we
  compute $\OccTwo{X}{\Text}$ as follows:
  \begin{enumerate}
  \item Initialize the output set $\mathcal{R} := \emptyset$.
  \item Using the lookup table $L_{\rm blocks}$, in $\bigO(1)$ time obtain
    the pointer to head of a linked list containing all the elements
    of the set ${\rm blocks}'(X)$.
    If the list is empty, then observe that, by definition of
    the sequence $(c_i)_{i \in [1 \dd m]}$ and the
    sets ${\rm blocks}(X)$ and ${\rm blocks}'(X)$,
    we then have $\OccTwo{X}{\Text} = \emptyset$.
    Thus, in this case we skip the rest of this step.
    Let us thus assume that ${\rm blocks}'(X) \neq \emptyset$.
    In $\bigO(1 + |{\rm blocks}'(X)|)$ time, we then
    iterate over all elements of ${\rm blocks}'(X)$. For every
    $(i,V) \in {\rm blocks}'(X)$, we proceed as follows:
    \begin{enumerate}
    \item Compute the set containing the positions of $\one$-bits in $V$,
      i.e., the set $\Delta_i := \OccTwo{X}{B(c_i)} \cap [1 \dd \ell]$.
      Note that $\Delta_i \neq \emptyset$.
      With the help of $L_{\rm bit}$, computing $\Delta_i$ takes
      $\bigO(|\Delta_i|)$ time.
    \item By scanning the array $A_{\rm blk}$ from index $i$, compute
      \[
        i' = \min\{j \in (i \dd m] : B(c_j) \neq B(c_i)\} \cup \{m+1\}.
      \]
      Denote $\mathcal{J}_i = [i \dd i')$ and note that $\mathcal{J}_i \neq \emptyset$.
      Denote also $\mathcal{C}_i = \{c_j : j \in [i \dd i')\}$.
      The computation of $i'$ takes $\bigO(|\mathcal{J}_i|) = \bigO(|\mathcal{C}_i|)$ time,
      since given any $p \in [1 \dd \Textlen]$, and the packed representation of
      $\Text$, we can in $\bigO(1)$ time obtain the packed representation of $B(p)$.
    \item In $\bigO(|\mathcal{C}_i| \cdot |\Delta_i|)$ time add the set
      \[
        \mathcal{P}_i := \{A_{\rm blk}[t] + \delta - 1 : (t,\delta) \in \mathcal{J}_i \times \Delta_i\}
        = \{c + \delta - 1 : (c,\delta) \in \mathcal{C}_i \times \Delta_i\}
      \]
      to $\mathcal{R}$.
      Note that since $\Delta_i \subseteq [1 \dd \ell]$, and
      every position $c \in \mathcal{C}_i$ is of the form $c = 1 + t \cdot \ell$, where $t \in \Zn$,
      it holds $|\mathcal{P}_i| = |\mathcal{C}_i| \cdot |\Delta_i|$,
    \end{enumerate}
    In total, we spend $\bigO(|\Delta_i| + |\mathcal{C}_i| + |\mathcal{C}_i| \cdot |\Delta_i|) =
    \bigO(|\mathcal{C}_i| \cdot |\Delta_i|) = \bigO(|\mathcal{P}_i|)$ time. Over all $(i,V) \in {\rm blocks}'(X)$, we spend
    $\bigO(\sum_{i \in {\rm blocks}(X)}|\mathcal{P}_i|)$ time (recall that $|{\rm blocks}(X)| = |{\rm blocks}'(X)|$, and for every
    $(i,V) \in {\rm blocks}'(X)$, it holds $i \in {\rm blocks}(X)$). By definition of the sequence
    $(c_j)_{j \in [1 \dd m]}$ and sets ${\rm blocks}(X)$ and ${\rm blocks}'(X)$, the set $\OccTwo{X}{\Text}$ is a disjoint
    union
    \[
      \OccTwo{X}{\Text} = \bigcup_{i \in {\rm blocks}(X)} \mathcal{P}_i.
    \]
    Thus, at the end of the algorithm, we have $\mathcal{R} = \OccTwo{X}{\Text}$, and the total
    time of this step is $\bigO(\sum_{i \in {\rm blocks}(X)} |\mathcal{P}_i|) = \bigO(|\OccTwo{X}{\Text}|)$.
  \item We return the set $\mathcal{R} = \OccTwo{X}{\Text}$ as the answer in
    $\bigO(1 + |\OccTwo{X}{\Text}|)$ time.
  \end{enumerate}
  In total, the query takes $\bigO(1 + |\OccTwo{X}{\Text}|)$ time.

  \DSConstruction
  The components of the structure are constructed as follows:
  \begin{enumerate}
  \item The construction of the array $A_{\rm blk}[1 \dd m]$ proceeds as follows:
    \begin{enumerate}
    \item In $\bigO(\tfrac{\Textlen}{\ell}) = \bigO(\Textlen / \log_{\AlphabetSize} \Textlen)$ time
      construct the array $A_{\rm sort}[1 \dd m]$ such that, for every
      $i \in [1 \dd m]$, $A_{\rm sort}[i] = (\Int{2\ell}{\AlphabetSize}{B(1+(i-1)\ell)},1+(i-1)\ell)$ (see \cref{def:int}).
    \item Sort the array $A_{\rm sort}[1 \dd m]$ lexicographically. By \cref{def:int}, the first coordinate
      is in the range $[0 \dd \AlphabetSize^{4\ell}) \subseteq [0 \dd \AlphabetSize^{12\tau})
      \subseteq [0 \dd \Textlen^{12\mu}) \subseteq [0 \dd \Textlen)$, and the second coordinate is in the range $[1 \dd \Textlen]$.
      We use a 4-round radix sort to achieve $\bigO(m + \sqrt{\Textlen}) = \bigO(\Textlen / \log_{\AlphabetSize} \Textlen)$
      sorting time. By \cref{lm:int}, the resulting array contains the sequence $(c_j)_{j \in [1 \dd m]}$ on the second coordinate.
      With one more scan, in $\bigO(m) = \bigO(\Textlen / \log_{\AlphabetSize} \Textlen)$ time we construct
      $A_{\rm blk}[1 \dd m]$.
    \end{enumerate}
    In total, constructing $A_{\rm blk}[1 \dd m]$ takes $\bigO(\Textlen / \log_{\AlphabetSize} \Textlen)$ time.
  \item The construction of the second component proceeds as follows:
    \begin{enumerate}
    \item First, we compute an array $A_{\mathcal{F}}[1 \dd m]$ such that for every $i \in [1 \dd m]$, $A_{\mathcal{F}}[i] = 1$
      holds if and only if $i \in \mathcal{F}$. With the help of the array $A_{\rm blk}[1 \dd m]$ (computed above), and the
      packed representation of $\Text$, computing $A_{\mathcal{F}}[1 \dd m]$ takes
      $\bigO(m)$ time.
    \item Next, we compute a linked list containing the elements of the set
      ${\rm blocks}'(\emptystring)$. To this end, it suffices to iterate over $A_{\mathcal{F}}$
      and add the pair $(i,\one^{\ell})$ (where $\one^{\ell}$ is represented as integer $2^{\ell}-1$) to ${\rm blocks}'(\emptystring)$
      for every $i \in [1 \dd m]$ such that $A_{\mathcal{F}}[i] = 1$. Lastly, we store the pointer to the head of the list
      in $L_{\rm blocks}[\emptystring]$. This takes $\bigO(m)$ time.
    \item To compute lists containing sets ${\rm blocks}'(X)$ for all $X \in \IntegerAlphabet^{\leq 3\tau-1} \setminus \{\emptystring\}$,
      we first initialize $L_{\rm blocks}[X]$ for all such $X$ to empty lists. For every $i \in [1 \dd m]$ such that
      $A_{\mathcal{F}}[i] = 1$, we then perform the following steps:
      \begin{enumerate}
      \item Denote $B = B(c_i) = B(A_{\rm blk}[i])$. We compute the
        set
        \[
          \mathcal{S}_i := \{X \in \IntegerAlphabet^{\leq \ell} \setminus \{\emptystring\} :
          \OccTwo{X}{B} \cap [1 \dd \ell] \neq \emptyset\},
        \]
        with each of the strings $X$ in the set
        represented as $\Int{\ell}{\AlphabetSize}{X}$ (\cref{def:int}). To this end,
        first create a multiset containing all substrings $B[s \dd s + t)$,
        where $1 \leq s < s + t \leq |B| + 1$, $t \leq \ell$, and $s \leq \ell$,
        and then remove the duplicates.
        This is easily implemented in $\bigO(\ell^3)$ time.
        Note that $|\mathcal{S}_i| \leq \ell^2$.
      \item For every $X \in \mathcal{S}_i$, compute the bitvector
        $V = \IndicatorVector{\ell}{\OccTwo{X}{B} \cap [1 \dd \ell]}$ (\cref{def:indicator-vector}), with
        $V$ represented as an integer in $[0 \dd 2^{\ell})$. Then, add the pair $(X,V)$ to the list
        whose head is currently stored in $L_{\rm blocks}[X]$.
        Using a naive algorithm, the computation of $V$ takes $\bigO(\ell^2)$.
        Over all $X \in \mathcal{S}_i$, we thus spend $\bigO(\ell^4)$ time.
      \end{enumerate}
      In total, the computation of ${\rm blocks}'(X)$ for all
      $X \in \IntegerAlphabet^{\leq 3\tau-1} \setminus \{\emptystring\}$
      takes $\bigO(m + |\mathcal{F}| \cdot \ell^4) = \bigO(m + \sqrt{\Textlen} \log^4 \Textlen)$ time.
    \end{enumerate}
    In total, the construction of the second component takes
    $\bigO(m + \sqrt{\Textlen} \log^4 \Textlen) =
    \bigO(\Textlen / \log_{\AlphabetSize} \Textlen)$ time.
  \item Computing $L_{\rm bit}[x]$ for any $x \in [1 \dd 2^{\ell})$
    takes $\bigO(\ell)$ time. Thus, over all $x \in [1 \dd 2^{\ell})$, we spend
    $\bigO(2^{\ell} \cdot \ell) = \bigO(\AlphabetSize^{\ell} \cdot \ell) = \bigO(\sqrt{\Textlen} \log \Textlen)
    = \bigO(\Textlen / \log_{\AlphabetSize} \Textlen)$ time.
  \end{enumerate}
  In total, the construction takes
  $\bigO(\Textlen / \log_{\AlphabetSize} \Textlen)$ time.
\end{proof}

\begin{proposition}\label{pr:lex-range-reporting-short}
  Let $\Text \in \IntegerAlphabet^{\Textlen}$ be such that $2 \leq \AlphabetSize < \Textlen^{1/13}$ and $\Text[\Textlen]$ does not
  occur in $\Text[1 \dd \Textlen)$. Let $\tau = \lfloor \mu\log_{\AlphabetSize} \Textlen \rfloor$, where $\mu$ is a positive
  constant smaller than $\tfrac{1}{12}$ such that $\tau \geq 1$.
  Given the packed representation of $\Text$, we can in $\bigO(\Textlen / \log_{\AlphabetSize} \Textlen)$ time
  construct a data structure that answers the following queries:
  \begin{enumerate}
  \item\label{pr:lex-range-reporting-short-it-1}
    Given the packed representation of any $X_1, X_2 \in \IntegerAlphabet^{\leq 3\tau-1}$,
    return $\mathcal{J} := \LexRange{X_1}{X_2}{\Text}$
    (see \cref{def:lex-range}) in $\bigO(1 + |\mathcal{J}|)$ time.
  \item\label{pr:lex-range-reporting-short-it-2}
    Given the packed representation of any $X \in \IntegerAlphabet^{\leq 3\tau-1}$,
    return $\mathcal{J} := \LexRange{X}{X c^{\infty}}{\Text}$
    (where $c = \AlphabetSize - 1$) in $\bigO(1 + |\mathcal{J}|)$ time.
  \item\label{pr:lex-range-reporting-short-it-3}
    Given the packed representation of any $X_1, X_2 \in \IntegerAlphabet^{\leq 3\tau-1}$,
    return $\mathcal{J} := \LexRange{X_1 c^{\infty}}{X_2}{\Text}$
    (where $c = \AlphabetSize - 1$) in $\bigO(1 + |\mathcal{J}|)$ time.
  \end{enumerate}
\end{proposition}
\begin{proof}

  We use the following definitions.
  Let $B_{3\tau-1}[1 \dd \Textlen]$ be a bitvector defined so that
  for every $i \in [1 \dd \Textlen]$, $B_{3\tau-1}[i] = \one$ holds
  if and only if $i = \Textlen$, or $i < \Textlen$ and $\LCE{\Text}{\SA{\Text}[i]}{\SA{\Text}[i+1]} < 3\tau-1$.
  Let $k = \Rank{B_{3\tau-1}}{\Textlen}{\one}$. Let $A_{\rm str}[1 \dd k]$ be an array defined
  such that, for every $i \in [1 \dd k]$,
  it holds $A_{\rm str}[i] = \Text[\SA{\Text}[i'] \dd \min(\Textlen + 1, \SA{\Text}[i'] + 3\tau - 1))$,
  where $i' = \Select{B_{3\tau-1}}{i}{\one}$.
  Observe that $k = \bigO(\AlphabetSize^{3\tau-1}) = \bigO(\sqrt{\Textlen})$.

  \DSComponents
  The data structure consists of the following components:
  \begin{enumerate}
  \item The data structure from \cref{pr:nav-index-short}. It needs
    $\bigO(\Textlen / \log_{\AlphabetSize} \Textlen)$ space.
  \item The bitvector $B_{3\tau-1}[1 \dd \Textlen]$ augmented with the support for $\bigO(1)$-time rank and select queries
    using \cref{th:bin-rank-select}. The bitvector together with the augmentation of \cref{th:bin-rank-select}
    needs $\bigO(\Textlen / \log \Textlen) = \bigO(\Textlen / \log_{\AlphabetSize} \Textlen)$ space.
  \item The array $A_{\rm str}[1 \dd k]$ stored in plain form using
    $\bigO(k) = \bigO(\sqrt{\Textlen}) = \bigO(\Textlen / \log_{\AlphabetSize} \Textlen)$ space.
  \item The structure from \cref{pr:short-occ-reporting} for text $\Text$. It needs
    $\bigO(\Textlen / \log_{\AlphabetSize} \Textlen)$ space.
  \end{enumerate}
  In total, the structure needs $\bigO(\Textlen / \log_{\AlphabetSize} \Textlen)$ space.

  \DSQueries
  The queries are answered as follows:
  \begin{enumerate}
  \item Let $X_1, X_2 \in \IntegerAlphabet^{\leq 3\tau-1}$.
    Given the packed representation of strings $X_1$ and $X_2$, we compute the set
    $\LexRange{X_1}{X_2}{\Text}$ as follows:
    \begin{enumerate}
    \item Using \cref{pr:nav-index-short}, in $\bigO(1)$ time compute
      $b_k = \RangeBegTwo{X_k}{\Text}$ for $k \in \{1,2\}$. Observe that
      then
      \[
        \LexRange{X_1}{X_2}{\Text} = \{\SA{\Text}[i]\}_{i \in (b_1 \dd b_2]}
      \]
      (see \cref{rm:lex-range}).
      If $b_1 \geq b_2$, then $\LexRange{X_1}{X_2}{\Text} = \emptyset$.
      Let us thus assume that $b_1 < b_2$.
      Note that, by definition of $B_{3\tau-1}$,
      if $b_k > 0$, then $B_{3\tau-1}[b_k] = \one$ holds for $k \in \{1,2\}$.
    \item If $b_1 = 0$, then we set $i_1 = 0$. Otherwise, using \cref{th:bin-rank-select}, in $\bigO(1)$ time
      we compute $i_1 = \Rank{B_{3\tau-1}}{b_1}{\one}$. Analogously we compute $i_2$.
      Observe that we then have
      \[
        \LexRange{X_1}{X_2}{\Text} = \bigcup_{i \in (i_1 \dd i_2]} \OccTwo{A_{\rm str}[i]}{\Text}.
      \]
    \item Initialize the set $\mathcal{P} := \emptyset$.
    \item For every $i \in (i_1 \dd i_2]$, using \cref{pr:short-occ-reporting}, compute the set
      $\mathcal{J}_i := \OccTwo{A_{\rm str}[i]}{\Text}$ in $\bigO(1 + |\mathcal{J}_i|)$ time,
      and add to $\mathcal{P}$. In total, this takes
      $\bigO((i_2 - i_1) + \sum_{i \in (i_1 \dd i_2]} |\mathcal{J}_i|)
      = \bigO(1 + |\LexRange{X_1}{X_2}{\Text}|)$ time, where we used that since for every $i \in (i_1 \dd i_2]$, it holds
      $\mathcal{J}_i \neq \emptyset$, it follows that
      $i_2 - i_1 = \bigO(\sum_{i \in (i_1 \dd i_2]} |\mathcal{J}_i|)$.
      Note also that the set of strings in $A_{\rm str}[1 \dd k]$ is prefix-free (i.e., no string is a prefix of another),
      which implies that for every $i,j \in [1 \dd k]$, $i \neq j$ implies that
      $\OccTwo{A_{\rm str}[i]}{\Text} \cap \OccTwo{A_{\rm str}[j]}{\Text} = \emptyset$, and hence
      $|\LexRange{X_1}{X_2}{\Text}| = \sum_{i \in (i_1 \dd i_2]} |\mathcal{J}_i|$.
      We then return $\mathcal{P}$ as the answer.
    \end{enumerate}
    In total, we spend $\bigO(1 + |\LexRange{X_1}{X_2}{\Text}|)$ time.
  \item Let us now consider $X \in \IntegerAlphabet^{\leq 3\tau-1}$,
    where $c = \AlphabetSize - 1$. Given the packed representation of $X$, we compute
    $\LexRange{X}{X c^{\infty}}{\Text}$
    by setting $X_1 = X$ and $X_2 = X c^{\infty}$, and
    using the same algorithm as above, except
    to compute $b_2 = \RangeBegTwo{X c^{\infty}}{\Text}$, we observe that
    $\RangeBegTwo{X c^{\infty}}{\Text} = \RangeEndTwo{X}{\Text}$. Thus, we can determine $b_2$ using
    \cref{pr:nav-index-short} in $\bigO(1)$ time.
  \item Let $X_1, X_2 \in \IntegerAlphabet^{\leq 3\tau-1}$.
    Given the packed representation of $X_1$ and $X_2$, we compute the set
    $\LexRange{X_1 c^{\infty}}{X_2}{\Text}$ (where $c = \AlphabetSize - 1$)
    similarly as above, except to compute the value $b_1 = \RangeBegTwo{X_1 c^{\infty}}{\Text}$, we use the fact
    that $\RangeBegTwo{X_1 c^{\infty}}{\Text} = \RangeEndTwo{X_1}{\Text}$. Thus, we can
    compute $b_1$ using \cref{pr:nav-index-short} in $\bigO(1)$ time.
  \end{enumerate}

  \DSConstruction
  The components of the data structure are constructed as follows:
  \begin{enumerate}
  \item The first component is constructed using \cref{pr:nav-index-short}
    in $\bigO(\Textlen / \log_{\AlphabetSize} \Textlen)$ time.
  \item To construct the second component, we first compute the bitvector
    $B_{3\tau-1}[1 \dd \Textlen]$ as described in~\cite[Proposition~5.3]{breaking}
    in $\bigO(\Textlen / \log_{\AlphabetSize} \Textlen)$ time.
    We then augment $B_{3\tau-1}$ with support for $\bigO(1)$-time rank/select queries
    using \cref{th:bin-rank-select} in
    $\bigO(\Textlen / \log \Textlen) = \bigO(\Textlen / \log_{\AlphabetSize} \Textlen)$ time.
  \item The array
    $A_{\rm str}[1 \dd k]$ (with all strings stored in the packed representation) is constructed
    as described in~\cite[Proposition~5.3]{breaking} in $\bigO(\Textlen / \log_{\AlphabetSize} \Textlen)$ time.
  \item The last component is constructed using \cref{pr:short-occ-reporting} in
    $\bigO(\Textlen / \log_{\AlphabetSize} \Textlen)$ time.
  \end{enumerate}
  In total, the construction takes
  $\bigO(\Textlen / \log_{\AlphabetSize} \Textlen)$ time.
\end{proof}

\paragraph{The Nonperiodic Patterns}

\subparagraph{Preliminaries}

\begin{definition}[Sequence of positions in lex-order]\label{def:lex-sorted}
  Let $\Text \in \Sigma^{\Textlen}$, $Q \subseteq [1 \dd \Textlen]$, and $q = |Q|$. By
  $\LexSorted{Q}{\Text}$, we denote a sequence $(a_i)_{1 \in [1 \dd q]}$
  containing all positions from $Q$ such that for every
  $i, j \in [1 \dd q]$, $i < j$ implies
  $\Text[a_i \dd \Textlen] \prec \Text[a_j \dd \Textlen]$.
\end{definition}

\begin{definition}[Sequence of positions in text-order]\label{def:sort}
  Consider any finite set $Q \subseteq \Z$, and let $q = |Q|$. By
  $\Sort{Q}$, we denote a sequence $(a_i)_{1 \in [1 \dd q]}$
  containing all positions from $Q$ such that for every
  $i, j \in [1 \dd q]$, $i < j$ implies $a_i < a_j$.
\end{definition}

\begin{definition}[Successor]\label{def:succ}
  Consider any totally ordered set $\mathcal{U}$, and let $A \subseteq \mathcal{U}$ be a nonempty subset of $\mathcal{U}$.
  For every $x \in \mathcal{U}$ satisfying $x \leq \max A$,
  we denote $\Successor{A}{x} = \min\{x' \in A : x' \succeq x\}$.
\end{definition}

\begin{definition}[Distinguishing prefix of a suffix]\label{def:dist-prefixes}
  Let $\Text \in \Sigma^{\Textlen}$.
  Let $\tau \in [1 \dd \lfloor \tfrac{\Textlen}{2} \rfloor]$, and let
  $\SSS$ be a $\tau$-synchronizing set of $\Text$.
  For every $j \in [1 \dd \Textlen - 3\tau + 2] \setminus \RTwo{\tau}{\Text}$,
  we denote (see \cref{def:succ})
  \[
    \DistPrefixPos{j}{\tau}{\Text}{\SSS} := \Text[j \dd \Successor{\SSS}{j} + 2\tau).
  \]
  We then let
  \[
    \DistPrefixes{\tau}{\Text}{\SSS}
      := \{\DistPrefixPos{j}{\tau}{\Text}{\SSS} : j \in [1 \dd \Textlen -
          3\tau + 2] \setminus \RTwo{\tau}{\Text}\}.
  \]
\end{definition}

\begin{remark}\label{rm:dist-prefixes}
  Note that $\Successor{\SSS}{j}$ in \cref{def:dist-prefixes}
  is well-defined for every $j \in [1 \dd \Textlen - 3\tau + 2]
  \setminus \RTwo{\tau}{\Text}$, because by
  \cref{def:sss}\eqref{def:sss-density}, for such $j$
  we have $[j \dd j + \tau) \cap \SSS \neq \emptyset$.
\end{remark}

\begin{lemma}[{\cite{sublinearlz}}]\label{lm:dist-prefixes}
  Let $\Text \in \Sigma^{\Textlen}$.
  Let $\tau \in [1 \dd \lfloor \tfrac{\Textlen}{2} \rfloor]$
  and let $\SSS$ be a $\tau$-synchronizing set of $\Text$. Then:
  \begin{enumerate}
  \item\label{lm:dist-prefixes-it-1}
    It holds $\DistPrefixes{\tau}{\Text}{\SSS} \subseteq \IntegerAlphabet^{\leq 3\tau - 1}$.
  \item\label{lm:dist-prefixes-it-2}
    $\DistPrefixes{\tau}{\Text}{\SSS}$ is prefix-free, i.e.,
    for $D, D' \in \DistPrefixes{\tau}{\Text}{\SSS}$,
    $D \neq D'$ implies that $D$ is not a prefix~of~$D'$.
  \end{enumerate}
\end{lemma}

\begin{lemma}\label{lm:dist-prefix-existence}
  Let $\Text \in \Sigma^{\Textlen}$.
  Let $\tau \in [1 \dd \lfloor \tfrac{\Textlen}{2} \rfloor]$ and let
  $\SSS$ be a $\tau$-synchronizing set of $\Text$. Let $\Pat \in \Sigma^{m}$ be
  a $\tau$-nonperiodic pattern (\cref{def:periodic-pattern}) such
  that $m \geq 3\tau - 1$ and $\OccTwo{\Pat[1 \dd 3\tau-1]}{\Text} \neq \emptyset$.
  Then, there exists a unique $D \in \DistPrefixes{\tau}{\Text}{\SSS}$ (\cref{def:dist-prefixes})
  that is a prefix of $\Pat$.
\end{lemma}
\begin{proof}
  Denote $\Pat' = \Pat[1 \dd 3\tau - 1]$.
  Consider any $j \in \OccTwo{\Pat'}{\Text}$ (such position exists by
  the assumption $\OccTwo{\Pat'}{\Text} \neq \emptyset$). Since $\Pat'$
  is $\tau$-nonperiodic, it follows by $|\Pat'| = 3\tau - 1$ and \cref{def:periodic-pattern}
  that $j \in [1 \dd \Textlen - 3\tau + 2]$ and
  $\per{\Text[j \dd j + 3\tau - 1)} = \per{\Pat'} > \tfrac{1}{3}\tau$.
  By \cref{def:sss}, we thus have $j \in [1 \dd \Textlen - 3\tau + 2] \setminus \RTwo{\tau}{\Text}$
  and $[j \dd j + \tau) \cap \SSS \neq \emptyset$.
  Consequently, letting $j' = \Successor{\SSS}{j}$ (\cref{def:succ}), it holds
  $j' - j < \tau$. Therefore, letting $D = \Text[j \dd j' + 2\tau)$, it
  holds $|D| = j' - j + 2\tau \leq 3\tau - 1 = |\Pat'|$, and hence $D$ is a prefix of $\Pat'$ (and thus
  also $\Pat$).
  On the other hand, by \cref{def:dist-prefixes}, we have $D \in \DistPrefixes{\tau}{\Text}{\SSS}$.
  It remains to observe that 
  since $\DistPrefixes{\tau}{\Text}{\SSS}$ is prefix-free (\cref{lm:dist-prefixes}\eqref{lm:dist-prefixes-it-2}),
  no other string from $\DistPrefixes{\tau}{\Text}{\SSS}$ can be a prefix of $\Pat$.
\end{proof}

\begin{definition}[Distinguishing prefix of a pattern]\label{def:dist-prefix-pat}
  Let $\Text \in \Sigma^{\Textlen}$.
  Let $\tau \in [1 \dd \lfloor \tfrac{\Textlen}{2} \rfloor]$, and let $\SSS$ be
  a $\tau$-synchronizing set of $\Text$. For every $\tau$-nonperiodic
  pattern $\Pat \in \Sigma^{m}$ satisfying $m \geq 3\tau - 1$
  and $\OccTwo{\Pat[1 \dd 3\tau-1]}{\Text} \neq \emptyset$, by $\DistPrefixPat{\Pat}{\tau}{\Text}{\SSS}$
  we denote the unique string $D \in \DistPrefixes{\tau}{\Text}{\SSS}$ (\cref{def:dist-prefixes})
  that is a prefix of $\Pat$ (such $D$ exists by \cref{lm:dist-prefix-existence}).
\end{definition}

\subparagraph{Basic Navigational Primitives}

\begin{proposition}[{\cite[Proposition~5.13]{sublinearlz}}]\label{pr:nav-index-nonperiodic}
  Let $\Text \in \IntegerAlphabet^{\Textlen}$ be such that $2 \leq \AlphabetSize < \Textlen^{1/13}$
  and $\Text[\Textlen]$ does not occur in $\Text[1 \dd \Textlen)$.
  Let $\tau = \lfloor \mu\log_{\AlphabetSize} \Textlen \rfloor$, where
  $\mu$ is a positive constant smaller than $\tfrac{1}{12}$ such that $\tau \geq 1$.
  Let $\SSS$ be a $\tau$-synchronizing
  set of $\Text$ satisfying $|\SSS| = \bigO(\tfrac{\Textlen}{\tau})$.
  Denote $(s_i)_{i \in [1 \dd n']} = \LexSorted{\SSS}{\Text}$ (\cref{def:lex-sorted}).
  Given the set $\SSS$ and the packed representation of the text
  $\Text$, we can in $\bigO(\Textlen / \log_{\AlphabetSize} \Textlen)$ time
  construct a data structure, denoted $\NavNonperiodic{\SSS}{\tau}{\Text}$,
  that supports the following queries:
  \begin{enumerate}
  \item\label{pr:nav-index-nonperiodic-it-1}
    Given the packed representation of any string $X \in \IntegerAlphabet^{\leq 3\tau - 1}$,
    in $\bigO(1)$ time return the packed representation of string $\revstr{X}$ (see \cref{sec:prelim}).
  \item\label{pr:nav-index-nonperiodic-it-2}
    Let $\Pat \in \Sigma^{m}$ be a $\tau$-nonperiodic pattern satisfying
    $m \geq 3\tau - 1$ and $\OccTwo{\Pat[1 \dd 3\tau-1]}{\Text} \neq \emptyset$. Denote
    $D = \DistPrefixPat{\Pat}{\tau}{\Text}{\SSS}$ (\cref{def:dist-prefix-pat}), $\deltatext = |D| - 2\tau$, and
    $\Pat' = \Pat(\deltatext \dd m]$.
    \begin{enumerate}
    \item\label{pr:nav-index-nonperiodic-it-2a}
      Given the packed representation of $\Pat$, in $\bigO(1)$ time
      compute the packed representation~of~$D$.
    \item\label{pr:nav-index-nonperiodic-it-2b}
      Given the packed representation of $\Pat$, in $\bigO(m / \log_{\AlphabetSize} \Textlen + \log \log \Textlen)$ time
      compute a pair of integers $(b,e)$ such that:
      \begin{itemize}
      \item $b = |\{i \in [1 \dd n'] : \Text[s_i \dd \Textlen] \prec \Pat'\}|$ and
      \item $e = |\{i \in [1 \dd n'] : \Text[s_i \dd \Textlen] \prec \Pat'c^{\infty}\}|$, where $c = \AlphabetSize - 1$.
      \end{itemize}
    \end{enumerate}
  \end{enumerate}
\end{proposition}
\begin{proof}
  Note that in~\cite[Proposition~5.13]{sublinearlz}, we assumed that $\OccTwo{\Pat}{\Text} \neq \emptyset$.
  It is easy to check that this is not in fact required, and the structure works even if only
  $\OccTwo{\Pat[1 \dd 3\tau-1]}{\Text} \neq \emptyset$ holds. Note also that in the original claim,
  the value $e$ is defined so that it holds
  $e-b = |\{i \in [1 \dd n'] : \Pat'\text{ is a prefix of }\Text[s_i \dd \Textlen]\}|$.
  Here we use a slightly different but equivalent definition.
\end{proof}

\begin{theorem}[{\cite[Theorem~4.3]{sss}}]\label{th:sss-lex-sort}
  Given the packed representation of text $\Text \in \IntegerAlphabet^{\Textlen}$
  and its $\tau$-synchronizing set $\SSS$ of
  size $|\SSS| = \bigO(\tfrac{\Textlen}{\tau})$ for $\tau = \bigO(\log_{\AlphabetSize} \Textlen)$,
  we can compute the sequence $\LexSorted{\SSS}{\Text}$
  (\cref{def:lex-sorted}) in $\bigO(\tfrac{\Textlen}{\tau})$ time.
\end{theorem}

\subparagraph{Combinatorial Properties}

\begin{lemma}\label{lm:lex-range-nonperiodic}
  Let $\Text \in \Sigma^{\Textlen}$ and $\tau \in [1 \dd \lfloor \tfrac{\Textlen}{2} \rfloor]$.
  Let $\SSS$ be a $\tau$-synchronizing set of $\Text$. Denote
  $(s_i)_{i \in [1 \dd n']} = \LexSorted{\SSS}{\Text}$ (\cref{def:lex-sorted}).
  Let $D \in \DistPrefixes{\tau}{\Text}{\SSS}$ (\cref{def:dist-prefixes})
  and let $\Pat_1, \Pat_2 \in \Sigma^{+}$ be $\tau$-nonperiodic patterns (\cref{def:periodic-pattern})
  both having $D$ as a prefix.
  Denote $\deltatext = |D| - 2\tau$. For $k \in \{1,2\}$, let
  $\Pat'_k = \Pat_k(\deltatext \dd |\Pat_k|]$, and
  $b_k = |\{i \in [1 \dd n'] : \Text[s_i \dd \Textlen] \prec \Pat'_k\}|$.
  Then, it holds (see \cref{def:lex-range})
  \begin{align*}
    \LexRange{\Pat_1}{\Pat_2}{\Text} =
      \{s_i - \deltatext :
        i \in (b_1 \dd b_2]\text{ and }
        s_i - \deltatext \in \OccTwo{D}{\Text}\}.
  \end{align*}
\end{lemma}
\begin{proof}

  Denote $A = \{s_i - \deltatext : i \in (b_1 \dd b_2]\text{ and }s_i - \deltatext \in \OccTwo{D}{\Text}\}$.
  Observe that since $D$ is a prefix of $\Pat_1$ and $\Pat_2$, for every $j \in \LexRange{\Pat_1}{\Pat_2}{\Text}$,
  it holds $j \in \OccTwo{D}{\Text}$. Note also that, by definition of $b_k$ for $k \in \{1,2\}$, it follows
  that for every $i \in [1 \dd n']$, $s_i \in \LexRange{\Pat'_1}{\Pat'_2}{\Text}$ holds if and only if $i \in (b_1 \dd b_2]$.

  First, we prove that $\LexRange{\Pat_1}{\Pat_2}{\Text} \subseteq A$. Let $j \in \LexRange{\Pat_1}{\Pat_2}{\Text}$
  and let $j' = j + \deltatext$.
  \begin{itemize}
  \item First, observe that by the above discussion, we have $j \in \OccTwo{D}{\Text}$.
  \item Second, note that since $\lcp{\Pat_1}{\Pat_2} \geq |D| \geq \deltatext$,
    and since for $k \in \{1,2\}$ it holds $\deltatext + |\Pat'_k| = |\Pat_k|$, it follows by
    $j \in \LexRange{\Pat_1}{\Pat_2}{\Text}$ that $j' \in \LexRange{\Pat'_1}{\Pat'_2}{\Text}$.
  \item Finally, we prove that $j' \in \SSS$. By \cref{def:dist-prefixes}, there exists
    $i \in [1 \dd \Textlen - 3\tau + 2] \setminus \RTwo{\tau}{\Text}$ such that
    $i \in \OccTwo{D}{\Text}$ and $\Successor{\SSS}{i} = i + |D| - 2\tau = i + \deltatext$. Thus,
    $i + \deltatext \in \SSS$. Since above we proved that $j \in \OccTwo{D}{\Text}$, it follows
    that $\Text[j + \deltatext \dd j + |D|) = \Text[i + \deltatext \dd i + |D|)$, and hence by
    the consistency of $\SSS$ (\cref{def:sss}\eqref{def:sss-consistency}), $j' = j + \deltatext \in \SSS$.
  \end{itemize}
  By $j' \in \SSS$, there exists $i \in [1 \dd n']$ such that $s_i = j'$.
  Above we proved that $j' \in \LexRange{\Pat'_1}{\Pat'_2}{\Text}$.
  Thus, by $s_i = j'$ and the earlier characterization of the set $\{s_t\}_{t \in (b_1 \dd b_2]}$,
  we obtain $i \in (b_1 \dd b_2]$. It remains to note that above we also proved that
  $j = j' - \deltatext = s_i - \deltatext \in \OccTwo{D}{\Text}$.
  Putting everything together, we have thus proved that
  there exists $i \in (b_1 \dd b_2]$ such that $j = s_i - \deltatext$ and
  $s_i - \deltatext \in \OccTwo{D}{\Text}$, i.e., $j \in A$.

  We now prove that $A \subseteq \LexRange{\Pat_1}{\Pat_2}{\Text}$. Let $j \in A$.
  To prove $j \in \LexRange{\Pat_1}{\Pat_2}{\Text}$, by \cref{def:lex-range}, we need
  to show that $j \in [1 \dd \Textlen]$ and $\Pat_1 \preceq \Text[j \dd \Textlen] \prec \Pat_2$.
  We proceed as follows:
  \begin{itemize}
  \item First, note that by $j \in A$, there exists $i \in (b_1 \dd b_2]$ such that
    $j = s_i - \deltatext$ and $s_i - \deltatext \in \OccTwo{D}{\Text}$. In particular, by
    \cref{def:occ}, this implies $j = s_i - \deltatext \in [1 \dd \Textlen]$.
  \item Next, observe that by $i \in (b_1 \dd b_2]$ and the above characterization
    of $\{s_t\}_{t \in (b_1 \dd b_2]}$, it follows that
    $s_i \in \LexRange{\Pat'_1}{\Pat'_2}{\Text}$, i.e.,
    $\Pat'_1 \preceq \Text[s_i \dd \Textlen] \prec \Pat'_2$. By
    $s_i = j + \deltatext$, we thus have
    $\Pat'_1 \preceq \Text[j + \deltatext \dd \Textlen] \prec \Pat'_2$.
    Recall now that we assumed that both $\Pat_1$ and $\Pat_2$ are prefixed with
    $D$. Moreover, above we observed that $j \in \OccTwo{D}{\Text}$. Since
    $|D| \geq \deltatext$, we thus have $\Pat_1[1 \dd \deltatext] =
    \Pat_2[1 \dd \deltatext] = \Text[j \dd j + \deltatext) = D[1 \dd \deltatext]$.
    Combining with $\Pat'_1 \preceq \Text[j + \deltatext \dd \Textlen] \prec \Pat'_2$,
    we thus obtain $\Pat_1 \preceq \Text[j \dd \Textlen] \prec \Pat_2$.
  \end{itemize}
  This concludes the proof of $j \in \LexRange{\Pat_1}{\Pat_2}{\Text}$.
\end{proof}

\begin{lemma}\label{lm:lex-range-nonperiodic-array}
  Let $\Text \in \Sigma^{\Textlen}$, $\tau \in [1 \dd \lfloor \tfrac{\Textlen}{2} \rfloor]$, and
  assume that $\Text[\Textlen]$ does not occur in $\Text[1 \dd \Textlen)$.
  Let $\SSS$ be a $\tau$-synchronizing set of $\Text$. Denote
  $(s_i)_{i \in [1 \dd n']} = \LexSorted{\SSS}{\Text}$ (\cref{def:lex-sorted}).
  Let $A_{\SSS}[1 \dd n']$ and $A_{\rm str}[1 \dd n']$ be defined by
  \begin{itemize}
  \item $A_{\SSS}[i] = s_i$,
  \item $A_{\rm str}[i] = \revstr{D_i}$, where $D_i = \Textinf[s_i - \tau \dd s_i + 2\tau)$.
  \end{itemize}
  Let $D \in \DistPrefixes{\tau}{\Text}{\SSS}$ (\cref{def:dist-prefixes})
  and let $\Pat_1, \Pat_2 \in \Sigma^{+}$ be $\tau$-nonperiodic patterns (\cref{def:periodic-pattern})
  both having $D$ as a prefix.
  Denote $\deltatext = |D| - 2\tau$. For $k \in \{1,2\}$, let
  $\Pat'_k = \Pat_k(\deltatext \dd |\Pat_k|]$, and
  $b_k = |\{i \in [1 \dd n'] : \Text[s_i \dd \Textlen] \prec \Pat'_k\}|$.
  Then, it holds (see \cref{def:lex-range})
  \begin{align*}
    \LexRange{\Pat_1}{\Pat_2}{\Text} =
      \{A_{\SSS}[i] - \deltatext :
        i \in (b_1 \dd b_2]\text{ and }
        \revstr{D}\text{ is a prefix of }A_{\rm str}[i]\}.
  \end{align*}
\end{lemma}
\begin{proof}

  Observe that for every $i \in [1 \dd n']$, $s_i - \deltatext \in \OccTwo{D}{\Text}$ holds if and only
  if $\revstr{D}$ is a prefix of $A_{\rm str}[i]$. The proof of this equivalence follows as
  in~\cite[Lemma~5.15]{sublinearlz} (full version).

  By putting together the above equivalence and \cref{lm:lex-range-nonperiodic}, we obtain
  \begin{align*}
    \LexRange{\Pat_1}{\Pat_2}{\Text}
      &= \{s_i - \deltatext : i \in (b_1 \dd b_2]\text{ and }s_i - \deltatext \in \OccTwo{D}{\Text}\}\\
      &= \{A_{\SSS}[i] - \deltatext : i \in (b_1 \dd b_2]\text{ and }\revstr{D}\text{ is a prefix of }A_{\rm str}[i]\}.
      \qedhere
  \end{align*}
\end{proof}

\subparagraph{Algorithms}

\begin{proposition}\label{pr:lex-range-reporting-nonperiodic}
  Consider a data structure answering prefix range reporting queries
  that, given any sequence $W[1 \dd k]$ strings of length
  $\ell$ over alphabet $\IntegerAlphabet$
  achieves the following complexities (where the input strings
  during construction are given in the packed representation,
  and at query time we are given any $b,e \in [0 \dd k]$ and the packed
  representation of any $X \in \IntegerAlphabet^{\leq \ell}$,
  and we return the set
  $\mathcal{I} := \{i \in (b \dd e] : X\text{ is a prefix of }W[i]\}$):
  \begin{itemize}
  \item space usage $S(k,\ell,\AlphabetSize)$,
  \item preprocessing time $P_t(k,\ell,\AlphabetSize)$,
  \item preprocessing space $P_s(k,\ell,\AlphabetSize)$,
  \item query time $Q_{\rm base}(k,\ell,\AlphabetSize) + |\mathcal{I}| \cdot Q_{\rm elem}(k,\ell,\AlphabetSize)$.
  \end{itemize}
  Let $\Text \in \IntegerAlphabet^{\Textlen}$ be such that
  $2 \leq \AlphabetSize < \Textlen^{1/13}$ and $\Text[\Textlen]$
  does not occur in $\Text[1 \dd \Textlen)$.
  Let $\tau = \lfloor \mu\log_{\AlphabetSize} \Textlen \rfloor$, where $\mu$ is a positive
  constant smaller than $\tfrac{1}{12}$ such that $\tau \geq 1$. There exist positive integers
  $k = \Theta(\Textlen / \log_{\AlphabetSize} \Textlen)$ and
  $\ell \leq (1 + \lfloor \log k \rfloor) / \lceil \log \AlphabetSize \rceil$
  such that, given the packed representation of $\Text$, we can
  in
  $\bigO(\Textlen / \log_{\AlphabetSize} \Textlen + P_t(k,\ell,\AlphabetSize))$ time and using
  $\bigO(\Textlen / \log_{\AlphabetSize} \Textlen + P_s(k,\ell,\AlphabetSize))$ working space construct a data structure of size
  $\bigO(\Textlen / \log_{\AlphabetSize} \Textlen + S(k,\ell,\AlphabetSize))$ that answers the following queries:
  \begin{enumerate}
  \item\label{pr:lex-range-reporting-nonperiodic-it-1}
    Given the packed representation of
    a $\tau$-nonperiodic (\cref{def:periodic-pattern}) patterns $\Pat_1, \Pat_2 \in \IntegerAlphabet^{+}$ such that
    $\lcp{\Pat_1}{\Pat_2} \geq 3\tau - 1$,
    computes the set $\mathcal{J} := \LexRange{\Pat_1}{\Pat_2}{\Text}$ (see \cref{def:lex-range}) in
    $\bigO(\log \log \Textlen +
    Q_{\rm base}(k,\ell,\AlphabetSize) + |\mathcal{J}| \cdot Q_{\rm elem}(k,\ell,\AlphabetSize) +
    |\Pat_1| / \log_{\AlphabetSize} \Textlen +
    |\Pat_2| / \log_{\AlphabetSize} \Textlen)$
    time.
  \item\label{pr:lex-range-reporting-nonperiodic-it-2}
    Given the packed representation of a $\tau$-nonperiodic pattern
    $\Pat \in \IntegerAlphabet^{+}$ satisfying $|\Pat| \geq 3\tau-1$,
    computes the set $\mathcal{J} := \LexRange{\Pat}{\Pat' c^{\infty}}{\Text}$
    (where $\Pat' = \Pat[1 \dd 3\tau-1]$ and $c = \AlphabetSize - 1$) in
    $\bigO(\log \log \Textlen +
    Q_{\rm base}(k,\ell,\AlphabetSize) + |\mathcal{J}| \cdot Q_{\rm elem}(k,\ell,\AlphabetSize) +
    |\Pat| / \log_{\AlphabetSize} \Textlen)$
    time.
  \end{enumerate}
\end{proposition}
\begin{proof}

  We use the following definitions.
  Let $\SSS$ be a $\tau$-synchronizing set of $\Text$ of size $|\SSS| = \bigO(\frac{\Textlen}{\tau})$ constructed
  using \cref{th:sss-packed-construction}. Denote $\Textlen' = |\SSS|$.
  Denote $(s_i)_{i \in [1 \dd \Textlen']} = \LexSorted{\SSS}{\Text}$ (\cref{def:lex-sorted}).
  Let $A_{\SSS}[1 \dd \Textlen']$ be an array defined by $A_{\SSS}[i] = s_i$.
  Let $k = \max(\Textlen', k') = \Theta(\Textlen / \log_{\AlphabetSize} \Textlen)$,
  where $k' = \lceil \Textlen / \log_{\AlphabetSize} \Textlen \rceil$, and let
  $A_{\rm str}[1 \dd k]$ be an array defined so that for every $i \in [1 \dd \Textlen']$,
  $A_{\rm str}[i] = \revstr{D_i}$, where $D_i = \Textinf[s_i - \tau \dd s_i + 2\tau)$.
  We leave the remaining element initialized arbitrarily.
  Denote $\ell = 3\tau$, and note that
  by the same analysis as in \cref{th:sa-to-prefix-select-nonbinary}, it holds
  $\ell \leq (1 + \lfloor \log k \rfloor) / \lceil \log \AlphabetSize \rceil$.

  \DSComponents
  The data structure consists of the following components:
  \begin{enumerate}
  \item The structure from \cref{pr:nav-index-short} for text $\Text$. It needs $\bigO(\Textlen / \log_{\AlphabetSize} \Textlen)$ space.
  \item The structure $\NavNonperiodic{\SSS}{\tau}{\Text}$ from \cref{pr:nav-index-nonperiodic}. It needs
    $\bigO(\Textlen / \log_{\AlphabetSize} \Textlen)$ space.
  \item The structure from the claim (answering prefix range reporting queries) for the sequence $A_{\rm str}[1 \dd k]$.
    It needs $\bigO(S(k,\ell,\AlphabetSize))$ space.
  \item The array $A_{\SSS}[1 \dd \Textlen']$ in plain form using
    $\bigO(\Textlen') = \bigO(\Textlen / \log_{\AlphabetSize} \Textlen)$ space.
  \end{enumerate}
  In total, the structure needs
  $\bigO(\Textlen / \log_{\AlphabetSize} \Textlen + S(k,\ell,\AlphabetSize))$ space.

  \DSQueries
  The queries are answered as follows.
  \begin{enumerate}
  \item Let $\Pat_1, \Pat_2 \in \IntegerAlphabet^{+}$ be $\tau$-nonperiodic patterns satisfying
    $\lcp{\Pat_1}{\Pat_2} \geq 3\tau - 1$.
    Denote $\Pat' = \Pat_1[1 \dd 3\tau-1] = \Pat_2[1 \dd 3\tau-1]$.
    Given the packed representation of $\Pat_1$ and $\Pat_2$, we compute
    $\LexRange{\Pat_1}{\Pat_2}{\Text}$ as follows:
    \begin{enumerate}
    \item First, using \cref{pr:nav-index-short}, in $\bigO(1)$ time we compute
      $\RangeBegTwo{\Pat'}{\Text}$ and $\RangeEndTwo{\Pat'}{\Text}$. This lets us determine $|\OccTwo{\Pat'}{\Text}|$.
      If $\OccTwo{\Pat'}{\Text} = \emptyset$, then it holds $\LexRange{\Pat_1}{\Pat_2}{\Text} = \emptyset$ (since any element of
      this set would have $\Pat'$ as a prefix), and we finish the query algorithm. Let us now assume that
      $\OccTwo{\Pat'}{\Text} \neq \emptyset$.
    \item Using \cref{pr:nav-index-nonperiodic}\eqref{pr:nav-index-nonperiodic-it-2a}, in $\bigO(1)$ time
      compute the packed representation of the string
      $D = \DistPrefixPat{\Pat_1}{\tau}{\Text}{\SSS}$ (\cref{def:dist-prefix-pat}).
      Note that since $\lcp{\Pat_1}{\Pat_2} \geq 3\tau - 1 \geq |D|$,
      $D$ is also a prefix of $\Pat_2$
      (and by \cref{lm:dist-prefix-existence} no other string in $\DistPrefixes{\tau}{\Text}{\SSS}$ is a prefix of $\Pat_2$).
      In $\bigO(1)$ time we then calculate $\deltatext = |D| - 2\tau$.
    \item Using \cref{pr:nav-index-nonperiodic}\eqref{pr:nav-index-nonperiodic-it-1}, in $\bigO(1)$ time
      compute the packed representation of $\revstr{D}$.
    \item Denote $P'_k = \Pat_k(\deltatext \dd |\Pat_k|]$, where $k \in \{1,2\}$.
      Using \cref{pr:nav-index-nonperiodic}\eqref{pr:nav-index-nonperiodic-it-2b},
      compute the value
      $b_k = |\{i \in [1 \dd \Textlen'] : \Text[s_i \dd \Textlen] \prec \Pat'_k\}|$
      for $k \in \{1,2\}$.
      This takes
      $\bigO(\log \log \Textlen + |\Pat'_1| / \log_{\AlphabetSize} \Textlen +
      |\Pat'_2| / \log_{\AlphabetSize} \Textlen) = \bigO(\log \log \Textlen + |\Pat_1| / \log_{\AlphabetSize} \Textlen +
      |\Pat_2| / \log_{\AlphabetSize} \Textlen)$ time.
    \item Using the data structure from the claim, compute the elements of
      the set $\mathcal{I} = \{i \in (b_1 \dd b_2] : \revstr{D}\text{ is a prefix of }A_{\rm str}[i]\}$
      in $\bigO(Q_{\rm base}(k,\ell,\AlphabetSize) + |\mathcal{I}| \cdot Q_{\rm elem}(k,\ell,\AlphabetSize))$ time.
      In $\bigO(1 + |\mathcal{I}|)$ time we then compute and return the set
      $\mathcal{J} = \{A_{\SSS}[i] - \deltatext : i \in \mathcal{I}\}$ as the answer.
      By \cref{lm:lex-range-nonperiodic-array}, it holds
      $\mathcal{J} = \LexRange{\Pat_1}{\Pat_2}{\Text}$.
      Note also that $|\mathcal{I}| = |\mathcal{J}|$.
      Finally, observe that the array $A_{\rm str}$ defined here is padded with
      extra strings compared to the array in \cref{lm:lex-range-nonperiodic-array}, but this does not
      affect the result of the query, since $b_1, b_2 \leq \Textlen'$.
    \end{enumerate}
    In total, we spend
    $\bigO(\log \log \Textlen +
      Q_{\rm base}(k,\ell,\AlphabetSize) +
      |\mathcal{J}| \cdot Q_{\rm elem}(k,\ell,\AlphabetSize) +
      |\Pat_1| / \log_{\AlphabetSize} \Textlen +
      |\Pat_2| / \log_{\AlphabetSize} \Textlen)$ time,
    where $\mathcal{J} = \LexRange{\Pat_1}{\Pat_2}{\Text}$.
  \item Let $\Pat \in \IntegerAlphabet^{+}$ be a $\tau$-nonperiodic pattern satisfying $|\Pat| \geq 3\tau-1$.
    To compute the elements of the set $\LexRange{\Pat}{\Pat' c^{\infty}}{\Text}$
    (where $\Pat' = \Pat[1 \dd 3\tau-1]$ and $c = \AlphabetSize - 1$), we
    set $\Pat_1 = \Pat$ and $\Pat_2 = \Pat' c^{\infty}$, and then proceed similarly as above, except
    we compute $b_2 = |\{i \in [1 \dd \Textlen'] : \Text[s_i \dd \Textlen] \prec \Pat'_2\}|$
    using the second integer computed in \cref{pr:nav-index-nonperiodic}\eqref{pr:nav-index-nonperiodic-it-2b}
    (which requires only the packed representation of $\Pat'$).
    The whole query takes
    $\bigO(\log \log \Textlen +
      Q_{\rm base}(k,\ell,\AlphabetSize) +
      |\mathcal{J}| \cdot Q_{\rm elem}(k,\ell,\AlphabetSize) +
      |\Pat| / \log_{\AlphabetSize} \Textlen)$ time,
    where $\mathcal{J} = \LexRange{\Pat}{\Pat' c^{\infty}}{\Text}$.
  \end{enumerate}

  \DSConstruction
  The components of the structure are constructed as follows:
  \begin{enumerate}
  \item We apply \cref{pr:nav-index-short}. This uses $\bigO(\Textlen / \log_{\AlphabetSize} \Textlen)$ time and working space.
  \item Using \cref{th:sss-packed-construction}, we compute the $\tau$-synchronizing set $\SSS$ satisfying
    $|\SSS| = \bigO(\tfrac{\Textlen}{\tau}) = \bigO(\Textlen / \log_{\AlphabetSize} \Textlen)$ in
    $\bigO(\Textlen / \log_{\AlphabetSize} \Textlen)$ time. Then, using $\SSS$ and the packed representation of $\Text$ as input,
    we construct $\NavNonperiodic{\SSS}{\tau}{\Text}$ in $\bigO(\Textlen / \log_{\AlphabetSize} \Textlen)$ time using
    \cref{pr:nav-index-nonperiodic}.
  \item To construct the next component of the structure, we proceed as follows:
    \begin{enumerate}
    \item Using \cref{th:sss-lex-sort}, in
      $\bigO(\tfrac{\Textlen}{\tau}) = \bigO(\Textlen / \log_{\AlphabetSize} \Textlen)$ time we first compute
      the sequence $(s_i)_{i \in [1 \dd \Textlen']} = \LexSorted{\SSS}{\Text}$ (\cref{def:lex-sorted}).
    \item Using \cref{pr:nav-index-nonperiodic}\eqref{pr:nav-index-nonperiodic-it-1}, in $\bigO(\Textlen') =
      \bigO(\Textlen / \log_{\AlphabetSize} \Textlen)$ time we compute the elements of the array $A_{\rm str}$ at
      indexes $i \in [1 \dd \Textlen']$, i.e., $A_{\rm str}[i] = \revstr{D_i}$, where $D_i = \Textinf[s_i - \tau \dd s_i + 2\tau)$.
      In $\bigO(k - \Textlen') = \bigO(k) = \bigO(\Textlen / \log_{\AlphabetSize} \Textlen)$ time we then pad
      the remaining elements of $A_{\rm str}$ at indexes $i \in (\Textlen' \dd k]$.
    \item Finally, we apply the preprocessing
      from the claim to the array $A_{\rm str}$.
      This takes $\bigO(P_t(k,\ell,\AlphabetSize))$ time and uses
      $\bigO(P_s(k,\ell,\AlphabetSize))$ working space.
    \end{enumerate}
    In total, constructing this component of the structure
    takes $\bigO(\Textlen / \log_{\AlphabetSize} \Textlen + P_t(k,\ell,\AlphabetSize))$ time and uses
    $\bigO(\Textlen / \log_{\AlphabetSize} \Textlen + P_s(k,\ell,\AlphabetSize))$ working space.
  \item To construct the last component, we simply store the sequence
    $(s_i)_{i \in [1 \dd \Textlen']} = \LexSorted{\SSS}{\Text}$ (computed above
    in $\bigO(\Textlen / \log_{\AlphabetSize} \Textlen)$ time) in array $A_{\SSS}[1 \dd \Textlen']$.
  \end{enumerate}
  In total, the construction takes
  $\bigO(\Textlen / \log_{\AlphabetSize} \Textlen + P_t(k,\ell,\AlphabetSize))$ time and uses
  $\bigO(\Textlen / \log_{\AlphabetSize} \Textlen + P_s(k,\ell,\AlphabetSize))$ working space.
\end{proof}

\paragraph{The Periodic Patterns}

\subparagraph{Preliminaries}

\begin{definition}[Basic properties of $\tau$-periodic patterns]\label{def:pat-root}
  Let $\tau \geq 1$ and let $\Pat \in \Sigma^{+}$ be a $\tau$-periodic
  pattern (\cref{def:periodic-pattern}). Denote $p = \per{\Pat[1 \dd 3\tau - 1]}$.
  We then define
  \begin{itemize}
  \item $\RootPat{\Pat}{\tau}
    := \min\{\Pat[1 + t \dd 1 + t + p) : t \in [0 \dd p)\}$,
  \item $\RunEndPat{\Pat}{\tau} := 1 + p +
    \lcp{\Pat[1 \dd |\Pat|]}{\Pat[1 + p \dd |\Pat|]}$.
  \end{itemize}
  Furthermore, we then let
  \[
    \TypePat{\Pat}{\tau} =
      \begin{cases}
       +1 & \text{if }\RunEndPat{\Pat}{\tau} \leq |\Pat|\text{ and }
            \Pat[\RunEndPat{\Pat}{\tau}] \succ \Pat[\RunEndPat{\Pat}{\tau} - p],\\
       -1 & \text{otherwise}.
      \end{cases}
  \]
\end{definition}

\begin{remark}\label{rm:pat-root}
  Observe that in \cref{def:pat-root}, we can
  write $\Pat[1 \dd \RunEndPat{\Pat}{\tau}) = H' H^{k} H''$, where $H =
  \RootPat{\Pat}{\tau}$, and $H'$ (resp.\ $H''$) is a proper
  suffix (resp.\ proper prefix) of $H$. This factorization is unique, since the
  opposite would contradict the synchronization property of primitive
  strings~\cite[Lemma~1.11]{AlgorithmsOnStrings}.
\end{remark}

\begin{definition}[Run properties in a $\tau$-periodic pattern]\label{def:pat-head}
  Let $\tau \geq 1$ and $\Pat \in \Sigma^{+}$ be a $\tau$-periodic pattern.
  Let $\Pat[1 \dd \RunEndPat{\Pat}{\tau}) = H' H^{k} H''$ be such that
  $H = \RootPat{\Pat}{\tau}$, and $H'$ (resp.\ $H''$) is a proper
  suffix (resp.\ proper prefix) of $H$ (see \cref{rm:pat-root}).
  We then define:
  \begin{itemize}
  \item $\HeadPat{\Pat}{\tau} := |H'|$,
  \item $\ExpPat{\Pat}{\tau} := k$,
  \item $\TailPat{\Pat}{\tau} := |H''|$,
  \item $\RunEndFullPat{\Pat}{\tau} :=
    1 + |H'| + k \cdot |H| =
    \RunEndPat{\Pat}{\tau} - \TailPat{\Pat}{\tau}$.
  \end{itemize}
\end{definition}

\begin{lemma}[{\cite{collapsing}}]\label{lm:periodic-pat-lce}
  Let $\tau \geq 1$ and $\Pat \in \Sigma^{+}$ be a $\tau$-periodic
  pattern. For every $\Pat'
  \in \Sigma^{+}$, $\lcp{\Pat}{\Pat'} \geq 3\tau - 1$ holds if and
  only if $\Pat'$ is $\tau$-periodic, $\RootPat{\Pat'}{\tau} =
  \RootPat{\Pat}{\tau}$, and $\HeadPat{\Pat'}{\tau} =
  \HeadPat{\Pat}{\tau}$.  Moreover, if $\RunEndPat{\Pat}{\tau}
  \leq |\Pat|$ and $\lcp{\Pat}{\Pat'} \geq \RunEndPat{\Pat}{\tau}$
  (which holds, in particular, when $\Pat$ is a prefix of $\Pat'$),
  then:
  \begin{itemize}
  \item $\RunEndPat{\Pat'}{\tau} = \RunEndPat{\Pat}{\tau}$,
  \item $\TailPat{\Pat'}{\tau} = \TailPat{\Pat}{\tau}$,
  \item $\RunEndFullPat{\Pat'}{\tau} =
    \RunEndFullPat{\Pat}{\tau}$,
  \item $\ExpPat{\Pat'}{\tau} = \ExpPat{\Pat}{\tau}$,
  \item $\TypePat{\Pat'}{\tau} = \TypePat{\Pat}{\tau}$.
  \end{itemize}
\end{lemma}

\begin{lemma}[{\cite{collapsing}}]\label{lm:pat-lex}
  Let $\tau \geq 1$ and
  $\Pat_1, \Pat_2 \in \Sigma^{+}$ be $\tau$-periodic patterns such
  that $\RootPat{\Pat_1}{\tau} = \RootPat{\Pat_2}{\tau}$
  and $\HeadPat{\Pat_1}{\tau} =
  \HeadPat{\Pat_2}{\tau}$. Denote $t_1 =
  \RunEndPat{\Pat_1}{\tau} - 1$ and $t_2 = \RunEndPat{\Pat_2}{\tau}
  - 1$. Then, it holds $\lcp{\Pat_1}{\Pat_2} \geq \min(t_1,
  t_2)$. Moreover:
  \begin{enumerate}
  \item\label{lm:pat-lex-it-1} If $\TypePat{\Pat_1}{\tau} \neq
    \TypePat{\Pat_2}{\tau}$ or $t_1 \neq t_2$, then $\Pat_1 \neq
    \Pat_2$ and $\lcp{\Pat_1}{\Pat_2} = \min(t_1, t_2)$,
  \item\label{lm:pat-lex-it-2} If $\TypePat{\Pat_1}{\tau} \neq
    \TypePat{\Pat_2}{\tau}$, then $\Pat_1 \prec \Pat_2$ if and only if
    $\TypePat{\Pat_1}{\tau} < \TypePat{\Pat_2}{\tau}$,
  \item\label{lm:pat-lex-it-3} If $\TypePat{\Pat_1}{\tau} = -1$,
    then $t_1 < t_2$ implies $\Pat_1 \prec \Pat_2$,
  \item\label{lm:pat-lex-it-4} If $\TypePat{\Pat_1}{\tau} =
    +1$, then $t_1 < t_2$ implies $\Pat_1 \succ \Pat_2$,
  \item\label{lm:pat-lex-it-5} If $\TypePat{\Pat_1}{\tau} =
    \TypePat{\Pat_2}{\tau}  = -1$ and $t_1 \neq
    t_2$, then $t_1 < t_2$ if and only if $\Pat_1 \prec \Pat_2$,
  \item\label{lm:pat-lex-it-6} If $\TypePat{\Pat_1}{\tau} =
    \TypePat{\Pat_2}{\tau} = +1$ and $t_1 \neq t_2$, then
    $t_1 < t_2$ if and only if $\Pat_1 \succ \Pat_2$.
  \end{enumerate}
\end{lemma}

\begin{definition}[Properties of $\tau$-periodic positions]\label{def:pos-root}
  Let $\Text \in \Sigma^{\Textlen}$,
  $\tau \in [1 \dd \floor{\frac{\Textlen}{2}}]$,
  and $j \in \RTwo{\tau}{\Text}$.
  Letting $\Pat = \Text[j \dd \Textlen]$, we
  define:
  \begin{itemize}
  \item $\RootPos{j}{\tau}{\Text} := \RootPat{\Pat}{\tau}$,
  \item $\HeadPos{j}{\tau}{\Text} := \HeadPat{\Pat}{\tau}$,
  \item $\ExpPos{j}{\tau}{\Text} := \ExpPat{\Pat}{\tau}$,
  \item $\TailPos{j}{\tau}{\Text} := \TailPat{\Pat}{\tau}$,
  \item $\RunEndPos{j}{\tau}{\Text} := j + \RunEndPat{\Pat}{\tau} - 1$,
  \item $\RunEndFullPos{j}{\tau}{\Text} := j + \RunEndFullPat{\Pat}{\tau} - 1$,
  \item $\TypePos{j}{\tau}{\Text} := \TypePat{\Pat}{\tau}$.
  \end{itemize}
\end{definition}

\begin{remark}\label{rm:pos-root}
  Note that applying the notation for $\tau$-periodic patterns (\cref{def:periodic-pattern}) to $\Pat$ in
  \cref{def:pos-root} is well-defined since it is easy to see that for
  every $j \in \RTwo{\tau}{\Text}$ (\cref{def:sss}), the string $\Text[j \dd \Textlen]$
  is $\tau$-periodic. Note also that, letting
  $s = \HeadPos{j}{\tau}{\Text}$,
  $H = \RootPos{j}{\tau}{\Text}$,
  $p = |H|$, and
  $k = \ExpPos{j}{\tau}{\Text}$, it holds:
  \begin{itemize}
  \item $\RunEndPos{j}{\tau}{\Text} = j + p + \LCE{\Text}{j}{j+p}$,
  \item $\RunEndFullPos{j}{\tau}{\Text} = j + s + kp = \RunEndPos{j}{\tau}{\Text} - \TailPos{j}{\tau}{\Text}$.
  \end{itemize}
\end{remark}

\begin{definition}[Subsets of $\RTwo{\tau}{\Text}$]\label{def:R-subsets}
  Let $\Text \in \Sigma^{\Textlen}$ and
  $\tau \in [1 \dd \floor{\frac{\Textlen}{2}}]$.
  For every $H \in \Sigma^{+}$, $s \in \Zz$, and $k \in \Zp$,
  we define the following subsets of $\RTwo{\tau}{\Text}$:
  \begin{itemize}
  \item $\RMinusTwo{\tau}{\Text} := \{j \in \RTwo{\tau}{\Text} : \TypePos{j}{\tau}{\Text} = -1\}$,
  \item $\RPlusTwo{\tau}{\Text} := \RTwo{\tau}{\Text} \setminus \RMinusTwo{\tau}{\Text}$,
  \item $\RThree{H}{\tau}{\Text} := \{j \in \RTwo{\tau}{\Text} : \RootPos{j}{\tau}{\Text} = H\}$,
  \item $\RMinusThree{H}{\tau}{\Text} := \RMinusTwo{\tau}{\Text} \cap \RThree{H}{\tau}{\Text}$,
  \item $\RPlusThree{H}{\tau}{\Text} := \RPlusTwo{\tau}{\Text} \cap \RThree{H}{\tau}{\Text}$,
  \item $\RFour{s}{H}{\tau}{\Text} := \{j \in \RThree{H}{\tau}{\Text} : \HeadPos{j}{\tau}{\Text} = s\}$,
  \item $\RMinusFour{s}{H}{\tau}{\Text} := \RMinusTwo{\tau}{\Text} \cap \RFour{s}{H}{\tau}{\Text}$,
  \item $\RPlusFour{s}{H}{\tau}{\Text} := \RPlusTwo{\tau}{\Text} \cap \RFour{s}{H}{\tau}{\Text}$,
  \item $\RFive{s}{k}{H}{\tau}{\Text} := \{j \in \RFour{s}{H}{\tau}{\Text} : \ExpPos{j}{\tau}{\Text} = k\}$,
  \item $\RMinusFive{s}{k}{H}{\tau}{\Text} := \RMinusTwo{\tau}{\Text} \cap \RFive{s}{k}{H}{\tau}{\Text}$,
  \item $\RPlusFive{s}{k}{H}{\tau}{\Text} := \RPlusTwo{\tau}{\Text} \cap \RFive{s}{k}{H}{\tau}{\Text}$.
  \end{itemize}
\end{definition}

\begin{definition}[Starting positions of $\tau$-runs]\label{def:R-prim}
  Let $\Text \in \Sigma^{\Textlen}$ and
  $\tau \in [1 \dd \floor{\frac{\Textlen}{2}}]$.
  The set of starting positions of maximal blocks in $\RTwo{\tau}{\Text}$
  is denoted
  \[
    \RPrimTwo{\tau}{\Text} := \{j \in \RTwo{\tau}{\Text} : j-1 \not\in \RTwo{\tau}{\Text}\}.
  \]
  For every $H \in \Sigma^{+}$, we then define:
  \begin{itemize}
  \item $\RPrimMinusTwo{\tau}{\Text} := \RPrimTwo{\tau}{\Text} \cap \RMinusTwo{\tau}{\Text}$,
  \item $\RPrimPlusTwo{\tau}{\Text} := \RPrimTwo{\tau}{\Text} \cap \RPlusTwo{\tau}{\Text}$,
  \item $\RPrimMinusThree{H}{\tau}{\Text} := \RPrimTwo{\tau}{\Text} \cap \RMinusThree{H}{\tau}{\Text}$,
  \item $\RPrimPlusThree{H}{\tau}{\Text} := \RPrimTwo{\tau}{\Text} \cap \RPlusThree{H}{\tau}{\Text}$.
  \end{itemize}
\end{definition}

\begin{lemma}[{\cite{collapsing}}]\label{lm:R-text-block}
  Let $\Text \in \Sigma^{\Textlen}$ and $\tau \in [1 \dd \floor{\frac{\Textlen}{2}}]$. For every $j \in
  \RTwo{\tau}{\Text}$ such that $j-1 \in \RTwo{\tau}{\Text}$, it holds
  \begin{itemize}
  \item $\RootPos{j-1}{\tau}{\Text} = \RootPos{j}{\tau}{\Text}$,
  \item $\RunEndPos{j-1}{\tau}{\Text} = \RunEndPos{j}{\tau}{\Text}$,
  \item $\TailPos{j-1}{\tau}{\Text} = \TailPos{j}{\tau}{\Text}$,
  \item $\RunEndFullPos{j-1}{\tau}{\Text} = \RunEndFullPos{j}{\tau}{\Text}$,
  \item $\TypePos{j-1}{\tau}{\Text} = \TypePos{j}{\tau}{\Text}$.
  \end{itemize}
\end{lemma}

\begin{lemma}[{\cite{collapsing}}]\label{lm:end}
  Let $\Text \in \Sigma^{\Textlen}$ and $\tau \in [1 \dd \floor{\frac{\Textlen}{2}}]$. For every $j \in
  \RTwo{\tau}{\Text}$, it holds
  \[
    \RunEndPos{j}{\tau}{\Text} =
      \max\{j' \in [j \dd \Textlen] : [j \dd j'] \subseteq \RTwo{\tau}{\Text}\} +
      3\tau - 1.
  \]
\end{lemma}

\begin{lemma}[{\cite{collapsing}}]\label{lm:periodic-pos-lce}
  Let $\Text \in \Sigma^{\Textlen}$, $\tau \in [1 \dd \lfloor \tfrac{\Textlen}{2} \rfloor]$, and
  $j \in [1 \dd \Textlen]$.
  \begin{enumerate}
  \item\label{lm:periodic-pos-lce-it-1}
    Let $\Pat \in \Sigma^{+}$ be a $\tau$-periodic pattern. Then,
    the following conditions are equivalent:
    \begin{itemize}
    \item $\lcp{\Pat}{\Text[j \dd \Textlen]} \geq 3\tau - 1$,
    \item $j \in \RTwo{\tau}{\Text}$, $\RootPos{j}{\tau}{\Text} =
      \RootPat{\Pat}{\tau}$, and $\HeadPos{j}{\tau}{\Text} =
      \HeadPat{\Pat}{\tau}$.
    \end{itemize}
    Moreover, if, letting $t = \RunEndPat{\Pat}{\tau} - 1$, it holds
    $\lcp{\Pat}{\Text[j \dd \Textlen]} > t$, then:
    \begin{itemize}
    \item $\RunEndPat{\Pat}{\tau} - 1 = \RunEndPos{j}{\tau}{\Text} - j$,
    \item $\TailPat{\Pat}{\tau} = \TailPos{j}{\tau}{\Text}$,
    \item $\RunEndFullPat{\Pat}{\tau} - 1 =
      \RunEndFullPos{j}{\tau}{\Text} - j$,
    \item $\ExpPat{\Pat}{\tau} = \ExpPos{j}{\tau}{\Text}$,
    \item $\TypePat{\Pat}{\tau} = \TypePos{j}{\tau}{\Text}$.
    \end{itemize}
  \item\label{lm:periodic-pos-lce-it-2}
    Let $j' \in \RTwo{\tau}{\Text}$. Then, the following conditions
    are equivalent:
    \begin{itemize}
    \item $\LCE{\Text}{j'}{j} \geq 3\tau - 1$,
    \item $j \in \RTwo{\tau}{\Text}$,
      $\RootPos{j}{\tau}{\Text} = \RootPos{j'}{\tau}{\Text}$, and
      $\HeadPos{j'}{\tau}{\Text} = \HeadPos{j}{\tau}{\Text}$.
    \end{itemize}
    Moreover, if letting $t = \RunEndPos{j}{\tau}{\Text} - j$, it holds
    $\LCE{\Text}{j}{j'} > t$, then:
    \begin{itemize}
    \item $\RunEndPos{j'}{\tau}{\Text} - j' = \RunEndPos{j}{\tau}{\Text} - j$,
    \item $\TailPos{j'}{\tau}{\Text} = \TailPos{j}{\tau}{\Text}$,
    \item $\RunEndFullPos{j'}{\tau}{\Text} - j' =
      \RunEndFullPos{j}{\tau}{\Text} - j$,
    \item $\ExpPos{j'}{\tau}{\Text} = \ExpPos{j}{\tau}{\Text}$,
    \item $\TypePos{j'}{\tau}{\Text} = \TypePos{j}{\tau}{\Text}$.
    \end{itemize}
  \end{enumerate}
\end{lemma}

\begin{lemma}[{\cite{collapsing}}]\label{lm:R-lex-block-pos}
  Let $\Text \in \Sigma^{\Textlen}$, $\tau \in [1 \dd \floor{\frac{\Textlen}{2}}]$,
  and $j \in \RTwo{\tau}{\Text}$. Let
  $j' \in \RTwo{\tau}{\Text}$ be such that $\RootPos{j'}{\tau}{\Text} =
  \RootPos{j}{\tau}{\Text}$ and $\HeadPos{j'}{\tau}{\Text} =
  \HeadPos{j}{\tau}{\Text}$.  Then, letting $t_1 =
  \RunEndPos{j}{\tau}{\Text} - j$ and $t_2 = \RunEndPos{j'}{\tau}{\Text} - j'$,
  it holds $\LCE{\Text}{j}{j'} \geq \min(t_1, t_2)$ and:
  \begin{enumerate}
  \item\label{lm:R-lex-block-pos-it-1}
    If $\TypePos{j}{\tau}{\Text} \neq \TypePos{j'}{\tau}{\Text}$ or $t_1
    \neq t_2$, then $\LCE{\Text}{j}{j'} = \min(t_1, t_2)$,
  \item\label{lm:R-lex-block-pos-it-2}
    If $\TypePos{j}{\tau}{\Text} \neq \TypePos{j'}{\tau}{\Text}$, then
    $\Text[j \dd \Textlen] \prec \Text[j' \dd \Textlen]$ iff %
    $\TypePos{j}{\tau}{\Text} <
    \TypePos{j'}{\tau}{\Text}$,
  \item\label{lm:R-lex-block-pos-it-3}
    If $\TypePos{j}{\tau}{\Text} = \TypePos{j'}{\tau}{\Text} = -1$ and
    $t_1 \neq t_2$, then $t_1 < t_2$ iff %
    $\Text[j \dd \Textlen] \prec \Text[j'
    \dd \Textlen]$,
  \item\label{lm:R-lex-block-pos-it-4}
    If $\TypePos{j}{\tau}{\Text} = \TypePos{j'}{\tau}{\Text} = +1$ and
    $t_1 \neq t_2$, then $t_1 < t_2$ iff %
    $\Text[j \dd \Textlen] \succ \Text[j'
    \dd \Textlen]$.
  \end{enumerate}
\end{lemma}

\begin{lemma}[{\cite{collapsing}}]\label{lm:run-end}
  Let $\Text \in \Sigma^{\Textlen}$ and
  $\tau \in [1 \dd \floor{\frac{\Textlen}{2}}]$.  Let $j, j',
  j'' \in [1 \dd \Textlen]$ be such that $j, j'' \in \RTwo{\tau}{\Text}$, $j'
  \not\in \RTwo{\tau}{\Text}$, and $j < j' < j''$. Then, it holds
  $\RunEndPos{j}{\tau}{\Text} \leq j'' + \tau - 1$.
\end{lemma}

\begin{lemma}[{\cite[Section~5.3.2]{breaking}}]\label{lm:runs}
  For every $\Text \in \Sigma^{\Textlen}$ and
  $\tau \in [1 \dd \lfloor \tfrac{\Textlen}{2} \rfloor]$,
  it holds $|\RPrimTwo{\tau}{\Text}| \leq \tfrac{2\Textlen}{\tau}$
  and $\sum_{i \in \RPrimTwo{\tau}{\Text}} \RunEndPos{i}{\tau}{\Text} - i \leq 2\Textlen$.
\end{lemma}

\begin{lemma}[{\cite{collapsing}}]\label{lm:efull}
  Let $\Text \in \Sigma^{\Textlen}$ and
  $\tau \in [1 \dd \floor{\frac{\Textlen}{2}}]$.
  For any $j, j' \in
  \RPrimTwo{\tau}{\Text}$, $j \neq j'$ implies $\RunEndFullPos{j}{\tau}{\Text} \neq
  \RunEndFullPos{j'}{\tau}{\Text}$.
\end{lemma}

\begin{lemma}[{\cite{collapsing}}]\label{lm:RskH}
  Let $\Text \in \Sigma^{\Textlen}$ and
  $\tau \in [1 \dd \lfloor \tfrac{\Textlen}{2} \rfloor]$.
  Let $H \in \Sigma^{+}$, $p = |H|$, $s \in [0 \dd p)$, and
  $k_{\min} = \lceil \tfrac{3\tau-1-s}{p} \rceil - 1$.
  For every $k \in (k_{\min} \dd \Textlen]$, it holds
  \[
    \RMinusFive{s}{k}{H}{\tau}{\Text} =
      \{\RunEndFullPos{j}{\tau}{\Text} - s - kp :
      j \in \RPrimMinusThree{H}{\tau}{\Text}
      \text{ and }
      s + kp \leq \RunEndFullPos{j}{\tau}{\Text} - j\}.
  \]
\end{lemma}

\begin{lemma}[{\cite{hierarchy}}]\label{lm:RskH-size}
  Let $\Text \in \Sigma^{\Textlen}$ and
  $\tau \in [1 \dd \lfloor \tfrac{\Textlen}{2} \rfloor]$.
  Let $H \in \Sigma^{+}$, $p = |H|$, $s \in [0 \dd p)$, and
  $k_{\min} = \lceil \tfrac{3\tau-1-s}{p} \rceil - 1$.
  For every $k \in (k_{\min} \dd \Textlen)$, it holds
  \[
    |\RFive{s}{k}{H}{\tau}{\Text}| \geq |\RFive{s}{k+1}{H}{\tau}{\Text}|.
  \]
\end{lemma}

\begin{definition}[Anchored lex-orderings of $\tau$-runs]\label{def:runs-minus-lex-sorted}
  Let $\Text \in \Sigma^{\Textlen}$ and
  $\tau \in [1 \dd \lfloor \tfrac{\Textlen}{2} \rfloor]$.
  \begin{itemize}
  \item Letting $m = |\RPrimMinusTwo{\tau}{\Text}|$,
    by $\RunsMinusLexSortedTwo{\tau}{\Text}$, we denote
    a sequence $(a_i)_{i \in [1 \dd m]}$
    containing all elements of $\RPrimMinusTwo{\tau}{\Text}$ such
    that for every $i, j \in [1 \dd m]$, $i < j$ implies
    $\RootPos{a_i}{\tau}{\Text} \prec \RootPos{a_j}{\tau}{\Text}$,
    or $\RootPos{a_i}{\tau}{\Text} = \RootPos{a_j}{\tau}{\Text}$ and
    $\Text[\RunEndFullPos{a_i}{\tau}{\Text} \dd \Textlen] \prec
    \Text[\RunEndFullPos{a_j}{\tau}{\Text} \dd \Textlen]$.
  \item For
    every $H \in \Sigma^{+}$, letting
    $m_H = |\RPrimMinusThree{H}{\tau}{\Text}|$, by
    $\RunsMinusLexSortedThree{H}{\tau}{\Text}$, we denote
    a sequence $(a_i)_{i \in [1 \dd m_{H}]}$
    containing all elements of $\RPrimMinusThree{H}{\tau}{\Text}$ such
    that for every $i, j \in [1 \dd m_{H}]$, $i < j$ implies
    $\Text[\RunEndFullPos{a_i}{\tau}{\Text} \dd \Textlen] \prec
    \Text[\RunEndFullPos{a_j}{\tau}{\Text} \dd \Textlen]$.
  \end{itemize}
\end{definition}

\begin{definition}[Text orderings of $\tau$-runs]\label{def:runs-minus-text-sorted}
  Let $\Text \in \Sigma^{\Textlen}$ and
  $\tau \in [1 \dd \lfloor \tfrac{\Textlen}{2} \rfloor]$.
  \begin{itemize}
  \item Letting $m = |\RPrimMinusTwo{\tau}{\Text}|$,
    by $\RunsMinusTextSortedTwo{\tau}{\Text}$, we denote
    a sequence $(a_i)_{i \in [1 \dd m]}$
    containing all elements of $\RPrimMinusTwo{\tau}{\Text}$ such
    that for every $i, j \in [1 \dd m]$, $i < j$ implies
    $\RootPos{a_i}{\tau}{\Text} \prec \RootPos{a_j}{\tau}{\Text}$,
    or $\RootPos{a_i}{\tau}{\Text} = \RootPos{a_j}{\tau}{\Text}$ and $a_i < a_j$.
  \item For every $H \in \Sigma^{+}$, letting
    $m_{H} = |\RPrimMinusThree{H}{\tau}{\Text}|$,
    by $\RunsMinusTextSortedThree{H}{\tau}{\Text}$, we denote
    a sequence $(a_i)_{i \in [1 \dd m_{H}]}$
    containing all elements of $\RPrimMinusThree{H}{\tau}{\Text}$ such
    that for every $i, j \in [1 \dd m_{H}]$, $i < j$ implies $a_i < a_j$.
  \end{itemize}
\end{definition}

\begin{definition}[Length-based ordering of $\tau$-runs]\label{def:runs-minus-length-sorted}
  Let $\Text \in \Sigma^{\Textlen}$ and
  $\tau \in [1 \dd \lfloor \tfrac{\Textlen}{2} \rfloor]$.
  Let $m = |\RPrimMinusTwo{\tau}{\Text}|$. By
  $\RunsMinusLengthSortedTwo{\tau}{\Text}$, we denote
  a sequence $(a_i)_{i \in [1 \dd m]}$
  containing all elements of $\RPrimMinusTwo{\tau}{\Text}$ such
  that for every $i, j \in [1 \dd m]$, $i < j$ implies
  that one of the following three conditions hold:
  \begin{itemize}
  \item
    $\RootPos{a_i}{\tau}{\Text} \prec \RootPos{a_j}{\tau}{\Text}$,
  \item
    $\RootPos{a_i}{\tau}{\Text} = \RootPos{a_j}{\tau}{\Text}$ and
    $\RunEndFullPos{a_i}{\tau}{\Text} - a_i < \RunEndFullPos{a_j}{\tau}{\Text} - a_j$,
  \item
    $\RootPos{a_i}{\tau}{\Text} = \RootPos{a_j}{\tau}{\Text}$,
    $\RunEndFullPos{a_i}{\tau}{\Text} - a_i = \RunEndFullPos{a_j}{\tau}{\Text} - a_j$, and
    $a_i < a_j$.
  \end{itemize}
\end{definition}

\begin{definition}[Exponent-block indicator bitvector]\label{def:exp-block-bitvector-minus}
  Let $\Text \in \Sigma^{\Textlen}$ and
  $\tau \in [1 \dd \floor{\tfrac{\Textlen}{2}}]$.
  By \[\ExpBlockBitvectorMinusTwo{\tau}{\Text},\] we denote a bitvector $B[1 \dd \Textlen]$
  defined such that for every $i \in [1 \dd \Textlen]$, $B[i] = \zero$ holds if and only if
  $\SA{\Text}[i] \in [1 \dd \Textlen] \setminus \RMinusTwo{\tau}{\Text}$, or
  $i < \Textlen$ and the positions $j = \SA{\Text}[i]$ and $j' = \SA{\Text}[i+1]$ satisfy
  $j,j' \in \RMinusFive{s}{k}{H}{\tau}{\Text}$ for some $s \in \Zn$, $k \in \Zp$, and $H \in \Sigma^{+}$.
\end{definition}

\begin{definition}[Leftmost representatives of $\tau$-run occurrences]\label{def:rmin}
  For every $\Text \in \Sigma^{\Textlen}$ and
  $\tau \in [1 \dd \lfloor \tfrac{\Textlen}{2} \rfloor]$, we
  define
  \[
    \RMinMinusTwo{\tau}{\Text} := \{j \in \RMinusTwo{\tau}{\Text}
      : j = \min \OccTwo{\Text[j \dd \RunEndPos{j}{\tau}{\Text})}{\Text}
                 \cap \RMinusTwo{\tau}{\Text}\}.
  \]
\end{definition}

\begin{definition}[Leftmost $\tau$-run indicator bitvectors]\label{def:min-pos-bitvector-minus}
  Let $\Text \in \Sigma^{\Textlen}$ and
  $\tau \in [1 \dd \floor{\tfrac{\Textlen}{2}}]$. By
  \[\MinPosBitvectorMinusTwo{\tau}{\Text},\] we denote a bitvector $B[1 \dd \Textlen]$
  defined such that for
  every $i \in [1 \dd \Textlen]$,
  \[
    B[i] =
      \begin{cases}
       1 & \text{if }\SA{\Text}[i] \in \RMinMinusTwo{\tau}{\Text},\\
       0 & \text{otherwise}.\\
   \end{cases}
  \]
  Moreover, for every $H \in \Sigma^{+}$, $s \in [0 \dd |H|)$, and $k \in \Zp$,
  we then denote
  \begin{align*}
    \MinPosBitvectorMinusFour{s}{H}{\tau}{\Text}
      &= B(x \dd y],\\
    \MinPosBitvectorMinusFive{s}{k}{H}{\tau}{\Text}
      &= B(x' \dd y'],\\
  \end{align*}
  where $x, y, x', y' \in [0 \dd \Textlen]$ are such that
  \begin{align*}
    \{\SA{\Text}[i] : i \in (x \dd y]\}
      &= \RMinusFour{s}{H}{\tau}{\Text},\\
    \{\SA{\Text}[i] : i \in (x' \dd y']\}
      &= \RMinusFive{s}{k}{H}{\tau}{\Text}.
  \end{align*}
\end{definition}

\begin{remark}\label{rm:min-pos-bitvector-minus}
  To show that $x$ and $y$ in \cref{def:min-pos-bitvector-minus} are well-defined, note
  that by \cref{lm:periodic-pos-lce}\eqref{lm:periodic-pos-lce-it-2},
  the set $\RFour{s}{H}{\tau}{\Text}$ occupies a contiguous block
  of positions in $\SA{\Text}$. Moreover, by \cref{lm:R-lex-block-pos}\eqref{lm:R-lex-block-pos-it-2}, all
  positions $j \in \RFour{s}{H}{\tau}{\Text}$ satisfying
  $\TypePos{j}{\tau}{\Text} = -1$ precede positions
  $j' \in \RFour{s}{H}{\tau}{\Text}$ satisfying
  $\TypePos{j'}{\tau}{\Text} = +1$. In order words, positions in
  $\RMinusFour{s}{H}{\tau}{\Text}$ occupy a contiguous
  block in $\SA{\Text}$. Thus, $x,y$ are indeed well-defined.

  To show that $x'$ and $y'$ in \cref{def:min-pos-bitvector-minus} are well-defined,
  observe now that letting
  $a,b \in [0 \dd \Textlen]$ be such that $\{\SA{\Text}[i] : i \in (a \dd b]\} = \RMinusFour{s}{H}{\tau}{\Text}$,
  it follows by \cref{lm:R-lex-block-pos}\eqref{lm:R-lex-block-pos-it-3} that for
  every $i \in (a \dd b)$,
  it holds $\RunEndPos{\SA{\Text}[i]}{\tau}{\Text} - \SA{\Text}[i] \leq \RunEndPos{\SA{\Text}[i+1]}{\tau}{\Text} - \SA{\Text}[i+1]$.
  By $\HeadPos{\SA{\Text}[i]}{\tau}{\Text} = \HeadPos{\SA{\Text}[i+1]}{\tau}{\Text} = s$, we thus obtain
  \begin{align*}
    \ExpPos{\SA{\Text}[i]}{\tau}{\Text}
      &= \lfloor \tfrac{\RunEndPos{\SA{\Text}[i]}{\tau}{\Text} - \SA{\Text}[i] - s}{p} \rfloor\\
      &\leq \lfloor \tfrac{\RunEndPos{\SA{\Text}[i+1]}{\tau}{\Text} - \SA{\Text}[i+1] - s}{p} \rfloor\\
      &= \ExpPos{\SA{\Text}[i+1]}{\tau}{\Text}.
  \end{align*}
  Thus, all positions from $\RMinusFive{s}{k}{H}{\tau}{\Text}$ occupy
  a contiguous block in $\SA{\Text}$. The
  values $x',y'$ are thus indeed well-defined.
\end{remark}

\begin{lemma}[{\cite{sublinearlz}}]\label{lm:bmin-bit}
  Let $\Text \in \Sigma^{\Textlen}$ and
  $\tau \in [1 \dd \floor{\tfrac{\Textlen}{2}}]$. Let
  $H \in \Sigma^{+}$, $s \in \Zn$, and $b, e \in [0 \dd \Textlen]$ be
  such that $b < e$ and
  $\{\SA{\Text}[i] : i \in (b \dd e]\} = \RMinusFour{s}{H}{\tau}{\Text}$.
  Let $\BitvectorMin[1 \dd \Textlen] = \MinPosBitvectorMinusTwo{\tau}{\Text}$
  (\cref{def:min-pos-bitvector-minus}). If $\BitvectorMin[i] = 1$ for
  some $i \in (b \dd e]$, then for every $i' \in (i \dd e]$, it holds
  \[\SA{\Text}[i'] > \SA{\Text}[i].\]
\end{lemma}

\subparagraph{Basic Navigational Primitives}

\begin{proposition}[{\cite{breaking,sublinearlz}}]\label{pr:periodic-constructions}
  Let $\Text \in \IntegerAlphabet^{\Textlen}$ be such that
  $2 \leq \AlphabetSize < \Textlen^{1/13}$ and $\Text[\Textlen]$ does not
  occur in $\Text[1 \dd \Textlen)$. Let $\tau = \lfloor \mu\log_{\AlphabetSize} \Textlen \rfloor$, where
  $\mu$ is a positive constant smaller than $\tfrac{1}{12}$ such that $\tau \geq 1$. Given
  the packed representation of $\Text$, we can in
  $\bigO(\Textlen / \log_{\AlphabetSize} \Textlen)$ time
  construct:
  \begin{enumerate}
  \item\label{pr:periodic-constructions-it-1}
    the packed representation of bitvector $\ExpBlockBitvectorMinusTwo{\tau}{\Text}$ (\cref{def:exp-block-bitvector-minus}),
  \item\label{pr:periodic-constructions-it-2}
    the packed representation of bitvector $\MinPosBitvectorMinusTwo{\tau}{\Text}$ (\cref{def:min-pos-bitvector-minus}),
  \item\label{pr:periodic-constructions-it-3}
    the sequence $\RunsMinusLexSortedTwo{\tau}{\Text}$ (\cref{def:runs-minus-lex-sorted}),
  \end{enumerate}
\end{proposition}
\begin{proof}
  The construction proceeds as follows:
  \begin{enumerate}
  \item This follows by combining Proposition~5.3 and Proposition~5.15 from~\cite{breaking}.
  \item This follows by combining Proposition~5.40 and Proposition~5.71 from~\cite{sublinearlz}.
  \item This follows again by combining Proposition~5.3 and Proposition~5.15 from~\cite{breaking}.
  \qedhere
  \end{enumerate}
\end{proof}

\begin{proposition}\label{pr:nav-index-periodic}
  Let $\Text \in \IntegerAlphabet^{\Textlen}$ be such that
  $2 \leq \AlphabetSize < \Textlen^{1/13}$ and $\Text[\Textlen]$ does not
  occur in $\Text[1 \dd \Textlen)$. Let $\tau = \lfloor \mu\log_{\AlphabetSize} \Textlen \rfloor$, where
  $\mu$ is a positive constant smaller than $\tfrac{1}{12}$ such that $\tau \geq 1$.
  Denote $q = |\RPrimMinusTwo{\tau}{\Text}|$ and
  $(a_i)_{i \in [1 \dd q]} = \RunsMinusLexSortedTwo{\tau}{\Text}$ (\cref{def:runs-minus-lex-sorted}).
  Let also $A_{\rm pos}[1 \dd q]$ and $A_{\rm len}[1 \dd q]$ be arrays defined by
  \begin{itemize}
  \item $A_{\rm pos}[i] = \RunEndFullPos{a_i}{\tau}{\Text}$,
  \item $A_{\rm len}[i] = \RunEndFullPos{a_i}{\tau}{\Text} - a_i$.
  \end{itemize}
  Given the packed representation of $\Text$, we can in
  $\bigO(\Textlen / \log_{\AlphabetSize} \Textlen)$ time
  construct a data structure, denoted $\NavPeriodic{\tau}{\Text}$,
  that answers the following queries:
  \begin{enumerate}
  \item\label{pr:nav-index-periodic-it-1}
    Given any $i \in [1 \dd \Textlen]$ such that $\SA{\Text}[i] \in \RTwo{\tau}{\Text}$, compute
    $\SA{\Text}[i]$ in $\bigO(\log \log \Textlen)$ time.
  \item\label{pr:nav-index-periodic-it-2}
    Given any $j \in \RTwo{\tau}{\Text}$, in $\bigO(1)$ time compute:
    \begin{itemize}
    \item $\HeadPos{j}{\tau}{\Text}$,
    \item $|\RootPos{j}{\tau}{\Text}|$,
    \item $\ExpPos{j}{\tau}{\Text}$,
    \item $\RunEndFullPos{j}{\tau}{\Text}$,
    \item integers $x,y,z \in [0 \dd \Textlen]$ satisfying
      $x \leq y \leq z$, 
      $\{\SA{\Text}[i]\}_{i \in (x \dd y]} = \RMinusFour{s}{H}{\tau}{\Text}$, and
      $\{\SA{\Text}[i]\}_{i \in (y \dd z]} = \RPlusFour{s}{H}{\tau}{\Text}$,
      where $s = \HeadPos{j}{\tau}{\Text}$ and $H = \RootPos{j}{\tau}{\Text}$.
    \end{itemize}
  \item\label{pr:nav-index-periodic-it-3}
    Given any $j \in \RMinusTwo{\tau}{\Text}$, in $\bigO(1)$ time compute
    \begin{itemize}
    \item index $i \in [1 \dd q]$ satisfying $\RunEndFullPos{j}{\tau}{\Text} = \RunEndFullPos{a_i}{\tau}{\Text}$,
    \item integers $x,y \in [0 \dd q]$ satisfying
      $x = \textstyle\sum_{H' \in \IntegerAlphabet^{+} : H' \prec H} |\RPrimMinusThree{H'}{\tau}{\Text}|$ and
      $y = x + |\RPrimMinusThree{H}{\tau}{\Text}|$, where $H = \RootPos{j}{\tau}{\Text}$.
    \end{itemize}
  \item\label{pr:nav-index-periodic-it-pat-range}
    Given the packed representation of any $\tau$-periodic (\cref{def:periodic-pattern})
    pattern $\Pat \in \IntegerAlphabet^{+}$, return the pair
    \[
      (\RangeBegTwo{\Pat}{\Text}, \RangeEndTwo{\Pat}{\Text})
    \]
    in $\bigO(\log \log \Textlen + |\Pat| / \log_{\AlphabetSize} \Textlen)$ time.
  \item\label{pr:nav-index-periodic-it-rmq}
    Given any $b,e \in [0 \dd q]$ and $v \geq 0$ such that $\ThreeSidedRMQ{A_{\rm pos}}{A_{\rm len}}{b}{e}{v} \neq \infty$
    (see \cref{def:three-sided-rmq}), return
    \[
      A_{\rm pos}[\ThreeSidedRMQ{A_{\rm pos}}{A_{\rm len}}{b}{e}{v}]
    \]
    in $\bigO(\log \log \Textlen)$ time.
  \item\label{pr:nav-index-periodic-it-reporting}
    Given any $b,e \in [0 \dd q]$ and $v \geq 0$, return
    \[
      \mathcal{I} = \{i \in (b \dd e] : \RunEndFullPos{a_i}{\tau}{\Text} - a_i \geq v\}
    \]
    in $\bigO(1 + |\mathcal{I}|)$ time.
  \end{enumerate}
\end{proposition}
\begin{proof}

  Let $L_{\rm runs}$ be a mapping defined such that for every
  $H \in \IntegerAlphabet^{\leq \tau}$ satisfying $\RMinusThree{H}{\tau}{\Text} \neq \emptyset$,
  $L_{\rm runs}$ maps $X$ to the pair of integers $(x,y)$ defined by
  $x = \sum_{H' \in \IntegerAlphabet^{+} : H' \prec H} |\RPrimMinusThree{H'}{\tau}{\Text}|$ and
  $y = x + |\RPrimMinusThree{H}{\tau}{\Text}|$. Denote $q_{\max} = 2\Textlen / \tau$.

  \DSComponents
  The structure consists of the following components:
  \begin{enumerate}
  \item The components of the indexes (CSA, CST, and pattern matching index) from
    Sections 5.3.2, 6.3.2, 7.3.1\ in~\cite{breaking}. All these structures need
    $\bigO(\Textlen / \log_{\AlphabetSize} \Textlen)$ space.
  \item The lookup table $L_{\rm runs}$. When accessing $L_{\rm runs}$, each $H \in \IntegerAlphabet^{\leq \tau}$
    is converted to the integer $\Int{\tau}{\AlphabetSize}{H}$ (\cref{def:int}). By
    $\Int{\tau}{\AlphabetSize}{H} \in [0 \dd \AlphabetSize^{2\tau})$, this component
    needs $\bigO(\AlphabetSize^{2\tau}) = \bigO(\Textlen^{2\mu}) = \bigO(\Textlen^{1/6}) =
    \bigO(\Textlen / \log_{\AlphabetSize} \Textlen)$ space.
  \item The plain representations of arrays $A_{\rm pos}[1 \dd q]$ and $A_{\rm len}[1 \dd q]$
    augmented with the data structure from \cref{th:three-sided-rmq}.
    Observe that we can use \cref{th:three-sided-rmq} since, by \cref{lm:runs}, it holds
    $q \leq q_{\max}$,
    $\max_{i=1}^{q} A_{\rm pos}[i] \leq \Textlen = \bigO(q_{\max} \log q_{\max})$,
    and
    $\sum_{i=1}^{q} A_{\rm len}[i] \leq 2\Textlen = \bigO(q_{\max} \log q_{\max})$.
    The arrays need
    $\bigO(q) = \bigO(\tfrac{\Textlen}{\tau}) = \bigO(\Textlen / \log_{\AlphabetSize} \Textlen)$ space, and
    the structure from \cref{th:three-sided-rmq} needs
    $\bigO(q_{\max}) = \bigO(\Textlen / \log_{\AlphabetSize} \Textlen)$ space.
  \item The array $A_{\rm len}[1 \dd q]$ augmented with the data structure from
    \cref{cr:two-sided-range-reporting}. By the above analysis,
    we have $\sum_{i=1}^{q} A_{\rm len}[i] = \bigO(q_{\max} \log q_{\max})$,
    where $q \leq q_{\max} = \bigO(\Textlen / \tau)$. Thus, this component
    needs $\bigO(q_{\max}) = \bigO(\Textlen / \log_{\AlphabetSize} \Textlen)$ space.
  \end{enumerate}
  In total, the structure needs
  $\bigO(\Textlen / \log_{\AlphabetSize} \Textlen)$ space.

  \DSQueries
  The queries are answered as follows:
  \begin{enumerate}
  \item Given any $i \in [1 \dd \Textlen]$ such that $\SA{\Text}[i] \in \RTwo{\tau}{\Text}$,
    we can compute $\SA{\Text}[i]$ in $\bigO(\log \log \Textlen)$ time using the
    first component of the above structure (\cite[Proposition~5.14]{breaking}).
  \item The queries are standard navigational queries for periodic positions;
    see~\cite[Proposition~5.7]{breaking}.
    The indices $x,y,z$ are computed with the help of rank/select queries on the
    bitvector $\ExpBlockBitvectorMinusTwo{\tau}{\Text}$
    (\cref{def:exp-block-bitvector-minus}) and its symmetric version (adapted
    for positions in $\RPlusTwo{\tau}{\Text}$);
    see the proof of~\cite[Proposition~5.9]{breaking}.
  \item The first query is again a simple application of the first component
    (see the proof of~\cite[Proposition~5.10]{breaking} for
    a similar computation). The second query is
    answered using the lookup table $L_{\rm runs}$ (defined above).
  \item Given the packed representation of any $\tau$-periodic pattern
    $\Pat \in \IntegerAlphabet^{+}$, we compute the pair
    $(\RangeBegTwo{\Pat}{\Text}, \allowbreak\RangeEndTwo{\Pat}{\Text})$
    in $\bigO(\log \log \Textlen + |\Pat| / \log_{\AlphabetSize} \Textlen)$ time
    using the first component of the structure,
    as described in~\cite[Proposition~6.12]{breaking}.
  \item Given $b,e \in [0 \dd q]$ and $v \geq 0$ such that
    $\ThreeSidedRMQ{A_{\rm pos}}{A_{\rm len}}{b}{e}{v} \neq \infty$, we compute
    the value $A_{\rm pos}[\ThreeSidedRMQ{A_{\rm pos}}{A_{\rm len}}{b}{e}{v}]$ using
    \cref{th:three-sided-rmq}
    in $\bigO(\log \log q) = \bigO(\log \log \Textlen)$ time.
  \item Given any $b,e \in [0 \dd q]$ and $v \geq 0$, we compute
    the set $\mathcal{I} = \{i \in (b \dd e] : \RunEndFullPos{a_i}{\tau}{\Text} - a_i \geq v\}$
    using \cref{cr:two-sided-range-reporting} (applied to array $A_{\rm len}$)
    in $\bigO(1 + |\mathcal{I}|)$ time.
  \end{enumerate}

  \DSConstruction
  The components of the structure are constructed as follows:
  \begin{enumerate}
  \item The first component is constructed by combining Propositions
    5.3, 5.15, 6.3, 6.13, 7.7, and 7.23 from~\cite{breaking}. In total, this
    takes $\bigO(\Textlen / \log_{\AlphabetSize} \Textlen)$ time.
  \item The lookup table is constructed in $\bigO(\Textlen / \log_{\AlphabetSize} \Textlen)$ time
    similarly as the closely related lookup table $L_{\rm runs}$ in the proof
    of~\cite[Proposition~5.15]{breaking}.
  \item The third component is constructed as follows:
    \begin{enumerate}
    \item Using \cref{pr:periodic-constructions}\eqref{pr:periodic-constructions-it-3},
      compute the sequence $(a_i)_{i \in [1 \dd q]} = \RunsMinusLexSortedTwo{\tau}{\Text}$
      (\cref{def:runs-minus-lex-sorted}) in
      $\bigO(\Textlen / \log_{\AlphabetSize} \Textlen)$ time.
    \item With the help of the first component, we can then compute
      arrays $A_{\rm pos}[1 \dd q]$ and $A_{\rm len}[1 \dd q]$
      in $\bigO(q) = \bigO(\Textlen / \log_{\AlphabetSize} \Textlen)$ time.
    \item Apply \cref{th:three-sided-rmq} to arrays
      $A_{\rm pos}[1 \dd q]$ and $A_{\rm len}[1 \dd q]$
      in $\bigO(q_{\max}) = \bigO(\Textlen / \log_{\AlphabetSize} \Textlen)$ time.
    \end{enumerate}
    In total, the construction of the third component
    takes $\bigO(\Textlen / \log_{\AlphabetSize} \Textlen)$ time.
  \item First, compute the array $A_{\rm len}[1 \dd q]$ as described above.
    Then, apply \cref{cr:two-sided-range-reporting} to array
    $A_{\rm len}[1 \dd q]$ in $\bigO(q_{\max}) =
    \bigO(\Textlen / \log_{\AlphabetSize} \Textlen)$ time.
  \end{enumerate}
  In total, the construction takes
  $\bigO(\Textlen / \log_{\AlphabetSize} \Textlen)$ time.
\end{proof}

\subparagraph{Combinatorial Properties}

\begin{lemma}\label{lm:exp-subblock-characterization}
  Let $\Text \in \Sigma^{\Textlen}$ and $\tau \in [1 \dd \lfloor \tfrac{\Textlen}{2} \rfloor]$.
  Let $i_1,i_2 \in [1 \dd \Textlen]$ be such that $i_1 \leq i_2$ and
  $\{\SA{\Text}[i]\}_{i \in [i_1 \dd i_2]} \subseteq \RMinusFive{s}{k}{H}{\tau}{\Text}$
  for some $s \in \Zn$, $k \in \Zp$, and $H \in \Sigma^{+}$.
  Denote $p = |H|$, $\deltatext = s + kp$, and
  $(a_j)_{j \in [1 \dd q]} = \RunsMinusLexSortedTwo{\tau}{\Text}$ (\cref{def:runs-minus-lex-sorted}).
  Let $j_1,j_2 \in [1 \dd q]$ be such that
  for $k \in \{1,2\}$,
  $\RunEndFullPos{\SA{\Text}[i_k]}{\tau}{\Text} = \RunEndFullPos{a_{j_k}}{\tau}{\Text}$.
  Then, it holds
  \[
    \{\SA{\Text}[i]\}_{i \in [i_1 \dd i_2]} =
        \{\RunEndFullPos{a_j}{\tau}{\Text} - \deltatext :
          j \in [j_1 \dd j_2]\text{ and }
          \RunEndFullPos{a_j}{\tau}{\Text} - a_j \geq \deltatext\}.
  \]
\end{lemma}
\begin{proof}

  Denote $A = \{\RunEndFullPos{a_j}{\tau}{\Text} - \deltatext :
  j \in [j_1 \dd j_2]\text{ and }
  \RunEndFullPos{a_j}{\tau}{\Text} - a_j \geq \deltatext\}$.
  Before proving the claim, i.e., that $\{\SA{\Text}[i]\}_{i \in [i_1 \dd i_2]} = A$, we first establish
  some auxiliary properties:
  \begin{itemize}
  \item First, we show that for $k \in \{1,2\}$, it holds $a_{j_k} \leq \SA{\Text}[i_k]$ and
    $[a_{j_k} \dd \SA{\Text}[i_k]] \subseteq \RMinusTwo{\tau}{\Text}$. Let $k \in \{1,2\}$ and
    $t \in \RPrimMinusTwo{\tau}{\Text}$ be such that $[t \dd \SA{\Text}[i_k]] \subseteq \RMinusTwo{\tau}{\Text}$.
    If $t \neq a_{j_k}$, then by \cref{lm:efull}, we have
    $\RunEndFullPos{t}{\tau}{\Text} \neq \RunEndFullPos{a_{j_k}}{\tau}{\Text}$.
    On the other hand, by \cref{lm:R-text-block},
    $\RunEndFullPos{t}{\tau}{\Text} = \RunEndFullPos{\SA{\Text}[i_k]}{\tau}{\Text}$.
    Thus, we obtain $\RunEndFullPos{\SA{\Text}[i_k]}{\tau}{\Text} \neq \RunEndFullPos{a_{j_k}}{\tau}{\Text}$, which
    contradicts the assumption the claim. We thus must have $t = a_{j_k}$, i.e., $a_{j_k} \leq \SA{\Text}[i_k]$
    and
    \[
      [a_{j_k} \dd \SA{\Text}[i_k]] \subseteq \RMinusTwo{\tau}{\Text}.
    \]
  \item Second, observe that by the above property and
    \cref{lm:R-text-block}, for $k \in \{1,2\}$ it holds:
    \begin{itemize}
    \item $\RootPos{a_{j_k}}{\tau}{\Text} = \RootPos{\SA{\Text}[i_k]}{\tau}{\Text}$,
    \item $\RunEndPos{a_{j_1}}{\tau}{\Text} = \RunEndPos{\SA{\Text}[i_1]}{\tau}{\Text}$,
    \item $\RunEndFullPos{a_{j_1}}{\tau}{\Text} = \RunEndFullPos{\SA{\Text}[i_1]}{\tau}{\Text}$.
    \end{itemize}
  \item Next, observe that for every $i \in [i_1 \dd i_2]$, the assumption $\SA{\Text}[i] \in \RMinusFive{s}{k}{H}{\tau}{\Text}$
    implies that
    \[
      \RunEndFullPos{\SA{\Text}[i]}{\tau}{\Text} = \SA{\Text}[i] + s + kp = \SA{\Text}[i] + \deltatext.
    \]
  \item Next, we prove that it holds $\TailPos{a_{j_1}}{\tau}{\Text} + \deltatext \geq 3\tau - 1$.
    First, observe that, by definition, for every $y \in \RTwo{\tau}{\Text}$, it holds
    $\RunEndPos{y}{\tau}{\Text} - y \geq 3\tau - 1$. In particular,
    $\RunEndPos{\SA{\Text}[i_1]}{\tau}{\Text} - \SA{\Text}[i_1] \geq 3\tau - 1$.
    By the above,
    we can rewrite this as $\RunEndPos{a_{j_1}}{\tau}{\Text} - \SA{\Text}[i_1] \geq 3\tau - 1$.
    Substituting $\SA{\Text}[i_1] = \RunEndFullPos{\SA{\Text}[i_1]}{\tau}{\Text} - \deltatext =
    \RunEndFullPos{a_{j_1}}{\tau}{\Text} - \deltatext$ (see above), we thus obtain
    $\RunEndPos{a_{j_1}}{\tau}{\Text} - (\RunEndFullPos{a_{j_1}}{\tau}{\Text} - \deltatext) \geq 3\tau - 1$,
    which simplifies to
    \[
      \TailPos{a_{j_1}}{\tau}{\Text} + \deltatext \geq 3\tau - 1.
    \]
  \item Finally, we prove that
    $\TailPos{\SA{\Text}[i_1]}{\tau}{\Text} \leq
    \TailPos{\SA{\Text}[i_1+1]}{\tau}{\Text} \leq \dots \leq
    \TailPos{\SA{\Text}[i_2]}{\tau}{\Text}$.
    By \cref{lm:R-lex-block-pos}\eqref{lm:R-lex-block-pos-it-3},
    $\RunEndPos{\SA{\Text}[i_1]}{\tau}{\Text}-\SA{\Text}[i_1] \leq
    \RunEndPos{\SA{\Text}[i_1+1]}{\tau}{\Text}-\SA{\Text}[i_1+1]) \leq \dots \leq
    \RunEndPos{\SA{\Text}[i_2]}{\tau}{\Text}-\SA{\Text}[i_2]$.
    On the other hand, note that for every $i \in [i_1 \dd i_2]$,
    $\SA{\Text}[i] \in \RMinusFive{s}{k}{H}{\tau}{\Text}$ implies that 
    $\TailPos{\SA{\Text}[i]}{\tau}{\Text} =
    (\RunEndPos{\SA{\Text}[i]}{\tau}{\Text} - \SA{\Text}[i]) - \deltatext$.
    This yields
    \[
      \TailPos{\SA{\Text}[i_1]}{\tau}{\Text} \leq
      \TailPos{\SA{\Text}[i_1+1]}{\tau}{\Text} \leq \dots \leq
      \TailPos{\SA{\Text}[i_2]}{\tau}{\Text}.
    \]
  \end{itemize}

  We now return to the proof of the main claim.
  First, we prove that $\{\SA{\Text}[i]\}_{i \in [i_1 \dd i_2]} \subseteq A$.
  Let $x \in \{\SA{\Text}[i]\}_{i \in [i_1 \dd i_2]}$, and let $i \in [i_1 \dd i_2]$
  be such that $x = \SA{\Text}[i]$. Let $x' \in \RPrimMinusTwo{\tau}{\Text}$ be such
  that $[x' \dd x] \subseteq \RMinusTwo{\tau}{\Text}$. Since $(a_j)_{j \in [1 \dd q]}$ contains
  all elements of $\RPrimMinusTwo{\tau}{\Text}$ (see \cref{def:runs-minus-lex-sorted}), there
  exists $j \in [1 \dd q]$ such that $x' = a_j$.
  We will now prove that $j \in [j_1 \dd j_2]$. We proceed in two steps:
  \begin{enumerate}

  \item Above we observed that
    $\RootPos{a_{j_1}}{\tau}{\Text} = \RootPos{\SA{\Text}[i_1]}{\tau}{\Text} = H =
    \RootPos{\SA{\Text}[i_2]}{\tau}{\Text} = \RootPos{a_{j_2}}{\tau}{\Text}$.
    On the other hand, $[a_j \dd \SA{\Text}[i]] \subseteq \RMinusTwo{\tau}{\Text}$ and
    $\SA{\Text}[i] \in \RMinusFive{s}{k}{H}{\tau}{\Text}$ imply by \cref{lm:R-text-block} that
    $\RootPos{a_j}{\tau}{\Text} = \RootPos{\SA{\Text}[i]}{\tau}{\Text} = H$. Thus,
    \[
      \RootPos{a_{j_1}}{\tau}{\Text} = \RootPos{a_j}{\tau}{\Text} = \RootPos{a_{j_2}}{\tau}{\Text} = H.
    \]

  \item Second, we prove that
    $\Text[\RunEndFullPos{a_{j_1}}{\tau}{\Text} \dd \Textlen] \preceq
    \Text[\RunEndFullPos{a_j}{\tau}{\Text} \dd \Textlen] \preceq
    \Text[\RunEndFullPos{a_{j_2}}{\tau}{\Text} \dd \Textlen]$.
    To this end, first observe that by $\SA{\Text}[i_1],\SA{\Text}[i],\SA{\Text}[i_2] \in \RMinusFive{s}{k}{H}{\tau}{\Text}$,
    it follows that suffixes
    $\Text[\SA{\Text}[i_1] \dd \Textlen]$,
    $\Text[\SA{\Text}[i] \dd \Textlen]$, and
    $\Text[\SA{\Text}[i_2] \dd \Textlen]$ share a common
    prefix of length $s + kp = \deltatext$. Thus, by
    $i_1 \leq i \leq i_2$ and the definition of the suffix array, we have
    \[\Text[\SA{\Text}[i_1]+\deltatext \dd \Textlen] \preceq
    \Text[\SA{\Text}[i]+\deltatext \dd \Textlen] \preceq
    \Text[\SA{\Text}[i_2]+\deltatext \dd \Textlen].\]
    On the other hand, note that $\SA{\Text}[i_1]+\deltatext = \RunEndFullPos{\SA{\Text}[i_1]}{\tau}{\Text}$,
    and similarly,
    $\SA{\Text}[i]+\deltatext = \RunEndFullPos{\SA{\Text}[i]}{\tau}{\Text}$, and
    $\SA{\Text}[i_2]+\deltatext = \RunEndFullPos{\SA{\Text}[i_2]}{\tau}{\Text}$.
    Combining this with the assumptions
    $\RunEndFullPos{\SA{\Text}[i_k]}{\tau}{\Text} = \RunEndFullPos{a_{j_k}}{\tau}{\Text}$ ($k \in \{1,2\}$)
    and the fact that
    $\RunEndFullPos{\SA{\Text}[i]}{\tau}{\Text} =
    \RunEndFullPos{a_j}{\tau}{\Text}$ (following from \cref{lm:R-text-block}),
    we obtain that
    $\SA{\Text}[i_k]+\deltatext = \RunEndFullPos{a_{j_k}}{\tau}{\Text}$ ($k \in \{1,2\}$) and
    $\SA{\Text}[i]+\deltatext = \RunEndFullPos{a_j}{\tau}{\Text}$.
    Thus, the above inequalities can be equivalent rewritten as
    $\Text[\RunEndFullPos{a_{j_1}}{\tau}{\Text} \dd \Textlen] \preceq
    \Text[\RunEndFullPos{a_j}{\tau}{\Text} \dd \Textlen] \preceq
    \Text[\RunEndFullPos{a_{j_2}}{\tau}{\Text} \dd \Textlen]$.
  \end{enumerate}
  Combining the above properties with the definition of $(a_j)_{j \in [1 \dd q]}$
  (see \cref{def:runs-minus-lex-sorted}), we immediately obtain that $j_1 \leq j \leq j_2$.
  It remains to recall that by $a_j \leq \SA{\Text}[i]$,
  $\RunEndFullPos{\SA{\Text}[i]}{\tau}{\Text} - \SA{\Text}[i] = \deltatext$, it follows that
  $\RunEndFullPos{a_j}{\tau}{\Text} - a_j \geq \RunEndFullPos{a_j}{\tau}{\Text} - \SA{\Text}[i]
  = \RunEndFullPos{\SA{\Text}[i]}{\tau}{\Text} - \SA{\Text}[i] = \deltatext$.
  Thus, by $\RunEndFullPos{a_j}{\tau}{\Text} - \deltatext = \RunEndFullPos{\SA{\Text}[i]}{\tau}{\Text} - \deltatext
  = \SA{\Text}[i] = x$, we obtain $x \in A$. This concludes the proof of the inclusion
  $\{\SA{\Text}[i]\}_{i \in [i_1 \dd i_2]} \subseteq A$.

  We now prove that $A \subseteq \{\SA{\Text}[i]\}_{i \in [i_1 \dd i_2]}$.
  Let $x \in A$, and let $j \in [j_1 \dd j_2]$ be such that $\RunEndFullPos{a_j}{\tau}{\Text} - a_j \geq \deltatext$
  and $x = \RunEndFullPos{a_j}{\tau}{\Text} - \deltatext$. We proceed as follows:
  \begin{enumerate}

  \item In the first step, we prove that $a_j \leq x$ and that it holds
    $[a_j \dd x] \subseteq \RTwo{\tau}{\Text}$. To this end, first note that by
    \cref{lm:end}, the maximal block of positions from the set $\RTwo{\tau}{\Text}$
    containing any $y \in \RPrimTwo{\tau}{\Text}$ is $[y \dd \RunEndPos{y}{\tau}{\Text} - (3\tau - 1)]$.
    Thus, to show $[a_j \dd x] \subseteq \RTwo{\tau}{\Text}$, we need to prove that $a_j \leq x$ and
    $x \leq \RunEndPos{a_j}{\tau}{\Text} - (3\tau - 1)$. Recall that
    $x = \RunEndFullPos{a_j}{\tau}{\Text} - \deltatext$,
    and that $a_j$ satisfies $\RunEndFullPos{a_j}{\tau}{\Text} - \deltatext \geq a_j$. This immediately yields
    $x \geq a_j$. To show $x \leq \RunEndPos{a_j}{\tau}{\Text} - (3\tau - 1)$, recall that above we proved that
    $\TailPos{a_j}{\tau}{\Text} \geq \TailPos{a_{j_1}}{\tau}{\Text}$, and
    $\TailPos{a_{j_1}}{\tau}{\Text} + \deltatext \geq 3\tau - 1$.
    Putting these inequalities together with $x = \RunEndFullPos{a_j}{\tau}{\Text} - \deltatext$, we obtain
      $\RunEndPos{a_j}{\tau}{\Text} - x
        = \RunEndPos{a_j}{\tau}{\Text} - (\RunEndFullPos{a_j}{\tau}{\Text} - \deltatext)
        = \TailPos{a_j}{\tau}{\Text} + \deltatext
        \geq \TailPos{a_{j_1}}{\tau}{\Text} + \deltatext
        \geq 3\tau - 1$
    or equivalently, $x \leq \RunEndPos{a_j}{\tau}{\Text} - (3\tau - 1)$.
    Thus, we indeed obtain
    \[
      [a_j \dd x] \subseteq
      [a_j \dd \RunEndPos{a_j}{\tau}{\Text} - (3\tau - 1)] \subseteq
      \RTwo{\tau}{\Text}.
    \]

  \item In this step, we prove that $x \in \RMinusFive{s}{k}{H}{\tau}{\Text}$.
    Above we showed that $a_j \leq x$ and $[a_j \dd x] \subseteq \RTwo{\tau}{\Text}$.
    By \cref{lm:R-text-block}, this implies that
    $\RunEndFullPos{x}{\tau}{\Text} = \RunEndFullPos{a_j}{\tau}{\Text}$,
    $\RootPos{x}{\tau}{\Text} = \RootPos{a_j}{\tau}{\Text} = H$, and
    $\TypePos{x}{\tau}{\Text} = \TypePos{a_j}{\tau}{\Text} = -1$.
    Moreover, we then have
    $\HeadPos{x}{\tau}{\Text} = (\RunEndFullPos{x}{\tau}{\Text} - x) \bmod |H| =
    (\RunEndFullPos{a_j}{\tau}{\Text} - x) \bmod |H| = \deltatext = s$ and
    $\ExpPos{x}{\tau}{\Text} = ((\RunEndFullPos{x}{\tau}{\Text} - x) - \HeadPos{x}{\tau}{\Text})/|H|
    = (\deltatext - s)/|H| = k$. We thus obtain
    \[
      x \in \RMinusFive{s}{k}{H}{\tau}{\Text}.
    \]

  \item In the next step, we show that
    $\Text[\SA{\Text}[i_1] \dd \Textlen] \preceq \Text[x \dd \Textlen] \preceq \Text[\SA{\Text}[i_2] \dd \Textlen]$.
    Let us focus on showing $\Text[\SA{\Text}[i_1] \dd \Textlen] \preceq \Text[x \dd \Textlen]$.
    Observe that by $x,\SA{\Text}[i_1] \in \RMinusFive{s}{k}{H}{\tau}{\Text}$,
    we have $\RunEndFullPos{x}{\tau}{\Text} - x = \RunEndFullPos{\SA{\Text}[i_1]}{\tau}{\Text} - \SA{\Text}[i_1]$
    and
    \[
      \Text[\SA{\Text}[i_1] \dd \RunEndFullPos{\SA{\Text}[i_1]}{\tau}{\Text}) =
      \Text[x \dd \RunEndFullPos{x}{\tau}{\Text}).
    \]
    On the other hand, observe that the assumption
    $j_1 \leq j$ combined with the above observation
    $\RootPos{a_{j_1}}{\tau}{\Text} = \RootPos{a_j}{\tau}{\Text}$ and
    the definition of the sequence $(a_j)_{j \in [1 \dd q]}$ (see \cref{def:runs-minus-lex-sorted}),
    imply that
    $\Text[\RunEndFullPos{a_{j_1}}{\tau}{\Text} \dd \Textlen] \preceq
    \Text[\RunEndFullPos{a_j}{\tau}{\Text} \dd \Textlen]$.
    Recalling that we have $\RunEndFullPos{a_{j_1}}{\tau}{\Text} = \RunEndFullPos{\SA{\Text}[i_1]}{\tau}{\Text}$
    and $\RunEndFullPos{a_j}{\tau}{\Text} = \RunEndFullPos{x}{\tau}{\Text}$, we can equivalently write this as
    \[
      \Text[\RunEndFullPos{\SA{\Text}[i_1]}{\tau}{\Text} \dd \Textlen] \preceq
      \Text[\RunEndFullPos{x}{\tau}{\Text} \dd \Textlen].
    \]
    Putting the above two observations together yields
    $\Text[\SA{\Text}[i_1] \dd \Textlen] \preceq \Text[x \dd \Textlen]$.
    The proof of $\Text[x \dd \Textlen] \preceq \Text[\SA{\Text}[i_2] \dd \Textlen]$
    is analogous.

  \item We are now ready to finish the proof. By the previous step and the definition
    of the suffix array, there exists $i \in [i_1 \dd i_2]$ such that $x = \SA{\Text}[i]$.
    Thus, $x \in \{\SA{\Text}[i]\}_{i \in [i_1 \dd i_2]}$. This concludes
    the proof of the inclusion $A \subseteq \{\SA{\Text}[i]\}_{i \in [i_1 \dd i_2]}$.
    \qedhere
  \end{enumerate}
\end{proof}

\subparagraph{Algorithms}

\begin{proposition}\label{pr:RskH-minus-report}
  Let $\Text \in \IntegerAlphabet^{\Textlen}$ be such that
  $2 \leq \AlphabetSize < \Textlen^{1/13}$ and $\Text[\Textlen]$ does not
  occur in $\Text[1 \dd \Textlen)$. Let
  $\tau = \lfloor \mu\log_{\AlphabetSize} \Textlen \rfloor$,
  where $\mu$ is a positive constant smaller than $\tfrac{1}{12}$
  such that $\tau \geq 1$.
  Given the packed representation of $\Text$, we can in
  $\bigO(\Textlen / \log_{\AlphabetSize} \Textlen)$ time construct
  a data structure that given any $j \in \RMinusFour{s}{H}{\tau}{\Text}$
  (where $H \in \IntegerAlphabet^{+}$ and $s \in [0 \dd |H|)$) and some
  $k > k_{\min}$, where $k_{\min} = \lceil \tfrac{3\tau - 1 - s}{|H|} \rceil - 1$,
  returns the set
  \[
    \RMinusFive{s}{k}{H}{\tau}{\Text}
  \]
  in
  $\bigO(1 + |\RMinusFive{s}{k}{H}{\tau}{\Text}|)$ time.
\end{proposition}
\begin{proof}

  Let $q = |\RPrimMinusTwo{\tau}{\Text}|$ and
  $(b_i)_{i \in [1 \dd q]} = \RunsMinusLengthSortedTwo{\tau}{\Text}$
  (\cref{def:runs-minus-length-sorted}). Let $A_{\rm runs}[1 \dd q]$ be an
  array defined by $A_{\rm runs}[i] = b_i$.

  \DSComponents
  The data structure consists of the following components:
  \begin{enumerate}
  \item The data structure from \cref{pr:nav-index-periodic} for text $\Text$.
    It needs $\bigO(\Textlen / \log_{\AlphabetSize} \Textlen)$ space.
  \item The array $A_{\rm runs}[1 \dd q]$ in plain form. By
    \cref{lm:runs}, it needs
    $\bigO(\Textlen / \tau) = \bigO(\Textlen / \log_{\AlphabetSize} \Textlen)$ space.
  \end{enumerate}
  In total, the structure needs $\bigO(\Textlen / \log_{\AlphabetSize} \Textlen)$ space.

  \DSQueries
  The queries are answered as follows.
  Let $j \in \RMinusFour{s}{H}{\tau}{\Text}$, where $H \in \IntegerAlphabet^{+}$, $p = |H|$, and $s \in [0 \dd p)$.
  Denote $k_{\min} = \lceil \tfrac{3\tau - 1 - s}{p} \rceil - 1$. Given $j$ and any $k > k_{\min}$, we compute
  $\RMinusFive{s}{k}{H}{\tau}{\Text}$ as follows:
  \begin{enumerate}
  \item Using \cref{pr:nav-index-periodic}\eqref{pr:nav-index-periodic-it-3}, in $\bigO(1)$ time
    compute integers $x, z \in [0 \dd q]$ such that it holds
    $x = \sum_{H' \in \IntegerAlphabet^{+} : H' \prec H} |\RPrimMinusThree{H'}{\tau}{\Text}|$
    and $z = x + |\RPrimMinusThree{H}{\tau}{\Text}|$. Note that we then have
    $\{b_i\}_{i \in (x \dd z]} = \RPrimMinusThree{H}{\tau}{\Text}$.
  \item Denote $\deltatext = s + kp$. In the next step, we
    compute an index $y \in [x \dd z]$ such that, for every $i \in (x \dd z]$,
    $\RunEndFullPos{b_i}{\tau}{\Text} - b_i \geq \deltatext$ holds if and only if
    $i \in (y \dd z]$. To this end, we scan the array $A_{\rm runs}(x \dd z]$ right-to-left.
    For each $i \in (x \dd z]$, in $\bigO(1)$ time we first compute $e = \RunEndFullPos{A_{\rm runs}[i]}{\tau}{\Text}$
    using \cref{pr:nav-index-periodic}\eqref{pr:nav-index-periodic-it-2}. We then check if $e - A_{\rm runs}[i] \geq \deltatext$.
    Once we reach $i = x$, or we find $i \in (x \dd z]$ such that $e - A_{\rm runs}[i] < \deltatext$, we stop the scan.
    The correctness of this algorithm follows by \cref{def:runs-minus-length-sorted}, i.e., we use that for all
    $i_1, i_2 \in (x \dd z]$ such that $i_1 < i_2$, it holds
    $\RunEndFullPos{b_{i_1}}{\tau}{\Text} - b_{i_1} \leq
    \RunEndFullPos{b_{i_2}}{\tau}{\Text} - b_{i_2}$.
    The computation of $y$ takes $\bigO(1 + (z-y))$ time. By
    \cref{lm:RskH}, it holds
    \[
      \RMinusFive{s}{k}{H}{\tau}{\Text} = \{\RunEndFullPos{b_i}{\tau}{\Text} - \deltatext : i \in (y \dd z]\}.
    \]
    By \cref{lm:efull}, $z - y = |\RMinusFive{s}{k}{H}{\tau}{\Text}|$, and hence this step
    takes $\bigO(1 + |\RMinusFive{s}{k}{H}{\tau}{\Text}|)$ time.
  \item We perform one more scan of $A_{\rm runs}(y \dd z]$, and with the help of
    \cref{pr:nav-index-periodic}\eqref{pr:nav-index-periodic-it-2}, we compute and return the
    set $\mathcal{P} = \{\RunEndFullPos{A_{\rm runs}[i]}{\tau}{\Text} - \deltatext : i \in (y \dd z]\}$.
    This takes $\bigO(1 + (z-y)) = \bigO(1 + |\RMinusFive{s}{k}{H}{\tau}{\Text}|)$ time, and by the above
    discussion, we have $\mathcal{P} = \RMinusFive{s}{k}{H}{\tau}{\Text}$.
  \end{enumerate}
  In total, we spend $\bigO(1 + |\RMinusFive{s}{k}{H}{\tau}{\Text}|)$ time.

  \DSConstruction
  The components of the structure are constructed as follows:
  \begin{enumerate}
  \item The construction of the first component takes
    $\bigO(\Textlen / \log_{\AlphabetSize} \Textlen)$ time
    using \cref{pr:nav-index-periodic}.
  \item To construct the second component, we proceed as follows:
    \begin{enumerate}
    \item First, using \cref{pr:periodic-constructions}\eqref{pr:periodic-constructions-it-3},
      compute the sequence
      $(a_i)_{i \in [1 \dd q]} = \RunsMinusLexSortedTwo{\tau}{\Text}$ (\cref{def:runs-minus-lex-sorted}) in
      $\bigO(\Textlen / \log_{\AlphabetSize} \Textlen)$ time.
    \item With the help of \cref{pr:nav-index-periodic}\eqref{pr:nav-index-periodic-it-2}, create an array
      $A_{\rm sort}[1 \dd q]$ of triples defined by $A_{\rm sort}[i] =
      (\Int{\tau}{\AlphabetSize}{\RootPos{a_i}{\tau}{\Text}}, \RunEndFullPos{a_i}{\tau}{\Text} - a_i, a_i)$
      (see \cref{def:int}). This takes $\bigO(q) = \bigO(\Textlen / \log_{\AlphabetSize} \Textlen)$ time.
    \item We sort the array $A_{\rm sort}[1 \dd q]$ lexicographically. Since the first coordinate is
      an integer smaller than $\AlphabetSize^{2\tau} = \bigO(n^{1/6})$, and the other two coordinates
      are numbers in $[1 \dd \Textlen]$, it suffices to use a 5-round radix sort to achieve
      $\bigO(q + \sqrt{\Textlen}) = \bigO(\Textlen / \log_{\AlphabetSize} \Textlen)$ time.
      The last coordinate of the resulting array contains the sequence
      $(b_i)_{i \in [1 \dd q]} = \RunsMinusLengthSortedTwo{\tau}{\Text}$ (\cref{def:runs-minus-length-sorted}),
      i.e., the elements of the array $A_{\rm runs}[1 \dd q]$. The correctness of this step follows by
      \cref{lm:int} and \cref{def:runs-minus-length-sorted}.
    \end{enumerate}
    In total, construction of the second component takes
    $\bigO(\Textlen / \log_{\AlphabetSize} \Textlen)$ time.
  \end{enumerate}
  In total, the construction takes
  $\bigO(\Textlen / \log_{\AlphabetSize} \Textlen)$ time.
\end{proof}

\begin{proposition}\label{pr:periodic-lex-range-minus-reporting-subblock}
  Let $\Text \in \IntegerAlphabet^{\Textlen}$ be such that
  $2 \leq \AlphabetSize < \Textlen^{1/13}$ and $\Text[\Textlen]$ does not
  occur in $\Text[1 \dd \Textlen)$. Let
  $\tau = \lfloor \mu\log_{\AlphabetSize} \Textlen \rfloor$,
  where $\mu$ is a positive constant smaller than $\tfrac{1}{12}$
  such that $\tau \geq 1$.
  Given the packed representation of $\Text$, we can in
  $\bigO(\Textlen / \log_{\AlphabetSize} \Textlen)$ time construct
  a data structure that given any $b,e \in [0 \dd \Textlen]$ such
  that $b < e$ and
  $\{\SA{\Text}[i]\}_{i \in (b \dd e]} \subseteq \RMinusFive{s}{k}{H}{\tau}{\Text}$ holds
  for some $s \in \Zn$, $H \in \IntegerAlphabet^{+}$, and $k \in \Zp$,
  returns the set
  \[
    \{\SA{\Text}[i]\}_{i \in (b \dd e]}
  \]
  in $\bigO((e-b) + \log \log \Textlen)$ time.
\end{proposition}
\begin{proof}

  Let $q = |\RPrimMinusTwo{\tau}{\Text}|$ and
  $(a_i)_{i \in [1 \dd q]} = \RunsMinusLexSortedTwo{\tau}{\Text}$
  (\cref{def:runs-minus-lex-sorted}). Let $A_{\rm runs}[1 \dd q]$ be an
  array defined by $A_{\rm runs}[i] = a_i$.

  \DSComponents
  The data structure consists of the following components:
  \begin{enumerate}
  \item The data structure from \cref{pr:nav-index-periodic} for text $\Text$.
    It needs $\bigO(\Textlen / \log_{\AlphabetSize} \Textlen)$ space.
  \item The array $A_{\rm runs}[1 \dd q]$ in plain form. By
    \cref{lm:runs}, it needs
    $\bigO(\Textlen / \tau) = \bigO(\Textlen / \log_{\AlphabetSize} \Textlen)$ space.
  \end{enumerate}
  In total, the structure needs $\bigO(\Textlen / \log_{\AlphabetSize} \Textlen)$ space.

  \DSQueries
  The queries are answered as follows. Let $b,e \in [0 \dd \Textlen]$ be such
  that $b < e$ and $\{\SA{\Text}[i]\}_{i \in (b \dd e]} \subseteq
  \RMinusFive{s}{k}{H}{\tau}{\Text}$ for some $s \in \Zn$, $H \in \IntegerAlphabet^{+}$,
  and $k \in \Zp$. Given $b,e$, we compute the set
  $\{\SA{\Text}[i]\}_{i \in (b \dd e]}$ as follows:
  \begin{enumerate}
  \item Denote $i_1 = b + 1$ and $i_2 = e$.
    Using \cref{pr:nav-index-periodic}\eqref{pr:nav-index-periodic-it-1},
    in $\bigO(\log \log \Textlen)$ time compute
    $p_k = \SA{\Text}[i_k]$ for $k \in \{1,2\}$.
  \item Using \cref{pr:nav-index-periodic}\eqref{pr:nav-index-periodic-it-2},
    in $\bigO(1)$ time compute
    $s = \HeadPos{p_1}{\tau}{\Text}$,
    $p = |\RootPos{p_1}{\tau}{\Text}|$, and
    $k = \ExpPos{p_1}{\tau}{\Text}$.
    We then set $\deltatext = s + kp$.
  \item Using \cref{pr:nav-index-periodic}\eqref{pr:nav-index-periodic-it-3},
    in $\bigO(1)$ time compute $j_k \in [1 \dd q]$ such that
    $\RunEndFullPos{p_k}{\tau}{\Text} = \RunEndFullPos{a_{j_k}}{\tau}{\Text}$ for
    $k \in \{1,2\}$.
  \item Using \cref{pr:nav-index-periodic}\eqref{pr:nav-index-periodic-it-reporting},
    compute the set
    $\mathcal{J} = \{j \in [j_1 \dd j_2] : \RunEndFullPos{a_j}{\tau}{\Text} - a_j \geq \deltatext\}$
    in $\bigO(1 + |\mathcal{J}|)$ time. Observe now that by
    \cref{lm:exp-subblock-characterization}, it holds
    \[
      \{\SA{\Text}[i]\}_{i \in (b \dd e]} =
      \{\SA{\Text}[i]\}_{i \in [i_1 \dd i_2]} =
      \{\RunEndFullPos{a_j}{\tau}{\Text} - \deltatext : j \in \mathcal{J}\}.
    \]
    By \cref{lm:efull}, $|\mathcal{J}| = i_2 - i_1 + 1 = e - b$.
    Hence, this step takes
    $\bigO(1 + |\mathcal{J}|) = \bigO(1 + (e-b))$ time.
  \item Using \cref{pr:nav-index-periodic}\eqref{pr:nav-index-periodic-it-2},
    compute and return the set
    $\{\RunEndFullPos{A_{\rm runs}[j]}{\tau}{\Text} - \deltatext : j \in \mathcal{J}\}$
    in $\bigO(1 + |\mathcal{J}|) = \bigO(1 + (e-b))$ time. By the above formula, this
    is precisely $\{\SA{\Text}[i]\}_{i \in (b \dd e]}$.
  \end{enumerate}
  In total, we spend $\bigO((e-b) + \log \log \Textlen)$ time.

  \DSConstruction
  The components of the structure are constructed as follows:
  \begin{enumerate}
  \item The construction of the first component takes
    $\bigO(\Textlen / \log_{\AlphabetSize} \Textlen)$ time
    using \cref{pr:nav-index-periodic}.
  \item The second component is constructed
    using \cref{pr:periodic-constructions}\eqref{pr:periodic-constructions-it-3}
    in $\bigO(\Textlen / \log_{\AlphabetSize} \Textlen)$ time.
  \end{enumerate}
  In total, the construction takes
  $\bigO(\Textlen / \log_{\AlphabetSize} \Textlen)$ time.
\end{proof}

\begin{proposition}\label{pr:periodic-lex-range-minus-reporting}
  Let $\Text \in \IntegerAlphabet^{\Textlen}$ be such that
  $2 \leq \AlphabetSize < \Textlen^{1/13}$ and $\Text[\Textlen]$ does not
  occur in $\Text[1 \dd \Textlen)$. Let
  $\tau = \lfloor \mu\log_{\AlphabetSize} \Textlen \rfloor$,
  where $\mu$ is a positive constant smaller than $\tfrac{1}{12}$
  such that $\tau \geq 1$.
  Given the packed representation of $\Text$, we can in
  $\bigO(\Textlen / \log_{\AlphabetSize} \Textlen)$ time construct
  a data structure that given any $b,e \in [0 \dd \Textlen]$ such
  that $b < e$ and $\{\SA{\Text}[i]\}_{i \in (b \dd e]} \subseteq \RMinusFour{s}{H}{\tau}{\Text}$ holds
  for some $s \in \Zn$ and $H \in \IntegerAlphabet^{+}$,
  returns the set
  \[
    \{\SA{\Text}[i]\}_{i \in (b \dd e]}
  \]
  in $\bigO((e-b) + \log \log \Textlen)$ time.
\end{proposition}
\begin{proof}

  We use the following definitions.
  Denote
  $(a_i)_{i \in [1 \dd q]} = \RunsMinusLexSortedTwo{\tau}{\Text}$ (\cref{def:runs-minus-lex-sorted}).
  Let also $B_{\rm exp} = \ExpBlockBitvectorMinusTwo{\tau}{\Text}$ (\cref{def:exp-block-bitvector-minus}).

  \DSComponents
  The data structure consists of the following components:
  \begin{enumerate}
  \item The data structure from \cref{pr:nav-index-periodic} for text $\Text$. It needs
    $\bigO(\Textlen / \log_{\AlphabetSize} \Textlen)$ space.
  \item The data structure from \cref{pr:RskH-minus-report} for text $\Text$. It needs
    $\bigO(\Textlen / \log_{\AlphabetSize} \Textlen)$ space.
  \item The data structure from \cref{pr:periodic-lex-range-minus-reporting-subblock}
    for text $\Text$. It also needs $\bigO(\Textlen / \log_{\AlphabetSize} \Textlen)$ space.
  \item The bitvector $B_{\rm exp}$ augmented with the support for $\bigO(1)$-time rank and select queries
    using \cref{th:bin-rank-select}. The bitvector together with the augmentation of \cref{th:bin-rank-select}
    needs $\bigO(\Textlen / \log \Textlen) = \bigO(\Textlen / \log_{\AlphabetSize} \Textlen)$ space.
  \end{enumerate}
  In total, the structure needs $\bigO(\Textlen / \log_{\AlphabetSize} \Textlen)$ space.

  \DSQueries
  The queries are answered as follows. Let $b, e \in [0 \dd \Textlen]$ be such that $b < e$ and
  $\{\SA{\Text}[i]\}_{i \in (b \dd e]} \subseteq \RMinusFour{s}{H}{\tau}{\Text}$ holds for some $s \in \Zn$
  and $H \in \IntegerAlphabet^{+}$. Given $b,e$, we compute the set $\{\SA{\Text}[i]\}_{i \in (b \dd e]}$ as follows.
  \begin{enumerate}
  \item In $\bigO(1)$ time compute $x = \Select{B_{\rm exp}}{\Rank{B_{\rm exp}}{b}{\one} + 1}{\one}$.
    If $x \geq e$, then by \cref{def:exp-block-bitvector-minus}, there exists $k \in \Zp$ such that
    $\{\SA{\Text}[i]\}_{i \in (b \dd e]} \subseteq \RMinusFive{s}{k}{H}{\tau}{\Text}$.
    In this case, we compute $\{\SA{\Text}[i]\}_{i \in (b \dd e]}$ using
    \cref{pr:periodic-lex-range-minus-reporting-subblock} in $\bigO((e-b) + \log \log \Textlen)$ time, and
    this completes the query algorithm.
    Let us now assume that $x < e$.
  \item In $\bigO(1)$ time, compute $y = \Select{B_{\rm exp}}{\Rank{B_{\rm exp}}{e-1}{\one}}{\one}$.
    Note that then $b < x \leq y < e$ and,
    by \cref{def:exp-block-bitvector-minus}, there exist $k_1, k_2 \in \Zp$ such that
    $k_1 < k_2$, and it holds:
    \begin{itemize}
    \item $\{\SA{\Text}[i]\}_{i \in (b \dd x]} \subseteq \RMinusFive{s}{k_1}{H}{\tau}{\Text}$,
    \item $\{\SA{\Text}[i]\}_{i \in (x \dd y]} = \bigcup_{k \in (k_1 \dd k_2)} \RMinusFive{s}{k}{H}{\tau}{\Text}$,
    \item $\{\SA{\Text}[i]\}_{i \in (y \dd e]} \subseteq \RMinusFive{s}{k_2}{H}{\tau}{\Text}$.
    \end{itemize}
    We enumerate the elements of each of the above sets as follows:
    \begin{itemize}
    \item The set $\{\SA{\Text}[i]\}_{i \in (b \dd x]}$ is computed
      using \cref{pr:periodic-lex-range-minus-reporting-subblock} in
      $\bigO((x-b) + \log \log \Textlen)$ time.
    \item Next, we check if $x = y$. If so, we skip this step. Let us thus assume that $x < y$. Then,
      it holds $k_1 + 1 < k_2$. Moreover, note that $k_1 + 1 > k_{\min}$, where
      $k_{\min} = \lceil \tfrac{3\tau - 1 - s}{|H|} \rceil - 1$.
      To see this, note that by $b < x$, it follows that $\RMinusFive{s}{k_1}{H}{\tau}{\Text} \neq \emptyset$.
      For any element $j$ of this set, it holds, by definition,
      \[
        k_1 =
        \ExpPos{j}{\tau}{\Text} =
        \big\lfloor \tfrac{\RunEndPos{j}{\tau}{\Text} - j - s}{|H|} \big\rfloor \geq
        \big\lfloor \tfrac{3\tau - 1 - s}{|H|} \big\rfloor \geq
        \big\lceil \tfrac{3\tau - 1 - s}{|H|} \big\rceil - 1 =
        k_{\min}.
      \]
      Thus, $k_1 + 1 > k_{\min}$. To compute
      $\{\SA{\Text}[i]\}_{i \in (x \dd y]} = \bigcup_{k \in (k_1 \dd k_2)} \RMinusFive{s}{k}{H}{\tau}{\Text}$,
      we proceed as follows:
      \begin{enumerate}
      \item Using \cref{pr:nav-index-periodic}\eqref{pr:nav-index-periodic-it-1}, in $\bigO(\log \log \Textlen)$ time
        compute $j_{\rm first} = \SA{\Text}[x+1]$ and $j_{\rm last} = \SA{\Text}[y]$.
      \item Using \cref{pr:nav-index-periodic}\eqref{pr:nav-index-periodic-it-2}, in
        $\bigO(1)$ time compute integers
        $k_{\rm first} = \ExpPos{j_{\rm first}}{\tau}{\Text}$ and
        $k_{\rm last} = \ExpPos{j_{\rm last}}{\tau}{\Text}$.
        Observe that it holds $k_1 + 1 = k_{\rm first}$ and $k_2 - 1 = k_{\rm last}$.
      \item We apply \cref{pr:RskH-minus-report} to position $j_1$ and $k = k_{\rm first}, k_{\rm first} + 1, \dots, k_{\rm last}$
        to compute all elements of the set $\bigcup_{k \in (k_1 \dd k_2)} \RMinusFive{s}{k}{H}{\tau}{\Text}$.
        In total, this takes $\bigO((k_2 - k_1) + (y-x))$. To simplify this complexity, note that by
        \cref{lm:RskH-size}, it holds
        \[
          |\RMinusFive{s}{k_1+1}{H}{\tau}{\Text}| \geq \ldots \geq |\RMinusFive{s}{k_2-1}{H}{\tau}{\Text}|.
        \]
        Since by $j_{\rm last} \in \RMinusFive{s}{k_2-1}{H}{\tau}{\Text}$, we have
        $\RMinusFive{s}{k_2-1}{H}{\tau}{\Text} \neq \emptyset$,
        we thus obtain that
        $k_2 - k_1 = \bigO(\sum_{k \in (k_1 \dd k_2)} |\RMinusFive{s}{k}{H}{\tau}{\Text}|) = \bigO(y-x)$,
        we can thus simplify the above runtime to $\bigO(y-x)$.
      \end{enumerate}
      In total, computing
      $\{\SA{\Text}[i]\}_{i \in (x \dd y]} = \bigcup_{k \in (k_1 \dd k_2)} \RMinusFive{s}{k}{H}{\tau}{\Text}$
      takes $\bigO((y-x) + \log \log \Textlen)$ time.
    \item The set $\{\SA{\Text}[i]\}_{i \in (y \dd e]}$ is computed
      using \cref{pr:periodic-lex-range-minus-reporting-subblock} in
      $\bigO((e-y) + \log \log \Textlen)$ time.
    \end{itemize}
    In total, we spend $\bigO((x-b) + (y-x) + (e-y) + \log \log \Textlen) = \bigO((e-b) + \log \log \Textlen)$ time.
  \end{enumerate}
  In total, the query takes $\bigO((e-b) + \log \log \Textlen)$ time.

  \DSConstruction
  The components of the structure are constructed as follows:
  \begin{enumerate}
  \item By \cref{pr:nav-index-periodic}, the construction of the first
    component takes $\bigO(\Textlen / \log_{\AlphabetSize} \Textlen)$ time.
  \item The second component is constructed using \cref{pr:RskH-minus-report} in
    $\bigO(\Textlen / \log_{\AlphabetSize} \Textlen)$ time.
  \item The third component is constructed using \cref{pr:periodic-lex-range-minus-reporting-subblock} in
    $\bigO(\Textlen / \log_{\AlphabetSize} \Textlen)$ time.
  \item To construct the last component, first, in $\bigO(\Textlen / \log_{\AlphabetSize} \Textlen)$ time
    we construct the bitvector $B_{\rm exp} = \ExpBlockBitvectorMinusTwo{\tau}{\Text}$ (\cref{def:exp-block-bitvector-minus})
    using \cref{pr:periodic-constructions}\eqref{pr:periodic-constructions-it-1}.
    We then augment $B_{\rm exp}$ using \cref{th:bin-rank-select} in $\bigO(\Textlen / \log \Textlen)
    = \bigO(\Textlen / \log_{\AlphabetSize} \Textlen)$ time.
  \end{enumerate}
  In total, the construction takes
  $\bigO(\Textlen / \log_{\AlphabetSize} \Textlen)$ time.
\end{proof}

\begin{proposition}\label{pr:periodic-lex-range-reporting}
  Let $\Text \in \IntegerAlphabet^{\Textlen}$ be such that
  $2 \leq \AlphabetSize < \Textlen^{1/13}$ and $\Text[\Textlen]$ does not
  occur in $\Text[1 \dd \Textlen)$. Let
  $\tau = \lfloor \mu\log_{\AlphabetSize} \Textlen \rfloor$,
  where $\mu$ is a positive constant smaller than $\tfrac{1}{12}$
  such that $\tau \geq 1$.
  Given the packed representation of $\Text$, we can in
  $\bigO(\Textlen / \log_{\AlphabetSize} \Textlen)$ time construct
  a data structure that given any $b,e \in [0 \dd \Textlen]$ such
  that $b < e$ and $\{\SA{\Text}[i]\}_{i \in (b \dd e]} \subseteq \RFour{s}{H}{\tau}{\Text}$
  for some $s \in \Zn$ and $H \in \IntegerAlphabet^{+}$, 
  returns the set 
  \[
    \{\SA{\Text}[i]\}_{i \in (b \dd e]}.
  \]
  in $\bigO((e - b) + \log \log \Textlen)$ time.
\end{proposition}
\begin{proof}

  \DSComponents
  The data structure consists of the following components:
  \begin{enumerate}
  \item The structure from \cref{pr:nav-index-periodic} for text $\Text$. It needs
    $\bigO(\Textlen / \log_{\AlphabetSize} \Textlen)$ space.
  \item The structure from \cref{pr:periodic-lex-range-minus-reporting} for text $\Text$. It also
    needs $\bigO(\Textlen / \log_{\AlphabetSize} \Textlen)$ space.
  \item The symmetric version of the structure from \cref{pr:periodic-lex-range-minus-reporting} adapted to positions
    in $\RPlusTwo{\tau}{\Text}$ (see \cref{def:R-subsets}). It also needs
    $\bigO(\Textlen / \log_{\AlphabetSize} \Textlen)$ space.
  \end{enumerate}
  In total, the structure needs $\bigO(\Textlen / \log_{\AlphabetSize} \Textlen)$ space.

  \DSQueries
  The queries are answered as follows. Let
  $b, e \in [0 \dd \Textlen]$ be such that $b < e$ and
  $\{\SA{\Text}[i]\}_{i \in (b \dd e]} \subseteq \RFour{s}{H}{\tau}{\Text}$
  for some $s \in \Zn$ and $H \in \IntegerAlphabet^{+}$. Given $b, e$, we
  compute $\{\SA{\Text}[i]\}_{i \in (b \dd e]}$ as follows:
  \begin{enumerate}
  \item Using \cref{pr:nav-index-periodic}\eqref{pr:nav-index-periodic-it-1}, in
    $\bigO(\log \log \Textlen)$ time we compute $j = \SA{\Text}[e]$.
  \item Using \cref{pr:nav-index-periodic}\eqref{pr:nav-index-periodic-it-2}, in
    $\bigO(1)$ time we then compute $x,y,z \in [0 \dd \Textlen]$ such that $x \leq y \leq z$,
    $\{\SA{\Text}[i]\}_{i \in (x \dd y]} = \RMinusFour{s}{H}{\tau}{\Text}$, and
    $\{\SA{\Text}[i]\}_{i \in (y \dd z]} = \RPlusFour{s}{H}{\tau}{\Text}$,
    where $s = \HeadPos{j}{\tau}{\Text}$ and $H = \RootPos{j}{\tau}{\Text}$.
    Note that then $x \leq b < e \leq z$.
  \item
    In $\bigO(1)$ time initialize $\mathcal{M} = \emptyset$.
  \item If $b < y$, then using \cref{pr:periodic-lex-range-minus-reporting}, in
    $\bigO((\min(e,y)-b) + \log \log \Textlen)$ time compute
    \[
      \{\SA{\Text}[i]\}_{i \in (b \dd \min(e,y)]},
    \]
    and add to $\mathcal{M}$.
    Note that $\{\SA{\Text}[i]\}_{i \in (b \dd \min(e,y)]} \subseteq \RMinusFour{s}{H}{\tau}{\Text}$.
  \item
    If $e > y$, then using the symmetric version of the data structure from \cref{pr:periodic-lex-range-minus-reporting},
    in $\bigO((e-\max(b,y)) + \log \log \Textlen)$ time compute
    \[
      \{\SA{\Text}[i]\}_{i \in (\max(b,y), e]},
    \]
    and add to $\mathcal{M}$.
    Note that $\{\SA{\Text}[i]\}_{i \in (\max(b,y) \dd e]} \subseteq \RPlusFour{s}{H}{\tau}{\Text}$.
  \item
    Note that at this point, we have
    $\{\SA{\Text}[i]\}_{i \in (b \dd e]} = \mathcal{M}$.
    Thus, we return $\mathcal{M}$ as the answer.
  \end{enumerate}
  In total, answering the query takes $\bigO((e-b) + \log \log \Textlen)$ time.

  \DSConstruction
  The components of the data structure are constructed as follows:
  \begin{enumerate}
  \item By \cref{pr:nav-index-periodic}, the construction of the first component takes
    $\bigO(\Textlen / \log_{\AlphabetSize} \Textlen)$ time.
  \item By \cref{pr:periodic-lex-range-minus-reporting}, the construction of the second component takes
    $\bigO(\Textlen / \log_{\AlphabetSize} \Textlen)$ time.
  \item Since the third component is a symmetric version of the second component, its construction also
    takes $\bigO(\Textlen / \log_{\AlphabetSize} \Textlen)$ time.
  \end{enumerate}
  In total, the construction takes $\bigO(\Textlen / \log_{\AlphabetSize} \Textlen)$ time.
\end{proof}

\begin{proposition}\label{pr:lex-range-reporting-periodic}
  Let $\Text \in \IntegerAlphabet^{\Textlen}$ be such that $2 \leq \AlphabetSize < \Textlen^{1/13}$ and $\Text[\Textlen]$ does not
  occur in $\Text[1 \dd \Textlen)$. Let $\tau = \lfloor \mu\log_{\AlphabetSize} \Textlen \rfloor$, where $\mu$ is a positive
  constant smaller than $\tfrac{1}{12}$ such that $\tau \geq 1$.
  Given the packed representation of $\Text$, we can in
  $\bigO(\Textlen / \log_{\AlphabetSize} \Textlen)$ time construct a data structure
  that answers the following queries:
  \begin{enumerate}
  \item\label{pr:lex-range-reporting-periodic-it-1}
    Given the packed representation of
    $\tau$-periodic (\cref{def:periodic-pattern}) patterns $\Pat_1, \Pat_2 \in \IntegerAlphabet^{+}$
    such that $\lcp{\Pat_1}{\Pat_2} \geq 3\tau - 1$,
    computes the set
    $\mathcal{J} := \LexRange{\Pat_1}{\Pat_2}{\Text}$ (\cref{def:lex-range}) in
    $\bigO(\log \log \Textlen +
    |\mathcal{J}| + 
    |\Pat_1| / \log_{\AlphabetSize} \Textlen +
    |\Pat_2| / \log_{\AlphabetSize} \Textlen)$
    time.
  \item\label{pr:lex-range-reporting-periodic-it-2}
    Given the packed representation of a $\tau$-periodic pattern
    $\Pat \in \IntegerAlphabet^{+}$ satisfying
    $|\Pat| \geq 3\tau - 1$,
    computes the set
    $\mathcal{J} := \LexRange{\Pat}{\Pat' c^{\infty}}{\Text}$
    (where $\Pat' = \Pat[1 \dd 3\tau-1]$ and $c = \AlphabetSize - 1$) in
    $\bigO(\log \log \Textlen +
    |\mathcal{J}| +
    |\Pat| / \log_{\AlphabetSize} \Textlen)$
    time.
  \end{enumerate}
\end{proposition}
\begin{proof}

  \DSComponents
  The data structure consists of the following components:
  \begin{enumerate}
  \item The data structure from \cref{pr:nav-index-periodic}
    using $\bigO(\Textlen / \log_{\AlphabetSize} \Textlen)$ space.
  \item The data structure from \cref{pr:periodic-lex-range-reporting}
    using $\bigO(\Textlen / \log_{\AlphabetSize} \Textlen)$ space.
  \end{enumerate}
  In total, the structure needs
  $\bigO(\Textlen / \log_{\AlphabetSize} \Textlen)$ space.

  \DSQueries
  The queries are answered as follows:
  \begin{enumerate}
  \item Let $\Pat_1, \Pat_2 \in \IntegerAlphabet^{+}$ be $\tau$-periodic
    patterns such that $\lcp{\Pat_1}{\Pat_2} \geq 3\tau - 1$.
    Given the packed representation of $\Pat_1$ and $\Pat_2$, we compute
    $\LexRange{\Pat_1}{\Pat_2}{\Text}$ as follows:
    \begin{enumerate}
    \item Using \cref{pr:nav-index-periodic}\eqref{pr:nav-index-periodic-it-pat-range}, in
      $\bigO(\log \log \Textlen + |\Pat_1| / \log_{\AlphabetSize} \Textlen + |\Pat_2| / \log_{\AlphabetSize} \Textlen)$
      time we compute $b_k = \RangeBegTwo{\Pat_k}{\Text}$, where $k \in \{1,2\}$.
      If $b_1 \geq b_2$, then $\LexRange{\Pat_1}{\Pat_2}{\Text} = \emptyset$ (see \cref{rm:lex-range}).
      Let us now assume that $b_1 < b_2$. By \cref{rm:lex-range}, we then have
      \[
        \LexRange{\Pat_1}{\Pat_2}{\Text} = \{\SA{\Text}[i]\}_{i \in (b_1 \dd b_2]}.
      \]
      By $\lcp{\Pat_1}{\Pat_2} \geq 3\tau - 1$, for every $j_1, j_2 \in \LexRange{\Pat_1}{\Pat_2}{\Text}$,
      it holds $\LCE{\Text}{j_1}{j_2} \geq 3\tau - 1$. Thus, by
      \cref{lm:periodic-pos-lce}\eqref{lm:periodic-pos-lce-it-2}, there exists
      $s \in \Zn$ and $H \in \IntegerAlphabet^{+}$ such that
      $\{\SA{\Text}[i]\}_{i \in (b_1 \dd b_2]} \subseteq \RFour{s}{H}{\tau}{\Text}$.
    \item Using \cref{pr:periodic-lex-range-reporting}, in
      $\bigO((b_2 - b_1) + \log \log \Textlen)$ time compute
      and return all elements of the set $\{\SA{\Text}[i]\}_{i \in (b_1 \dd b_2]}$.
    \end{enumerate}
    In total, we spend
    \begin{align*}
      & \bigO(\log \log \Textlen +
        (b_2 - b_1) +
        |\Pat_1| / \log_{\AlphabetSize} \Textlen +
        |\Pat_2| / \log_{\AlphabetSize} \Textlen) = \\
      & \bigO(\log \log \Textlen +
        |\LexRange{\Pat_1}{\Pat_2}{\Text}| +
        |\Pat_1| / \log_{\AlphabetSize} \Textlen +
        |\Pat_2| / \log_{\AlphabetSize} \Textlen)
    \end{align*}
    time.
  \item Let us now consider a $\tau$-periodic pattern $\Pat \in \IntegerAlphabet^{+}$
    that satisfies $|\Pat| \geq 3\tau - 1$,
    where $\Pat' = \Pat[1 \dd 3\tau - 1]$ and $c = \AlphabetSize - 1$. Given the packed representation of
    $\Pat$, we compute $\LexRange{\Pat}{\Pat' c^{\infty}}{\Text}$ similarly as above, letting
    $\Pat_1 = \Pat$ and $\Pat_2 = \Pat' c^{\infty}$. The main difference is that to compute
    $b_2 = \RangeBegTwo{\Pat_2}{\Text}$, we observe that
    $\RangeBegTwo{\Pat' c^{\infty}}{\Text} = \RangeEndTwo{\Pat'}{\Text}$.
    Thus, we can determine the value of $b_2$ using \cref{pr:nav-index-periodic}\eqref{pr:nav-index-periodic-it-pat-range}
    in $\bigO(\log \log \Textlen + |\Pat'| / \log_{\AlphabetSize} \Textlen) = \bigO(\log \log \Textlen)$ time.
  \end{enumerate}

  \DSConstruction
  The components of the structure are constructed as follows:
  \begin{enumerate}
  \item The first component is constructed using \cref{pr:nav-index-periodic} in
    $\bigO(\Textlen / \log_{\AlphabetSize} \Textlen)$ time.
  \item The second component is constructed in the same time using
    \cref{pr:periodic-lex-range-reporting}.
  \end{enumerate}
  In total, the construction takes
  $\bigO(\Textlen / \log_{\AlphabetSize} \Textlen)$ time.
\end{proof}

\paragraph{Summary}\label{sec:lex-range-reporting-to-pref-range-reporting-summary}

\begin{proposition}\label{pr:lex-range-reporting-to-prefix-range-reporting-nonbinary}
  Consider a data structure answering prefix range reporting queries
  (see \cref{sec:prefix-range-reporting-and-lex-range-reporting-problem-def}) that, for any
  sequence $W$ of $k$ strings of length $\ell$ over alphabet $\IntegerAlphabet$
  achieves the following complexities (where the input strings
  during construction are given in the packed representation, and at query time we are given any $b,e \in [0 \dd k]$ and the packed
  representation of any $X \in \IntegerAlphabet^{\leq \ell}$, and we return the set
  $\mathcal{I} := \{i \in (b \dd e] : X\text{ is a prefix of }W[i]\}$):
  \begin{itemize}
  \item space usage $S(k,\ell,\AlphabetSize)$,
  \item preprocessing time $P_t(k,\ell,\AlphabetSize)$,
  \item preprocessing space $P_s(k,\ell,\AlphabetSize)$,
  \item query time $Q_{\rm base}(k,\ell,\AlphabetSize) + |\mathcal{I}| \cdot Q_{\rm elem}(k,\ell,\AlphabetSize)$.
  \end{itemize}
  Let $\Text \in \IntegerAlphabet^{\Textlen}$ be such that $2 \leq \AlphabetSize < \Textlen^{1/13}$ and $\Text[\Textlen]$ does not
  occur in $\Text[1 \dd \Textlen)$. There exist positive integers
  $k = \Theta(\Textlen / \log_{\AlphabetSize} \Textlen)$ and
  $\ell \leq (1 + \lfloor \log k \rfloor) / \lceil \log \AlphabetSize \rceil$
  such that, given the packed representation of $\Text$, we can
  in
  $\bigO(\Textlen / \log_{\AlphabetSize} \Textlen + P_t(k,\ell,\AlphabetSize))$ time and using
  $\bigO(\Textlen / \log_{\AlphabetSize} \Textlen + P_s(k,\ell,\AlphabetSize))$ working space construct a data structure of size
  $\bigO(\Textlen / \log_{\AlphabetSize} \Textlen + S(k,\ell,\AlphabetSize))$ that, given the packed representation
  of any patterns $\Pat_1, \Pat_2 \in \IntegerAlphabet^{*}$, computes the set
  $\mathcal{J} := \LexRange{\Pat_1}{\Pat_2}{\Text}$ (see \cref{def:lex-range}) in
  $\bigO(\log \log \Textlen +
  Q_{\rm base}(k,\ell,\AlphabetSize) + |\mathcal{J}| \cdot Q_{\rm elem}(k,\ell,\AlphabetSize) +
  |\Pat_1| / \log_{\AlphabetSize} \Textlen +
  |\Pat_2| / \log_{\AlphabetSize} \Textlen)$
  time.
\end{proposition}
\begin{proof}

  We use the following definitions.
  Let $\tau = \lfloor \mu\log_{\AlphabetSize} \Textlen \rfloor$, where $\mu$
  is a positive constant smaller than $\tfrac{1}{12}$ such that $\tau \geq 1$ 
  (such $\tau$ exists by the assumption $\AlphabetSize < \Textlen^{1/13}$).
  Let $k$ and $\ell$ be integers resulting from
  the application of \cref{pr:lex-range-reporting-nonperiodic} to text $\Text$,
  with the structure from the above claim to answer prefix range reporting queries.
  Note that they satisfy $k = \Theta(\Textlen / \log_{\AlphabetSize} \Textlen)$ and
  $\ell \leq (1 + \lfloor \log k \rfloor) / \lceil \log \AlphabetSize \rceil$.

  \DSComponents
  The data structure consists of the following components:
  \begin{enumerate}
  \item The structure from \cref{pr:per} for $\AlphabetSize$ and $\tau$. If needs
    $\bigO(\AlphabetSize^{3\tau} \cdot \tau^2) = \bigO(\Textlen / \log_{\AlphabetSize} \Textlen)$ space.
  \item The structure from \cref{pr:nav-index-short} for $\tau$ and $\Text$.
    It needs $\bigO(\Textlen / \log_{\AlphabetSize} \Textlen)$ space.
  \item The structure from \cref{pr:lex-range-reporting-short} for $\tau$ and $\Text$.
    It uses $\bigO(\Textlen / \log_{\AlphabetSize} \Textlen)$ space.
  \item The structure from \cref{pr:lex-range-reporting-nonperiodic} for $\tau$ and $\Text$, and
    using the structure from the above claim to
    answer prefix range reporting queries. It uses
    $\bigO(\Textlen / \log_{\AlphabetSize} \Textlen + S(k,\ell,\AlphabetSize))$ space, where $k$ and $\ell$ are defined above.
  \item The structure from \cref{pr:lex-range-reporting-periodic} for $\tau$ and $\Text$.
    It uses $\bigO(\Textlen / \log_{\AlphabetSize} \Textlen)$ space.
  \end{enumerate}
  In total, the structure needs $\bigO(\Textlen / \log_{\AlphabetSize} \Textlen + S(k,\ell,\AlphabetSize))$ space.

  \DSQueries
  The queries are answered as follows. Let $\Pat_1, \Pat_2 \in \IntegerAlphabet^{*}$. Given the packed
  representation of $\Pat_1$ and $\Pat_2$, we compute $\LexRange{\Pat_1}{\Pat_2}{\Text}$ as follows.
  First, in $\bigO(1 + |\Pat_1| / \log_{\AlphabetSize} \Textlen + |\Pat_2| / \log_{\AlphabetSize} \Textlen)$ time
  we check if $\Pat_1 \prec \Pat_2$. If not, then we have $\LexRange{\Pat_1}{\Pat_2}{\Text} = \emptyset$
  (see \cref{def:lex-range}), and the query algorithm is complete.
  Let us thus assume that it holds $\Pat_1 \prec \Pat_2$.
  In $\bigO(1)$ time we use the packed representation of $\Pat_1$ and $\Pat_2$ to compute the
  packed representation of prefixes $X_1 = \Pat_1[1 \dd \min(3\tau-1, |\Pat_1|)]$ and $X_2 = \Pat_2[1 \dd \min(3\tau-1, |\Pat_2|)]$.
  Note that the assumption $\Pat_1 \prec \Pat_2$ implies that $X_1 \preceq X_2$.
  In $\bigO(1)$ time we check if $X_1 = X_2$.
  In $\bigO(1)$ time we also check if $X_1$ is a prefix of $X_2$.
  We then consider three cases:
  \begin{itemize}
  \item First, assume that $X_1 = X_2$. Note that $|X_1| = |X_2| = 3\tau - 1$, since otherwise
    $\Pat_1 = X_1 = X_2 = \Pat_2$, and the algorithm would have already finished.
    Using \cref{pr:per}, in $\bigO(1)$ time we check if
    $X_1$ is $\tau$-periodic (\cref{def:periodic-pattern}).
    Consider two cases:
    \begin{itemize}
    \item If $X_1$ is $\tau$-periodic, then so are $X_2$, $\Pat_1$, and $\Pat_2$.
      We compute $\mathcal{J} := \LexRange{\Pat_1}{\Pat_2}{\Text}$ using
      \cref{pr:lex-range-reporting-periodic}\eqref{pr:lex-range-reporting-periodic-it-1} in
      $\bigO(\log \log \Textlen +
          |\mathcal{J}| +
          |\Pat_1| / \log_{\AlphabetSize} \Textlen +
          |\Pat_2| / \log_{\AlphabetSize} \Textlen)$
      time.
    \item Otherwise, $\Pat_1$ and $\Pat_2$ are $\tau$-nonperiodic. Since $\lcp{\Pat_1}{\Pat_2} \geq 3\tau - 1$, we then
      compute $\mathcal{J} := \LexRange{\Pat_1}{\Pat_2}{\Text}$ in
      $\bigO(\log \log \Textlen +
          Q_{\rm base}(k,\ell,\AlphabetSize) + |\mathcal{J}| \cdot Q_{\rm elem}(k,\ell,\AlphabetSize) +
          |\Pat_1| / \log_{\AlphabetSize} \Textlen +
          |\Pat_2| / \log_{\AlphabetSize} \Textlen)$
      time
      using \cref{pr:lex-range-reporting-nonperiodic}\eqref{pr:lex-range-reporting-nonperiodic-it-1}.
    \end{itemize}
    In total, we spend
    $\bigO(\log \log \Textlen +
        |\mathcal{J}| +
        Q_{\rm base}(k,\ell,\AlphabetSize) + |\mathcal{J}| \cdot Q_{\rm elem}(k,\ell,\AlphabetSize) +
        |\Pat_1| / \log_{\AlphabetSize} \Textlen +
        |\Pat_2| / \log_{\AlphabetSize} \Textlen) =
    \bigO(\log \log \Textlen +
        Q_{\rm base}(k,\ell,\AlphabetSize) + |\mathcal{J}| \cdot Q_{\rm elem}(k,\ell,\AlphabetSize) +
        |\Pat_1| / \log_{\AlphabetSize} \Textlen +
        |\Pat_2| / \log_{\AlphabetSize} \Textlen)$
    time.
  \item Let us now assume that $X_1 \neq X_2$ and $X_1$ is a prefix of $X_2$.
    Note that then $X_1$ must be a proper prefix of $X_2$, and hence $|X_1| < |X_2|$.
    Thus, we must have $\Pat_1 = X_1$. Combining this with $X_1 \preceq X_2$ (see above), we obtain
    $\Pat_1 \preceq X_2 \preceq \Pat_2$. Consequently, we can write
    $\LexRange{\Pat_1}{\Pat_2}{\Text}$ as a disjoint union:
    \[
      \LexRange{\Pat_1}{\Pat_2}{\Text} = \LexRange{\Pat_1}{X_2}{\Text} \cup \LexRange{X_2}{\Pat_2}{\Text}.
    \]
    We will separately enumerate each of the two sets.
    Using \cref{pr:lex-range-reporting-short}\eqref{pr:lex-range-reporting-short-it-1},
    we compute $\mathcal{J}_1 := \LexRange{X_1}{X_2}{\Text}$ (recall that $\Pat_1 = X_1$) in
    $\bigO(1 + |\mathcal{J}_1|)$ time.
    Let us now focus on computing $\mathcal{J}_2 := \LexRange{X_2}{\Pat_2}{\Text}$.
    If $|\Pat_2| < 3\tau - 1$, then again we compute $\mathcal{J}_2$ using
    \cref{pr:lex-range-reporting-short}\eqref{pr:lex-range-reporting-short-it-1} in $\bigO(1 + |\mathcal{J}_2|)$ time.
    Let us now assume that $|\Pat_2| \geq 3\tau - 1$.
    Using \cref{pr:per}, in $\bigO(1)$ time we check if
    $\Pat_2$ is $\tau$-periodic (\cref{def:periodic-pattern}).
    Consider two cases:
    \begin{itemize}
    \item If $\Pat_2$ is $\tau$-periodic, then using \cref{pr:lex-range-reporting-periodic}\eqref{pr:lex-range-reporting-periodic-it-1},
      we compute $\mathcal{J}_2$ in
      $\bigO(\log \log \Textlen +
          |\mathcal{J}_2| +
          |\Pat_2| / \log_{\AlphabetSize} \Textlen)$
      time. Note that we can use this because
      $X_2 = \Pat_2[1 \dd 3\tau - 1]$.
    \item If $\Pat_2$ is $\tau$-nonperiodic, then using
      \cref{pr:lex-range-reporting-nonperiodic}\eqref{pr:lex-range-reporting-nonperiodic-it-1}, we compute
      $\mathcal{J}_2$ in
      $\bigO(\log \log \Textlen +
          Q_{\rm base}(k,\ell,\AlphabetSize) + |\mathcal{J}_2| \cdot Q_{\rm elem}(k,\ell,\AlphabetSize) +
          |\Pat_2| / \log_{\AlphabetSize} \Textlen)$
      time.
    \end{itemize}
    In total, we spend
    $\bigO(\log \log \Textlen +
        |\mathcal{J}_1| +
        |\mathcal{J}_2| +
        Q_{\rm base}(k,\ell,\AlphabetSize) + |\mathcal{J}_2| \cdot Q_{\rm elem}(k,\ell,\AlphabetSize) +
        |\Pat_2| / \log_{\AlphabetSize} \Textlen) =
    \bigO(\log \log \Textlen +
        Q_{\rm base}(k,\ell,\AlphabetSize) + |\mathcal{J}| \cdot Q_{\rm elem}(k,\ell,\AlphabetSize) +
        |\Pat_2| / \log_{\AlphabetSize} \Textlen)$
    time, where
    \[
      \mathcal{J} := \LexRange{\Pat_1}{\Pat_2}{\Text} = \mathcal{J}_1 \cup \mathcal{J}_2.
    \]
  \item Let us now assume that $X_1 \neq X_2$ and $X_1$ is not a prefix of $X_2$. Observe
    that, letting $c = \AlphabetSize - 1$, we then have $X_1 c^{\infty} \prec X_2$.
    Combining this with $\Pat_1 \preceq X_1 c^{\infty}$ and $X_2 \preceq \Pat_2$, we can thus write
    $\LexRange{\Pat_1}{\Pat_2}{\Text}$ as a disjoint union:
    \begin{align*}
      \LexRange{\Pat_1}{\Pat_2}{\Text}
        =\,& \LexRange{\Pat_1}{X_1 c^{\infty}}{\Text} \,\cup\\
         \,& \LexRange{X_1 c^{\infty}}{X_2}{\Text} \,\cup\\
         \,& \LexRange{X_2}{\Pat_2}{\Text}.
    \end{align*}
    We will separately enumerate the elements of each of these three sets.
    We begin with the set $\mathcal{J}_1 := \LexRange{\Pat_1}{X_1 c^{\infty}}{\Text}$.
    If $|\Pat_1| \leq 3\tau - 1$, then $\Pat_1 = X_1$, and hence we compute $\mathcal{J}_1$ in
    $\bigO(1 + |\mathcal{J}_1|)$ time using \cref{pr:lex-range-reporting-short}\eqref{pr:lex-range-reporting-short-it-2}.
    Let us now assume that $|\Pat_1| > 3\tau - 1$. In $\bigO(1)$ time we check if $\Pat_1$ is $\tau$-periodic
    using \cref{pr:per}. We then consider two cases:
    \begin{itemize}
    \item If $\Pat_1$ is $\tau$-periodic, then we compute $\mathcal{J}_1$
      using \cref{pr:lex-range-reporting-periodic}\eqref{pr:lex-range-reporting-periodic-it-2} in
      $\bigO(\log \log \Textlen +
          |\mathcal{J}_1| +
          |\Pat_1| / \log_{\AlphabetSize} \Textlen)$ time.
    \item Otherwise (i.e., if $\Pat_1$ is $\tau$-nonperiodic), we compute $\mathcal{J}_1$
      using \cref{pr:lex-range-reporting-nonperiodic}\eqref{pr:lex-range-reporting-nonperiodic-it-2} in
      $\bigO(\log \log \Textlen +
          Q_{\rm base}(k,\ell,\AlphabetSize) + |\mathcal{J}_1| \cdot Q_{\rm elem}(k,\ell,\AlphabetSize) +
          |\Pat_1| / \log_{\AlphabetSize} \Textlen)$ time.
    \end{itemize}
    Next, we compute $\mathcal{J}_2 := \LexRange{X_1 c^{\infty}}{X_2}{\Text}$. Note that
    $\RangeBegTwo{X_1 c^{\infty}}{\Text} = \RangeEndTwo{X_1}{\Text}$. We thus proceed as follows. Using
    \cref{pr:nav-index-short}, in $\bigO(1)$ time we compute 
    $e_1 = \RangeEndTwo{X_1}{\Text} = \RangeBegTwo{X_1 c^{\infty}}{\Text}$ and $b_2 = \RangeBegTwo{X_2}{\Text}$.
    If $e_1 \geq b_2$, then by \cref{rm:lex-range}, we have $\mathcal{J}_2 = \emptyset$. Let us thus assume
    that $e_1 < b_2$. Then, we compute $\mathcal{J}_2$ using
    \cref{pr:lex-range-reporting-short}\eqref{pr:lex-range-reporting-short-it-3}
    in $\bigO(1 + |\mathcal{J}_2|)$ time.
    Finally, we address the problem of computing $\mathcal{J}_3 := \LexRange{X_2}{\Pat_2}{\Text}$.
    If $|\Pat_2| \leq 3\tau - 1$, then
    $\Pat_2 = X_2$, and hence we have $\mathcal{J}_3 = \emptyset$.
    Let us now assume that $|\Pat_2| > 3\tau - 1$. In $\bigO(1)$ time we check if $\Pat_2$ is $\tau$-periodic
    using \cref{pr:per}. We then consider two cases:
    \begin{itemize}
    \item If $\Pat_2$ is $\tau$-nonperiodic, then we compute $\mathcal{J}_3$
      using \cref{pr:lex-range-reporting-nonperiodic}\eqref{pr:lex-range-reporting-nonperiodic-it-1} in
      $\bigO(\log \log \Textlen +
          Q_{\rm base}(k,\ell,\AlphabetSize) + |\mathcal{J}_3| \cdot Q_{\rm elem}(k,\ell,\AlphabetSize) +
          |\Pat_2| / \log_{\AlphabetSize} \Textlen)$ time.
    \item Otherwise (i.e., if $\Pat_2$ is $\tau$-periodic), we compute the elements of the set $\mathcal{J}_3$
      using \cref{pr:lex-range-reporting-periodic}\eqref{pr:lex-range-reporting-periodic-it-1} in
      $\bigO(\log \log \Textlen +
          |\mathcal{J}_3| +
          |\Pat_2| / \log_{\AlphabetSize} \Textlen)$ time.
    \end{itemize}
    In total, we spend
    $\bigO(\log \log \Textlen +
        |\mathcal{J}_1| + |\mathcal{J}_2| + |\mathcal{J}_3| +
        Q_{\rm base}(k,\ell,\AlphabetSize) + (|\mathcal{J}_1| + |\mathcal{J}_3|) \cdot Q_{\rm elem}(k,\ell,\AlphabetSize) +
        |\Pat_1| / \log_{\AlphabetSize} \Textlen +
        |\Pat_2| / \log_{\AlphabetSize} \Textlen) =
    \bigO(\log \log \Textlen +
        Q_{\rm base}(k,\ell,\AlphabetSize) + |\mathcal{J}| \cdot Q_{\rm elem}(k,\ell,\AlphabetSize) +
        |\Pat_1| / \log_{\AlphabetSize} \Textlen +
        |\Pat_2| / \log_{\AlphabetSize} \Textlen)$
    time in the above step, where
    $\mathcal{J} = \mathcal{J}_1 \cup \mathcal{J}_2 \cup \mathcal{J}_3$.
  \end{itemize}
  In total, we spend
  $\bigO(\log \log \Textlen +
      Q_{\rm base}(k,\ell,\AlphabetSize) + |\mathcal{J}| \cdot Q_{\rm elem}(k,\ell,\AlphabetSize) +
      |\Pat_1| / \log_{\AlphabetSize} \Textlen +
      |\Pat_2| / \log_{\AlphabetSize} \Textlen)$ time,
  where $\mathcal{J} = \LexRange{\Pat_1}{\Pat_2}{\Text}$.

  \DSConstruction
  The components of the structure are constructed as follows:
  \begin{enumerate}
  \item In $\bigO(\AlphabetSize^{3\tau} \cdot \tau^2) = \bigO(\Textlen / \log_{\AlphabetSize} \Textlen)$ time
    we apply \cref{pr:per}.
  \item In $\bigO(\Textlen / \log_{\AlphabetSize} \Textlen)$ time we apply \cref{pr:nav-index-short} to text $\Text$.
  \item In the same time as above we apply \cref{pr:lex-range-reporting-short} to text $\Text$.
  \item In $\bigO(\Textlen / \log_{\AlphabetSize} \Textlen + P_t(k,\ell,\AlphabetSize))$ time and using
    $\bigO(\Textlen / \log_{\AlphabetSize} \Textlen + P_s(k,\ell,\AlphabetSize))$ working space we apply
    \cref{pr:lex-range-reporting-nonperiodic} to text $\Text$.
  \item In $\bigO(\Textlen / \log_{\AlphabetSize} \Textlen)$ time we apply
    \cref{pr:lex-range-reporting-periodic} to text $\Text$.
  \end{enumerate}
  In total, the construction takes
  $\bigO(\Textlen / \log_{\AlphabetSize} \Textlen + P_t(k,\ell,\AlphabetSize))$ time uses
  $\bigO(\Textlen / \log_{\AlphabetSize} \Textlen + P_s(k,\ell,\AlphabetSize))$ working space.
\end{proof}

\subsubsection{Alphabet Reduction for Prefix Range Reporting Queries}\label{sec:prefix-range-reporting-alphabet-reduction}

\begin{proposition}\label{pr:prefix-range-reporting-alphabet-reduction}
  Consider a data structure answering prefix range reporting queries
  (see \cref{sec:prefix-range-reporting-and-lex-range-reporting-problem-def}) that, for
  and any sequence $W$ of $m$ binary strings of length $\ell = 1 + \lfloor \log m \rfloor$ (where $m \geq 1$),
  achieves the following complexities
  (where both the input strings
  during construction are given in the packed representation, and at query time, we are given
  some $b,e \in [0 \dd m]$, and the packed representation of some $X \in \BinaryAlphabet^{\leq \ell}$, and we return
  the set $\mathcal{I} := \{i \in (b \dd e] : X\text{ is a prefix of }W[i]\}$):
  \begin{itemize}
  \item space usage $S(m)$,
  \item preprocessing time $P_t(m)$,
  \item preprocessing space $P_s(m)$,
  \item query time $Q_{\rm base}(m) + |\mathcal{I}| \cdot Q_{\rm elem}(m)$.
  \end{itemize}
  Let $m' \geq 1$ and let
  $W'[1 \dd m']$ be a sequence of $m'$ equal-length strings over alphabet
  $\IntegerAlphabet$ (where $\AlphabetSize \geq 2$) of length
  $\ell' \leq (1 + \lfloor \log m' \rfloor) / \lceil \log \AlphabetSize \rceil$.
  Given the array $W'$, with all strings represented in the packed form, we can in
  $\bigO(m' + P_t(m'))$ time and using $\bigO(m' + P_s(m'))$ working space
  construct a data structure of size $\bigO(m' + S(m'))$ that, given the packed
  representation of any $X' \in \IntegerAlphabet^{\leq \ell'}$
  and any $b,e \in [0 \dd m']$, computes
  \[
    \mathcal{I}' := \{i \in (b \dd e] : X'\text{ is a prefix of }W'[i]\}
  \]
  in $\bigO(Q_{\rm base}(m') + |\mathcal{I}'| \cdot Q_{\rm elem}(m'))$ time.
\end{proposition}
\begin{proof}
  The above reduction is essentially the same as the one presented in
  \cref{pr:prefix-select-alphabet-reduction}, except we only need
  \cref{lm:prefix-queries-alphabet-reduction}\eqref{lm:prefix-queries-alphabet-reduction-it-1c}.
\end{proof}

\subsubsection{Summary}\label{sec:lex-range-reporting-to-prefix-range-reporting-summary}

\begin{theorem}\label{th:lex-range-reporting-to-prefix-range-reporting}
  Consider a data structure answering prefix range reporting queries
  (see \cref{sec:prefix-range-reporting-and-lex-range-reporting-problem-def}) that, for any
  sequence $W$ of $m \geq 1$ binary strings of length
  $\ell = 1 + \lfloor \log m \rfloor$, achieves the following
  complexities
  (where the input strings during construction are given in the packed representation,
  and at query time we are given any $b,e \in [0 \dd m]$, and the packed representation
  of any $X \in \BinaryAlphabet^{\leq \ell}$, and we return
  $\mathcal{I} := \{i \in (b \dd e] : X\text{ is a prefix of }W[i]\}$):
  \begin{itemize}
  \item space usage $S(m)$,
  \item preprocessing time $P_t(m)$,
  \item preprocessing space $P_s(m)$,
  \item query time $Q_{\rm base}(m) + |\mathcal{I}| \cdot Q_{\rm elem}(m)$.
  \end{itemize}
  For every $\Text \in \BinaryAlphabet^{\Textlen}$, there exists
  $m = \Theta(\Textlen / \log \Textlen)$ such that, given the packed
  representation of $\Text$, we can in
  $\bigO(\Textlen / \log \Textlen + P_t(m))$ time and using
  $\bigO(\Textlen / \log \Textlen + P_s(m))$ working space construct
  a data structure of size $\bigO(\Textlen / \log \Textlen + S(m))$ that,
  given the packed representation of any $\Pat_1, \Pat_2 \in \BinaryAlphabet^{*}$,
  computes $\mathcal{J} := \LexRange{\Pat_1}{\Pat_2}{\Text}$ (see \cref{def:lex-range}) in
  $\bigO(\log \log \Textlen +
    Q_{\rm base}(m) + |\mathcal{J}| \cdot Q_{\rm elem}(m) +
    |\Pat_1| / \log \Textlen +
    |\Pat_2| / \log \Textlen)$
  time.
\end{theorem}
\begin{proof}

  Assume that $\Textlen > 3^{13}-1$ (otherwise, the claim holds trivially).

  We use the following definitions. Let $\AlphabetSize = 3$ and
  let $\Text' \in \IntegerAlphabet^{\Textlen + 1}$ be a string defined so that $\Text'[1 \dd \Textlen] = \Text$ and
  $\Text'[\Textlen + 1] = \two$. Denote $\Textlen' = |\Text'|$. Note that $\Text'[\Textlen']$ does not occur
  in $\Text'[1 \dd \Textlen')$ and it holds $2 \leq \AlphabetSize < (\Textlen')^{1/13}$.
  Let $D$ denote the data structure resulting from applying \cref{pr:prefix-range-reporting-alphabet-reduction} with alphabet size
  $\AlphabetSize$
  to the structure (answering prefix range reporting queries on binary strings) form the claim. Given any sequence $W[1 \dd m']$
  of $m' \geq 1$ equal-length strings over alphabet $\IntegerAlphabet$ of length
  $\ell' \leq (1 + \lfloor \log m' \rfloor) / \lceil \log \AlphabetSize \rceil$ as input (with all strings represented in the packed
  form), the structure $D$ achieves the following complexities (where at query time it is given
  the packed representation of any $X \in \IntegerAlphabet^{\leq \ell'}$ and any
  $b,e \in [0 \dd m']$, and computes $\mathcal{I} := \{i \in (b \dd e] : X\text{ is a prefix of }W[i]\}$):
  \begin{itemize}
  \item space usage $S'(m',\ell',\AlphabetSize) = \bigO(m' + S(m'))$,
  \item preprocessing time $P'_t(m',\ell',\AlphabetSize) = \bigO(m' + P_t(m'))$,
  \item preprocessing space $P'_s(m',\ell',\AlphabetSize) = \bigO(m' + P_s(m'))$,
  \item query time $Q'_{\rm base}(m',\ell',\AlphabetSize) = \bigO(Q_{\rm base}(m'))$ and
    $Q'_{\rm elem}(m',\ell',\AlphabetSize) = \bigO(Q_{\rm elem}(m'))$,
  \end{itemize}
  where the complexities $S(m')$, $P_t(m')$, $P_s(m')$, $Q_{\rm base}(m')$, and $Q_{\rm elem}(m')$
  refer to the structure from the claim (answering
  prefix range reporting queries for binary strings).

  \DSComponents
  The data structure answering lex-range reporting queries
  consists of a single component: the data structure from \cref{pr:lex-range-reporting-to-prefix-range-reporting-nonbinary}
  applied for text $\Text'$ and with the structure $D$ as the structure answering prefix range reporting queries.
  Note that we can use $D$, since it supports prefix range reporting queries for the combination of parameters required
  in \cref{pr:lex-range-reporting-to-prefix-range-reporting-nonbinary}, i.e., for sequences over alphabet $\IntegerAlphabet$, with the
  length $\ell$ of all strings satisfying
  $\ell \leq (1 + \lfloor \log k \rfloor) / \lceil \log \AlphabetSize \rceil$ (where $k$ is the
  length of the input sequence).
  To bound the space usage of this component, note that by
  \cref{pr:lex-range-reporting-to-prefix-range-reporting-nonbinary} and the above discussion, there exists
  $m = \Theta(\Textlen' / \log_{\AlphabetSize} \Textlen') = \Theta(\Textlen / \log \Textlen)$ (recall that $\AlphabetSize = 3$)
  such that the structure needs
  $\bigO(\Textlen' / \log_{\AlphabetSize} \Textlen' + S'(m,\ell,\AlphabetSize)) =
  \bigO(\Textlen' / \log_{\AlphabetSize} \Textlen' + m + S(m)) = \bigO(\Textlen / \log \Textlen + S(m))$ space.

  \DSQueries
  The lex-range reporting queries are answered as follows. Let $\Pat_1, \Pat_2 \in \BinaryAlphabet^{*}$.
  Note that by definition of
  $\Text'$, it follows that $\LexRange{\Pat_1}{\Pat_2}{\Text} = \LexRange{\Pat_1}{\Pat_2}{\Text'}$.
  By \cref{pr:lex-range-reporting-to-prefix-range-reporting-nonbinary} and the above
  discussion, enumerating the set $\mathcal{J} := \LexRange{\Pat_1}{\Pat_2}{\Text'}$ takes
  $\bigO(\log \log \Textlen' +
      Q'_{\rm base}(m,\ell,\AlphabetSize) + |\mathcal{J}| \cdot Q'_{\rm elem}(m,\ell,\AlphabetSize) +
      |\Pat_1| / \log_{\AlphabetSize} \Textlen' +
      |\Pat_2| / \log_{\AlphabetSize} \Textlen') =
  \bigO(\log \log \Textlen +
      Q_{\rm base}(m) + |\mathcal{J}| \cdot Q_{\rm elem}(m) +
      |\Pat_1| / \log \Textlen +
      |\Pat_2| / \log \Textlen)$
  time.

  \DSConstruction
  By \cref{pr:lex-range-reporting-to-prefix-range-reporting-nonbinary}
  and the above discussion, construction of the above data structure
  answering lex-range reporting queries on $\Text'$ (and hence also on $\Text$) takes
  $\bigO(\Textlen' / \log_{\AlphabetSize} \Textlen' + P'_t(m,\ell,\AlphabetSize)) =
  \bigO(\Textlen' / \log_{\AlphabetSize} \Textlen' + m + P_t(m)) =
  \bigO(\Textlen / \log \Textlen + P_t(m))$
  time and uses
  $\bigO(\Textlen' / \log_{\AlphabetSize} \Textlen' + P'_s(m,\ell,\AlphabetSize)) =
  \bigO(\Textlen' / \log_{\AlphabetSize} \Textlen' + m + P_s(m)) =
  \bigO(\Textlen / \log \Textlen + P_s(m))$ working space.
\end{proof}

\section{Equivalence of Lex-Range Emptiness and Prefix Range Emptiness}\label{sec:prefix-range-emptiness}

\subsection{Problem Definitions}\label{sec:prefix-range-emptiness-and-lex-range-emptiness-problem-def}
\vspace{-1.5ex}

\begin{framed}
  \noindent
  \probname{Indexing for Lex-Range Emptiness Queries}
  \begin{bfdescription}
  \item[Input:]
    The packed representation of a string
    $\Text \in \BinaryAlphabet^{\Textlen}$.
  \item[Output:]
    A data structure that, given the packed representation of any
    $\Pat_1, \Pat_2 \in \BinaryAlphabet^{*}$, returns a
    $\texttt{YES}/\texttt{NO}$ answer indicating whether
    the set $\LexRange{\Pat_1}{\Pat_2}{\Text}$ (\cref{def:lex-range}) is empty.
  \end{bfdescription}
\end{framed}

\begin{framed}
  \noindent
  \probname{Indexing for Prefix Range Emptiness Queries}
  \begin{bfdescription}
  \item[Input:]
    A sequence $W[1 \dd m]$ of $m \geq 1$ binary strings of length $\ell = 1 + \lfloor \log m \rfloor$, with all
    strings represented in the packed form.
  \item[Output:]
    A data structure that, given
    the packed representation of any $X \in \BinaryAlphabet^{\leq \ell}$ and any
    $b,e \in [0 \dd m]$, returns a $\texttt{YES}/\texttt{NO}$ answer indicating
    whether $\PrefixRank{W}{e}{X} - \PrefixRank{W}{b}{X} > 0$, i.e., whether
    there exists $i \in (b \dd e]$ such that $X$ is a prefix of~$W[i]$.
  \end{bfdescription}
\end{framed}
\vspace{2ex}

\subsection{Reducing Prefix Range Emptiness to Lex-Range Emptiness Queries}\label{sec:prefix-emptiness-to-lex-range-emptiness}

\subsubsection{Problem Reduction}\label{sec:prefix-range-emptiness-to-lex-range-emptiness-problem-reduction}

\begin{lemma}\label{lm:prefix-range-emptiness-to-lex-range-emptiness}
  Let $W[1 \dd m]$ be a sequence of $m \geq 1$ strings of length $\ell \geq 1$ over alphabet $\BinaryAlphabet$.
  Let $s(i)$ (where $i \in [0 \dd m]$) be as in \cref{def:seq-to-string}.
  Denote
  $\Text = \SeqToString{\ell}{W}$ (\cref{def:seq-to-string}).
  Consider a string $X \in \BinaryAlphabet^{\leq \ell}$ and let $b, e \in [0 \dd m]$ be such that $b < e$.
  Denote
  $\Pat_1 = \revstr{X} \cdot \four \cdot s(b)$ and
  $\Pat_2 = \revstr{X} \cdot \four \cdot s(e)$.
  Then, it holds (see \cref{def:lex-range})
  \[
    \PrefixRank{W}{e}{X} - \PrefixRank{W}{b}{X} =
    |\LexRange{\Pat_1}{\Pat_2}{\Text}|.
  \]
\end{lemma}
\begin{proof}
  By \cref{lm:prefix-rank-to-pattern-rank}, letting $\alpha = \RangeBegTwo{\revstr{X}\cdot\four}{\Text}$, it holds
  \begin{align*}
    \PrefixRank{W}{e}{X} - \PrefixRank{W}{b}{X}
      &= (\RangeBegTwo{\Pat_2}{\Text} - \alpha) - (\RangeBegTwo{\Pat_1}{\Text} - \alpha)\\
      &= \RangeBegTwo{\Pat_2}{\Text} - \RangeBegTwo{\Pat_1}{\Text}\\
      &= |\LexRange{\Pat_1}{\Pat_2}{\Text}|,
  \end{align*}
  where in the last equality we use that $\Pat_1 \prec \Pat_2$ (which holds due to $b < e$); see also
  \cref{rm:lex-range}.
\end{proof}

\subsubsection{Alphabet Reduction for Lex-Range Emptiness Queries}\label{sec:lex-range-emptiness-alphabet-reduction}

\begin{lemma}\label{lm:lex-range-emptiness-alphabet-reduction}
  Let $\AlphabetSize \geq 2$ and $\Text, \Pat_1, \Pat_2 \in \IntegerAlphabet^{*}$. Then
  $\LexRange{\Pat_1}{\Pat_2}{\Text} \neq \emptyset$ holds if and only if $\LexRange{\Pat'_1}{\Pat'_2}{\Text'} \neq \emptyset$,
  where $k = \lceil \log \AlphabetSize \rceil$, $\Pat'_1 = \ebin{k}{\Pat_1}$, $\Pat'_2 = \ebin{k}{\Pat_2}$,
  and $\Text' = \ebin{k}{\Text}$ (\cref{def:ebin}).
\end{lemma}
\begin{proof}
  The proof follows immediately by \cref{lm:lex-range-alphabet-reduction}.
\end{proof}

\begin{proposition}\label{pr:lex-range-emptiness-alphabet-reduction}
  Let $\Text \in \IntegerAlphabet^{\Textlen}$ be a nonempty string, where
  $2 \leq \AlphabetSize < \Textlen^{\bigO(1)}$. Given the packed representation of
  $\Text$, we can in $\bigO(\Textlen / \log_{\AlphabetSize} \Textlen)$ time
  construct the packed representation of a
  text $\Text_{\rm bin} \in \BinaryAlphabet^{*}$ satisfying
  $|\Text_{\rm bin}| = \Theta(\Textlen \log \AlphabetSize)$, and a data structure that,
  given the packed representation of any
  $\Pat_1, \Pat_2 \in \IntegerAlphabet^{\leq m}$, in
  $\bigO(1 + m / \log_{\AlphabetSize} \Textlen)$ time returns the
  packed representation of $\Pat'_1, \Pat'_2 \in \BinaryAlphabet^{*}$
  satisfying
  $|\Pat'_i| = |\Pat_i| \cdot (2\lceil \log \AlphabetSize \rceil + 3)$ (where $i \in \{1,2\}$)
  such that $\LexRange{\Pat_1}{\Pat_2}{\Text} \neq \emptyset$ holds if and only if
  $\LexRange{\Pat'_1}{\Pat'_2}{\Text_{\rm bin}} \neq \emptyset$.
\end{proposition}
\begin{proof}

  Let $k = \lceil \log \AlphabetSize \rceil$.
  To compute $\Text_{\rm bin}$, we proceed as follows:
  \begin{enumerate}
  \item In $\bigO(\Textlen / \log_{\AlphabetSize} \Textlen)$ time we construct the data structure from
    \cref{pr:ebin} for $u = \Textlen$.
    Given the packed representation of any $S \in \IntegerAlphabet^{*}$, we can then
    compute the packed representation of $\ebin{k}{S}$ (\cref{def:ebin}) in
    $\bigO(1 + |S| / \log_{\AlphabetSize} \Textlen)$ time.
  \item Compute the packed representation of
    $\Text_{\rm bin} = \ebin{k}{\Text}$ in $\bigO(\Textlen / \log_{\AlphabetSize} \Textlen)$ time.
  \end{enumerate}
  In total, the construction of the packed representation of $\Text_{\rm bin}$
  takes $\bigO(\Textlen / \log_{\AlphabetSize} \Textlen)$. By \cref{def:ebin}, it
  holds $|\Text_{\rm bin}| = \Textlen \cdot (2k + 3) = \Theta(\Textlen \log \AlphabetSize)$.

  Next, we address the construction of the data structure from the claim.
  It consists of a single component: the data structure from \cref{pr:ebin} for $u = \Textlen$.
  It uses $\bigO(\Textlen / \log_{\AlphabetSize} \Textlen)$ space.

  The queries are implemented as follows. Given the packed representation of
  $\Pat_1,\Pat_2 \in \IntegerAlphabet^{\leq m}$, we compute the packed representation
  of $\Pat'_1$ and $\Pat'_2$ defined as $\Pat'_i = \ebin{k}{\Pat_i}$.
  Using the structure from \cref{pr:ebin}, this takes $\bigO(1 + m / \log_{\AlphabetSize} \Textlen)$ time.
  By \cref{def:ebin}, it holds $|\Pat'_i| = |\Pat_i| \cdot (2k + 3)$ for $i \in \{1,2\}$,
  and by \cref{lm:lex-range-emptiness-alphabet-reduction},
  $\LexRange{\Pat_1}{\Pat_2}{\Text} \neq \emptyset$ holds if and only
  if $\LexRange{\Pat'_1}{\Pat'_2}{\Text_{\rm bin}} \neq \emptyset$.

  By \cref{pr:ebin}, construction of the data structure takes
  $\bigO(\Textlen / \log_{\AlphabetSize} \Textlen)$ time.
\end{proof}

\subsubsection{Summary}\label{sec:prefix-range-emptiness-to-lex-range-emptiness-summary}

\begin{theorem}\label{th:prefix-range-emptiness-to-lex-range-emptiness}
  Consider a data structure answering lex-range emptiness queries
  (see \cref{sec:prefix-range-emptiness-and-lex-range-emptiness-problem-def})
  that, for any
  text $\Text \in \BinaryAlphabet^{\Textlen}$, achieves the following complexities
  (where in the preprocessing we assume that we are given as input the packed
  representation of $\Text$, and at query time we are given a packed
  representation of $\Pat_1,\Pat_2 \in \BinaryAlphabet^{\leq k}$):
  \begin{itemize}
  \item space usage $S(\Textlen)$,
  \item preprocessing time $P_t(\Textlen)$,
  \item preprocessing space $P_s(\Textlen)$,
  \item query time $Q(\Textlen, k)$.
  \end{itemize}
  For every sequence $W[1 \dd m]$ of $m \geq 1$ binary strings of length
  $\ell = 1 + \lfloor \log m \rfloor$, there exists
  $\Textlen = \bigO(m \log m)$ and $k = \bigO(\log m)$ such that, given the sequence $W$, with
  all strings represented in the packed form, we can in
  $\bigO(m + P_t(\Textlen))$ time and using $\bigO(m + P_s(\Textlen))$
  working space construct a data structure of size $\bigO(m + S(\Textlen))$
  that, given the packed representation of any $X \in \BinaryAlphabet^{\leq \ell}$
  and any pair of integers $b,e \in [0 \dd m]$, determines if
  there exists $i \in (b \dd e]$ such that $X$ is a prefix of $W[i]$
  (i.e., if $\PrefixRank{W}{e}{X} - \PrefixRank{W}{b}{X} > 0$)
  in $\bigO(Q(\Textlen, k))$ time.
\end{theorem}
\begin{proof}

  We use the following definitions.
  Let $\Text_{\rm aux} = \SeqToString{\ell}{W}$ (\cref{def:seq-to-string}).
  Denote $\Textlen_{\rm aux} =
  |\Text_{\rm aux}| = m \cdot (2\ell+1) = \Theta(m \log m)$. Let
  $\Text_{\rm bin}$ denote the text obtained
  by applying \cref{pr:lex-range-emptiness-alphabet-reduction} to text
  $\Text_{\rm aux}$. Denote $\Textlen = |\Text_{\rm bin}|$.
  Since $\Text_{\rm aux}$ is over alphabet $\{\zero, \dots, \four\}$,
  by \cref{pr:lex-range-emptiness-alphabet-reduction}, we have
  $\Textlen = \Theta(|\Text_{\rm aux}|) = \Theta(m \log m)$.
  Denote $k = 18 \lfloor \log m \rfloor + 27 = \bigO(\log m)$.

  \DSComponents
  The data structure consists of the following components:
  \begin{enumerate}
  \item The structure from \cref{pr:lex-range-emptiness-alphabet-reduction} applied
    to text $\Text_{\rm aux}$. Since $\Text_{\rm aux}$ is over alphabet
    $\{\zero, \dots, \four\}$, the structure needs
    $\bigO(|\Text_{\rm aux}| / \log |\Text_{\rm aux}|) =
    \bigO(m)$ space.
  \item The data structure from the claim (i.e., answering lex-range emptiness
    queries) for the text $\Text_{\rm bin}$. The structure needs
    $\bigO(S(\Textlen))$ space.
  \item A lookup table $L_{\rm rev}$ defined as in
    \cref{th:prefix-rank-to-pattern-ranking}. It needs
    $\bigO(\sqrt{\Textlen}) = \bigO(m)$ space.
  \item A lookup table $L_{\rm map}$ defined as in
    \cref{th:prefix-rank-to-pattern-ranking}. It also needs
    $\bigO(\sqrt{\Textlen}) = \bigO(m)$ space.
  \end{enumerate}
  In total, the structure needs $\bigO(m + S(\Textlen))$ space.

  \DSQueries
  The queries are answered as follows. Assume that we are given
  the packed representation of some $X \in \BinaryAlphabet^{\leq \ell}$ and
  two integers $b,e \in [0 \dd m]$.
  If $b \geq e$, then we immediately return that $\PrefixRank{W}{e}{X} - \PrefixRank{W}{b}{X} \leq 0$.
  Let us thus assume that $b < e$.
  The query algorithm proceeds as follows:
  \begin{enumerate}
  \item Using $L_{\rm rev}$ and $L_{\rm map}$, in $\bigO(1)$ time we compute the packed
    representation of strings $\Pat_1 = \revstr{X} \cdot \four \cdot s(b)$
    and $\Pat_2 = \revstr{X} \cdot \four \cdot s(e)$
    (over alphabet $\{\zero, \dots, \four\}$).
    Note that $|\Pat_1| = |\Pat_2| = |X| + 1 + \ell \leq 2\ell + 1 \leq 2\lfloor \log m \rfloor + 3$.
  \item Using \cref{pr:lex-range-emptiness-alphabet-reduction}, in
    $\bigO(1 + |\Pat_1| / \log \Textlen_{\rm aux} + |\Pat_2| / \log \Textlen_{\rm aux}) =
    \bigO(1 + \ell / \log m) = \bigO(1)$ time, we compute
    the packed representation of strings $\Pat'_1,\Pat'_2 \in \BinaryAlphabet^{*}$
    of length $|\Pat'_1| = |\Pat'_2| = 9 \cdot |\Pat_1| \leq 18 \lfloor \log m \rfloor + 27 \leq k$
    such that $\LexRange{\Pat_1}{\Pat_2}{\Text} \neq \emptyset$ if and only if
    $\LexRange{\Pat'_1}{\Pat'_2}{\Text_{\rm bin}} \neq \emptyset$.
  \item Using the structure from the claim, we determine whether it holds
    $\LexRange{\Pat'_1}{\Pat'_2}{\Text_{\rm bin}} \neq \emptyset$.
    By $|\Pat'_1|, |\Pat'_2| \leq k$, this takes $\bigO(Q(\Textlen, k))$ time.
    By the observation in the previous step and \cref{lm:prefix-range-emptiness-to-lex-range-emptiness}, this
    answer also determines whether $\PrefixRank{W}{e}{X} - \PrefixRank{W}{b}{X} > 0$. Thus we return
    that bit as the output of the query.
  \end{enumerate}
  In total, the query takes $\bigO(Q(\Textlen, k))$ time.

  \DSConstruction
  The components of the data structure are constructed as follows:
  \begin{enumerate}
  \item The first component of the structure is computed as follows:
    \begin{enumerate}
    \item In $\bigO(m)$ time we apply \cref{pr:seq-to-string} to
      the sequence $W$ to compute the packed representation of the string
      $\Text_{\rm aux} = \SeqToString{\ell}{W}$ (\cref{def:seq-to-string}).
    \item We apply \cref{pr:lex-range-emptiness-alphabet-reduction}
      to $\Text_{\rm aux}$. Since the string $\Text_{\rm aux}$ is over alphabet
      $\{\zero, \dots, \four\}$, this takes
      $\bigO(|\Text_{\rm aux}| / \log |\Text_{\rm aux}|) = \bigO(m)$
      time. Note that \cref{pr:lex-range-emptiness-alphabet-reduction}, in addition
      to the data structure, also returns
      the packed representation of the string $\Text_{\rm bin}$ (defined above).
    \end{enumerate}
  \item We apply the preprocessing from the claim to the string $\Text_{\rm bin}$.
    This takes $\bigO(P_t(|\Text_{\rm bin}|)) = \bigO(P_t(\Textlen))$ time
    and uses $\bigO(P_s(|\Text_{\rm bin}|)) = \bigO(P_s(\Textlen))$ working space.
  \item The lookup table $L_{\rm rev}$ is computed in $\bigO(m)$ time similarly
    as in the proof of \cref{pr:seq-to-string}.
  \item Next, we compute the lookup table $L_{\rm map}$. Similarly as above,
    we proceed as in \cref{pr:seq-to-string}, and spend $\bigO(m)$ time.
  \end{enumerate}
  In total, the construction takes
  $\bigO(m + P_t(\Textlen))$ time and uses
  $\bigO(m + P_s(\Textlen))$ working space.
\end{proof}

\subsection{Reducing Lex-Range Emptiness to Prefix Range Emptiness Queries}\label{sec:lex-range-emptiness-to-pref-range-emptiness}

\subsubsection{Problem Reduction}

\paragraph{The Short Patterns}\label{sec:lex-range-emptiness-to-pref-range-emptiness-short}

\begin{proposition}\label{pr:lex-range-emptiness-short}
  Let $\Text \in \IntegerAlphabet^{\Textlen}$ be such that $2 \leq \AlphabetSize < \Textlen^{1/13}$ and $\Text[\Textlen]$ does not
  occur in $\Text[1 \dd \Textlen)$. Let $\tau = \lfloor \mu\log_{\AlphabetSize} \Textlen \rfloor$, where $\mu$ is a positive
  constant smaller than $\tfrac{1}{12}$ such that $\tau \geq 1$.
  Given the packed representation of $\Text$, we can in $\bigO(\Textlen / \log_{\AlphabetSize} \Textlen)$ time
  construct a data structure that answers the following queries:
  \begin{enumerate}
  \item\label{pr:lex-range-emptiness-short-it-1}
    Given the packed representation of any $X_1, X_2 \in \IntegerAlphabet^{\leq 3\tau-1}$,
    check whether $\LexRange{X_1}{X_2}{\Text} \neq \emptyset$ in $\bigO(1)$ time.
  \item\label{pr:lex-range-emptiness-short-it-2}
    Given the packed representation of any $X \in \IntegerAlphabet^{\leq 3\tau-1}$,
    check whether $\LexRange{X}{X c^{\infty}}{\Text} \neq \emptyset$ (where $c = \AlphabetSize - 1$) in $\bigO(1)$ time.
  \end{enumerate}
\end{proposition}
\begin{proof}

  \DSComponents
  The data structure consists of a single component: the structure from \cref{pr:nav-index-short}. It needs
  $\bigO(\Textlen / \log_{\AlphabetSize} \Textlen)$ space.

  \DSQueries
  The queries are answered as follows:
  \begin{enumerate}
  \item Let $X_1, X_2 \in \IntegerAlphabet^{\leq 3\tau - 1}$.
    Using \cref{pr:nav-index-short} and the packed representations of $X_1$ and $X_2$,
    in $\bigO(1)$ time we compute the values
    $b_1 = \RangeBegTwo{X_1}{\Text}$ and $b_2 = \RangeBegTwo{X_2}{\Text}$.
    If $b_1 < b_2$, we have $\LexRange{X_1}{X_2}{\Text} \neq \emptyset$.
    Otherwise, we return that $\LexRange{X_1}{X_2}{\Text} = \emptyset$. The correctness
    follows by \cref{rm:lex-range}. In total, we spend $\bigO(1)$ time.
  \item Let $X \in \IntegerAlphabet^{\leq 3\tau - 1}$. Observe that it holds
    $\RangeBegTwo{X c^{\infty}}{\Text} = \RangeEndTwo{X}{\Text}$. Thus, the query is
    answered as above, except $b_2$ is computed as $\RangeEndTwo{X}{\Text}$ using
    \cref{pr:nav-index-short}.
    In total, we again spend $\bigO(1)$ time.
  \end{enumerate}

  \DSConstruction
  The construction of the structure takes $\bigO(\Textlen / \log_{\AlphabetSize} \Textlen)$ time by \cref{pr:nav-index-short}.
\end{proof}

\paragraph{The Nonperiodic Patterns}\label{sec:lex-range-emptiness-to-pref-range-emptiness-nonperiodic}

\subparagraph{Combinatorial Properties}

\begin{lemma}\label{lm:lex-range-emptiness-nonperiodic}
  Let $\Text \in \Sigma^{\Textlen}$, $\tau \in [1 \dd \lfloor \tfrac{\Textlen}{2} \rfloor]$, and assume that
  $\Text[\Textlen]$ does not occur in $\Text[1 \dd \Textlen)$.
  Let $\SSS$ be a $\tau$-synchronizing set of $\Text$. Denote
  $(s_i)_{i \in [1 \dd n']} = \LexSorted{\SSS}{\Text}$ (\cref{def:lex-sorted}).
  Let $A_{\SSS}[1 \dd n']$ and $A_{\rm str}[1 \dd n']$ be defined by
  \begin{itemize}
  \item $A_{\SSS}[i] = s_i$,
  \item $A_{\rm str}[i] = \revstr{D_i}$, where $D_i = \Textinf[s_i - \tau \dd s_i + 2\tau)$.
  \end{itemize}
  Let $D \in \DistPrefixes{\tau}{\Text}{\SSS}$ (\cref{def:dist-prefixes})
  and let $\Pat_1, \Pat_2 \in \Sigma^{+}$ be $\tau$-nonperiodic patterns (\cref{def:periodic-pattern})
  both having $D$ as a prefix.
  Denote $\deltatext = |D| - 2\tau$. For $k \in \{1,2\}$, let
  $\Pat'_k = \Pat_k(\deltatext \dd |\Pat_k|]$, and
  $b_k = |\{i \in [1 \dd n'] : \Text[s_i \dd \Textlen] \prec \Pat'_k\}|$.
  Then, the following conditions are equivalent:
  \begin{itemize}
  \item $\LexRange{\Pat_1}{\Pat_2}{\Text} \neq \emptyset$,
  \item $\{i \in (b_1 \dd b_2] : \revstr{D}\text{ is a prefix of }A_{\rm str}[i]\} \neq \emptyset$.
  \end{itemize}
\end{lemma}
\begin{proof}
  The result follows from \cref{lm:lex-range-nonperiodic-array}.
\end{proof}

\subparagraph{Algorithms}

\begin{proposition}\label{pr:lex-range-emptiness-nonperiodic}
  Consider a data structure answering prefix range emptiness queries
  (see \cref{sec:prefix-range-emptiness-and-lex-range-emptiness-problem-def})
  that, given any sequence $W$ of $k$ strings of length $\ell$ over alphabet $\IntegerAlphabet$
  achieves the following complexities (where the input strings
  during construction are given in the packed representation, and at query time we are given any $b,e \in [0 \dd k]$ and the packed
  representation of any $X \in \IntegerAlphabet^{\leq \ell}$, and we return a bit indicating whether
  $\{i \in (b \dd e] : X\text{ is a prefix of }W[i]\} \neq \emptyset$):
  \begin{itemize}
  \item space usage $S(k,\ell,\AlphabetSize)$,
  \item preprocessing time $P_t(k,\ell,\AlphabetSize)$,
  \item preprocessing space $P_s(k,\ell,\AlphabetSize)$,
  \item query time $Q(k,\ell,\AlphabetSize)$.
  \end{itemize}
  Let $\Text \in \IntegerAlphabet^{\Textlen}$ be such that $2 \leq \AlphabetSize < \Textlen^{1/13}$ and $\Text[\Textlen]$ does not
  occur in $\Text[1 \dd \Textlen)$. Let $\tau = \lfloor \mu\log_{\AlphabetSize} \Textlen \rfloor$, where $\mu$ is a positive
  constant smaller than $\tfrac{1}{12}$ such that $\tau \geq 1$. There exist positive integers
  $k = \Theta(\Textlen / \log_{\AlphabetSize} \Textlen)$ and
  $\ell \leq (1 + \lfloor \log k \rfloor) / \lceil \log \AlphabetSize \rceil$
  such that, given the packed representation of $\Text$, we can
  in
  $\bigO(\Textlen / \log_{\AlphabetSize} \Textlen + P_t(k,\ell,\AlphabetSize))$ time and using
  $\bigO(\Textlen / \log_{\AlphabetSize} \Textlen + P_s(k,\ell,\AlphabetSize))$ working space construct a data structure of size
  $\bigO(\Textlen / \log_{\AlphabetSize} \Textlen + S(k,\ell,\AlphabetSize))$ that answers the following queries:
  \begin{enumerate}
  \item\label{pr:lex-range-emptiness-nonperiodic-it-1}
    Given the packed representation of
    any $\tau$-nonperiodic (\cref{def:periodic-pattern}) patterns $\Pat_1, \Pat_2 \in \IntegerAlphabet^{+}$ satisfying
    $\lcp{\Pat_1}{\Pat_2} \geq 3\tau - 1$,
    determines whether $\LexRange{\Pat_1}{\Pat_2}{\Text} \neq \emptyset$ (\cref{def:lex-range}) in
    $\bigO(\log \log \Textlen +
    Q(k,\ell,\AlphabetSize) +
    |\Pat_1| / \log_{\AlphabetSize} \Textlen +
    |\Pat_2| / \log_{\AlphabetSize} \Textlen)$
    time.
  \item\label{pr:lex-range-emptiness-nonperiodic-it-2}
    Given the packed representation of a $\tau$-nonperiodic pattern
    $\Pat \in \IntegerAlphabet^{+}$ satisfying $|\Pat| \geq 3\tau-1$, determines whether
    $\LexRange{\Pat}{\Pat' c^{\infty}}{\Text} \neq \emptyset$
    (where $\Pat' = \Pat[1 \dd 3\tau-1]$ and $c = \AlphabetSize - 1$) in
    $\bigO(\log \log \Textlen +
    Q(k,\ell,\AlphabetSize) +
    |\Pat| / \log_{\AlphabetSize} \Textlen)$
    time.
  \end{enumerate}
\end{proposition}
\begin{proof}

  We use the following definitions.
  Let $\SSS$ be a $\tau$-synchronizing set of $\Text$ of size $|\SSS| = \bigO(\frac{\Textlen}{\tau})$ constructed
  using \cref{th:sss-packed-construction}. Denote $\Textlen' = |\SSS|$.
  Denote $(s_i)_{i \in [1 \dd \Textlen']} = \LexSorted{\SSS}{\Text}$ (\cref{def:lex-sorted}).
  Let $k = \max(\Textlen', k') = \Theta(\Textlen / \log_{\AlphabetSize} \Textlen)$,
  where $k' = \lceil \Textlen / \log_{\AlphabetSize} \Textlen \rceil$, and let
  $A_{\rm str}[1 \dd k]$ be an array defined so that for every $i \in [1 \dd \Textlen']$,
  $A_{\rm str}[i] = \revstr{D_i}$, where $D_i = \Textinf[s_i - \tau \dd s_i + 2\tau)$.
  We leave the remaining element initialized arbitrarily.
  Denote $\ell = 3\tau$, and note that
  by the same analysis as in \cref{th:sa-to-prefix-select-nonbinary}, it holds
  $\ell \leq (1 + \lfloor \log k \rfloor) / \lceil \log \AlphabetSize \rceil$.

  \DSComponents
  The data structure consists of the following components:
  \begin{enumerate}
  \item The structure from \cref{pr:nav-index-short} for text $\Text$. It needs $\bigO(\Textlen / \log_{\AlphabetSize} \Textlen)$ space.
  \item The structure $\NavNonperiodic{\SSS}{\tau}{\Text}$ from \cref{pr:nav-index-nonperiodic}. It needs
    $\bigO(\Textlen / \log_{\AlphabetSize} \Textlen)$ space.
  \item The structure from the claim (answering prefix range emptiness queries) for the sequence $A_{\rm str}[1 \dd k]$.
    It needs $\bigO(S(k,\ell,\AlphabetSize))$ space.
  \end{enumerate}
  In total, the structure needs
  $\bigO(\Textlen / \log_{\AlphabetSize} \Textlen + S(k,\ell,\AlphabetSize))$ space.

  \DSQueries
  The queries are answered as follows.
  \begin{enumerate}
  \item Let $\Pat_1, \Pat_2 \in \IntegerAlphabet^{+}$ be $\tau$-nonperiodic patterns satisfying
    $\lcp{\Pat_1}{\Pat_2} \geq 3\tau - 1$.
    Denote $\Pat' = \Pat_1[1 \dd 3\tau-1] = \Pat_2[1 \dd 3\tau-1]$.
    Given the packed representation of $\Pat_1$ and $\Pat_2$, we determine if
    $\LexRange{\Pat_1}{\Pat_2}{\Text} \neq \emptyset$ as follows:
    \begin{enumerate}
    \item First, using \cref{pr:nav-index-short}, in $\bigO(1)$ time we compute
      the values of
      $\RangeBegTwo{\Pat'}{\Text}$ and $\RangeEndTwo{\Pat'}{\Text}$. This lets us determine $|\OccTwo{\Pat'}{\Text}|$.
      If $\OccTwo{\Pat'}{\Text} = \emptyset$, then we return that $\LexRange{\Pat_1}{\Pat_2}{\Text} = \emptyset$ (since any element of
      this set would have $\Pat'$ as a prefix), and finish the query algorithm. Let us now assume that
      $\OccTwo{\Pat'}{\Text} \neq \emptyset$.
    \item Using \cref{pr:nav-index-nonperiodic}\eqref{pr:nav-index-nonperiodic-it-2a}, in $\bigO(1)$ time
      compute the packed representation of the string
      $D = \DistPrefixPat{\Pat_1}{\tau}{\Text}{\SSS}$ (\cref{def:dist-prefix-pat}).
      Note that since $\lcp{\Pat_1}{\Pat_2} \geq 3\tau - 1 \geq |D|$,
      $D$ is also a prefix of $\Pat_2$
      (and by \cref{lm:dist-prefix-existence} no other string in $\DistPrefixes{\tau}{\Text}{\SSS}$ is a prefix of $\Pat_2$).
      In $\bigO(1)$ time we then calculate $\deltatext = |D| - 2\tau$.
    \item Using \cref{pr:nav-index-nonperiodic}\eqref{pr:nav-index-nonperiodic-it-1}, in $\bigO(1)$ time
      compute the packed representation of $\revstr{D}$.
    \item Denote $P'_k = \Pat_k(\deltatext \dd |\Pat_k|]$, where $k \in \{1,2\}$.
      Using \cref{pr:nav-index-nonperiodic}\eqref{pr:nav-index-nonperiodic-it-2b},
      compute the value
      $b_k = |\{i \in [1 \dd \Textlen'] : \Text[s_i \dd \Textlen] \prec \Pat'_k\}|$
      for $k \in \{1,2\}$.
      This takes
      $\bigO(\log \log \Textlen + |\Pat'_1| / \log_{\AlphabetSize} \Textlen +
      |\Pat'_2| / \log_{\AlphabetSize} \Textlen) = \bigO(\log \log \Textlen + |\Pat_1| / \log_{\AlphabetSize} \Textlen +
      |\Pat_2| / \log_{\AlphabetSize} \Textlen)$ time.
    \item Using the structure from the claim, in $\bigO(Q(k,\ell,\AlphabetSize))$ time we check
      if there exists $i \in (b_1 \dd b_2]$ such that $\revstr{D}$ is a prefix of $A_{\rm str}[i]$.
      By \cref{lm:lex-range-emptiness-nonperiodic}, this is equivalent to checking if
      $\LexRange{\Pat_1}{\Pat_2}{\Text} \neq \emptyset$. Note that the array defined here is padded with
      extra strings compared to the array in \cref{lm:lex-range-emptiness-nonperiodic}, but this does not
      affect the result of the above query, since $b_1, b_2 \leq \Textlen'$.
    \end{enumerate}
    In total, we spend $\bigO(\log \log \Textlen + Q(k,\ell,\AlphabetSize) +
    |\Pat_1| / \log_{\AlphabetSize} \Textlen + |\Pat_2| / \log_{\AlphabetSize} \Textlen)$ time.
  \item Let $\Pat \in \IntegerAlphabet^{+}$ be a $\tau$-nonperiodic pattern satisfying $|\Pat| \geq 3\tau-1$.
    To determine whether it holds $\LexRange{\Pat}{\Pat' c^{\infty}}{\Text} \neq \emptyset$
    (where $\Pat' = \Pat[1 \dd 3\tau-1]$ and $c = \AlphabetSize - 1$), we
    set $\Pat_1 = \Pat$ and $\Pat_2 = \Pat' c^{\infty}$, and then proceed similarly as above, except
    we compute $b_2 = |\{i \in [1 \dd \Textlen'] : \Text[s_i \dd \Textlen] \prec \Pat'_2\}|$
    using the second integer computed in \cref{pr:nav-index-nonperiodic}\eqref{pr:nav-index-nonperiodic-it-2b}
    (which requires only the packed representation of $\Pat'$).
    The whole query takes
    $\bigO(\log \log \Textlen + Q(k,\ell,\AlphabetSize) + |\Pat| / \log_{\AlphabetSize} \Textlen)$ time.
  \end{enumerate}

  \DSConstruction
  The components of the structure are constructed as follows:
  \begin{enumerate}
  \item We apply \cref{pr:nav-index-short}. This uses $\bigO(\Textlen / \log_{\AlphabetSize} \Textlen)$ time and working space.
  \item Using \cref{th:sss-packed-construction}, we compute the $\tau$-synchronizing set $\SSS$ satisfying
    $|\SSS| = \bigO(\tfrac{\Textlen}{\tau}) = \bigO(\Textlen / \log_{\AlphabetSize} \Textlen)$ in
    $\bigO(\Textlen / \log_{\AlphabetSize} \Textlen)$ time. Then, using $\SSS$ and the packed representation of $\Text$ as input,
    we construct $\NavNonperiodic{\SSS}{\tau}{\Text}$ in $\bigO(\Textlen / \log_{\AlphabetSize} \Textlen)$ time using
    \cref{pr:nav-index-nonperiodic}.
  \item To construct the next component of the structure, we proceed as follows:
    \begin{enumerate}
    \item Using \cref{th:sss-lex-sort}, in
      $\bigO(\tfrac{\Textlen}{\tau}) = \bigO(\Textlen / \log_{\AlphabetSize} \Textlen)$ time we first compute
      the sequence $(s_i)_{i \in [1 \dd \Textlen']} = \LexSorted{\SSS}{\Text}$ (\cref{def:lex-sorted}).
    \item Using \cref{pr:nav-index-nonperiodic}\eqref{pr:nav-index-nonperiodic-it-1}, in $\bigO(\Textlen') =
      \bigO(\Textlen / \log_{\AlphabetSize} \Textlen)$ time we compute the elements of the array $A_{\rm str}$ at
      indexes $i \in [1 \dd \Textlen']$, i.e., $A_{\rm str}[i] = \revstr{D_i}$, where $D_i = \Textinf[s_i - \tau \dd s_i + 2\tau)$.
      In $\bigO(k - \Textlen') = \bigO(k) = \bigO(\Textlen / \log_{\AlphabetSize} \Textlen)$ time we then pad
      the remaining elements of $A_{\rm str}$ at indexes $i \in (\Textlen' \dd k]$.
    \item Finally, we apply the preprocessing
      from the claim to the array $A_{\rm str}$.
      This takes $\bigO(P_t(k,\ell,\AlphabetSize))$ time and uses
      $\bigO(P_s(k,\ell,\AlphabetSize))$ working space.
    \end{enumerate}
    In total, constructing this component of the structure
    takes $\bigO(\Textlen / \log_{\AlphabetSize} \Textlen + P_t(k,\ell,\AlphabetSize))$ time and uses
    $\bigO(\Textlen / \log_{\AlphabetSize} \Textlen + P_s(k,\ell,\AlphabetSize))$ working space.
  \end{enumerate}
  In total, the construction takes
  $\bigO(\Textlen / \log_{\AlphabetSize} \Textlen + P_t(k,\ell,\AlphabetSize))$ time and uses
  $\bigO(\Textlen / \log_{\AlphabetSize} \Textlen + P_s(k,\ell,\AlphabetSize))$ working space.
\end{proof}

\paragraph{The Periodic Patterns}\label{sec:lex-range-emptiness-to-pref-range-emptiness-periodic}

\begin{proposition}\label{pr:lex-range-emptiness-periodic}
  Let $\Text \in \IntegerAlphabet^{\Textlen}$ be such that $2 \leq \AlphabetSize < \Textlen^{1/13}$ and $\Text[\Textlen]$ does not
  occur in $\Text[1 \dd \Textlen)$. Let $\tau = \lfloor \mu\log_{\AlphabetSize} \Textlen \rfloor$, where $\mu$ is a positive
  constant smaller than $\tfrac{1}{12}$ such that $\tau \geq 1$.
  Given the packed representation of $\Text$, we can in
  $\bigO(\Textlen / \log_{\AlphabetSize} \Textlen)$ time construct a data structure
  that answers the following queries:
  \begin{enumerate}
  \item\label{pr:lex-range-emptiness-periodic-it-1}
    Given the packed representation of
    $\tau$-periodic (\cref{def:periodic-pattern}) patterns $\Pat_1, \Pat_2 \in \IntegerAlphabet^{+}$
    satisfying $\lcp{\Pat_1}{\Pat_2} \geq 3\tau - 1$,
    determines if
    $\LexRange{\Pat_1}{\Pat_2}{\Text} \neq \emptyset$ (\cref{def:lex-range}) in
    $\bigO(\log \log \Textlen +
    |\Pat_1| / \log_{\AlphabetSize} \Textlen +
    |\Pat_2| / \log_{\AlphabetSize} \Textlen)$
    time.
  \item\label{pr:lex-range-emptiness-periodic-it-2}
    Given the packed representation of a $\tau$-periodic pattern
    $\Pat \in \IntegerAlphabet^{+}$, determines whether it holds
    $\LexRange{\Pat}{\Pat' c^{\infty}}{\Text} \neq \emptyset$
    (where $\Pat' = \Pat[1 \dd 3\tau-1]$ and $c = \AlphabetSize - 1$) in
    $\bigO(\log \log \Textlen +
    |\Pat| / \log_{\AlphabetSize} \Textlen)$
    time.
  \end{enumerate}
\end{proposition}
\begin{proof}

  \DSComponents
  The data structure consists of a single component: the structure from \cref{pr:nav-index-periodic}. It needs
  $\bigO(\Textlen / \log_{\AlphabetSize} \Textlen)$ space.

  \DSQueries
  The queries are answered as follows:
  \begin{enumerate}
  \item Let $\Pat_1, \Pat_2 \in \IntegerAlphabet^{+}$ be
    $\tau$-periodic patterns.
    Using \cref{pr:nav-index-periodic}\eqref{pr:nav-index-periodic-it-pat-range}
    and the packed representation of $\Pat_1$ and $\Pat_2$,
    in $\bigO(\log \log \Textlen +
    |\Pat_1| / \log_{\AlphabetSize} \Textlen + |\Pat_2| / \log_{\AlphabetSize} \Textlen)$ time we compute the values
    $b_1 = \RangeBegTwo{\Pat_1}{\Text}$ and $b_2 = \RangeBegTwo{\Pat_2}{\Text}$.
    If $b_1 < b_2$, we have $\LexRange{\Pat_1}{\Pat_2}{\Text} \neq \emptyset$.
    Otherwise, we return that $\LexRange{\Pat_1}{\Pat_2}{\Text} = \emptyset$. The correctness
    follows by \cref{rm:lex-range}. In total, we spend
    $\bigO(\log \log \Textlen +
    |\Pat_1| / \log_{\AlphabetSize} \Textlen + |\Pat_2| / \log_{\AlphabetSize} \Textlen)$ time.
  \item Let $\Pat \in \IntegerAlphabet^{+}$ be a $\tau$-periodic pattern. Observe that it holds
    $\RangeBegTwo{\Pat' c^{\infty}}{\Text} = \RangeEndTwo{\Pat'}{\Text}$. Thus, the query is
    answered as above, except the value $b_2$ is computed as $\RangeEndTwo{\Pat'}{\Text}$ using
    \cref{pr:nav-index-periodic}\eqref{pr:nav-index-periodic-it-pat-range}.
    In total, we spend
    $\bigO(\log \log \Textlen +
    |\Pat| / \log_{\AlphabetSize} \Textlen + |\Pat'| / \log_{\AlphabetSize} \Textlen) =
    \bigO(\log \log \Textlen +
    |\Pat| / \log_{\AlphabetSize} \Textlen)$ time.
  \end{enumerate}

  \DSConstruction
  The construction of the structure takes $\bigO(\Textlen / \log_{\AlphabetSize} \Textlen)$ time by \cref{pr:nav-index-periodic}.
\end{proof}

\paragraph{Summary}\label{sec:lex-range-emptiness-to-pref-range-emptiness-summary}

\begin{proposition}\label{pr:lex-range-emptiness-to-prefix-range-emptiness-nonbinary}
  Consider a data structure answering prefix range emptiness queries
  (see \cref{sec:prefix-range-emptiness-and-lex-range-emptiness-problem-def}) that, for any
  sequence $W$ of $k$ strings of length $\ell$ over alphabet $\IntegerAlphabet$
  achieves the following complexities (where the input strings
  during construction are given in the packed representation, and at query time we are given any $b,e \in [0 \dd k]$ and the packed
  representation of any $X \in \IntegerAlphabet^{\leq \ell}$, and we return a bit indicating whether
  $\{i \in (b \dd e] : X\text{ is a prefix of }W[i]\} \neq \emptyset$):
  \begin{itemize}
  \item space usage $S(k,\ell,\AlphabetSize)$,
  \item preprocessing time $P_t(k,\ell,\AlphabetSize)$,
  \item preprocessing space $P_s(k,\ell,\AlphabetSize)$,
  \item query time $Q(k,\ell,\AlphabetSize)$.
  \end{itemize}
  Let $\Text \in \IntegerAlphabet^{\Textlen}$ be such that $2 \leq \AlphabetSize < \Textlen^{1/13}$ and $\Text[\Textlen]$ does not
  occur in $\Text[1 \dd \Textlen)$. There exist positive integers
  $k = \Theta(\Textlen / \log_{\AlphabetSize} \Textlen)$ and
  $\ell \leq (1 + \lfloor \log k \rfloor) / \lceil \log \AlphabetSize \rceil$
  such that, given the packed representation of $\Text$, we can
  in
  $\bigO(\Textlen / \log_{\AlphabetSize} \Textlen + P_t(k,\ell,\AlphabetSize))$ time and using
  $\bigO(\Textlen / \log_{\AlphabetSize} \Textlen + P_s(k,\ell,\AlphabetSize))$ working space construct a data structure of size
  $\bigO(\Textlen / \log_{\AlphabetSize} \Textlen + S(k,\ell,\AlphabetSize))$ that, given the packed representation
  of any patterns $\Pat_1, \Pat_2 \in \IntegerAlphabet^{*}$, determines if
  $\LexRange{\Pat_1}{\Pat_2}{\Text} \neq \emptyset$ (see \cref{def:lex-range}) in
  $\bigO(\log \log \Textlen +
  Q(k,\ell,\AlphabetSize) +
  |\Pat_1| / \log_{\AlphabetSize} \Textlen +
  |\Pat_2| / \log_{\AlphabetSize} \Textlen)$
  time.
\end{proposition}
\begin{proof}

  We use the following definitions.
  Let $\tau = \lfloor \mu\log_{\AlphabetSize} \Textlen \rfloor$, where $\mu$
  is a positive constant smaller than $\tfrac{1}{12}$ such that $\tau \geq 1$ 
  (such $\tau$ exists by the assumption $\AlphabetSize < \Textlen^{1/13}$).
  Let $k$ and $\ell$ be integers resulting from
  the application of \cref{pr:lex-range-emptiness-nonperiodic} to text $\Text$,
  with the structure from the above claim to answer prefix range emptiness queries.
  Note that they satisfy $k = \Theta(\Textlen / \log_{\AlphabetSize} \Textlen)$ and
  $\ell \leq (1 + \lfloor \log k \rfloor) / \lceil \log \AlphabetSize \rceil$.

  \DSComponents
  The data structure consists of the following components:
  \begin{enumerate}
  \item The structure from \cref{pr:per} for $\AlphabetSize$ and $\tau$.
    It needs $\bigO(\AlphabetSize^{3\tau} \cdot \tau^2) =
    \bigO(\Textlen / \log_{\AlphabetSize} \Textlen)$ space.
  \item The structure from \cref{pr:nav-index-short} for $\tau$ and $\Text$.
    It needs $\bigO(\Textlen / \log_{\AlphabetSize} \Textlen)$ space.
  \item The structure from \cref{pr:lex-range-emptiness-short} for $\tau$ and $\Text$.
    It uses $\bigO(\Textlen / \log_{\AlphabetSize} \Textlen)$ space.
  \item The structure from \cref{pr:lex-range-emptiness-nonperiodic} for $\tau$ and $\Text$, and
    using the structure from the above claim to
    answer prefix range emptiness queries. It uses
    $\bigO(\Textlen / \log_{\AlphabetSize} \Textlen + S(k,\ell,\AlphabetSize))$ space, where $k$ and $\ell$ are defined above.
  \item The structure from \cref{pr:lex-range-emptiness-periodic} for $\tau$ and $\Text$.
    It uses $\bigO(\Textlen / \log_{\AlphabetSize} \Textlen)$ space.
  \end{enumerate}
  In total, the structure needs $\bigO(\Textlen / \log_{\AlphabetSize} \Textlen + S(k,\ell,\AlphabetSize))$ space.

  \DSQueries
  The queries are answered as follows. Let $\Pat_1, \Pat_2 \in \IntegerAlphabet^{*}$. Given the packed
  representation of $\Pat_1$ and $\Pat_2$, we determine if $\LexRange{\Pat_1}{\Pat_2}{\Text} \neq \emptyset$ as follows.
  First, in $\bigO(1 + |\Pat_1| / \log_{\AlphabetSize} \Textlen + |\Pat_2| / \log_{\AlphabetSize} \Textlen)$ time
  we check if $\Pat_1 \prec \Pat_2$. If not, then we return that $\LexRange{\Pat_1}{\Pat_2}{\Text} = \emptyset$
  (see \cref{def:lex-range}), and the query algorithm is complete.
  Let us thus assume that it holds $\Pat_1 \prec \Pat_2$.
  In $\bigO(1)$ time we use the packed representation of $\Pat_1$ and $\Pat_2$ to compute the
  packed representation of prefixes $X_1 = \Pat_1[1 \dd \min(3\tau-1, |\Pat_1|)]$ and $X_2 = \Pat_2[1 \dd \min(3\tau-1, |\Pat_2|)]$.
  Note that the assumption $\Pat_1 \prec \Pat_2$ implies that $X_1 \preceq X_2$.
  In $\bigO(1)$ time we check if $X_1 = X_2$.
  In $\bigO(1)$ time we also check if $X_1$ is a prefix of $X_2$.
  We then consider three cases:
  \begin{itemize}
  \item First, assume that $X_1 = X_2$. Note that $|X_1| = |X_2| = 3\tau - 1$, since otherwise
    $\Pat_1 = X_1 = X_2 = \Pat_2$, and the algorithm would have already finished.
    Using \cref{pr:per}, in $\bigO(1)$ time we check if
    $X_1$ is $\tau$-periodic (\cref{def:periodic-pattern}).
    Consider two cases:
    \begin{itemize}
    \item If $X_1$ is $\tau$-periodic, then so are $X_2$, $\Pat_1$, and $\Pat_2$.
      We then check if $\LexRange{\Pat_1}{\Pat_2}{\Text} \neq \emptyset$ using
      \cref{pr:lex-range-emptiness-periodic}\eqref{pr:lex-range-emptiness-periodic-it-1}
      in $\bigO(\log \log \Textlen + |\Pat_1| / \log_{\AlphabetSize} \Textlen + |\Pat_2| / \log_{\AlphabetSize} \Textlen)$ time.
    \item Otherwise, $\Pat_1$ and $\Pat_2$ are $\tau$-nonperiodic. Since $\lcp{\Pat_1}{\Pat_2} \geq 3\tau - 1$, we then
      check whether it holds $\LexRange{\Pat_1}{\Pat_2}{\Text} \neq \emptyset$ in
      $\bigO(\log \log \Textlen + Q(k,\ell,\AlphabetSize) +
      |\Pat_1| / \log_{\AlphabetSize} \Textlen + |\Pat_2| / \log_{\AlphabetSize} \Textlen)$ time
      using \cref{pr:lex-range-emptiness-nonperiodic}\eqref{pr:lex-range-emptiness-nonperiodic-it-1}.
    \end{itemize}
    In total, we spend
    $\bigO(\log \log \Textlen + Q(k,\ell,\AlphabetSize) +
    |\Pat_1| / \log_{\AlphabetSize} \Textlen +
    |\Pat_2| / \log_{\AlphabetSize} \Textlen)$ time.
  \item Let us now assume that $X_1 \neq X_2$ and $X_1$ is a prefix of $X_2$.
    Note that then $X_1$ must be a proper prefix of $X_2$, and hence $|X_1| < |X_2|$.
    Thus, we must have $\Pat_1 = X_1$. Combining this with $X_1 \preceq X_2$ (see above), we obtain
    $\Pat_1 \preceq X_2 \preceq \Pat_2$. Consequently, we can write
    $\LexRange{\Pat_1}{\Pat_2}{\Text}$ as a disjoint union:
    \[
      \LexRange{\Pat_1}{\Pat_2}{\Text} = \LexRange{\Pat_1}{X_2}{\Text} \cup \LexRange{X_2}{\Pat_2}{\Text}.
    \]
    We will separately check the emptiness of each of the two sets. Based on this, we can easily determine
    if $\LexRange{\Pat_1}{\Pat_2}{\Text} \neq \emptyset$.
    Using \cref{pr:lex-range-emptiness-short}\eqref{pr:lex-range-emptiness-short-it-1},
    we check if $\LexRange{X_1}{X_2}{\Text} \neq \emptyset$ (recall that $\Pat_1 = X_1$) in $\bigO(1)$ time.
    Let us now focus on checking if $\LexRange{X_2}{\Pat_2}{\Text} \neq \emptyset$. If $|\Pat_2| < 3\tau - 1$, then
    again we check if $\LexRange{X_2}{\Pat_2}{\Text} \neq \emptyset$ using
    \cref{pr:lex-range-emptiness-short}\eqref{pr:lex-range-emptiness-short-it-1} in $\bigO(1)$ time.
    Let us now assume that $|\Pat_2| \geq 3\tau - 1$.
    Using \cref{pr:per}, in $\bigO(1)$ time we check if
    $\Pat_2$ is $\tau$-periodic (\cref{def:periodic-pattern}).
    Consider two cases:
    \begin{itemize}
    \item If $\Pat_2$ is $\tau$-periodic, then using \cref{pr:lex-range-emptiness-periodic}\eqref{pr:lex-range-emptiness-periodic-it-1},
      we check if $\LexRange{X_2}{\Pat_2}{\Text} \neq \emptyset$ in
      $\bigO(\log \log \Textlen + |\Pat_2| / \log_{\AlphabetSize} \Textlen)$ time. Note that we can use this because
      $X_2 = \Pat_2[1 \dd 3\tau - 1]$.
    \item If $\Pat_2$ is $\tau$-nonperiodic, then using
      \cref{pr:lex-range-emptiness-nonperiodic}\eqref{pr:lex-range-emptiness-nonperiodic-it-1}, we check if
      $\LexRange{X_2}{\Pat_2}{\Text} \neq \emptyset$ in
      $\bigO(\log \log \Textlen + Q(k,\ell,\AlphabetSize) + |\Pat_2| / \log_{\AlphabetSize} \Textlen)$ time.
    \end{itemize}
    In total, we spend
    $\bigO(\log \log \Textlen + Q(k,\ell,\AlphabetSize) + |\Pat_2| / \log_{\AlphabetSize} \Textlen)$ time.
  \item Let us now assume that $X_1 \neq X_2$ and $X_1$ is not a prefix of $X_2$. Observe
    that, letting $c = \AlphabetSize - 1$, we then have $X_1 c^{\infty} \prec X_2$.
    Combining this with $\Pat_1 \preceq X_1 c^{\infty}$ and $X_2 \preceq \Pat_2$, we can thus write
    $\LexRange{\Pat_1}{\Pat_2}{\Text}$ as a disjoint union:
    \begin{align*}
      \LexRange{\Pat_1}{\Pat_2}{\Text}
        =\,& \LexRange{\Pat_1}{X_1 c^{\infty}}{\Text} \,\cup\\
         \,& \LexRange{X_1 c^{\infty}}{X_2}{\Text} \,\cup\\
         \,& \LexRange{X_2}{\Pat_2}{\Text}.
    \end{align*}
    We will separately check the emptiness of each of the three sets. Based on this, we can easily determine
    if $\LexRange{\Pat_1}{\Pat_2}{\Text} \neq \emptyset$.
    We begin with the set $\LexRange{\Pat_1}{X_1 c^{\infty}}{\Text}$. If $|\Pat_1| \leq 3\tau - 1$, then
    $\Pat_1 = X_1$, and hence we check if $\LexRange{\Pat_1}{X_1 c^{\infty}}{\Text} \neq \emptyset$ in
    $\bigO(1)$ time using \cref{pr:lex-range-emptiness-short}\eqref{pr:lex-range-emptiness-short-it-2}.
    Let us now assume that $|\Pat_1| > 3\tau - 1$. In $\bigO(1)$ time we check if $\Pat_1$ is $\tau$-periodic
    using \cref{pr:per}. We then consider two cases:
    \begin{itemize}
    \item If $\Pat_1$ is $\tau$-periodic, then we check if $\LexRange{\Pat_1}{X_1 c^{\infty}}{\Text} \neq \emptyset$
      using \cref{pr:lex-range-emptiness-periodic}\eqref{pr:lex-range-emptiness-periodic-it-2} in
      $\bigO(\log \log \Textlen + |\Pat_1| / \log_{\AlphabetSize} \Textlen)$ time.
    \item Otherwise (i.e., if $\Pat_1$ is $\tau$-nonperiodic), we check if $\LexRange{\Pat_1}{X_1 c^{\infty}}{\Text} \neq \emptyset$
      using \cref{pr:lex-range-emptiness-nonperiodic}\eqref{pr:lex-range-emptiness-nonperiodic-it-2} in
      $\bigO(\log \log \Textlen + Q(k,\ell,\AlphabetSize) + |\Pat_1| / \log_{\AlphabetSize} \Textlen)$ time.
    \end{itemize}
    Next, we check if $\LexRange{X_1 c^{\infty}}{X_2}{\Text} \neq \emptyset$. To this end, note that
    $\RangeBegTwo{X_1 c^{\infty}}{\Text} = \RangeEndTwo{X_1}{\Text}$. We thus proceed as follows. Using
    \cref{pr:nav-index-short}, in $\bigO(1)$ time we compute 
    $e_1 = \RangeEndTwo{X_1}{\Text} = \RangeBegTwo{X_1 c^{\infty}}{\Text}$ and $b_2 = \RangeBegTwo{X_2}{\Text}$.
    If $e_1 < b_2$, we return that $\LexRange{X_1 c^{\infty}}{X_2}{\Text} \neq \emptyset$. Otherwise, we have
    $\LexRange{X_1 c^{\infty}}{X_2}{\Text} = \emptyset$. The correctness of this follows by \cref{rm:lex-range}.
    Finally, we address the problem of checking if $\LexRange{X_2}{\Pat_2}{\Text} \neq \emptyset$.
    If $|\Pat_2| \leq 3\tau - 1$, then
    $\Pat_2 = X_2$, and hence we have $\LexRange{X_2}{\Pat_2}{\Text} = \emptyset$.
    Let us now assume that $|\Pat_2| > 3\tau - 1$. In $\bigO(1)$ time we check if $\Pat_2$ is $\tau$-periodic
    using \cref{pr:per}. We then consider two cases:
    \begin{itemize}
    \item If $\Pat_2$ is $\tau$-nonperiodic, then we check if $\LexRange{X_2}{\Pat_2}{\Text} \neq \emptyset$
      using \cref{pr:lex-range-emptiness-nonperiodic}\eqref{pr:lex-range-emptiness-nonperiodic-it-1} in
      $\bigO(\log \log \Textlen + Q(k,\ell,\AlphabetSize) + |\Pat_2| / \log_{\AlphabetSize} \Textlen)$ time.
    \item Otherwise (i.e., if $\Pat_2$ is $\tau$-periodic), we check if $\LexRange{X_2}{\Pat_2}{\Text} \neq \emptyset$
      using \cref{pr:lex-range-emptiness-periodic}\eqref{pr:lex-range-emptiness-periodic-it-1} in
      $\bigO(\log \log \Textlen + |\Pat_2| / \log_{\AlphabetSize} \Textlen)$ time.
    \end{itemize}
    We spend
    $\bigO(\log \log \Textlen + Q(k,\ell,\AlphabetSize) +
    |\Pat_1| / \log_{\AlphabetSize} \Textlen +
    |\Pat_2| / \log_{\AlphabetSize} \Textlen)$ time in the above step.
  \end{itemize}
  Over all steps, we spend $\bigO(\log \log \Textlen + Q(k,\ell,\AlphabetSize) +
  |\Pat_1| / \log_{\AlphabetSize} \Textlen +
  |\Pat_2| / \log_{\AlphabetSize} \Textlen)$ time.

  \DSConstruction
  The components of the structure are constructed as follows:
  \begin{enumerate}
  \item In $\bigO(\AlphabetSize^{3\tau} \cdot \tau^2) = \bigO(\Textlen / \log_{\AlphabetSize} \Textlen)$ time
    we apply \cref{pr:per}.
  \item In $\bigO(\Textlen / \log_{\AlphabetSize} \Textlen)$ time we apply \cref{pr:nav-index-short} to text $\Text$.
  \item In the same time as above we apply \cref{pr:lex-range-emptiness-short} to text $\Text$.
  \item In $\bigO(\Textlen / \log_{\AlphabetSize} \Textlen + P_t(k,\ell,\AlphabetSize))$ time and using
    $\bigO(\Textlen / \log_{\AlphabetSize} \Textlen + P_s(k,\ell,\AlphabetSize))$ working space we apply
    \cref{pr:lex-range-emptiness-nonperiodic} to text $\Text$.
  \item In $\bigO(\Textlen / \log_{\AlphabetSize} \Textlen)$ time we apply
    \cref{pr:lex-range-emptiness-periodic} to text $\Text$.
  \end{enumerate}
  In total, the construction takes
  $\bigO(\Textlen / \log_{\AlphabetSize} \Textlen + P_t(k,\ell,\AlphabetSize))$ time uses
  $\bigO(\Textlen / \log_{\AlphabetSize} \Textlen + P_s(k,\ell,\AlphabetSize))$ working space.
\end{proof}

\subsubsection{Alphabet Reduction for Prefix Range Emptiness Queries}\label{sec:prefix-range-emptiness-alphabet-reduction}

\begin{proposition}\label{pr:prefix-range-emptiness-alphabet-reduction}
  Consider a data structure answering prefix range emptiness queries
  (see \cref{sec:prefix-range-emptiness-and-lex-range-emptiness-problem-def}) that, for any
  sequence of $m \geq 1$ binary strings of length
  $1 + \lfloor \log m \rfloor$, achieves the following
  complexities (where both the string at query time as well as
  input strings during construction are given in the packed representation):
  \begin{itemize}
  \item space usage $S(m)$,
  \item preprocessing time $P_t(m)$,
  \item preprocessing space $P_s(m)$,
  \item query time $Q(m)$.
  \end{itemize}
  Let $W[1 \dd m']$ be a sequence of $m' \geq 1$ equal-length strings over alphabet
  $\IntegerAlphabet$ (where $\AlphabetSize \geq 2$) of length
  $\ell \leq (1 + \lfloor \log m' \rfloor) / \lceil \log \AlphabetSize \rceil$.
  Given the
  sequence $W$, with all strings represented in the packed form, we can in
  $\bigO(m' + P_t(m'))$ time and using $\bigO(m' + P_s(m'))$ working space
  construct a data structure of size $\bigO(m' + S(m'))$ that, given the packed
  representation of any $X \in \IntegerAlphabet^{\leq \ell}$
  and any $b,e \in [0 \dd m']$, checks if there exists $i \in (b \dd e]$ such
  that $X$ is a prefix of $W[i]$, in $\bigO(Q(m'))$ time.
\end{proposition}
\begin{proof}
  The above reduction is essentially the same as the one presented in
  \cref{pr:prefix-select-alphabet-reduction}, except we only need
  \cref{lm:prefix-queries-alphabet-reduction}\eqref{lm:prefix-queries-alphabet-reduction-it-1c}.
\end{proof}

\subsubsection{Summary}\label{sec:lex-range-emptiness-to-prefix-range-emptiness-summary}

\begin{theorem}\label{th:lex-range-emptiness-to-prefix-range-emptiness}
  Consider a data structure answering prefix range emptiness queries
  (see \cref{sec:prefix-range-emptiness-and-lex-range-emptiness-problem-def}) that, for any
  sequence of $m \geq 1$ binary strings of length
  $1 + \lfloor \log m \rfloor$, achieves the following
  complexities
  (where both the string at query time as well as
  input strings during construction are given in the packed representation):
  \begin{itemize}
  \item space usage $S(m)$,
  \item preprocessing time $P_t(m)$,
  \item preprocessing space $P_s(m)$,
  \item query time $Q(m)$.
  \end{itemize}
  For every $\Text \in \BinaryAlphabet^{\Textlen}$, there exists
  $m = \Theta(\Textlen / \log \Textlen)$ such that, given the packed
  representation of $\Text$, we can in
  $\bigO(\Textlen / \log \Textlen + P_t(m))$ time and using
  $\bigO(\Textlen / \log \Textlen + P_s(m))$ working space construct
  a data structure of size $\bigO(\Textlen / \log \Textlen + S(m))$ that,
  given the packed representation of any $\Pat_1, \Pat_2 \in \BinaryAlphabet^{*}$,
  determines whether $\LexRange{\Pat_1}{\Pat_2}{\Text} \neq \emptyset$ (see \cref{def:lex-range})
  in $\bigO(\log \log \Textlen + Q(m) + |\Pat_1| / \log \Textlen + |\Pat_2| / \log \Textlen)$ time.
\end{theorem}
\begin{proof}

  Assume that $\Textlen > 3^{13}-1$ (otherwise, the claim holds trivially).

  We use the following definitions. Let $\AlphabetSize = 3$ and
  let $\Text' \in \IntegerAlphabet^{\Textlen + 1}$ be a string defined so that $\Text'[1 \dd \Textlen] = \Text$ and
  $\Text'[\Textlen + 1] = \two$. Denote $\Textlen' = |\Text'|$. Note that $\Text'[\Textlen']$ does not occur
  in $\Text'[1 \dd \Textlen')$ and it holds $2 \leq \AlphabetSize < (\Textlen')^{1/13}$.
  Let $D$ denote the data structure resulting from applying \cref{pr:prefix-range-emptiness-alphabet-reduction} with alphabet size
  $\AlphabetSize$
  to the structure (answering prefix range emptiness queries on binary strings) form the claim. Given any sequence $W[1 \dd m']$
  of $m' \geq 1$ equal-length strings over alphabet $\IntegerAlphabet$ of length
  $\ell' \leq (1 + \lfloor \log m' \rfloor) / \lceil \log \AlphabetSize \rceil$ as input (with all strings represented in the packed
  form), the structure $D$ achieves the following complexities (where at query time it is given
  the packed representation of any $X \in \IntegerAlphabet^{\leq \ell'}$ and any
  $b,e \in [0 \dd m']$, and determines whether there exists $i \in (b \dd e]$ such that $X$ is a prefix of $W[i]$):
  \begin{itemize}
  \item space usage $S'(m',\ell',\AlphabetSize) = \bigO(m' + S(m'))$,
  \item preprocessing time $P'_t(m',\ell',\AlphabetSize) = \bigO(m' + P_t(m'))$,
  \item preprocessing space $P'_s(m',\ell',\AlphabetSize) = \bigO(m' + P_s(m'))$,
  \item query time $Q'(m',\ell',\AlphabetSize) = \bigO(Q(m'))$,
  \end{itemize}
  where the complexities $S(m')$, $P_t(m')$, $P_s(m')$, and $Q(m')$ refer to the structure from the claim (answering
  prefix range emptiness queries for binary strings).

  \DSComponents
  The data structure answering lex-range emptiness queries
  consists of a single component: the data structure from \cref{pr:lex-range-emptiness-to-prefix-range-emptiness-nonbinary}
  applied for text $\Text'$ and with the structure $D$ as the structure answering prefix range emptiness queries.
  Note that we can use $D$, since it supports prefix range emptiness queries for the combination of parameters required
  in \cref{pr:lex-range-emptiness-to-prefix-range-emptiness-nonbinary}, i.e., for sequences over alphabet $\IntegerAlphabet$, with the
  length $\ell$ of all strings satisfying
  $\ell \leq (1 + \lfloor \log k \rfloor) / \lceil \log \AlphabetSize \rceil$ (where $k$ is the
  length of the input sequence).
  To bound the space usage of this component, note that by
  \cref{pr:lex-range-emptiness-to-prefix-range-emptiness-nonbinary} and the above discussion, there exists
  $m = \Theta(\Textlen' / \log_{\AlphabetSize} \Textlen') = \Theta(\Textlen / \log \Textlen)$ (recall that $\AlphabetSize = 3$)
  such that the structure needs
  $\bigO(\Textlen' / \log_{\AlphabetSize} \Textlen' + S'(m,\ell,\AlphabetSize)) =
  \bigO(\Textlen' / \log_{\AlphabetSize} \Textlen' + m + S(m)) = \bigO(\Textlen / \log \Textlen + S(m))$ space.

  \DSQueries
  The lex-range emptiness queries are answered as follows. Let $\Pat_1, \Pat_2 \in \BinaryAlphabet^{*}$.
  Note that by definition of
  $\Text'$, it follows that $\LexRange{\Pat_1}{\Pat_2}{\Text} \neq \emptyset$ if and only if
  $\LexRange{\Pat_1}{\Pat_2}{\Text'} \neq \emptyset$.
  By \cref{pr:lex-range-emptiness-to-prefix-range-emptiness-nonbinary} and the above
  discussion, determining if $\LexRange{\Pat_1}{\Pat_2}{\Text'} \neq \emptyset$ takes
  $\bigO(\log \log \Textlen' + Q'(m,\ell,\AlphabetSize) + |\Pat_1| / \log_{\AlphabetSize} \Textlen' +
  |\Pat_2| / \log_{\AlphabetSize} \Textlen') =
  \bigO(\log \log \Textlen + Q(m) + |\Pat_1| / \log \Textlen + |\Pat_2| / \log \Textlen)$ time.

  \DSConstruction
  By \cref{pr:lex-range-emptiness-to-prefix-range-emptiness-nonbinary}
  and the above discussion, construction of the above data structure
  answering lex-range emptiness queries on $\Text'$ (and hence also on $\Text$) takes
  $\bigO(\Textlen' / \log_{\AlphabetSize} \Textlen' + P'_t(m,\ell,\AlphabetSize)) =
  \bigO(\Textlen' / \log_{\AlphabetSize} \Textlen' + m + P_t(m)) =
  \bigO(\Textlen / \log \Textlen + P_t(m))$
  time and uses
  $\bigO(\Textlen' / \log_{\AlphabetSize} \Textlen' + P'_s(m,\ell,\AlphabetSize)) =
  \bigO(\Textlen' / \log_{\AlphabetSize} \Textlen' + m + P_s(m)) =
  \bigO(\Textlen / \log \Textlen + P_s(m))$ working space.
\end{proof}

\section{Equivalence of Lex-Range Minimum and Prefix RMQ}\label{sec:prefix-rmq}

\subsection{Problem Definitions}\label{sec:prefix-rmq-and-lex-range-minimum-problem-def}
\vspace{-1.5ex}

\begin{framed}
  \noindent
  \probname{Indexing for Lex-Range Minimum Queries}
  \begin{bfdescription}
  \item[Input:]
    The packed representation of a nonempty string
    $\Text \in \BinaryAlphabet^{\Textlen}$.
  \item[Output:]
    A data structure that, given the packed representation of any
    $\Pat_1,\Pat_2 \in \BinaryAlphabet^{*}$, returns
    the value (see \cref{def:lex-range})
    \[
      \min \LexRange{\Pat_1}{\Pat_2}{\Text} \cup \{\infty\}.
    \]
  \end{bfdescription}
\end{framed}

\begin{framed}
  \noindent
  \probname{Indexing for Prefix RMQ Queries}
  \begin{bfdescription}
  \item[Input:]
    A nonempty sequence $A[1 \dd m]$ of $m$ nonnegative integers
    (with $A[i] \in [1 \dd m]$ for $i \in [1 \dd m]$) and
    a sequence $W[1 \dd m]$ of $m$ binary strings of length
    $\ell = 1 + \lfloor \log m \rfloor$, with all
    strings represented in the packed form.
  \item[Output:]
    A data structure that, given
    the packed representation of any $X \in \BinaryAlphabet^{\leq \ell}$ and any
    indexes $b, e \in [0 \dd m]$,
    returns $\PrefixRMQ{A}{W}{b}{e}{X}$ (see \cref{def:prefix-rmq}).
  \end{bfdescription}
\end{framed}
\vspace{2ex}

\subsection{Reducing Prefix RMQ to Lex-Range Minimum Queries}\label{sec:from-prefix-rmq-to-lex-range-minimum}

\subsubsection{Preliminaries}\label{sec:from-prefix-rmq-to-lex-range-minimum-prelim}

\begin{definition}[Permuted sequence-to-string mapping]\label{def:perm-seq-to-string}
  Let $W[1 \dd m]$ be a sequence of $m \geq 1$ strings of length $\ell \geq 1$ over alphabet $\BinaryAlphabet$.
  Let $\pi[1 \dd m]$ be an array containing a permutation of $[1 \dd m]$. Denote
  $k = 1 + \lfloor \log m \rfloor$. For any $i \in [0 \dd m]$, let
  $s(i) = \Substitute{\Substitute{\bin{k}{i}}{\zero}{\two}}{\one}{\three}$ (see \cref{def:bin,def:sub}).
  We then define
  \[
    \PermSeqToString{\ell}{\pi}{W} := \textstyle\bigodot_{i=1}^{m} \revstr{W[\pi[i]]} \cdot \four \cdot s(\pi[i]-1).
  \]
\end{definition}

\begin{lemma}\label{lm:perm-seq-to-string-occ}
  Let $W[1 \dd m]$ be a sequence of $m \geq 1$ strings of length $\ell \geq 1$ over alphabet $\BinaryAlphabet$.
  Let $\pi[1 \dd m]$ be an array containing a permutation of $[1 \dd m]$, and let $\pi^{-1}[1 \dd m]$ be
  the inverse permutation (i.e., such that for every $i \in [1 \dd m]$, $\pi[\pi^{-1}[i]] = i$).
  Let $\Text = \PermSeqToString{\ell}{\pi}{W}$ (\cref{def:perm-seq-to-string}), $k = 1 + \lfloor \log m \rfloor$,
  and $\beta = \ell + k + 1$.
  Finally, let $X \in \BinaryAlphabet^{\leq \ell}$ and
  $\mathcal{P} = \{i \in [1 \dd m] : X\text{ is a prefix of }W[i]\}$.
  Then, it holds
  \[
    \OccTwo{\revstr{X} \cdot \four}{\Text} = \{\pi^{-1}[i] \cdot \beta - (k+|X|) : i \in \mathcal{P}\}.
  \]
\end{lemma}
\begin{proof}

  Denote $A = \{\pi^{-1}[i] \cdot \beta - (k+|X|) : i \in \mathcal{P}\}$.

  First, we show the inclusion $\OccTwo{\revstr{X} \cdot \four}{\Text} \subseteq A$.
  Let $j \in \OccTwo{\revstr{X} \cdot \four}{\Text}$. Note that $\OccTwo{\four}{\Text} = \{i \cdot \beta - k : i \in [1 \dd m]\}$.
  Thus, $j \in \OccTwo{\revstr{X} \cdot \four}{\Text}$ implies that for some $i \in [1 \dd m]$, it holds
  $j = i \cdot \beta - (k + |X|)$. Since $\Text[i \cdot \beta - k \dd |\Text|]$ is preceded
  in $\Text$ with $\revstr{W[\pi[i]]}$, it follows that $\revstr{X}$ is a suffix of $\revstr{W[\pi[i]]}$, or
  equivalently, $X$ is a prefix of $W[\pi[i]]$. This implies that $\pi[i] \in \mathcal{P}$, and hence, by definition
  of $A$, $\pi^{-1}[\pi[i]] \cdot \beta - (k+|X|) \in A$.
  It remains to note that $\pi^{-1}[\pi[i]] \cdot \beta - (k+|X|) = i \cdot \beta - (k+|X|) = j$.

  We now show the opposite inclusion. Let $j \in A$. Then, there exists $i \in \mathcal{P}$ such that
  $j = \pi^{-1}[i] \cdot \beta - (k+|X|)$. By $i \in \mathcal{P}$, it holds $i \in [1 \dd m]$ and $X$ is a prefix of $W[i]$.
  Note that, by \cref{def:perm-seq-to-string}, $\Text(\pi^{-1}[i] \cdot \beta \dd |\Text|]$
  is preceded in $\Text$ with $\revstr{W[\pi[\pi^{-1}[i]]]} \cdot \four \cdot s(\pi[\pi^{-1}[i]]-1)
  = \revstr{W[i]} \cdot \four \cdot s(i-1)$. Since $X$ is a prefix of $W[i]$, or equivalently, $\revstr{X}$ is a suffix
  of $\revstr{W[i]}$, we thus obtain that $\Text(\pi^{-1}[i] \cdot \beta \dd |\Text|]$ is preceded in $\Text$
  with $\revstr{X} \cdot \four \cdot s(i-1)$. Thus, $\pi^{-1}[i] \cdot \beta - (k+|X|) \in \OccTwo{\revstr{X} \cdot \four}{\Text}$.
  By $j = \pi^{-1}[i] \cdot \beta - (k+|X|)$, we thus obtain $j \in \OccTwo{\revstr{X} \cdot \four}{\Text}$.
\end{proof}

\begin{lemma}\label{lm:perm-seq-to-string}
  Let $W[1 \dd m]$ be a sequence of $m \geq 1$ strings of length $\ell \geq 1$ over alphabet $\BinaryAlphabet$.
  Let $\pi[1 \dd m]$ be an array containing a permutation of $[1 \dd m]$. Let $s(i)$ (where $i \in [0 \dd m]$) be
  as in \cref{def:perm-seq-to-string}.
  Denote
  $\Text = \PermSeqToString{\ell}{\pi}{W}$ (\cref{def:perm-seq-to-string}).
  Consider a string $X \in \BinaryAlphabet^{\leq \ell}$ and let $b, e \in [0 \dd m]$.
  Denote
  $\Pat_1 = \revstr{X} \cdot \four \cdot s(b)$,
  $\Pat_2 = \revstr{X} \cdot \four \cdot s(e)$,
  and $\beta = \ell + \lfloor \log m \rfloor + 2$. Finally, let
  \begin{itemize}
  \item $\mathcal{P} = \{i \in (b \dd e] : X\text{ is a prefix of }W[i]\}$,
  \item $\mathcal{Q} = \{\lceil j/\beta \rceil : j \in \LexRange{\Pat_1}{\Pat_2}{\Text}\}$ (\cref{def:lex-range}).
  \end{itemize}
  Then, it holds
  \[
    \mathcal{P} = \{\pi[q] : q \in \mathcal{Q}\}.
  \]
\end{lemma}
\begin{proof}

  Denote $k = 1 + \lfloor \log m \rfloor$.

  First, we prove the inclusion $\mathcal{P} \subseteq \{\pi[q] : q \in \mathcal{Q}\}$. Let $i \in \mathcal{P}$. Then,
  $i \in (b \dd e]$ and $X$ is a prefix of $W[i]$. Let $\pi^{-1}[1 \dd m]$ be an array containing the inverse permutation
  of $\pi$, i.e., such that for every $i \in [1 \dd m]$, it holds $\pi^{-1}[\pi[i]] = i$. Denote
  $i' = \pi^{-1}[i]$ and let $j = i' \cdot \beta - (k + |X|)$. Observe that, by \cref{def:perm-seq-to-string}, the
  string $\Text[i' \cdot \beta + 1 \dd |\Text|]$ is preceded in $\Text$ with $\revstr{W[\pi[i']]} \cdot \four \cdot s(\pi[i']-1)
  = \revstr{W[i]} \cdot \four \cdot s(i-1)$. Since $X$ is a prefix of $W[i]$, we thus obtain that 
  $\Text[j \dd |\Text|]$ has $\revstr{X} \cdot \four \cdot s(i-1)$ as a prefix. The assumption $b < i \leq e$, or equivalently,
  $b \leq i-1 < e$, implies that $s(b) \preceq s(i-1) \prec s(e)$. We thus obtain
  that $\Pat_1 \preceq \revstr{X} \cdot \four \cdot s(i-1) \prec \Pat_2$, and hence
  $\Pat_1 \preceq \Text[j \dd |\Text|] \prec \Pat_2$. We have thus proved that $j \in \LexRange{\Pat_1}{\Pat_2}{\Text}$.
  Since $\lceil j/\beta \rceil = i'$, we thus obtain $i' \in \mathcal{Q}$, and hence
  $\pi[i'] = \pi[\pi^{-1}[i]] = i \in \{\pi[q] : q \in \mathcal{Q}\}$.

  We now prove the second inclusion. Let $q' \in \{\pi[q] : q \in \mathcal{Q}\}$.
  Let $q \in \mathcal{Q}$ be such that $q' = \pi[q]$. Then, there exists $j \in \LexRange{\Pat_1}{\Pat_2}{\Text}$
  such that $\lceil j/\beta \rceil = q$. The assumption $j \in \LexRange{\Pat_1}{\Pat_2}{\Text}$ implies
  that $\Pat_1 \preceq \Text[j \dd |\Text|] \prec \Pat_2$. By definition of $\Pat_1$ and $\Pat_2$, this implies
  that $\Text[j \dd |\Text|]$ has the string $\revstr{X} \cdot \four$ as a prefix. Consequently, by
  \cref{lm:perm-seq-to-string-occ}, there exists $i \in [1 \dd m]$ such that $X$ is a prefix of $W[i]$ and
  $j = \pi^{-1}[i] \cdot \beta - (k + |X|)$. By \cref{def:perm-seq-to-string}, the suffix
  $\Text(\pi^{-1}[i] \cdot \beta - k \dd |\Text|]$ has the string $s(\pi[\pi^{-1}[i]]-1) = s(i-1)$ as a prefix.
  Putting together with the previous observation, we thus obtain that $\Text[j \dd |\Text|]$ has
  $\revstr{X} \cdot \four \cdot s(i-1)$ as a prefix. From the definition of $\Pat_1$ and $\Pat_2$, and the assumption
  $\Pat_1 \preceq \Text[j \dd |\Text|] \prec \Pat_2$, this implies that $s(b) \preceq s(i-1) \prec s(e)$, and hence
  $b \leq i-1 < e$, or equivalently, $b < i \leq e$. Combining this with $X$ being a prefix of $W[i]$ (see above),
  we obtain $i \in \mathcal{P}$. It remains to note that $j = \pi^{-1}[i] \cdot \beta - (k+|X|)$ implies that
  $q = \lceil j/\beta \rceil = \pi^{-1}[i]$. Hence, $q' = \pi[q] = \pi[\pi^{-1}[i]] = i$. We have thus proved
  that $q' \in \mathcal{P}$.
\end{proof}

\begin{proposition}\label{pr:perm-seq-to-string}
  Let $W[1 \dd m]$ be an array of $m \geq 1$ binary strings of
  length $\ell = 1 + \lfloor \log m \rfloor$ and let $\pi[1 \dd m]$ be an
  array containing a permutation of $[1 \dd m]$. Given arrays $W$ and $\pi$, with all
  strings in $W$ represented in the packed form, we can in $\bigO(m)$ time compute
  the packed representation of the string $\PermSeqToString{\ell}{\pi}{W}$ (\cref{def:perm-seq-to-string}).
\end{proposition}
\begin{proof}
  The construction proceeds as in \cref{pr:seq-to-string}, except in the last step,
  we additionally use the permutation $\pi$.
\end{proof}

\subsubsection{Problem Reduction}\label{sec:from-prefix-rmq-to-lex-range-minimum-problem-reduction}

\begin{lemma}\label{lm:prefix-rmq-to-lex-range-minimum}
  Let $A[1 \dd m]$ be an array of $m \geq 1$ integers and let $W[1 \dd m]$ be
  a sequence of $m$ strings of length $\ell \geq 1$ over alphabet $\BinaryAlphabet$.
  Let $s(i)$ (where $i \in [0 \dd m]$) be as in \cref{def:perm-seq-to-string}.
  Let $\pi[1 \dd m]$ be a permutation of $[1 \dd m]$ such that
  \[
    A[\pi[1]] \leq A[\pi[2]] \leq \cdots \leq A[\pi[m]].
  \]
  Denote $\Text = \PermSeqToString{\ell}{\pi}{W}$ (\cref{def:perm-seq-to-string}).
  Then, for every $X \in \BinaryAlphabet^{\leq \ell}$ and every $b, e \in [0 \dd m]$,
  letting 
  $\Pat_1 = \revstr{X} \cdot \four \cdot s(b)$ and
  $\Pat_2 = \revstr{X} \cdot \four \cdot s(e)$, the following two conditions are equivalent:
  \begin{itemize}
  \item $\PrefixRMQ{A}{W}{b}{e}{X} \neq \infty$ (\cref{def:prefix-rmq}),
  \item $\min \LexRange{\Pat_1}{\Pat_2}{\Text} \cup \{\infty\} \neq \infty$ (\cref{def:lex-range}).
  \end{itemize}
  Moreover, if
  $\min \LexRange{\Pat_1}{\Pat_2}{\Text} \cup \{\infty\} \neq \infty$,
  then
  \[
    \PrefixRMQ{A}{W}{b}{e}{X} = \pi\Bigg[\left\lceil\frac{\min \LexRange{\Pat_1}{\Pat_2}{\Text}}{\beta}\right\rceil\Bigg],
  \]
  where $\beta = \ell + \lfloor \log m \rfloor + 2$.
\end{lemma}
\begin{proof}

  Denote
  \begin{itemize}
  \item $\mathcal{P} = \{i \in (b \dd e] : X\text{ is a prefix of }W[i]\}$,
  \item $\mathcal{Q} = \{\lceil j/\beta \rceil : j \in \LexRange{\Pat_1}{\Pat_2}{\Text}\}$.
  \end{itemize}
  To show the first claim, observe that the following four conditions are equivalent.
  \begin{enumerate}
  \item\label{lm:prefix-rmq-to-lex-range-minimum-condition-1}
    $\PrefixRMQ{A}{W}{b}{e}{X} \neq \infty$,
  \item\label{lm:prefix-rmq-to-lex-range-minimum-condition-2}
    $\mathcal{P} \neq \emptyset$,
  \item\label{lm:prefix-rmq-to-lex-range-minimum-condition-3}
    $\mathcal{Q} \neq \emptyset$,
  \item\label{lm:prefix-rmq-to-lex-range-minimum-condition-4}
    $\min \LexRange{\Pat_1}{\Pat_2}{\Text} \cup \{\infty\} \neq \infty$.
  \end{enumerate}
  The equivalence of
  Conditions~\ref{lm:prefix-rmq-to-lex-range-minimum-condition-1} and~\ref{lm:prefix-rmq-to-lex-range-minimum-condition-2}
  follows by \cref{def:prefix-rmq}.
  The equivalence for~\ref{lm:prefix-rmq-to-lex-range-minimum-condition-2} and~\ref{lm:prefix-rmq-to-lex-range-minimum-condition-3}
  follows by \cref{lm:perm-seq-to-string}.
  Finally, the equivalence of Conditions~\ref{lm:prefix-rmq-to-lex-range-minimum-condition-3}
  and~\ref{lm:prefix-rmq-to-lex-range-minimum-condition-4} follows simply by definition of $Q$.
  Thus, we obtain the equivalence of Conditions~\ref{lm:prefix-rmq-to-lex-range-minimum-condition-1}
  and~\ref{lm:prefix-rmq-to-lex-range-minimum-condition-4}, i.e., the claim.

  We now show the second claim.
  Assume that
  it holds $\min \LexRange{\Pat_1}{\Pat_2}{\Text} \cup \{\infty\} \neq \infty$ and
  $\PrefixRMQ{A}{W}{b}{e}{X} \neq \infty$.
  Denote $j_{\min} = \min \LexRange{\Pat_1}{\Pat_2}{\Text}$ and let $i = \pi[\lceil j_{\min}/\beta\rceil]$.
  To show that it holds $i = \PrefixRMQ{A}{W}{b}{e}{X}$, we need to prove (see \cref{def:prefix-rmq})
  that $i \in (b \dd e]$, $X$ is a prefix of $W[i]$, and
  for that every $i' \in (b \dd e]$ such that $X$ is a prefix of $W[i']$, it holds $A[i] \leq A[i']$.
  \begin{itemize}
  \item By definition of $\mathcal{Q}$, we have
    $\lceil j_{\min} / \beta \rceil \in \mathcal{Q}$. Consequently, by \cref{lm:perm-seq-to-string}, it holds
    $\pi[\lceil j_{\min} / \beta \rceil] \in \mathcal{P}$, i.e., $i \in \mathcal{P}$. By definition of $\mathcal{P}$,
    we thus have $i \in (b \dd e]$ and $X$ is a prefix of $W[i]$.
  \item To show the remaining claim, consider any $i' \in (b \dd e]$ such that $X$ is a prefix of $W[i']$.
    By the definition of $\mathcal{P}$, we have $i' \in \mathcal{P}$.
    By \cref{lm:perm-seq-to-string}, there exist $j' \in \LexRange{\Pat_1}{\Pat_2}{\Text}$
    such that $\pi[\lceil j'/\beta \rceil] = i'$. By definition of $j_{\min}$, we have $j_{\min} \leq j'$.
    Thus, $\lceil j_{\min}/\beta \rceil \leq \lceil j'/\beta \rceil$. By the assumption
    $A[\pi[1]] \leq A[\pi[2]] \leq \dots \leq A[\pi[m]]$, we thus obtain that
    $A[\pi[\lceil j_{\min}/\beta \rceil]] \leq A[\pi[\lceil j'/\beta \rceil]]$.
    In other words, $A[i] \leq A[i']$.
    \qedhere
  \end{itemize}
\end{proof}

\subsubsection{Alphabet Reduction for Lex-Range Minimum Queries}\label{sec:lex-range-minimum-alphabet-reduction}

\begin{lemma}\label{lm:lex-range-min-alphabet-reduction}
  Let $\AlphabetSize \geq 2$, $\Text \in \IntegerAlphabet^{*}$,
  $\Pat_1 \in \IntegerAlphabet^{*}$, and $\Pat_2 \in \IntegerAlphabet^{*}$.
  Let
  $k = \lceil \log \AlphabetSize \rceil$,
  $\delta = 2k + 3$,
  $\Pat'_1 = \ebin{k}{\Pat_1}$,
  $\Pat'_2 = \ebin{k}{\Pat_2}$, and
  $\Text' = \ebin{k}{\Text}$ (\cref{def:ebin}).
  Then, the following conditions are equivalent (see \cref{def:lex-range}):
  \begin{itemize}
  \item $\LexRange{\Pat_1}{\Pat_2}{\Text} \neq \emptyset$,
  \item $\LexRange{\Pat'_1}{\Pat'_2}{\Text'} \neq \emptyset$.
  \end{itemize}
  Moreover, if $\LexRange{\Pat_1}{\Pat_2}{\Text} \neq \emptyset$, then
  \[
    \min \LexRange{\Pat_1}{\Pat_2}{\Text} =
      \frac{\min \LexRange{\Pat'_1}{\Pat'_2}{\Text'}-1}{\delta} + 1.
  \]
\end{lemma}
\begin{proof}
  The first claim (i.e., the equivalence) follows immediately by \cref{lm:lex-range-alphabet-reduction}.
  To prove the second claim, note that, letting $\LexRange{\Pat_1}{\Pat_2}{\Text} = \{a_1, a_2, \ldots, a_k\}$
  be such that $a_1 < a_2 < \dots < a_k$, and $(b_i)_{i \in [1 \dd k]}$ be a sequence defined by
  $b_i = (a_i-1)\delta + 1$, by \cref{lm:lex-range-alphabet-reduction} it holds
  $\LexRange{\Pat'_1}{\Pat'_2}{\Text'} = \{b_1, b_2, \ldots, b_k\}$. Clearly, we have $b_1 < b_2 < \dots < b_k$.
  Thus, $\min \LexRange{\Pat_1}{\Pat_2}{\Text} = a_1 = (b_1-1)/\delta + 1
  = (\min \LexRange{\Pat'_1}{\Pat'_2}{\Text'}-1)/\delta + 1$.
\end{proof}

\begin{proposition}\label{pr:lex-range-minimum-alphabet-reduction}
  Let $\Text \in \IntegerAlphabet^{\Textlen}$ be a nonempty string, where
  $2 \leq \AlphabetSize < \Textlen^{\bigO(1)}$. Given the packed representation of
  $\Text$, we can in $\bigO(\Textlen / \log_{\AlphabetSize} \Textlen)$ time
  construct the packed representation of a
  text $\Text_{\rm bin} \in \BinaryAlphabet^{*}$ satisfying
  $|\Text_{\rm bin}| = \Theta(\Textlen \log \AlphabetSize)$, integers
  $\alpha$, $\beta$, and $\gamma$, and a data structure that,
  given the packed representation of any
  $\Pat_1, \Pat_2 \in \IntegerAlphabet^{\leq m}$, in
  $\bigO(1 + m / \log_{\AlphabetSize} \Textlen)$ time returns the
  packed representation of $\Pat'_1, \Pat'_2 \in \BinaryAlphabet^{*}$
  satisfying
  $|\Pat'_i| = |\Pat_i| \cdot (2\lceil \log \AlphabetSize \rceil + 3)$ (where $i \in \{1,2\}$)
  and such that the following conditions are equivalent:
  \begin{itemize}
  \item $\LexRange{\Pat_1}{\Pat_2}{\Text} \neq \emptyset$,
  \item $\LexRange{\Pat'_1}{\Pat'_2}{\Text} \neq \emptyset$.
  \end{itemize}
  Moreover, if $\LexRange{\Pat_1}{\Pat_2}{\Text} \neq \emptyset$, then it holds
  \[
    \min \LexRange{\Pat_1}{\Pat_2}{\Text}
      = \frac{\min \LexRange{\Pat'_1}{\Pat'_2}{\Text_{\rm bin}} - \alpha}{\beta} + \gamma.
  \]
\end{proposition}
\begin{proof}
  The proof proceeds the same as in \cref{pr:lex-range-reporting-alphabet-reduction}, except
  the correctness of the formula follows by \cref{lm:lex-range-min-alphabet-reduction} (rather than
  \cref{lm:lex-range-alphabet-reduction}).
\end{proof}

\subsubsection{Summary}\label{sec:from-prefix-rmq-to-lex-range-minimum-summary}

\begin{theorem}\label{th:prefix-rmq-to-lex-range-minimum}
  Consider a data structure answering lex-range minimum queries
  (see \cref{sec:prefix-rmq-and-lex-range-minimum-problem-def})
  that, for any
  text $\Text \in \BinaryAlphabet^{\Textlen}$, achieves the following complexities
  (where in the preprocessing we assume that we are given as input the packed
  representation of $\Text$, and at query time we are given a packed
  representation of $\Pat_1,\Pat_2 \in \BinaryAlphabet^{\leq k}$):
  \begin{itemize}
  \item space usage $S(\Textlen)$,
  \item preprocessing time $P_t(\Textlen)$,
  \item preprocessing space $P_s(\Textlen)$,
  \item query time $Q(\Textlen, k)$.
  \end{itemize}
  For every sequence $W[1 \dd m]$ of $m \geq 1$ binary strings of length
  $\ell = 1 + \lfloor \log m \rfloor$ and any array $A[1 \dd m]$ of integers
  (where $A[i] \in [1 \dd m]$ for $i \in [1 \dd m]$), there exists
  $\Textlen = \bigO(m \log m)$ and $k = \bigO(\log m)$ such that, given the sequences $A$ and $W$, with
  all strings in $W$ represented in the packed form, we can in
  $\bigO(m + P_t(\Textlen))$ time and using $\bigO(m + P_s(\Textlen))$
  working space construct a data structure of size $\bigO(m + S(\Textlen))$
  that, given the packed representation of any $X \in \BinaryAlphabet^{\leq \ell}$
  and any pair of integers $b,e \in [0 \dd m]$, computes
  $\PrefixRMQ{A}{W}{b}{e}{X}$ in $\bigO(Q(\Textlen, k))$ time.
\end{theorem}
\begin{proof}

  We use the following definitions.
  Let $\pi[1 \dd m]$ be a permutation of $[1 \dd m]$ such that
  for every $i, j \in [1 \dd m]$ satisfying $i < j$, it holds
  $A[\pi[i]] < A[\pi[j]]$, or $A[\pi[i]] = A[\pi[j]]$ and $\pi[i] < \pi[j]$.
  Note that, in particular, we then have $A[\pi[1]] \leq A[\pi[2]] \leq \dots \leq A[\pi[m]]$.
  Let $\Text_{\rm aux} = \PermSeqToString{\ell}{\pi}{W}$ (\cref{def:perm-seq-to-string}).
  Denote $\Textlen_{\rm aux} =
  |\Text_{\rm aux}| = m \cdot (2\ell+1) = \Theta(m \log m)$. Let
  $\Text_{\rm bin}$ and $\alpha_{\rm bin}$, $\beta_{\rm bin}$, and $\gamma_{\rm bin}$
  denote the text and the three integers obtained
  by applying \cref{pr:lex-range-minimum-alphabet-reduction} to text
  $\Text_{\rm aux}$. Denote $\Textlen = |\Text_{\rm bin}|$.
  Since $\Text_{\rm aux}$ is over alphabet $\{\zero, \dots, \four\}$,
  by \cref{pr:lex-range-minimum-alphabet-reduction} we have
  $\Textlen = \Theta(|\Text_{\rm aux}|) = \Theta(m \log m)$.
  Denote $k = 18 \lfloor \log m \rfloor + 27 = \bigO(\log m)$.

  \DSComponents
  The data structure consists of the following components:
  \begin{enumerate}
  \item The structure from \cref{pr:lex-range-minimum-alphabet-reduction} applied
    to text $\Text_{\rm aux}$. Since $\Text_{\rm aux}$ is over alphabet
    $\{\zero, \dots, \four\}$, the structure needs
    $\bigO(|\Text_{\rm aux}| / \log |\Text_{\rm aux}|) =
    \bigO(m)$ space.
  \item The data structure from the claim (i.e., answering lex-range minimum
    queries) for the text $\Text_{\rm bin}$. The structure needs
    $\bigO(S(\Textlen))$ space.
  \item The array $\pi[1 \dd m]$ storing the permutation $\pi$ defined above. It needs $\bigO(m)$ space.
  \item A lookup table $L_{\rm rev}$ defined as in
    \cref{th:prefix-rank-to-pattern-ranking}. It needs
    $\bigO(\sqrt{\Textlen}) = \bigO(m)$ space.
  \item A lookup table $L_{\rm map}$ defined as in
    \cref{th:prefix-rank-to-pattern-ranking}. It also needs
    $\bigO(\sqrt{\Textlen}) = \bigO(m)$ space.
  \item The integers $\alpha_{\rm bin}$, $\beta_{\rm bin}$, and $\gamma_{\rm bin}$.
  \end{enumerate}
  In total, the structure needs $\bigO(m + S(\Textlen))$ space.

  \DSQueries
  The queries are answered as follows. Assume that we are given
  the packed representation of some $X \in \BinaryAlphabet^{\leq \ell}$ and
  two integers $b,e \in [0 \dd m]$. The algorithm to compute $\PrefixRMQ{A}{W}{b}{e}{X}$
  proceeds as follows:
  \begin{enumerate}
  \item Using $L_{\rm rev}$ and $L_{\rm map}$, in $\bigO(1)$ time we compute the packed
    representation of strings $\Pat_1 = \revstr{X} \cdot \four \cdot s(b)$
    and $\Pat_2 = \revstr{X} \cdot \four \cdot s(e)$
    (over alphabet $\{\zero, \dots, \four\}$).
    Note that $|\Pat_1| = |\Pat_2| = |X| + 1 + \ell \leq 2\ell + 1 \leq 2\lfloor \log m \rfloor + 3$.
  \item Using \cref{pr:lex-range-minimum-alphabet-reduction}, in
    $\bigO(1 + |\Pat_1| / \log \Textlen_{\rm aux} + |\Pat_2| / \log \Textlen_{\rm aux}) =
    \bigO(1 + \ell / \log m) = \bigO(1)$ time, we compute
    the packed representation of strings $\Pat'_1,\Pat'_2 \in \BinaryAlphabet^{*}$
    of length $|\Pat'_1| = |\Pat'_2| = 9 \cdot |\Pat_1| \leq 18 \lfloor \log m \rfloor + 27 \leq k$
    such that $\LexRange{\Pat_1}{\Pat_2}{\Text_{\rm aux}} \neq \emptyset$ holds if and only if
    $\LexRange{\Pat'_1}{\Pat'_2}{\Text_{\rm bin}} \neq \emptyset$, and moreover, if
    $\LexRange{\Pat_1}{\Pat_2}{\Text_{\rm aux}} \neq \emptyset$, then it holds
    \[
      \min \LexRange{\Pat_1}{\Pat_2}{\Text_{\rm aux}} =
        \frac{\min \LexRange{\Pat'_1}{\Pat'_2}{\Text_{\rm bin}} - \alpha_{\rm bin}}{\beta_{\rm bin}} + \gamma_{\rm bin}.
    \]
  \item Using the structure from the claim, we compute
    $p'_{\min} = \min \LexRange{\Pat'_1}{\Pat'_2}{\Text_{\rm bin}} \cup \{\infty\}$.
    By $|\Pat'_1|, |\Pat'_2| \leq k$, this takes $\bigO(Q(\Textlen, k))$ time.
  \item If $p'_{\min} = \infty$
    (which is equivalent to $\LexRange{\Pat'_1}{\Pat'_2}{\Text_{\rm bin}} = \emptyset$),
    then by the above, it holds $\LexRange{\Pat_1}{\Pat_2}{\Text_{\rm aux}} = \emptyset$, or equivalently,
    $\min \LexRange{\Pat_1}{\Pat_2}{\Text_{\rm aux}} \cup \{\infty\} = \infty$. By
    \cref{lm:prefix-rmq-to-lex-range-minimum}, this is equivalent to
    $\PrefixRMQ{A}{W}{b}{e}{X} = \infty$. We thus return $\infty$, and conclude the query algorithm.
    Let us now assume that $p'_{\min} \neq \infty$. We then have
    $\LexRange{\Pat'_1}{\Pat'_2}{\Text_{\rm bin}} \neq \emptyset$,
    $\LexRange{\Pat_1}{\Pat_2}{\Text_{\rm aux}} \neq \emptyset$, and
    $\min \LexRange{\Pat_1}{\Pat_2}{\Text_{\rm aux}} \cup \{\infty\} \neq \infty$.
  \item In $\bigO(1)$ time, we compute the value
    $p_{\rm min} = (p'_{\rm min} - \alpha_{\rm bin})/\beta_{\rm bin} + \gamma_{\rm bin}$.
    By the above, it holds $p_{\rm min} = \min \LexRange{\Pat_1}{\Pat_2}{\Text_{\rm aux}}$.
  \item In $\bigO(1)$ time compute $i = \lceil p_{\rm min} / \beta \rceil$, where $\beta = 2\ell + 1$.
    Then, compute and return the value $r = \pi[i]$.
    By \cref{lm:prefix-rmq-to-lex-range-minimum}, it holds $\PrefixRMQ{A}{W}{b}{e}{X} = r$.
  \end{enumerate}
  In total, the query takes $\bigO(Q(\Textlen, k))$ time.

  \DSConstruction
  The components of the data structure are constructed as follows:
  \begin{enumerate}
  \item The first component of the structure is computed as follows:
    \begin{enumerate}
    \item First, we compute the array containing the permutation $\pi$.
      To this end, in $\bigO(m)$ time we initialize an array $A_{\rm perm}[1 \dd m]$
      defined by $A_{\rm perm}[i] = (A[i],i)$. We then sort this array lexicographically.
      Utilizing that all numbers are in the range $[1 \dd m]$, we use a 2-round
      radix sort and achieve $\bigO(m)$ time. The resulting array contains the
      permutation $\pi[1 \dd m]$ on the second coordinate.
    \item In $\bigO(m)$ time we apply \cref{pr:perm-seq-to-string} to the
      permutation $\pi$ and the sequence $W$ to compute the packed representation of the string
      $\Text_{\rm aux} = \PermSeqToString{\ell}{\pi}{W}$ (\cref{def:perm-seq-to-string}).
    \item Lastly, we apply \cref{pr:lex-range-minimum-alphabet-reduction}
      to $\Text_{\rm aux}$. Since $\Text_{\rm aux}$ is over alphabet
      $\{\zero, \dots, \four\}$, this takes
      $\bigO(|\Text_{\rm aux}| / \log |\Text_{\rm aux}|) = \bigO(m)$
      time. Note that \cref{pr:lex-range-minimum-alphabet-reduction}, in addition
      to the data structure, also returns integers $\alpha_{\rm bin}$, $\beta_{\rm bin}$, and
      $\gamma_{\rm bin}$, and the packed representation of the string $\Text_{\rm bin}$ (defined above).
    \end{enumerate}
  \item We apply the preprocessing from the claim to the string $\Text_{\rm bin}$.
    This takes $\bigO(P_t(|\Text_{\rm bin}|)) = \bigO(P_t(\Textlen))$ time
    and uses $\bigO(P_s(|\Text_{\rm bin}|)) = \bigO(P_s(\Textlen))$ working space.
  \item We store the permutation $\pi$ (computed above) in $\bigO(m)$ time.
  \item The lookup table $L_{\rm rev}$ is computed in $\bigO(m)$ time similarly
    as in the proof of \cref{pr:seq-to-string}.
  \item Next, we compute the lookup table $L_{\rm map}$. Similarly as above,
    we proceed as in \cref{pr:seq-to-string}, and spend $\bigO(m)$ time.
  \item Lastly, we store integers $\alpha_{\rm bin}$, $\beta_{\rm bin}$,
    and $\gamma_{\rm bin}$, which were already computed above.
  \end{enumerate}
  In total, the construction takes
  $\bigO(m + P_t(\Textlen))$ time and uses
  $\bigO(m + P_s(\Textlen))$ working space.
\end{proof}

\subsection{Reducing Lex-Range Minimum to Prefix RMQ}\label{sec:lex-range-minimum-to-prefix-rmq}

\subsubsection{Problem Reduction}

\paragraph{The Short Patterns}

\begin{proposition}\label{pr:short-occ-min}
  Let $\Text \in \IntegerAlphabet^{\Textlen}$ be such that $2 \leq \AlphabetSize < \Textlen^{1/13}$ and $\Text[\Textlen]$ does not
  occur in $\Text[1 \dd \Textlen)$. Let $\tau = \lfloor \mu\log_{\AlphabetSize} \Textlen \rfloor$, where $\mu$ is a positive
  constant smaller than $\tfrac{1}{12}$ such that $\tau \geq 1$.
  Given the packed representation of $\Text$, we can in $\bigO(\Textlen / \log_{\AlphabetSize} \Textlen)$ time
  construct a data structure that given the packed representation of any $X \in \IntegerAlphabet^{\leq 3\tau-1}$
  satisfying $\OccTwo{X}{\Text} \neq \emptyset$, returns $\min \OccTwo{X}{\Text}$ in $\bigO(1)$ time.
\end{proposition}
\begin{proof}
  The data structure consists of a single component: the lookup table $L_{\rm minocc}$.
  Its definition, space requirement, usage, and construction is as
  described in Section~5.2 of~\cite{sublinearlz}.\footnote{The construction is originally described only for $X \in \IntegerAlphabet^{<3\tau - 1}$, but it is easy to check that without modifications it works for $X \in \IntegerAlphabet^{\leq 3\tau-1}$.}
\end{proof}

\begin{proposition}\label{pr:lex-range-minimum-short}
  Let $\Text \in \IntegerAlphabet^{\Textlen}$ be such that $2 \leq \AlphabetSize < \Textlen^{1/13}$ and $\Text[\Textlen]$ does not
  occur in $\Text[1 \dd \Textlen)$. Let $\tau = \lfloor \mu\log_{\AlphabetSize} \Textlen \rfloor$, where $\mu$ is a positive
  constant smaller than $\tfrac{1}{12}$ such that $\tau \geq 1$.
  Given the packed representation of $\Text$, we can in $\bigO(\Textlen / \log_{\AlphabetSize} \Textlen)$ time
  construct a data structure that answers the following queries:
  \begin{enumerate}
  \item\label{pr:lex-range-minimum-short-it-1}
    Given the packed representation of any $X_1, X_2 \in \IntegerAlphabet^{\leq 3\tau-1}$,
    return $\min \LexRange{X_1}{X_2}{\Text} \cup \{\infty\}$
    (see \cref{def:lex-range}) in $\bigO(1)$ time.
  \item\label{pr:lex-range-minimum-short-it-2}
    Given the packed representation of any $X \in \IntegerAlphabet^{\leq 3\tau-1}$,
    return $\min \LexRange{X}{X c^{\infty}}{\Text} \cup \{\infty\}$
    (where $c = \AlphabetSize - 1$) in $\bigO(1)$ time.
  \item\label{pr:lex-range-minimum-short-it-3}
    Given the packed representation of any $X_1, X_2 \in \IntegerAlphabet^{\leq 3\tau-1}$,
    return $\min \LexRange{X_1 c^{\infty}}{X_2}{\Text} \cup \{\infty\}$
    (where $c = \AlphabetSize - 1$) in $\bigO(1)$ time.
  \end{enumerate}
\end{proposition}
\begin{proof}

  We use the following definitions.
  Let $B_{3\tau-1}[1 \dd \Textlen]$ be a bitvector defined so that
  for every $i \in [1 \dd \Textlen]$, $B_{3\tau-1}[i] = \one$ holds
  if and only if $i = \Textlen$, or $i < \Textlen$ and $\LCE{\Text}{\SA{\Text}[i]}{\SA{\Text}[i+1]} < 3\tau-1$.
  Let $k = \Rank{B_{3\tau-1}}{\Textlen}{\one}$. Let $A_{\min}[1 \dd k]$ be an array of positive integers
  such that for every $i \in [1 \dd k]$, $A_{\min}[i] = \min \OccTwo{X}{\Text}$, where $i' = \Select{B_{3\tau-1}}{i}{\one}$
  and $X = \Text[\SA{\Text}[i'] \dd \min(\Textlen + 1, \SA{\Text}[i'] + 3\tau - 1))$. Observe that
  $k = \bigO(\AlphabetSize^{3\tau-1}) = \bigO(\sqrt{\Textlen})$.

  \DSComponents
  The data structure consists of the following components:
  \begin{enumerate}
  \item The data structure from \cref{pr:nav-index-short}. It needs
    $\bigO(\Textlen / \log_{\AlphabetSize} \Textlen)$ space.
  \item The bitvector $B_{3\tau-1}[1 \dd \Textlen]$ augmented with the support for $\bigO(1)$-time rank and select queries
    using \cref{th:bin-rank-select}. The bitvector together with the augmentation of \cref{th:bin-rank-select}
    needs $\bigO(\Textlen / \log \Textlen) = \bigO(\Textlen / \log_{\AlphabetSize} \Textlen)$ space.
  \item The array $A_{\min}[1 \dd k]$ augmented with support for $\bigO(1)$-time RMQ queries using
    \cref{th:rmq}. The array together with the augmentation of \cref{th:rmq} needs
    $\bigO(k) = \bigO(\sqrt{\Textlen}) = \bigO(\Textlen / \log_{\AlphabetSize} \Textlen)$ space.
  \end{enumerate}
  In total, the structure needs $\bigO(\Textlen / \log_{\AlphabetSize} \Textlen)$ space.

  \DSQueries
  The queries are answered as follows:
  \begin{enumerate}
  \item Let $X_1, X_2 \in \IntegerAlphabet^{\leq 3\tau-1}$.
    Given the packed representation of strings $X_1$ and $X_2$, we compute
    $\min \LexRange{X_1}{X_2}{\Text} \cup \{\infty\}$ as follows:
    \begin{enumerate}
    \item Using \cref{pr:nav-index-short}, in $\bigO(1)$ time compute
      $b_k = \RangeBegTwo{X_k}{\Text}$ for $k \in \{1,2\}$. Observe that
      then
      \[
        \LexRange{X_1}{X_2}{\Text} = \{\SA{\Text}[i]\}_{i \in (b_1 \dd b_2]}
      \]
      (see \cref{rm:lex-range}).
      If $b_1 \geq b_2$, then it holds $\LexRange{X_1}{X_2}{\Text} = \emptyset$, and hence
      we return $\min \LexRange{X_1}{X_2}{\Text} \cup \{\infty\} = \infty$.
      Let us thus assume that $b_1 < b_2$.
      Note that, by definition of $B_{3\tau-1}$,
      if $b_k > 0$, then $B_{3\tau-1}[b_k] = \one$ holds for $k \in \{1,2\}$.
    \item If $b_1 = 0$, then we set $i_1 = 0$. Otherwise, using \cref{th:bin-rank-select}, in $\bigO(1)$ time
      we compute $i_1 = \Rank{B_{3\tau-1}}{b_1}{\one}$. Analogously we compute $i_2$.
    \item Using \cref{th:rmq}, in $\bigO(1)$ time we compute $i_{\min} = \RMQ{A_{\min}}{i_1}{i_2}$. Observe that
      by definition of $B_{3\tau-1}$ and $A_{\min}$, we then have
      $\min \{\SA{\Text}[i]\}_{i \in (b_1 \dd b_2]} = A_{\min}[i_{\min}]$.
      Thus, we return $A_{\min}[i_{\min}]$ as the answer.
    \end{enumerate}
    In total, we spend $\bigO(1)$ time.
  \item Let us now consider $X \in \IntegerAlphabet^{\leq 3\tau-1}$,
    where $c = \AlphabetSize - 1$. Given the packed representation of $X$, we compute
    $\min \LexRange{X}{X c^{\infty}}{\Text} \cup \{\infty\}$
    by setting $X_1 = X$ and $X_2 = X c^{\infty}$, and
    using the same algorithm as above, except
    to compute $b_2 = \RangeBegTwo{X c^{\infty}}{\Text}$, we observe that
    $\RangeBegTwo{X c^{\infty}}{\Text} = \RangeEndTwo{X}{\Text}$. Thus, we can determine $b_2$ using
    \cref{pr:nav-index-short} in $\bigO(1)$ time.
  \item Let $X_1, X_2 \in \IntegerAlphabet^{\leq 3\tau-1}$.
    Given the packed representation of strings $X_1$ and $X_2$, we compute
    $\min \LexRange{X_1 c^{\infty}}{X_2}{\Text} \cup \{\infty\}$ (where $c = \AlphabetSize - 1$)
    similarly as above, except to compute $b_1 = \RangeBegTwo{X_1 c^{\infty}}{\Text}$, we use the fact
    that $\RangeBegTwo{X_1 c^{\infty}}{\Text} = \RangeEndTwo{X_1}{\Text}$. Thus, we can
    compute $b_1$ using \cref{pr:nav-index-short} in $\bigO(1)$ time.
  \end{enumerate}

  \DSConstruction
  The components of the data structure are constructed as follows:
  \begin{enumerate}
  \item The first component is constructed using \cref{pr:nav-index-short}
    in $\bigO(\Textlen / \log_{\AlphabetSize} \Textlen)$ time.
  \item To construct the second component, we first compute the bitvector
    $B_{3\tau-1}[1 \dd \Textlen]$ as described in~\cite[Proposition~5.3]{breaking}
    in $\bigO(\Textlen / \log_{\AlphabetSize} \Textlen)$ time.
    We then augment $B_{3\tau-1}$ with support for $\bigO(1)$-time rank/select queries
    using \cref{th:bin-rank-select} in
    $\bigO(\Textlen / \log \Textlen) = \bigO(\Textlen / \log_{\AlphabetSize} \Textlen)$ time.
  \item To describe the construction of $A_{\min}[1 \dd k]$, we first introduce as auxiliary
    array $A_{\rm str}[1 \dd k]$ defined such that, for every $i \in [1 \dd k]$,
    it holds $A_{\rm str}[i] = \Text[\SA{\Text}[i'] \dd \min(\Textlen + 1, \SA{\Text}[i'] + 3\tau - 1))$,
    where $i' = \Select{B_{3\tau-1}}{i}{\one}$. Observe that for every $i \in [1 \dd k]$, we then
    have $A_{\min}[i] = \min \OccTwo{A_{\rm str}[i]}{\Text}$. The construction of $A_{\min}[1 \dd k]$ proceeds
    as follows:
    \begin{enumerate}
    \item In $\bigO(\Textlen / \log_{\AlphabetSize} \Textlen)$ time construct the array
      $A_{\rm str}[1 \dd k]$ (with all strings stored in the packed representation)
      as described in~\cite[Proposition~5.3]{breaking}.
    \item In $\bigO(\Textlen / \log_{\AlphabetSize} \Textlen)$ time construct the data
      structure from \cref{pr:short-occ-min}.
    \item For every $i \in [1 \dd k]$, compute $A_{\min}[i] = \min \OccTwo{A_{\rm str}[i]}{\Text}$
      in $\bigO(1)$ time using \cref{pr:short-occ-min}. In total, we spend $\bigO(k)$ time.
    \end{enumerate}
    In total, the construction of $A_{\min}[1 \dd k]$ takes
    $\bigO(\Textlen / \log_{\AlphabetSize} \Textlen + k) = \bigO(\Textlen / \log_{\AlphabetSize} \Textlen)$ time.
  \end{enumerate}
  In total, the construction takes
  $\bigO(\Textlen / \log_{\AlphabetSize} \Textlen)$ time.
\end{proof}

\paragraph{The Nonperiodic Patterns}

\subparagraph{Combinatorial Properties}

\begin{lemma}\label{lm:lex-range-minimum-nonperiodic}
  Let $\Text \in \Sigma^{\Textlen}$, $\tau \in [1 \dd \lfloor \tfrac{\Textlen}{2} \rfloor]$, and assume that
  $\Text[\Textlen]$ does not occur in $\Text[1 \dd \Textlen)$.
  Let $\SSS$ be a $\tau$-synchronizing set of $\Text$. Denote
  $(s_i)_{i \in [1 \dd n']} = \LexSorted{\SSS}{\Text}$ (\cref{def:lex-sorted}).
  Let $A_{\SSS}[1 \dd n']$ and $A_{\rm str}[1 \dd n']$ be defined by
  \begin{itemize}
  \item $A_{\SSS}[i] = s_i$,
  \item $A_{\rm str}[i] = \revstr{D_i}$, where $D_i = \Textinf[s_i - \tau \dd s_i + 2\tau)$.
  \end{itemize}
  Let $D \in \DistPrefixes{\tau}{\Text}{\SSS}$ (\cref{def:dist-prefixes})
  and let $\Pat_1, \Pat_2 \in \Sigma^{+}$ be $\tau$-nonperiodic patterns (\cref{def:periodic-pattern})
  both having $D$ as a prefix.
  Denote $\deltatext = |D| - 2\tau$. For $k \in \{1,2\}$, let
  $\Pat'_k = \Pat_k(\deltatext \dd |\Pat_k|]$, and
  $b_k = |\{i \in [1 \dd n'] : \Text[s_i \dd \Textlen] \prec \Pat'_k\}|$.
  Then, the following conditions are equivalent:
  \begin{enumerate}
  \item $\LexRange{\Pat_1}{\Pat_2}{\Text} \neq \emptyset$ (see \cref{def:lex-range}),
  \item $\PrefixRMQ{A_{\SSS}}{A_{\rm str}}{b_1}{b_2}{\revstr{D}} \neq \infty$ (see \cref{def:prefix-rmq}).
  \end{enumerate}
  Moreover, if $\LexRange{\Pat_1}{\Pat_2}{\Text} \neq \emptyset$, then
  \[
    \min \LexRange{\Pat_1}{\Pat_2}{\Text} =
      A_{\SSS}[\PrefixRMQ{A_{\SSS}}{A_{\rm str}}{b_1}{b_2}{\revstr{D}}] - \deltatext.
  \]
\end{lemma}
\begin{proof}

  The first claim (i.e., the equivalence)
  follows by combining \cref{def:prefix-rmq} and \cref{lm:lex-range-emptiness-nonperiodic}.

  To prove the second claim, we combine \cref{lm:lex-range-nonperiodic-array} and \cref{def:prefix-rmq}
  to obtain:
  \begin{align*}
    \min \LexRange{\Pat_1}{\Pat_2}{\Text}
      &= \min \{A_{\SSS}[i] - \deltatext : i \in (b_1 \dd b_2]\text{ and }\revstr{D}\text{ is a prefix of }A_{\rm str}[i]\}\\
      &= \min \{A_{\SSS}[i] : i \in (b_1 \dd b_2]\text{ and }\revstr{D}\text{ is a prefix of }A_{\rm str}[i]\} - \deltatext\\
      &= A_{\SSS}[\PrefixRMQ{A_{\SSS}}{A_{\rm str}}{b_1}{b_2}{\revstr{D}}] - \deltatext.
      \qedhere
  \end{align*}
\end{proof}

\begin{lemma}\label{lm:lex-range-minimum-nonperiodic-permuted}
  Let $\Text \in \Sigma^{\Textlen}$, $\tau \in [1 \dd \lfloor \tfrac{\Textlen}{2} \rfloor]$, and assume that
  $\Text[\Textlen]$ does not occur in $\Text[1 \dd \Textlen)$.
  Let $\SSS$ be a $\tau$-synchronizing set of $\Text$. Denote
  $(s^{\rm lex}_i)_{i \in [1 \dd n']} = \LexSorted{\SSS}{\Text}$ (\cref{def:lex-sorted}),
  $(s^{\rm text}_i)_{i \in [1 \dd n']} = \Sort{\SSS}$ (\cref{def:sort}), and let
  $\pi[1 \dd n']$ be a permutation of $[1 \dd n']$ such that for every $i \in [1 \dd n']$, it holds
  $s^{\rm lex}_i = s^{\rm text}_{\pi[i]}$.
  Let $A_{\pi}[1 \dd n']$, $A_{\SSS}[1 \dd n']$, and $A_{\rm str}[1 \dd n']$ be defined by
  \begin{itemize}
  \item $A_{\pi}[i] = \pi[i]$,
  \item $A_{\SSS}[i] = s^{\rm text}_i$,
  \item $A_{\rm str}[i] = \revstr{D_i}$, where $D_i = \Textinf[s_i - \tau \dd s_i + 2\tau)$.
  \end{itemize}
  Let $D \in \DistPrefixes{\tau}{\Text}{\SSS}$ (\cref{def:dist-prefixes})
  and let $\Pat_1, \Pat_2 \in \Sigma^{+}$ be $\tau$-nonperiodic patterns (\cref{def:periodic-pattern})
  both having $D$ as a prefix.
  Denote $\deltatext = |D| - 2\tau$. For $k \in \{1,2\}$, let
  $\Pat'_k = \Pat_k(\deltatext \dd |\Pat_k|]$, and
  $b_k = |\{i \in [1 \dd n'] : \Text[s^{\rm lex}_i \dd \Textlen] \prec \Pat'_k\}|$.
  Then, the following conditions are equivalent:
  \begin{enumerate}
  \item $\LexRange{\Pat_1}{\Pat_2}{\Text} \neq \emptyset$ (see \cref{def:lex-range}),
  \item $\PrefixRMQ{A_{\pi}}{A_{\rm str}}{b_1}{b_2}{\revstr{D}} \neq \infty$ (see \cref{def:prefix-rmq}).
  \end{enumerate}
  Moreover, if $\LexRange{\Pat_1}{\Pat_2}{\Text} \neq \emptyset$, then
  \[
    \min \LexRange{\Pat_1}{\Pat_2}{\Text} =
      A_{\SSS}[A_{\pi}[\PrefixRMQ{A_{\pi}}{A_{\rm str}}{b_1}{b_2}{\revstr{D}}]] - \deltatext.
  \]
\end{lemma}
\begin{proof}

  To show the first claim, note that 
  whether $\PrefixRMQ{A}{S}{b}{e}{X} \neq \infty$ holds in \cref{def:prefix-rmq}
  does not depend on the array $A$.
  Thus, the first claim follows by \cref{lm:lex-range-minimum-nonperiodic}.

  To prove the second claim, observe that by definition of the permutation $\pi$,
  for every $i,j \in [1 \dd n']$ such that $i \neq j$,
  $s^{\rm lex}_i < s^{\rm lex}_j$ holds if and only if $s^{\rm text}_{\pi[i]} < s^{\rm text}_{\pi[j]}$.
  On the other hand, by \cref{def:sort}, $s^{\rm text}_{\pi[i]} < s^{\rm text}_{\pi[j]}$ holds if and only if
  $\pi[i] < \pi[j]$. Thus, $s^{\rm lex}_i < s^{\rm lex}_j$ holds if and only if $\pi[i] < \pi[j]$.
  This implies that for every $Q \subseteq [1 \dd n']$, $\argmin_{j \in Q} s^{\rm lex}_j =
  \argmin_{j \in Q} \pi[j]$.
  In particular, letting $i = \argmin_{j \in Q} \pi[j]$, it holds $\min \{s^{\rm lex}_j : j \in Q\} = s_i =
  s^{\rm text}_{\pi[i]}$.
  Consequently, combining \cref{lm:lex-range-nonperiodic} and \cref{def:prefix-rmq},
  we obtain:
  \begin{align*}
    \min \LexRange{\Pat_1}{\Pat_2}{\Text}
      &= \min \{s^{\rm lex}_i - \deltatext : i \in (b_1 \dd b_2]\text{ and }\revstr{D}\text{ is a prefix of }A_{\rm str}[i]\}\\
      &= \min \{s^{\rm lex}_i : i \in (b_1 \dd b_2]\text{ and }\revstr{D}\text{ is a prefix of }A_{\rm str}[i]\} - \deltatext\\
      &= A_{\SSS}[A_{\pi}[\PrefixRMQ{A_{\pi}}{A_{\rm str}}{b_1}{b_2}{\revstr{D}}]] - \deltatext.
      \qedhere
  \end{align*}
\end{proof}

\begin{remark}\label{rm:lex-range-minimum-nonperiodic-permuted}
  The reason for introducing the array $A_{\pi}$ in \cref{lm:lex-range-minimum-nonperiodic-permuted}
  is so that the first array used in prefix RMQ queries has values between $1$ and the size of the array.
  This is used in \cref{pr:lex-range-minimum-nonperiodic}.
\end{remark}

\subparagraph{Algorithms}

\begin{proposition}\label{pr:lex-range-minimum-nonperiodic}
  Consider a data structure that, given any array of integers $A[1 \dd k]$ (where $A[i] \in [1 \dd k]$ for $i \in [1 \dd k]$)
  and a sequence $W[1 \dd k]$ strings of length
  $\ell$ over alphabet $\IntegerAlphabet$
  achieves the following complexities (where the input strings
  during construction are given in the packed representation, and at query time we are given any $b,e \in [0 \dd k]$ and the packed
  representation of any $X \in \IntegerAlphabet^{\leq \ell}$, and we return the position
  $\PrefixRMQ{A}{W}{b}{e}{X}$; see \cref{def:prefix-rmq}):
  \begin{itemize}
  \item space usage $S(k,\ell,\AlphabetSize)$,
  \item preprocessing time $P_t(k,\ell,\AlphabetSize)$,
  \item preprocessing space $P_s(k,\ell,\AlphabetSize)$,
  \item query time $Q(k,\ell,\AlphabetSize)$.
  \end{itemize}
  Let $\Text \in \IntegerAlphabet^{\Textlen}$ be such that $2 \leq \AlphabetSize < \Textlen^{1/13}$ and $\Text[\Textlen]$ does not
  occur in $\Text[1 \dd \Textlen)$. Let $\tau = \lfloor \mu\log_{\AlphabetSize} \Textlen \rfloor$, where $\mu$ is a positive
  constant smaller than $\tfrac{1}{12}$ such that $\tau \geq 1$. There exist positive integers
  $k = \Theta(\Textlen / \log_{\AlphabetSize} \Textlen)$ and
  $\ell \leq (1 + \lfloor \log k \rfloor) / \lceil \log \AlphabetSize \rceil$
  such that, given the packed representation of $\Text$, we can
  in
  $\bigO(\Textlen / \log_{\AlphabetSize} \Textlen + P_t(k,\ell,\AlphabetSize))$ time and using
  $\bigO(\Textlen / \log_{\AlphabetSize} \Textlen + P_s(k,\ell,\AlphabetSize))$ working space construct a data structure of size
  $\bigO(\Textlen / \log_{\AlphabetSize} \Textlen + S(k,\ell,\AlphabetSize))$ that answers the following queries:
  \begin{enumerate}
  \item\label{pr:lex-range-minimum-nonperiodic-it-1}
    Given the packed representation of
    a $\tau$-nonperiodic (\cref{def:periodic-pattern}) patterns $\Pat_1, \Pat_2 \in \IntegerAlphabet^{+}$ such that
    $\lcp{\Pat_1}{\Pat_2} \geq 3\tau - 1$,
    computes $\min \LexRange{\Pat_1}{\Pat_2}{\Text} \cup \{\infty\}$ (see \cref{def:lex-range}) in
    $\bigO(\log \log \Textlen +
    Q(k,\ell,\AlphabetSize) +
    |\Pat_1| / \log_{\AlphabetSize} \Textlen +
    |\Pat_2| / \log_{\AlphabetSize} \Textlen)$
    time.
  \item\label{pr:lex-range-minimum-nonperiodic-it-2}
    Given the packed representation of a $\tau$-nonperiodic pattern
    $\Pat \in \IntegerAlphabet^{+}$ satisfying $|\Pat| \geq 3\tau-1$,
    computes $\min \LexRange{\Pat}{\Pat' c^{\infty}}{\Text} \cup \{\infty\}$
    (where $\Pat' = \Pat[1 \dd 3\tau-1]$ and $c = \AlphabetSize - 1$) in
    $\bigO(\log \log \Textlen +
    Q(k,\ell,\AlphabetSize) +
    |\Pat| / \log_{\AlphabetSize} \Textlen)$
    time.
  \end{enumerate}
\end{proposition}
\begin{proof}

  We use the following definitions.
  Let $\SSS$ be a $\tau$-synchronizing set of $\Text$ of size $|\SSS| = \bigO(\frac{\Textlen}{\tau})$ constructed
  using \cref{th:sss-packed-construction}. Denote $\Textlen' = |\SSS|$.
  Denote $(s^{\rm lex}_i)_{i \in [1 \dd \Textlen']} = \LexSorted{\SSS}{\Text}$ (\cref{def:lex-sorted}),
  $(s^{\rm text}_i)_{i \in [1 \dd \Textlen']} = \Sort{\SSS}$ (\cref{def:sort}),
  and let $\pi[1 \dd \Textlen']$ be a permutation of $[1 \dd \Textlen']$ such that for every $i \in [1 \dd \Textlen']$
  it holds $s^{\rm lex}_i = s^{\rm text}_{\pi[i]}$.
  Let $k = \max(\Textlen', k') = \Theta(\Textlen / \log_{\AlphabetSize} \Textlen)$,
  where $k' = \lceil \Textlen / \log_{\AlphabetSize} \Textlen \rceil$.
  Let $A_{\pi}[1 \dd \Textlen']$ be defined so that for every $i \in [1 \dd \Textlen']$,
  $A_{\pi}[i] = \pi[i]$.
  Let $A_{\SSS}[1 \dd k]$ and $A_{\rm str}[1 \dd k]$ be arrays defined so that for every $i \in [1 \dd \Textlen']$,
  $A_{\SSS}[i] = s^{\rm text}_i$ and $A_{\rm str}[i] = \revstr{D_i}$, where $D_i = \Textinf[s_i - \tau \dd s_i + 2\tau)$.
  The remaining elements of the arrays are initialized arbitrarily (they are not used).
  Denote $\ell = 3\tau$, and note that
  by the same analysis as in \cref{th:sa-to-prefix-select-nonbinary}, it holds
  $\ell \leq (1 + \lfloor \log k \rfloor) / \lceil \log \AlphabetSize \rceil$.

  \DSComponents
  The data structure consists of the following components:
  \begin{enumerate}
  \item The structure from \cref{pr:nav-index-short} for text $\Text$. It needs $\bigO(\Textlen / \log_{\AlphabetSize} \Textlen)$ space.
  \item The structure $\NavNonperiodic{\SSS}{\tau}{\Text}$ from \cref{pr:nav-index-nonperiodic}. It needs
    $\bigO(\Textlen / \log_{\AlphabetSize} \Textlen)$ space.
  \item The structure from the claim (answering prefix range minimum queries) for sequences $A_{\pi}[1 \dd k]$
    and $A_{\rm str}[1 \dd k]$. It needs $\bigO(S(k,\ell,\AlphabetSize))$ space.
  \item The plain representation of the array $A_{\SSS}[1 \dd k]$.
    It needs $\bigO(k) = \bigO(\Textlen / \log_{\AlphabetSize} \Textlen)$ space.
  \item The plain representation of the array $A_{\pi}[1 \dd \Textlen']$
    using $\bigO(\Textlen') = \bigO(\Textlen / \log_{\AlphabetSize} \Textlen)$ space.
  \end{enumerate}
  In total, the structure needs
  $\bigO(\Textlen / \log_{\AlphabetSize} \Textlen + S(k,\ell,\AlphabetSize))$ space.

  \DSQueries
  The queries are answered as follows.
  \begin{enumerate}
  \item Let $\Pat_1, \Pat_2 \in \IntegerAlphabet^{+}$ be $\tau$-nonperiodic patterns satisfying
    $\lcp{\Pat_1}{\Pat_2} \geq 3\tau - 1$. Denote $\Pat' = \Pat_1[1 \dd 3\tau-1] = \Pat_2[1 \dd 3\tau-1]$.
    Given the packed representation of $\Pat_1$ and $\Pat_2$, we compute
    $\min \LexRange{\Pat_1}{\Pat_2}{\Text} \cup \{\infty\}$ as follows:
    \begin{enumerate}
    \item First, using \cref{pr:nav-index-short}, in $\bigO(1)$ time we compute
      $\RangeBegTwo{\Pat'}{\Text}$ and $\RangeEndTwo{\Pat'}{\Text}$. This lets us determine $|\OccTwo{\Pat'}{\Text}|$.
      If $\OccTwo{\Pat'}{\Text} = \emptyset$, then we have $\LexRange{\Pat_1}{\Pat_2}{\Text} = \emptyset$
      (since any element of this set would have $\Pat'$ as a prefix), and hence
      we return $\min \LexRange{\Pat_1}{\Pat_2}{\Text} \cup \{\infty\} = \infty$, and finish
      the query algorithm.
      Let us now assume that $\OccTwo{\Pat'}{\Text} \neq \emptyset$.
    \item Using \cref{pr:nav-index-nonperiodic}\eqref{pr:nav-index-nonperiodic-it-2a}, in $\bigO(1)$ time
      compute the packed representation of the string
      $D = \DistPrefixPat{\Pat_1}{\tau}{\Text}{\SSS}$ (\cref{def:dist-prefix-pat}).
      Note that since $\lcp{\Pat_1}{\Pat_2} \geq 3\tau - 1 \geq |D|$,
      $D$ is also a prefix of $\Pat_2$
      (and by \cref{lm:dist-prefix-existence} no other string in $\DistPrefixes{\tau}{\Text}{\SSS}$ is a prefix of $\Pat_2$).
      In $\bigO(1)$ time we then calculate $\deltatext = |D| - 2\tau$.
    \item Using \cref{pr:nav-index-nonperiodic}\eqref{pr:nav-index-nonperiodic-it-1}, in $\bigO(1)$ time
      compute the packed representation of $\revstr{D}$.
    \item Denote $P'_k = \Pat_k(\deltatext \dd |\Pat_k|]$, where $k \in \{1,2\}$.
      Using \cref{pr:nav-index-nonperiodic}\eqref{pr:nav-index-nonperiodic-it-2b},
      compute the value
      $b_k = |\{i \in [1 \dd \Textlen'] : \Text[s_i \dd \Textlen] \prec \Pat'_k\}|$
      for $k \in \{1,2\}$.
      This takes
      $\bigO(\log \log \Textlen + |\Pat'_1| / \log_{\AlphabetSize} \Textlen +
      |\Pat'_2| / \log_{\AlphabetSize} \Textlen) = \bigO(\log \log \Textlen + |\Pat_1| / \log_{\AlphabetSize} \Textlen +
      |\Pat_2| / \log_{\AlphabetSize} \Textlen)$ time.
    \item Using the structure from the claim, we
      compute $j = \PrefixRMQ{A_{\pi}}{A_{\rm str}}{b_1}{b_2}{\revstr{D}}$ (\cref{def:prefix-rmq})
      in $\bigO(Q(k,\ell,\AlphabetSize))$ time.
      If $j = \infty$, then by \cref{lm:lex-range-minimum-nonperiodic-permuted},
      $\LexRange{\Pat_1}{\Pat_2}{\Text} = \emptyset$, and hence we return
      $\min \LexRange{\Pat_1}{\Pat_2}{\Text} \cup \{\infty\} = \infty$.
      Otherwise (i.e., if $j \neq \infty$),
      by \cref{lm:lex-range-minimum-nonperiodic-permuted} we
      have $A_{\SSS}[A_{\pi}[j]] = \min \LexRange{\Pat_1}{\Pat_2}{\Text}$.
      We thus return $A_{\SSS}[A_{\pi}[j]]$ as the answer.
      Note that the array defined here is padded with
      extra strings compared to the array in \cref{lm:lex-range-minimum-nonperiodic-permuted},
      but this does not affect the result of the above query, since $b_1, b_2 \leq \Textlen'$.
    \end{enumerate}
    In total, we spend $\bigO(\log \log \Textlen + Q(k,\ell,\AlphabetSize) +
    |\Pat_1| / \log_{\AlphabetSize} \Textlen + |\Pat_2| / \log_{\AlphabetSize} \Textlen)$ time.
  \item Let us now consider a $\tau$-nonperiodic pattern $\Pat \in \IntegerAlphabet^{+}$
    such that it holds $|\Pat| \geq 3\tau-1$.
    To compute $\min \LexRange{\Pat}{\Pat' c^{\infty}}{\Text} \cup \{\infty\}$
    (where $\Pat' = \Pat[1 \dd 3\tau-1]$ and $c = \AlphabetSize - 1$), we
    set $\Pat_1 = \Pat$ and $\Pat_2 = \Pat' c^{\infty}$, and then proceed similarly as above, except
    we compute $b_2 = |\{i \in [1 \dd \Textlen'] : \Text[s_i \dd \Textlen] \prec \Pat'_2\}|$
    using the second integer computed in \cref{pr:nav-index-nonperiodic}\eqref{pr:nav-index-nonperiodic-it-2b}
    (which requires only the packed representation of $\Pat'$).
    The whole query takes
    $\bigO(\log \log \Textlen + Q(k,\ell,\AlphabetSize) + |\Pat| / \log_{\AlphabetSize} \Textlen)$ time.
  \end{enumerate}

  \DSConstruction
  The components of the structure are constructed as follows:
  \begin{enumerate}
  \item We apply \cref{pr:nav-index-short}. This uses $\bigO(\Textlen / \log_{\AlphabetSize} \Textlen)$ time and working space.
  \item Using \cref{th:sss-packed-construction}, we compute the $\tau$-synchronizing set $\SSS$ satisfying
    $|\SSS| = \bigO(\tfrac{\Textlen}{\tau}) = \bigO(\Textlen / \log_{\AlphabetSize} \Textlen)$ in
    $\bigO(\Textlen / \log_{\AlphabetSize} \Textlen)$ time. Then, using $\SSS$ and the packed representation of $\Text$ as input,
    we construct $\NavNonperiodic{\SSS}{\tau}{\Text}$ in $\bigO(\Textlen / \log_{\AlphabetSize} \Textlen)$ time using
    \cref{pr:nav-index-nonperiodic}.
  \item To construct the next component of the structure, we proceed as follows:
    \begin{enumerate}
    \item Using \cref{th:sss-lex-sort}, in
      $\bigO(\tfrac{\Textlen}{\tau}) = \bigO(\Textlen / \log_{\AlphabetSize} \Textlen)$ time we first compute
      the sequence $(s_i)_{i \in [1 \dd \Textlen']} = \LexSorted{\SSS}{\Text}$ (\cref{def:lex-sorted}).
    \item Using \cref{pr:nav-index-nonperiodic}\eqref{pr:nav-index-nonperiodic-it-1}, in $\bigO(\Textlen') =
      \bigO(\Textlen / \log_{\AlphabetSize} \Textlen)$ time we compute the elements of the array $A_{\rm str}$ at
      indexes $i \in [1 \dd \Textlen']$, i.e., $A_{\rm str}[i] = \revstr{D_i}$, where $D_i = \Textinf[s_i - \tau \dd s_i + 2\tau)$.
    \item Finally, we apply the preprocessing
      from the claim to the array $A_{\rm str}$.
      This takes $\bigO(P_t(k,\ell,\AlphabetSize))$ time and uses
      $\bigO(P_s(k,\ell,\AlphabetSize))$ working space.
    \end{enumerate}
    In total, the construction of this component of the structure
    takes $\bigO(\Textlen / \log_{\AlphabetSize} \Textlen + P_t(k,\ell,\AlphabetSize))$ time and uses
    $\bigO(\Textlen / \log_{\AlphabetSize} \Textlen + P_s(k,\ell,\AlphabetSize))$ working space.
  \item The sequence $\Sort{\SSS}$ (and hence the array $A_{\SSS}$)
    is easily obtained from $\SSS$ (computed above) using a 2-round radix sort
    in $\bigO(|\SSS| + \sqrt{\Textlen}) = \bigO(\Textlen / \log_{\AlphabetSize} \Textlen)$ time.
  \item To compute the last component, first recall that the sequences $(s^{\rm lex}_i)_{i \in [1 \dd \Textlen']}$
    and $(s^{\rm text}_i)_{i \in [1 \dd \Textlen']}$ have already been constructed above. To compute
    the permutation $\pi[1 \dd \Textlen']$, we create an array $P[1 \dd \Textlen']$ of pairs defined by
    $P[i] = (s^{\rm lex}_i, i)$. We then sort the array $P$ lexicographically. Using a 4-round radix sort,
    we achieve $\bigO(\Textlen' + \sqrt{\Textlen}) = \bigO(\Textlen / \log_{\AlphabetSize} \Textlen)$ time.
    Let $(b_i)_{i \in [1 \dd \Textlen']}$ be the sequence containing the second coordinates in the resulting
    sequence of pairs. For every $i \in [1 \dd \Textlen']$, we set $A_{\pi}[b_i] = i$. In total, construction
    of $A_{\pi}[1 \dd \Textlen']$ takes $\bigO(\Textlen / \log_{\AlphabetSize} \Textlen)$ time.
  \end{enumerate}
  In total, the construction takes
  $\bigO(\Textlen / \log_{\AlphabetSize} \Textlen + P_t(k,\ell,\AlphabetSize))$ time and uses
  $\bigO(\Textlen / \log_{\AlphabetSize} \Textlen + P_s(k,\ell,\AlphabetSize))$ working space.
\end{proof}

\paragraph{The Periodic Patterns}\label{sec:lex-range-minimum-periodic}

\subparagraph{Combinatorial Properties}

\begin{lemma}\label{lm:exp-subblock-min}
  Let $\Text \in \Sigma^{\Textlen}$ and $\tau \in [1 \dd \lfloor \tfrac{\Textlen}{2} \rfloor]$.
  Let $(a_j)_{j \in [1 \dd q]} = \RunsMinusLexSortedTwo{\tau}{\Text}$ (\cref{def:runs-minus-lex-sorted}),
  and let $A_{\rm pos}[1 \dd q]$ and $A_{\rm len}[1 \dd q]$ be arrays defined by
  \begin{itemize}
  \item $A_{\rm pos}[i] = \RunEndFullPos{a_i}{\tau}{\Text}$,
  \item $A_{\rm len}[i] = \RunEndFullPos{a_i}{\tau}{\Text} - a_i$.
  \end{itemize}
  Let $i_1,i_2 \in [1 \dd \Textlen]$ be such that $i_1 \leq i_2$ and
  $\{\SA{\Text}[i]\}_{i \in [i_1 \dd i_2]} \subseteq \RMinusFive{s}{k}{H}{\tau}{\Text}$
  for some $s \in \Zn$, $k \in \Zp$, and $H \in \Sigma^{+}$.
  Denote $p = |H|$ and $\deltatext = s + kp$.
  Let $j_1,j_2 \in [1 \dd q]$ be such that
  for $k \in \{1,2\}$, it holds
  $\RunEndFullPos{\SA{\Text}[i_k]}{\tau}{\Text} = \RunEndFullPos{a_{j_k}}{\tau}{\Text}$.
  Then, there exists $j \in (j_1-1 \dd j_2]$ such that $A_{\rm len}[j] \geq \deltatext$. Moreover, 
  it holds (see \cref{def:three-sided-rmq}):
  \[
    \min \{\SA{\Text}[i]\}_{i \in [i_1 \dd i_2]} =
      A_{\rm pos}[\ThreeSidedRMQ{A_{\rm pos}}{A_{\rm len}}{j_1-1}{j_2}{\deltatext}] - \deltatext.
  \]
\end{lemma}
\begin{proof}
  The first claim follows by $i_1 \leq i_2$ and \cref{lm:exp-subblock-characterization}. To obtain
  the second claim, we apply \cref{def:three-sided-rmq} and \cref{lm:exp-subblock-characterization}, resulting in:
  \begin{align*}
    \min \{\SA{\Text}[i]\}_{i \in [i_1 \dd i_2]}
      &= \min \{A_{\rm pos}[j] - \deltatext : j \in [j_1 \dd j_2]\text{ and }A_{\rm len}[j] \geq \deltatext\}\\
      &= \min \{A_{\rm pos}[j] : j \in [j_1 \dd j_2]\text{ and }A_{\rm len}[j] \geq \deltatext\} - \deltatext\\
      &= A_{\rm pos}[\ThreeSidedRMQ{A_{\rm pos}}{A_{\rm len}}{j_1-1}{j_2}{\deltatext}] - \deltatext.
      \qedhere
  \end{align*}
\end{proof}

\begin{lemma}\label{lm:bmin-exp-block}
  Let $\Text \in \Sigma^{\Textlen}$ and $\tau \in [1 \dd \lfloor \tfrac{\Textlen}{2} \rfloor]$.
  Let $s \in \Zn$, $k \in \Zp$, and $H \in \Sigma^{+}$ be such that $\RMinusFive{s}{k}{H}{\tau}{\Text} \neq \emptyset$.
  Then, $\OccTwo{\one}{\MinPosBitvectorMinusFive{s}{k}{H}{\tau}{\Text}} = \emptyset$
  (see \cref{def:min-pos-bitvector-minus}) implies that
  $\RMinusFive{s}{k+1}{H}{\tau}{\Text} \neq \emptyset$ and
  \[
    \min \RMinusFive{s}{k+1}{H}{\tau}{\Text} < \min \RMinusFive{s}{k}{H}{\tau}{\Text}.
  \]
\end{lemma}
\begin{proof}
  Denote $p = |H|$.
  By \cref{def:min-pos-bitvector-minus}, the
  assumption $\OccTwo{\one}{\MinPosBitvectorMinusFive{s}{k}{H}{\tau}{\Text}} = \emptyset$
  implies that $\RMinusFive{s}{k}{H}{\tau}{\Text} \cap \RMinMinusTwo{\tau}{\Text} = \emptyset$.
  Denote $j = \min \RMinusFive{s}{k}{H}{\tau}{\Text}$, and let
  $X = \Text[j \dd \RunEndPos{j}{\tau}{\Text})$. It holds $j \in \OccTwo{X}{\Text} \cap \RMinusTwo{\tau}{\Text}$, but
  since $j \not\in \RMinMinusTwo{\tau}{\Text}$, the position $j' = \min \OccTwo{X}{\Text} \cap \RMinusTwo{\tau}{\Text}$
  must satisfy $j' < j$; see \cref{def:rmin}. Position $j'$ also satisfies the following properties:
  \begin{itemize}
  \item Note that $|X| \geq 3\tau - 1$, and hence by
    \cref{lm:periodic-pos-lce}\eqref{lm:periodic-pos-lce-it-2}, $j' \in \RMinusFour{s}{H}{\tau}{\Text}$.
    Moreover, $\RunEndPos{j'}{\tau}{\Text} - j' \geq |X|$, and hence
    $\ExpPos{j'}{\tau}{\Text} = \lfloor ((\RunEndPos{j'}{\tau}{\Text} - j') - s) / p \rfloor
    \geq \lfloor (|X| - s) / p \rfloor = k$.
  \item Next, we prove $j' \in \RMinMinusTwo{\tau}{\Text}$. Suppose this is not the case.
    Denote $X' = \Text[j' \dd \RunEndPos{j'}{\tau}{\Text})$. Since $j' \in \OccTwo{X'}{\Text}$ and $j' \in \RMinusTwo{\tau}{\Text}$,
    the assumption $j' \not\in \RMinMinusTwo{\tau}{\Text}$ implies that there exists $j'' < j'$ such that
    $j'' \in \OccTwo{X'}{\Text}$ and $j'' \in \RMinusTwo{\tau}{\Text}$. Since $X$ is a prefix of $X'$, we thus
    have $j'' \in \OccTwo{X}{\Text} \cap \RMinusTwo{\tau}{\Text}$.
    We thus obtain a contradiction with the assumption
    $j' = \min \OccTwo{X}{\Text} \cap \RMinusTwo{\tau}{\Text}$.
  \end{itemize}
  Denote $k' = \ExpPos{j'}{\tau}{\Text}$. Above we established that $k' \geq k$. However, since we also proved
  that $j' \in \RMinMinusTwo{\tau}{\Text}$ and $j' \in \RMinusFour{s}{H}{\tau}{\Text}$, and earlier we assumed that
  $\RMinusFive{s}{k}{H}{\tau}{\Text} \cap \RMinMinusTwo{\tau}{\Text} = \emptyset$, we must therefore have $k' > k$.
  Denote
  \[
    x = \RunEndFullPos{j'}{\tau}{\Text} - (s + (k+1) p).
  \]
  Observe that $x$ satisfies
  the following properties:
  \begin{itemize}
  \item First, we show that $j' \leq x$. To see this, recall that $j' \in \RMinusFive{s}{k'}{H}{\tau}{\Text}$,
    where $k' \geq (k+1)$. Thus,
    $\RunEndFullPos{j'}{\tau}{\Text} - j' = s + k' \cdot p \geq s + (k+1) \cdot p = \RunEndFullPos{j'}{\tau}{\Text} - x$.
    Equivalently, $j' \leq x$.
  \item Next, we prove that $j' \leq \RunEndPos{j'}{\tau}{\Text} - (3\tau - 1)$.
    First, note that by $j \in \RMinusFive{s}{k}{H}{\tau}{\Text}$, it holds
    \[
      s + (k+1)p > s + kp + \TailPos{j}{\tau}{\Text} = \RunEndPos{j}{\tau}{\Text} - j \geq 3\tau - 1.
    \]
    Thus, $x = \RunEndFullPos{j'}{\tau}{\Text} - (s + (k+1)p)
    \leq \RunEndPos{j'}{\tau}{\Text} - (s + (k+1)p) \leq \RunEndPos{j'}{\tau}{\Text} - (3\tau - 1)$.
  \item In this step, we prove that $x \in \RMinusFive{s}{k+1}{H}{\tau}{\Text}$.
    Recall that by \cref{lm:end}, $[j' \dd \RunEndPos{j'}{\tau}{\Text} - (3\tau - 1)]$ is a
    maximal block of positions in $\RTwo{\tau}{\Text}$ containing $j'$. Above we proved
    that $x \in [j' \dd \RunEndPos{j'}{\tau}{\Text} - (3\tau - 1)]$. Thus, $x \in \RTwo{\tau}{\Text}$.
    Moreover, by \cref{lm:R-text-block},
    $\RootPos{x}{\tau}{\Text} = \RootPos{j'}{\tau}{\Text} = H$,
    $\TypePos{x}{\tau}{\Text} = \TypePos{j'}{\tau}{\Text} = -1$, and
    $\RunEndFullPos{x}{\tau}{\Text} = \RunEndFullPos{j'}{\tau}{\Text}$. This, in turn, yields
    $\HeadPos{x}{\tau}{\Text} = (\RunEndFullPos{x}{\tau}{\Text} - x) \bmod p =
    (\RunEndFullPos{j'}{\tau}{\Text} - x) \bmod p = (s + (k+1)p) \bmod p = s$ and
    $\ExpPos{x}{\tau}{\Text} = \lfloor (\RunEndFullPos{x}{\tau}{\Text} - x) / p \rfloor
    = \lfloor (\RunEndFullPos{j'}{\tau}{\Text} - x) / p \rfloor = \lfloor (s + (k+1)p) / p \rfloor = k + 1$.
    Thus, we obtain $x \in \RMinusFive{s}{k+1}{H}{\tau}{\Text}$. In particular, this proves that
    $\RMinusFive{s}{k+1}{H}{\tau}{\Text} \neq \emptyset$, i.e., the first part of the claim.
  \item Lastly, we prove that $x < j$. Recall that $j' < j$. We consider two cases:
    \begin{itemize}
    \item First, assume that $[j' \dd j] \subseteq \RTwo{\tau}{\Text}$.
      By \cref{lm:R-text-block}, we have $\RunEndFullPos{j'}{\tau}{\Text} = \RunEndFullPos{j}{\tau}{\Text}$.
      On the other hand, $j \in \RMinusFive{s}{k}{H}{\tau}{\Text}$ implies that
      $\RunEndFullPos{j}{\tau}{\Text} - j = s + kp$. Equivalently, $j = \RunEndFullPos{j}{\tau}{\Text} - (s + kp)
      = \RunEndFullPos{j'}{\tau}{\Text} - (s + kp) = x + p$. Thus, $x < j$.
    \item Let us now assume that $[j' \dd j] \not\subseteq \RTwo{\tau}{\Text}$. Then, there exists
      $y \in [1 \dd \Textlen] \setminus \RTwo{\tau}{\Text}$ such that $j' < y < j$.
      By \cref{lm:run-end}, it holds $\RunEndPos{j'}{\tau}{\Text} - j < \tau$, or equivalently,
      $\RunEndPos{j'}{\tau}{\Text} - \tau < j$.
      On the other hand, recall that above we proved that $s + (k+1)p \geq 3\tau - 1$. Thus,
      we obtain
      \begin{align*}
        x
          &= \RunEndFullPos{j'}{\tau}{\Text} - (s + (k+1)p)\\
          &\leq \RunEndPos{j'}{\tau}{\Text} - (s + (k+1)p)\\
          &\leq \RunEndPos{j'}{\tau}{\Text} - (3\tau - 1)\\
          &< \RunEndPos{j'}{\tau}{\Text} - \tau < j.
      \end{align*}
    \end{itemize}
    In both cases, we thus obtain $x < j$.
  \end{itemize}
  To sum up, we proved that $x \in \RMinusFive{s}{k+1}{H}{\tau}{\Text}$ and $x < j$. Recalling that
  $j = \min \RMinusFive{s}{k}{H}{\tau}{\Text}$, we thus obtain the second part of the claim, i.e.,
  \[
    \min \RMinusFive{s}{k+1}{H}{\tau}{\Text} \leq x < j = \min \RMinusFive{s}{k}{H}{\tau}{\Text}.
    \qedhere
  \]
\end{proof}

\begin{lemma}\label{lm:exp-block-min}
  Let $\Text \in \Sigma^{\Textlen}$ and $\tau \in [1 \dd \lfloor \tfrac{\Textlen}{2} \rfloor]$. Let
  $b, e \in [0 \dd \Textlen]$ be such that $\{\SA{\Text}[i]\}_{i \in (b \dd e]}
  = \RMinusFive{s}{k}{H}{\tau}{\Text}$ for some $s \in \Zn$, $k \in \Zp$, and $H \in \Sigma^{+}$
  satisfying $\RMinusFive{s}{k}{H}{\tau}{\Text} \neq \emptyset$.
  Denote $B_{\min} = \MinPosBitvectorMinusTwo{\tau}{\Text}$ (\cref{def:min-pos-bitvector-minus}).
  If $\Rank{B_{\min}}{b}{\one} \neq \Rank{B_{\min}}{e}{\one}$, then, letting
  $z = \Select{B_{\min}}{\Rank{B_{\min}}{b}{\one} + 1}{\one}$, it holds
  \[
    \min \RMinusFive{s}{k}{H}{\tau}{\Text} = \SA{\Text}[z].
  \]
\end{lemma}
\begin{proof}
  Observe that by the assumption $\Rank{B_{\min}}{b}{\one} \neq \Rank{B_{\min}}{e}{\one}$ and the
  definition of rank and select queries, it holds $z \in (b \dd e]$ and $B_{\min}[z] = \one$.
  Consequently (see \cref{def:min-pos-bitvector-minus}),
  $\SA{\Text}[z] \in \RMinMinusTwo{\tau}{\Text}$ (\cref{def:rmin}), and hence,
  letting $Z = \Text[\SA{\Text}[z] \dd \RunEndPos{\SA{\Text}[z]}{\tau}{\Text})$, it holds
  \[
    \SA{\Text}[z] = \min \OccTwo{Z}{\Text} \cap \RMinusTwo{\tau}{\Text}.
  \]
  Note also that we then have $\OccTwo{\one}{B_{\min}(b \dd z)} = \emptyset$.
  The proof consists of two steps:
  \begin{itemize}
  \item First, we prove that for every $x \in (b \dd z)$, it holds $\SA{\Text}[x] > \SA{\Text}[z]$.
    Suppose that this is not the case, and there exists $x \in (b \dd z)$ such that $\SA{\Text}[x] < \SA{\Text}[z]$.
    We then proceed as follows.
    \begin{itemize}
    \item Denote $X = \Text[\SA{\Text}[x] \dd \RunEndPos{\SA{\Text}[x]}{\tau}{\Text})$. First, observe that
      it holds $3\tau - 1 \leq |X| < |Z|$.
      The inequality $3\tau - 1 \leq |X|$ follows by definition. To show the other inequality, note that if
      $|X| \geq |Z|$, then by $\SA{\Text}[x], \SA{\Text}[z] \in \RMinusFour{s}{H}{\tau}{\Text}$,
      $Z$ would be a prefix of $X$. Consequently, we would have $\SA{\Text}[x] \in \OccTwo{Z}{\Text} \cap \RMinusTwo{\tau}{\Text}$.
      Since $\SA{\Text}[x] < \SA{\Text}[z]$, this contradicts $\SA{\Text}[z] = \min \OccTwo{Z}{\Text} \cap \RMinusTwo{\tau}{\Text}$.
    \item Denote $j = \min \OccTwo{X}{\Text} \cap \RMinusTwo{\tau}{\Text}$ and
      $J = \Text[j \dd \RunEndPos{j}{\tau}{\Text})$. Next, we show that
      $|X| \leq |J| < |Z|$.
      On the one hand, by $3\tau - 1 \leq |X|$ and
      \cref{lm:periodic-pos-lce}\eqref{lm:periodic-pos-lce-it-2},
      we have $j \in \RMinusFour{s}{H}{\tau}{\Text}$.
      By $\SA{\Text}[x] \in \RMinusFour{s}{H}{\tau}{\Text}$, this implies
      that $|X| \leq |J|$. On the other hand, recall that we also have
      $\SA{\Text}[z] \in \RMinusFour{s}{H}{\tau}{\Text}$. Thus, by $j \leq \SA{\Text}[x] < \SA{\Text}[z]$, we
      must have $|J| < |Z|$, since otherwise we would obtain a contradiction with
      $\SA{\Text}[z] = \min \OccTwo{Z}{\Text} \cap \RMinusTwo{\tau}{\Text}$.
    \item Lastly, observe that it holds $j \in \RMinMinusTwo{\tau}{\Text}$. To see this i.e.,
      that we have $j = \min \OccTwo{J}{\Text} \cap \RMinusTwo{\tau}{\Text}$, note that a position
      $j' < j$ satisfying $j' \in \OccTwo{J}{\Text} \cap \RMinusTwo{\tau}{\Text}$ would contradict
      that $j = \min \OccTwo{X}{\Text} \cap \RMinusTwo{\tau}{\Text}$ (since $X$ is a prefix of $J$).
    \end{itemize}
    We have thus proved that the position $j$ satisfies
    $j \in \RMinusFour{s}{H}{\tau}{\Text}$, and
    \[
      \RunEndPos{\SA{\Text}[x]}{\tau}{\Text} - \SA{\Text}[x] \leq
      \RunEndPos{j}{\tau}{\Text} - j <
      \RunEndPos{\SA{\Text}[z]}{\tau}{\Text} - \SA{\Text}[z].
    \]
    Since $\SA{\Text}[x], \SA{\Text}[z] \in \RMinusFive{s}{k}{H}{\tau}{\Text}$, this
    implies that we also have $j \in \RMinusFive{s}{k}{H}{\tau}{\Text}$. Moreover, since by
    \cref{lm:R-lex-block-pos}\eqref{lm:R-lex-block-pos-it-3} it holds
    \[
      \RunEndPos{\SA{\Text}[b+1]}{\tau}{\Text} - \SA{\Text}[b+1] \leq
      \RunEndPos{\SA{\Text}[b+2]}{\tau}{\Text} - \SA{\Text}[b+2] \leq
      \cdots \leq
      \RunEndPos{\SA{\Text}[e]}{\tau}{\Text} - \SA{\Text}[e],
    \]
    it follows that there exists $x' \in (b \dd z)$ such that $j = \SA{\Text}[x']$.
    Since we also proved that $j \in \RMinMinusTwo{\tau}{\Text}$, we
    have $B_{\min}[x'] = \one$. This contradicts that $\OccTwo{\one}{B_{\min}(b \dd z)} = \emptyset$,
    which concludes the proof that for every $x \in (b \dd z)$, we have $\SA{\Text}[x] > \SA{\Text}[z]$.
  \item Second, observe that by \cref{lm:bmin-bit}, for every $x \in (z \dd e]$, it holds
    $\SA{\Text}[x] > \SA{\Text}[z]$.
  \end{itemize}
  We have proved that for every $x \in (b \dd e] \setminus \{z\}$, it holds $\SA{\Text}[x] > \SA{\Text}[z]$.
  This implies that
  \[
    \min \RMinusFive{s}{k}{H}{\tau}{\Text} = \min \{\SA{\Text}[i]\}_{i \in (b \dd e]} = \SA{\Text}[z].
    \qedhere
  \]
\end{proof}

\begin{lemma}\label{lm:exp-blocks-min}
  Consider a text $\Text \in \Sigma^{\Textlen}$ and let
  $\tau \in [1 \dd \lfloor \tfrac{\Textlen}{2} \rfloor]$. Let
  $b, e \in [0 \dd \Textlen]$ be such that $\{\SA{\Text}[i]\}_{i \in (b \dd e]}
  = \bigcup_{k \in [k_1 \dd k_2]} \RMinusFive{s}{k}{H}{\tau}{\Text}$ for some $s \in \Zn$,
  $k_1, k_2 \in \Zp$, and $H \in \Sigma^{+}$
  satisfying
  $k_1 \leq k_2$, $\RMinusFive{s}{k_1}{H}{\tau}{\Text} \neq \emptyset$, and
  $\RMinusFive{s}{k_2}{H}{\tau}{\Text} \neq \emptyset$.
  Let $B_{\min} = \MinPosBitvectorMinusTwo{\tau}{\Text}$ (\cref{def:min-pos-bitvector-minus}).
  \begin{enumerate}
  \item If $\Rank{B_{\min}}{b}{\one} \neq \Rank{B_{\min}}{e}{\one}$, then, letting
    $z = \Select{B_{\min}}{\Rank{B_{\min}}{b}{\one} + 1}{\one}$,
    \[
      \min \{\SA{\Text}[i]\}_{i \in (b \dd e]} = \SA{\Text}[z].
    \]
  \item If $\Rank{B_{\min}}{b}{\one} = \Rank{B_{\min}}{e}{\one}$, then
    \[
      \min \{\SA{\Text}[i]\}_{i \in (b \dd e]} = \min \RMinusFive{s}{k_2}{H}{\tau}{\Text}.
    \]
  \end{enumerate}
\end{lemma}
\begin{proof}
  We prove each of the claims separately:
  \begin{enumerate}
  \item The assumption $\Rank{B_{\min}}{b}{\one} \neq \Rank{B_{\min}}{e}{\one}$ and the definition
    of rank and select queries imply that $z \in (b \dd e]$. Thus, there exists $k' \in [k_1 \dd k_2]$ such
    that $\SA{\Text}[z] \in \RMinusFive{s}{k'}{H}{\tau}{\Text}$.
    Let $b', e' \in [b \dd e]$ be such that $\{\SA{\Text}[i]\}_{i \in (b' \dd e']} =
    \RMinusFive{s}{k'}{H}{\tau}{\Text}$ (see \cref{rm:min-pos-bitvector-minus}). Note that
    then $z \in (b' \dd e']$.
    By \cref{lm:periodic-pos-lce}, we also have
    $\{\SA{\Text}[i]\}_{i \in (b \dd b']} = \bigcup_{k \in [k_1 \dd k')} \RMinusFive{s}{k}{H}{\tau}{\Text}$ and
    $\{\SA{\Text}[i]\}_{i \in (e' \dd e]} = \bigcup_{k \in (k' \dd k_2]} \RMinusFive{s}{k}{H}{\tau}{\Text}$,
    and hence
    \begin{align*}
      B_{\min}(b \dd b'] &= \textstyle\bigodot_{k \in [k_1 \dd k')} \MinPosBitvectorMinusFive{s}{k}{H}{\tau}{\Text}\\
      B_{\min}(e' \dd e] &= \textstyle\bigodot_{k \in (k' \dd k_2]} \MinPosBitvectorMinusFive{s}{k}{H}{\tau}{\Text}.
    \end{align*}
    Finally, note that by definition of $z$, we have $B_{\min}[z] = \one$ and $\OccTwo{\one}{B_{\min}(b \dd z)} = \emptyset$.
    We are now ready to prove the claim.
    \begin{itemize}
    \item First, we prove that for every $i \in (b \dd b']$, it holds
      $\SA{\Text}[i] > \min \RMinusFive{s}{k'}{H}{\tau}{\Text}$. If $k' = k_1$, then $b' = b$, and the claim holds. Let us
      thus assume that $k_1 < k'$, and recall that $\RMinusFive{s}{k_1}{H}{\tau}{\Text} \neq \emptyset$.
      Observe that $z > b'$ and $\OccTwo{\one}{B_{\min}(b \dd z)} = \emptyset$ imply that
      $\OccTwo{\one}{\MinPosBitvectorMinusFive{s}{k_1}{H}{\tau}{\Text}} = \emptyset$.
      By \cref{lm:bmin-exp-block}, we then have $\RMinusFive{s}{k_1+1}{H}{\tau}{\Text} \neq \emptyset$ and
      \[
        \min \RMinusFive{s}{k_1+1}{H}{\tau}{\Text} < \min \RMinusFive{s}{k_1}{H}{\tau}{\Text}.
      \]
      If $k' = k_1 + 1$, then this proves the claim, because then we have
      $\{\SA{\Text}[i]\}_{i \in (b \dd b']} = \RMinusFive{s}{k_1}{H}{\tau}{\Text}$, and by the above, for every
      $x \in \RMinusFive{s}{k_1}{H}{\tau}{\Text} = \{\SA{\Text}[i]\}_{i \in (b \dd b']}$, it holds
      $x \geq \min \RMinusFive{s}{k_1}{H}{\tau}{\Text} > \min \RMinusFive{s}{k_1+1}{H}{\tau}{\Text} =
      \min \RMinusFive{s}{k'}{H}{\tau}{\Text}$. Let us thus assume that $k' > k_1 + 1$.
      Using again $z > b'$ and $\OccTwo{\one}{B_{\min}(b \dd z)} = \emptyset$, we obtain that
      $\OccTwo{\one}{\MinPosBitvectorMinusFive{s}{k_1+1}{H}{\tau}{\Text}} = \emptyset$. By \cref{lm:bmin-exp-block},
      we then obtain that it holds $\RMinusFive{s}{k_1+2}{H}{\tau}{\Text} \neq \emptyset$ and
      $\min \RMinusFive{s}{k_1+2}{H}{\tau}{\Text} < \min \RMinusFive{s}{k_1+1}{H}{\tau}{\Text}$.
      If $k' = k_1 + 2$, then we obtain the claim, because we now have
      \[
        \min \RMinusFive{s}{k_1+2}{H}{\tau}{\Text} <
        \min \RMinusFive{s}{k_1+1}{H}{\tau}{\Text} <
        \min \RMinusFive{s}{k_1}{H}{\tau}{\Text},
      \]
      which implies that for every $x \in \RMinusFive{s}{k_1}{H}{\tau}{\Text} \cup \RMinusFive{s}{k_1+1}{H}{\tau}{\Text}
      = \{\SA{\Text}[i]\}_{i \in (b \dd b']}$, we have
      $x \geq \min \RMinusFive{s}{k_1+1}{H}{\tau}{\Text} > \min \RMinusFive{s}{k_1+2}{H}{\tau}{\Text}
      = \min \RMinusFive{s}{k'}{H}{\tau}{\Text}$. Continuing analogously, we obtain that no matter the value of $k'$, for
      every $x \in \{\SA{\Text}[i]\}_{i \in (b \dd b']}$, it holds $x > \min \RMinusFive{s}{k'}{H}{\tau}{\Text}$, or
      equivalently, that for every $i \in (b \dd b']$, $\SA{\Text}[i] > \min \RMinusFive{s}{k'}{H}{\tau}{\Text}$.
    \item Next, we prove that for every $i \in (e' \dd e]$, it holds $\SA{\Text}[i] > \min \RMinusFive{s}{k'}{H}{\tau}{\Text}$.
      First, observe that since
      \[
        \RMinusFive{s}{k'}{H}{\tau}{\Text} \subseteq
        \textstyle\bigcup_{k \in [k_1 \dd k_2]} \RMinusFive{s}{k}{H}{\tau}{\Text} \subseteq
        \RMinusFour{s}{H}{\tau}{\Text},
      \]
      it follows that, letting $\hat{b},\hat{e} \in [0 \dd \Textlen]$ be such that
      $\{\SA{\Text}[i]\}_{i \in (\hat{b} \dd \hat{e}]} = \RMinusFour{s}{H}{\tau}{\Text}$ (see \cref{rm:min-pos-bitvector-minus}),
      it holds
      \[
        \hat{b} \leq b \leq b' < z \leq e' \leq e \leq \hat{e}.
      \]
      Recall that $z \in (b' \dd e']$ and $B_{\min}[z] = \one$. Thus, by \cref{lm:bmin-bit}, it follows that for
      every $i \in (z \dd \hat{e}]$, it holds $\SA{\Text}[i] > \SA{\Text}[z]$. In particular,
      this includes all $i \in (e' \dd e]$, i.e., for such $i$, we have
      $\SA{\Text}[i] > \SA{\Text}[z] \geq \min \RMinusFive{s}{k'}{H}{\tau}{\Text}$.
    \item Finally, we prove that $\min \RMinusFive{s}{k'}{H}{\tau}{\Text} = \SA{\Text}[z]$.
      First, note that
      $z \in (b' \dd e']$,
      $B_{\min}[z] = \one$, and
      $\{\SA{\Text}[i]\}_{i \in (b' \dd e']} = \RMinusFive{s}{k'}{H}{\tau}{\Text}$
      imply that $\RMinusFive{s}{k'}{H}{\tau}{\Text} \neq \emptyset$ and
      $\Rank{B_{\min}}{b'}{\one} \neq \Rank{B_{\min}}{e'}{\one}$. One the other hand, note that
      $\OccTwo{\one}{B_{\min}(b' \dd z)} = \emptyset$ implies that 
      $z = \Select{B_{\min}}{\Rank{B_{\min}}{b'}{\one} + 1}{\one}$. Thus, by \cref{lm:exp-block-min},
      we obtain $\min \RMinusFive{s}{k'}{H}{\tau}{\Text} = \SA{\Text}[z]$.
    \end{itemize}
    By the above, for every $i \in (b \dd e] \setminus (b' \dd e']$, it holds
    $\SA{\Text}[i] > \min \RMinusFive{s}{k'}{H}{\tau}{\Text} = \SA{\Text}[z]$.
    On the other hand, since we have $\{\SA{\Text}[i]\}_{i \in (b' \dd e']} = \RMinusFive{s}{k'}{H}{\tau}{\Text}$
    and $\min \RMinusFive{s}{k'}{H}{\tau}{\Text} = \SA{\Text}[z]$, we obtain that $\SA{\Text}[i] > \SA{\Text}[z]$ also
    for $i \in (b' \dd e'] \setminus \{z\}$.
    Thus, $\min \{\SA{\Text}[i]\}_{i \in (b \dd e]} = \SA{\Text}[z]$.
  \item The assumption $\Rank{B_{\min}}{b}{\one} = \Rank{B_{\min}}{e}{\one}$ implies
    that $\OccTwo{\one}{B_{\min}(b \dd e]} = \emptyset$. Consequently, for every
    $k \in [k_1 \dd k_2]$, it holds $\OccTwo{\one}{\MinPosBitvectorMinusFive{s}{k}{H}{\tau}{\Text}} = \emptyset$.
    Recall now that we assumed that $\RMinusFive{s}{k_1}{H}{\tau}{\Text} \neq \emptyset$.
    Thus, by repeatedly applying
    \cref{lm:bmin-exp-block} for $k=k_1,k_1+1,\ldots,k_2-1$, we obtain that
    for every $k \in [k_1 \dd k_2]$, it holds $\RMinusFive{s}{k}{H}{\tau}{\Text} \neq \emptyset$ and
    \[
      \min \RMinusFive{s}{k_1}{H}{\tau}{\Text} > \min \RMinusFive{s}{k_1+1}{H}{\tau}{\Text} > \dots >
      \min \RMinusFive{s}{k_2}{H}{\tau}{\Text}.
    \]
    This implies
    $\min \bigcup_{k \in [k_1 \dd k_2]} \RMinusFive{s}{k}{H}{\tau}{\Text} = \min \RMinusFive{s}{k_2}{H}{\tau}{\Text}$,
    or equivalently,
    $\min \{\SA{\Text}[i]\}_{i \in (b \dd e]} = \min \RMinusFive{s}{k_2}{H}{\tau}{\Text}$, i.e., the claim.
    \qedhere
  \end{enumerate}
\end{proof}

\subparagraph{Algorithms}

\begin{proposition}\label{pr:periodic-lex-range-minus-min}
  Let $\Text \in \IntegerAlphabet^{\Textlen}$ be such that
  $2 \leq \AlphabetSize < \Textlen^{1/13}$ and $\Text[\Textlen]$ does not
  occur in $\Text[1 \dd \Textlen)$. Let
  $\tau = \lfloor \mu\log_{\AlphabetSize} \Textlen \rfloor$,
  where $\mu$ is a positive constant smaller than $\tfrac{1}{12}$
  such that $\tau \geq 1$.
  Given the packed representation of $\Text$, we can in
  $\bigO(\Textlen / \log_{\AlphabetSize} \Textlen)$ time construct
  a data structure that given any $b,e \in [0 \dd \Textlen]$ such
  that $b < e$ and $\{\SA{\Text}[i]\}_{i \in (b \dd e]} \subseteq \RMinusFour{s}{H}{\tau}{\Text}$
  for some $s \in \Zn$ and $H \in \IntegerAlphabet^{+}$, 
  in $\bigO(\log \log \Textlen)$ time returns
  \[
    \min \{\SA{\Text}[i] : i \in (b \dd e]\}.
  \]
\end{proposition}
\begin{proof}

  Let $B_{\rm exp} = \ExpBlockBitvectorMinusTwo{\tau}{\Text}$ (\cref{def:exp-block-bitvector-minus}) and
  $B_{\min} = \MinPosBitvectorMinusTwo{\tau}{\Text}$ (\cref{def:min-pos-bitvector-minus}).
  Denote also
  $(a_i)_{i \in [1 \dd q]} = \RunsMinusLexSortedTwo{\tau}{\Text}$ (\cref{def:runs-minus-lex-sorted}).

  \DSComponents
  The data structure consists of the following components:
  \begin{enumerate}
  \item The data structure from \cref{pr:nav-index-periodic} for text $\Text$. It needs
    $\bigO(\Textlen / \log_{\AlphabetSize} \Textlen)$ space.
  \item The bitvector $B_{\rm exp}$ augmented with the support for $\bigO(1)$-time rank and select queries
    using \cref{th:bin-rank-select}. The bitvector together with the augmentation of \cref{th:bin-rank-select}
    needs $\bigO(\Textlen / \log \Textlen) = \bigO(\Textlen / \log_{\AlphabetSize} \Textlen)$ space.
  \item The bitvector $B_{\min}$ augmented with \cref{th:bin-rank-select} similarly as above.
    The bitvector together with the augmentation also needs
    $\bigO(\Textlen / \log \Textlen) = \bigO(\Textlen / \log_{\AlphabetSize} \Textlen)$ space.
  \end{enumerate}
  In total, the structure needs $\bigO(\Textlen / \log_{\AlphabetSize} \Textlen)$ space.

  \DSQueries
  The queries are answered as follows. Let $b, e \in [0 \dd \Textlen]$ be such that $b < e$ and
  $\{\SA{\Text}[i]\}_{i \in (b \dd e]} \subseteq \RMinusFour{s}{H}{\tau}{\Text}$ for some $s \in \Zn$
  and $H \in \IntegerAlphabet^{+}$. We gradually develop an algorithm that, given $b,e$, returns
  $\min \{\SA{\Text}[i]\}_{i \in (b \dd e]}$.

  First, let us assume that there exists $k \in \Zp$, such that $\{\SA{\Text}[i]\}_{i \in (b \dd e]}
  \subseteq \RMinusFive{s}{k}{H}{\tau}{\Text}$. In this case, we compute $\min \{\SA{\Text}[i]\}_{i \in (b \dd e]}$
  as follows:
  \begin{enumerate}
  \item Denote $i_1 = b+1$ and $i_2 = e$.
    Using \cref{pr:nav-index-periodic}\eqref{pr:nav-index-periodic-it-1}, in $\bigO(\log \log \Textlen)$ time
    compute $p_k = \SA{\Text}[i_k]$, where $k \in \{1,2\}$.
  \item Using \cref{pr:nav-index-periodic}\eqref{pr:nav-index-periodic-it-2}, in $\bigO(1)$ time compute
    $s = \HeadPos{p_1}{\tau}{\Text}$,
    $p = |\RootPos{p_1}{\tau}{\Text}|$,
    and $k = \ExpPos{p_1}{\tau}{\Text}$. We then set $\deltatext = s + pk$.
  \item Using \cref{pr:nav-index-periodic}\eqref{pr:nav-index-periodic-it-3}, in $\bigO(1)$ time
    compute indexes $j_k \in [1 \dd q]$ such that $\RunEndFullPos{p_k}{\tau}{\Text} = \RunEndFullPos{a_{j_k}}{\tau}{\Text}$,
    where $k \in \{1,2\}$. Note that then, by \cref{lm:exp-subblock-min}, there exists
    $j \in (j_1 - 1 \dd j_2]$ such that $A_{\rm len}[j] \geq \deltatext$.
  \item Using \cref{pr:nav-index-periodic}\eqref{pr:nav-index-periodic-it-rmq}, in $\bigO(\log \log \Textlen)$ time
    compute $x = A_{\rm pos}[\PrefixRMQ{A_{\rm pos}}{A_{\rm len}}{j_1-1}{j_2}{\deltatext}]$. By
    \cref{lm:exp-subblock-min}, it holds $\min\{\SA{\Text}[i]\}_{i \in (b \dd e]} = x - \deltatext$.
    Thus, we return $x - \deltatext$ as the answer.
  \end{enumerate}
  In total, we spend $\bigO(\log \log \Textlen)$ time.

  Let us now consider a general case, where $\{\SA{\Text}[i]\}_{i \in (b \dd e]} \subseteq \RMinusFour{s}{H}{\tau}{\Text}$.
  We then compute the value $\min \{\SA{\Text}[i]\}_{i \in (b \dd e]}$ as follows:
  \begin{enumerate}
  \item In $\bigO(1)$ time compute $x = \Select{B_{\rm exp}}{\Rank{B_{\rm exp}}{b}{\one} + 1}{\one}$.
    If $x \geq e$, then by \cref{def:exp-block-bitvector-minus}, there exists $k \in \Zp$ such that
    $\{\SA{\Text}[i]\}_{i \in (b \dd e]} \subseteq \RMinusFive{s}{k}{H}{\tau}{\Text}$. Thus, in this
    case we compute $\min \{\SA{\Text}[i]\}_{i \in (b \dd e]}$ using the algorithm presented above
    in $\bigO(\log \log \Textlen)$ time. Let us now assume that $x < e$.
  \item In $\bigO(1)$ time, compute $y = \Select{B_{\rm exp}}{\Rank{B_{\rm exp}}{e-1}{\one}}{\one}$.
    Note that then $b < x \leq y < e$ and,
    by \cref{def:exp-block-bitvector-minus}, there exist $k_1, k_2 \in \Zp$ such that
    $k_1 < k_2$, and it holds:
    \begin{itemize}
    \item $\{\SA{\Text}[i]\}_{i \in (b \dd x]} \subseteq \RMinusFive{s}{k_1}{H}{\tau}{\Text}$,
    \item $\{\SA{\Text}[i]\}_{i \in (x \dd y]} = \bigcup_{k \in (k_1 \dd k_2)} \RMinusFive{s}{k}{H}{\tau}{\Text}$,
    \item $\{\SA{\Text}[i]\}_{i \in (y \dd e]} \subseteq \RMinusFive{s}{k_2}{H}{\tau}{\Text}$.
    \end{itemize}
    Initialize the set $\mathcal{M} = \emptyset$, and then
    compute the minimum of each of the above sets as follows:
    \begin{itemize}
    \item The computation of $m_1 = \min \{\SA{\Text}[i]\}_{i \in (b \dd x]}$ proceeds using the algorithm described above
      and takes $\bigO(\log \log \Textlen)$ time. We then add $m_1$ to the set $\mathcal{M}$.
    \item Next, we check if $x = y$. If so, we skip this step. Let us thus assume that $x < y$. Then,
      it holds $k_1 + 1 < k_2$.
      In $\bigO(1)$ time we check if $\Rank{B_{\min}}{x}{\one} \neq \Rank{B_{\min}}{y}{\one}$. Consider two cases:
      \begin{itemize}
      \item First, assume that $\Rank{B_{\min}}{x}{\one} \neq \Rank{B_{\min}}{y}{\one}$. In
        $\bigO(1)$ time compute the value $z = \Select{B_{\min}}{\Rank{B_{\min}}{x}{\one} + 1}{\one}$.
        Using \cref{pr:nav-index-periodic}\eqref{pr:nav-index-periodic-it-1}, in $\bigO(\log \log \Textlen)$ time
        compute $m_2 = \SA{\Text}[z]$. By \cref{lm:exp-blocks-min},
        $\min \{\SA{\Text}[i]\}_{i \in (x \dd y]} = m_2$. Add $m_2$ to $\mathcal{M}$.
      \item Let us now assume that $\Rank{B_{\min}}{x}{\one} = \Rank{B_{\min}}{y}{\one}$.
        In $\bigO(1)$ time compute the position $y' = \Select{B_{\rm exp}}{\Rank{B_{\rm exp}}{y-1}{\one}}{\one}$.
        By \cref{def:exp-block-bitvector-minus}, we then have $x \leq y' < y$ and
        $\{\SA{\Text}[i]\}_{i \in (y' \dd y]} = \RMinusFive{s}{k_2-1}{H}{\tau}{\Text}$.
        Using the algorithm described above, in $\bigO(\log \log m)$ time we compute
        $m_2 = \min \{\SA{\Text}[i]\}_{i \in (y' \dd y]}$, and add $m_2$ to $\mathcal{M}$.
        Note that by \cref{lm:exp-blocks-min}, it holds $\min \{\SA{\Text}[i]\}_{i \in (x \dd y]}
        = \min \RMinusFive{s}{k_2-1}{H}{\tau}{\Text} = m_2$.
      \end{itemize}
    \item The computation of $m_3 = \min \{\SA{\Text}[i]\}_{i \in (y \dd e]}$ proceeds as above
      and takes $\bigO(\log \log \Textlen)$ time. We then add $m_3$ to the set $\mathcal{M}$.
    \end{itemize}
    By the above discussion, we have $2 \leq |\mathcal{M}| \leq 3$ and
    $\min \{\SA{\Text}[i]\}_{i \in (b \dd e]} = \min \mathcal{M}$.
    Thus, in $\bigO(1)$ time we compute and return $\min \mathcal{M}$.
  \end{enumerate}
  In total, the query in the general case takes $\bigO(\log \log \Textlen)$ time.

  \DSConstruction
  The components of the structure are constructed as follows:
  \begin{enumerate}
  \item By \cref{pr:nav-index-periodic}, the construction of the first
    component takes $\bigO(\Textlen / \log_{\AlphabetSize} \Textlen)$ time.
  \item To construct the second component, first, in $\bigO(\Textlen / \log_{\AlphabetSize} \Textlen)$ time
    we construct the bitvector $B_{\rm exp} = \ExpBlockBitvectorMinusTwo{\tau}{\Text}$ (\cref{def:exp-block-bitvector-minus})
    using \cref{pr:periodic-constructions}\eqref{pr:periodic-constructions-it-1}.
    We then augment $B_{\rm exp}$ using \cref{th:bin-rank-select} in $\bigO(\Textlen / \log \Textlen)
    = \bigO(\Textlen / \log_{\AlphabetSize} \Textlen)$ time.
  \item The third component of the data structure is constructed similarly as above, by combining
    \cref{pr:periodic-constructions}\eqref{pr:periodic-constructions-it-2} and
    \cref{th:bin-rank-select}.
  \end{enumerate}
  In total, the construction takes
  $\bigO(\Textlen / \log_{\AlphabetSize} \Textlen)$ time.
\end{proof}

\begin{proposition}\label{pr:periodic-lex-range-min}
  Let $\Text \in \IntegerAlphabet^{\Textlen}$ be such that
  $2 \leq \AlphabetSize < \Textlen^{1/13}$ and $\Text[\Textlen]$ does not
  occur in $\Text[1 \dd \Textlen)$. Let
  $\tau = \lfloor \mu\log_{\AlphabetSize} \Textlen \rfloor$,
  where $\mu$ is a positive constant smaller than $\tfrac{1}{12}$
  such that $\tau \geq 1$.
  Given the packed representation of $\Text$, we can in
  $\bigO(\Textlen / \log_{\AlphabetSize} \Textlen)$ time construct
  a data structure that given any $b,e \in [0 \dd \Textlen]$ such
  that $b < e$ and $\{\SA{\Text}[i]\}_{i \in (b \dd e]} \subseteq \RFour{s}{H}{\tau}{\Text}$
  for some $s \in \Zn$ and $H \in \IntegerAlphabet^{+}$, 
  in $\bigO(\log \log \Textlen)$ time returns
  \[
    \min \{\SA{\Text}[i] : i \in (b \dd e]\}.
  \]
\end{proposition}
\begin{proof}

  \DSComponents
  The data structure consists of the following components:
  \begin{enumerate}
  \item The structure from \cref{pr:nav-index-periodic} for text $\Text$. It needs
    $\bigO(\Textlen / \log_{\AlphabetSize} \Textlen)$ space.
  \item The structure from \cref{pr:periodic-lex-range-minus-min} for text $\Text$. It also
    needs $\bigO(\Textlen / \log_{\AlphabetSize} \Textlen)$ space.
  \item The symmetric version of the structure from \cref{pr:periodic-lex-range-minus-min} adapted to positions
    in $\RPlusTwo{\tau}{\Text}$ (see \cref{def:R-subsets}). It also needs
    $\bigO(\Textlen / \log_{\AlphabetSize} \Textlen)$ space.
  \end{enumerate}
  In total, the structure needs $\bigO(\Textlen / \log_{\AlphabetSize} \Textlen)$ space.

  \DSQueries
  The queries are answered as follows. Let
  $b, e \in [0 \dd \Textlen]$ be such that $b < e$ and
  $\{\SA{\Text}[i]\}_{i \in (b \dd e]} \subseteq \RFour{s}{H}{\tau}{\Text}$
  for some $s \in \Zn$ and $H \in \IntegerAlphabet^{+}$. Given $b, e$, we
  compute $\min \{\SA{\Text}[i]\}_{i \in (b \dd e]}$ as follows:
  \begin{enumerate}
  \item Using \cref{pr:nav-index-periodic}\eqref{pr:nav-index-periodic-it-1}, in
    $\bigO(\log \log \Textlen)$ time we compute $j = \SA{\Text}[e]$.
  \item Using \cref{pr:nav-index-periodic}\eqref{pr:nav-index-periodic-it-2}, in
    $\bigO(1)$ time we then compute $x,y,z \in [0 \dd \Textlen]$ such that $x \leq y \leq z$,
    $\{\SA{\Text}[i]\}_{i \in (x \dd y]} = \RMinusFour{s}{H}{\tau}{\Text}$, and
    $\{\SA{\Text}[i]\}_{i \in (y \dd z]} = \RPlusFour{s}{H}{\tau}{\Text}$,
    where $s = \HeadPos{j}{\tau}{\Text}$ and $H = \RootPos{j}{\tau}{\Text}$.
    Note that then $x \leq b < e \leq z$.
  \item
    In $\bigO(1)$ time initialize $\mathcal{M} = \emptyset$.
  \item If $b < y$, then using \cref{pr:periodic-lex-range-minus-min}, in $\bigO(\log \log \Textlen)$ time
    compute
    \[
      m^{-} = \min \{\SA{\Text}[i]\}_{i \in (b \dd \min(e,y)]},
    \]
    and add $m^{-}$ to $\mathcal{M}$.
    Note that $\{\SA{\Text}[i]\}_{i \in (b \dd \min(e,y)]} \subseteq \RMinusFour{s}{H}{\tau}{\Text}$.
  \item
    If $e > y$, then using the symmetric version of the structure from \cref{pr:periodic-lex-range-minus-min},
    in $\bigO(\log \log \Textlen)$ time compute
    \[
      m^{+} = \min \{\SA{\Text}[i]\}_{i \in (\max(b,y), e]},
    \]
    and add $m^{+}$ to $\mathcal{M}$.
    Note that $\{\SA{\Text}[i]\}_{i \in (\max(b,y) \dd e]} \subseteq \RPlusFour{s}{H}{\tau}{\Text}$.
  \item
    Note that at this point, we have $1 \leq |\mathcal{M}| \leq 2$, and
    $\min \{\SA{\Text}[i]\}_{i \in (b \dd e]} = \min \mathcal{M}$. Thus, in $\bigO(1)$ time we compute
    and return $\min \mathcal{M}$ as the answer.
  \end{enumerate}
  In total, answering the query takes $\bigO(\log \log \Textlen)$ time.

  \DSConstruction
  The components of the data structure are constructed as follows:
  \begin{enumerate}
  \item By \cref{pr:nav-index-periodic}, the construction of the first component takes
    $\bigO(\Textlen / \log_{\AlphabetSize} \Textlen)$ time.
  \item By \cref{pr:periodic-lex-range-minus-min}, the construction of the second component takes
    $\bigO(\Textlen / \log_{\AlphabetSize} \Textlen)$ time.
  \item Since the third component is a symmetric version of the second component, its construction also
    takes $\bigO(\Textlen / \log_{\AlphabetSize} \Textlen)$ time.
  \end{enumerate}
  In total, the construction takes $\bigO(\Textlen / \log_{\AlphabetSize} \Textlen)$ time.
\end{proof}

\begin{proposition}\label{pr:lex-range-minimum-periodic}
  Let $\Text \in \IntegerAlphabet^{\Textlen}$ be such that $2 \leq \AlphabetSize < \Textlen^{1/13}$ and $\Text[\Textlen]$ does not
  occur in $\Text[1 \dd \Textlen)$. Let $\tau = \lfloor \mu\log_{\AlphabetSize} \Textlen \rfloor$, where $\mu$ is a positive
  constant smaller than $\tfrac{1}{12}$ such that $\tau \geq 1$.
  Given the packed representation of $\Text$, we can in
  $\bigO(\Textlen / \log_{\AlphabetSize} \Textlen)$ time construct a data structure
  that answers the following queries:
  \begin{enumerate}
  \item\label{pr:lex-range-minimum-periodic-it-1}
    Given the packed representation of
    $\tau$-periodic (\cref{def:periodic-pattern}) patterns $\Pat_1, \Pat_2 \in \IntegerAlphabet^{+}$
    such that $\lcp{\Pat_1}{\Pat_2} \geq 3\tau - 1$,
    computes
    $\min \LexRange{\Pat_1}{\Pat_2}{\Text} \cup \{\infty\}$ (\cref{def:lex-range}) in
    $\bigO(\log \log \Textlen +
    |\Pat_1| / \log_{\AlphabetSize} \Textlen +
    |\Pat_2| / \log_{\AlphabetSize} \Textlen)$
    time.
  \item\label{pr:lex-range-minimum-periodic-it-2}
    Given the packed representation of a $\tau$-periodic pattern
    $\Pat \in \IntegerAlphabet^{+}$ satisfying
    $|\Pat| \geq 3\tau - 1$,
    computes
    $\min \LexRange{\Pat}{\Pat' c^{\infty}}{\Text} \cup \{\infty\}$
    (where $\Pat' = \Pat[1 \dd 3\tau-1]$ and $c = \AlphabetSize - 1$) in
    $\bigO(\log \log \Textlen +
    |\Pat| / \log_{\AlphabetSize} \Textlen)$
    time.
  \end{enumerate}
\end{proposition}
\begin{proof}

  \DSComponents
  The data structure consists of the following components:
  \begin{enumerate}
  \item The data structure from \cref{pr:nav-index-periodic}
    using $\bigO(\Textlen / \log_{\AlphabetSize} \Textlen)$ space.
  \item The data structure from \cref{pr:periodic-lex-range-min}
    using $\bigO(\Textlen / \log_{\AlphabetSize} \Textlen)$ space.
  \end{enumerate}
  In total, the structure needs
  $\bigO(\Textlen / \log_{\AlphabetSize} \Textlen)$ space.

  \DSQueries
  The queries are answered as follows:
  \begin{enumerate}
  \item Let $\Pat_1, \Pat_2 \in \IntegerAlphabet^{+}$ be $\tau$-periodic
    patterns such that $\lcp{\Pat_1}{\Pat_2} \geq 3\tau - 1$.
    Given the packed representation of $\Pat_1$ and $\Pat_2$, we compute
    $\min \LexRange{\Pat_1}{\Pat_2}{\Text} \cup \{\infty\}$ as follows:
    \begin{enumerate}
    \item Using \cref{pr:nav-index-periodic}\eqref{pr:nav-index-periodic-it-pat-range}, in
      $\bigO(\log \log \Textlen + |\Pat_1| / \log_{\AlphabetSize} \Textlen + |\Pat_2| / \log_{\AlphabetSize} \Textlen)$
      time we compute $b_k = \RangeBegTwo{\Pat_k}{\Text}$, where $k \in \{1,2\}$.
      If $b_1 \geq b_2$, then $\LexRange{\Pat_1}{\Pat_2}{\Text} = \emptyset$ (see \cref{rm:lex-range}), and hence
      we return $\min \LexRange{\Pat_1}{\Pat_2}{\Text} \cup \{\infty\} = \infty$. Let us now assume
      that $b_1 < b_2$. By \cref{rm:lex-range}, we then have
      \[
        \LexRange{\Pat_1}{\Pat_2}{\Text} = \{\SA{\Text}[i]\}_{i \in (b_1 \dd b_2]}.
      \]
      By $\lcp{\Pat_1}{\Pat_2} \geq 3\tau - 1$, for every $j_1, j_2 \in \LexRange{\Pat_1}{\Pat_2}{\Text}$,
      it holds $\LCE{\Text}{j_1}{j_2} \geq 3\tau - 1$. Thus, by
      \cref{lm:periodic-pos-lce}\eqref{lm:periodic-pos-lce-it-2}, there exists
      $s \in \Zn$ and $H \in \IntegerAlphabet^{+}$ such that
      $\{\SA{\Text}[i]\}_{i \in (b_1 \dd b_2]} \subseteq \RFour{s}{H}{\tau}{\Text}$.
    \item Using \cref{pr:periodic-lex-range-min}, in $\bigO(\log \log \Textlen)$ time compute
      and return $\min \{\SA{\Text}[i]\}_{i \in (b_1 \dd b_2]}$.
    \end{enumerate}
    In total, we spend
    $\bigO(\log \log \Textlen + |\Pat_1| / \log_{\AlphabetSize} \Textlen + |\Pat_2| / \log_{\AlphabetSize} \Textlen)$ time.
  \item Let us now consider a $\tau$-periodic pattern $\Pat \in \IntegerAlphabet^{+}$
    that satisfies $|\Pat| \geq 3\tau - 1$,
    where $\Pat' = \Pat[1 \dd 3\tau - 1]$ and $c = \AlphabetSize - 1$. Given the packed representation of
    $\Pat$, we compute $\min \LexRange{\Pat}{\Pat' c^{\infty}}{\Text} \cup \{\infty\}$ similarly as above, letting
    $\Pat_1 = \Pat$ and $\Pat_2 = \Pat' c^{\infty}$. The main difference is that to compute
    $b_2 = \RangeBegTwo{\Pat_2}{\Text}$, we observe that
    $\RangeBegTwo{\Pat' c^{\infty}}{\Text} = \RangeEndTwo{\Pat'}{\Text}$.
    Thus, we can determine the value of $b_2$ using \cref{pr:nav-index-periodic}\eqref{pr:nav-index-periodic-it-pat-range}
    in $\bigO(\log \log \Textlen + |\Pat'| / \log_{\AlphabetSize} \Textlen) = \bigO(\log \log \Textlen)$ time.
  \end{enumerate}

  \DSConstruction
  The components of the structure are constructed as follows:
  \begin{enumerate}
  \item The first component is constructed using \cref{pr:nav-index-periodic} in
    $\bigO(\Textlen / \log_{\AlphabetSize} \Textlen)$ time.
  \item The second component is constructed in the same time using
    \cref{pr:periodic-lex-range-min}.
  \end{enumerate}
  In total, the construction takes
  $\bigO(\Textlen / \log_{\AlphabetSize} \Textlen)$ time.
\end{proof}

\paragraph{Summary}

\begin{proposition}\label{pr:lex-range-minimum-to-prefix-rmq-nonbinary}
  Consider a data structure answering prefix RMQ
  (see \cref{sec:prefix-rmq-and-lex-range-minimum-problem-def}) that, for any
  array of integers $A[1 \dd k]$ (where $A[i] \in [1 \dd k]$ for $i \in [1 \dd k]$)
  and a sequence $W[1 \dd k]$ strings of length
  $\ell$ over alphabet $\IntegerAlphabet$
  achieves the following complexities (where the input strings at query time and
  during construction are given in the packed representation):
  \begin{itemize}
  \item space usage $S(k,\ell,\AlphabetSize)$,
  \item preprocessing time $P_t(k,\ell,\AlphabetSize)$,
  \item preprocessing space $P_s(k,\ell,\AlphabetSize)$,
  \item query time $Q(k,\ell,\AlphabetSize)$.
  \end{itemize}
  Let $\Text \in \IntegerAlphabet^{\Textlen}$ be such that $2 \leq \AlphabetSize < \Textlen^{1/13}$ and $\Text[\Textlen]$ does not
  occur in $\Text[1 \dd \Textlen)$. There exist positive integers
  $k = \Theta(\Textlen / \log_{\AlphabetSize} \Textlen)$ and
  $\ell \leq (1 + \lfloor \log k \rfloor) / \lceil \log \AlphabetSize \rceil$
  such that, given the packed representation of $\Text$, we can
  in
  $\bigO(\Textlen / \log_{\AlphabetSize} \Textlen + P_t(k,\ell,\AlphabetSize))$ time and using
  $\bigO(\Textlen / \log_{\AlphabetSize} \Textlen + P_s(k,\ell,\AlphabetSize))$ working space construct a data structure of size
  $\bigO(\Textlen / \log_{\AlphabetSize} \Textlen + S(k,\ell,\AlphabetSize))$ that, given the packed representation
  of any patterns $\Pat_1, \Pat_2 \in \IntegerAlphabet^{*}$,
  returns $\min \LexRange{\Pat_1}{\Pat_2}{\Text} \cup \{\infty\}$ (see \cref{def:lex-range}) in
  $\bigO(\log \log \Textlen +
  Q(k,\ell,\AlphabetSize) +
  |\Pat_1| / \log_{\AlphabetSize} \Textlen +
  |\Pat_2| / \log_{\AlphabetSize} \Textlen)$
  time.
\end{proposition}
\begin{proof}

  We use the following definitions.
  Let $\tau = \lfloor \mu\log_{\AlphabetSize} \Textlen \rfloor$, where $\mu$
  is a positive constant smaller than $\tfrac{1}{12}$ such that $\tau \geq 1$
  (such $\tau$ exists by the assumption $\AlphabetSize < \Textlen^{1/13}$).
  Let $k$ and $\ell$ be integers resulting from
  the application of \cref{pr:lex-range-minimum-nonperiodic} to text $\Text$,
  with the structure from the above claim to answer prefix RMQ.
  Note that they satisfy $k = \Theta(\Textlen / \log_{\AlphabetSize} \Textlen)$ and
  $\ell \leq (1 + \lfloor \log k \rfloor) / \lceil \log \AlphabetSize \rceil$.

  \DSComponents
  The data structure consists of the following components:
  \begin{enumerate}
  \item The structure from \cref{pr:per} for $\AlphabetSize$ and $\tau$. If needs
    $\bigO(\AlphabetSize^{3\tau} \cdot \tau^2) = \bigO(\Textlen / \log_{\AlphabetSize} \Textlen)$ space.
  \item The structure from \cref{pr:nav-index-short} for $\tau$ and $\Text$.
    It needs $\bigO(\Textlen / \log_{\AlphabetSize} \Textlen)$ space.
  \item The structure from \cref{pr:lex-range-minimum-short} for $\tau$ and $\Text$.
    It uses $\bigO(\Textlen / \log_{\AlphabetSize} \Textlen)$ space.
  \item The structure from \cref{pr:lex-range-minimum-nonperiodic} for $\tau$ and $\Text$, and
    using the structure from the above claim to
    answer prefix RMQ. It uses
    $\bigO(\Textlen / \log_{\AlphabetSize} \Textlen + S(k,\ell,\AlphabetSize))$ space, where $k$ and $\ell$ are defined above.
  \item The structure from \cref{pr:lex-range-minimum-periodic} for $\tau$ and $\Text$.
    It uses $\bigO(\Textlen / \log_{\AlphabetSize} \Textlen)$ space.
  \end{enumerate}
  In total, the structure needs $\bigO(\Textlen / \log_{\AlphabetSize} \Textlen + S(k,\ell,\AlphabetSize))$ space.

  \DSQueries
  The queries are answered as follows. Let $\Pat_1, \Pat_2 \in \IntegerAlphabet^{*}$. Given the packed
  representation of $\Pat_1$ and $\Pat_2$, we compute
  $\min \LexRange{\Pat_1}{\Pat_2}{\Text} \cup \{\infty\}$ as follows.
  First, in
  $\bigO(1 + |\Pat_1| / \log_{\AlphabetSize} \Textlen + |\Pat_2| / \log_{\AlphabetSize} \Textlen)$ time
  we check if $\Pat_1 \prec \Pat_2$. If not, then we have
  $\LexRange{\Pat_1}{\Pat_2}{\Text} = \emptyset$. We then return
  $\min \LexRange{\Pat_1}{\Pat_2}{\Text} \cup \{\infty\} = \infty$,
  and the query algorithm is complete.
  Let us thus assume that it holds $\Pat_1 \prec \Pat_2$.
  In $\bigO(1)$ time we use the packed representation of $\Pat_1$ and $\Pat_2$ to compute the
  packed representation of prefixes
  $X_1 = \Pat_1[1 \dd \min(3\tau-1, |\Pat_1|)]$ and
  $X_2 = \Pat_2[1 \dd \min(3\tau-1, |\Pat_2|)]$.
  Note that the assumption $\Pat_1 \prec \Pat_2$ implies that $X_1 \preceq X_2$.
  In $\bigO(1)$ time we check if $X_1 = X_2$.
  In $\bigO(1)$ time we also check if $X_1$ is a prefix of $X_2$.
  We then consider three cases:
  \begin{itemize}
  \item First, assume that $X_1 = X_2$. We then must have $|X_1| = |X_2| = 3\tau - 1$, since otherwise we would have
    $\Pat_1 = X_1 = X_2 = \Pat_2$, contradicting $\Pat_1 \prec \Pat_2$.
    Using \cref{pr:per}, in $\bigO(1)$ time we check if
    $X_1$ is $\tau$-periodic (\cref{def:periodic-pattern}).
    Consider two cases:
    \begin{itemize}
    \item If $X_1$ is $\tau$-periodic, then so are $\Pat_1$ and $\Pat_2$.
      The value $\min \LexRange{\Pat_1}{\Pat_2}{\Text} \cup \{\infty\}$ is computed using
      \cref{pr:lex-range-minimum-periodic}\eqref{pr:lex-range-minimum-periodic-it-1}
      in $\bigO(\log \log \Textlen + |\Pat_1| / \log_{\AlphabetSize} \Textlen + |\Pat_2| / \log_{\AlphabetSize} \Textlen)$ time.
    \item Otherwise, the strings $\Pat_1$ and $\Pat_2$ are $\tau$-nonperiodic. Since $\lcp{\Pat_1}{\Pat_2} \geq 3\tau - 1$, we then
      compute $\min \LexRange{\Pat_1}{\Pat_2}{\Text} \cup \{\infty\}$ in
      $\bigO(\log \log \Textlen + Q(k,\ell,\AlphabetSize) +
      |\Pat_1| / \log_{\AlphabetSize} \Textlen + |\Pat_2| / \log_{\AlphabetSize} \Textlen)$ time
      using \cref{pr:lex-range-minimum-nonperiodic}\eqref{pr:lex-range-minimum-nonperiodic-it-1}.
    \end{itemize}
    In total, we spend
    $\bigO(\log \log \Textlen + Q(k,\ell,\AlphabetSize) +
    |\Pat_1| / \log_{\AlphabetSize} \Textlen +
    |\Pat_2| / \log_{\AlphabetSize} \Textlen)$ time.
  \item Let us now assume that $X_1 \neq X_2$ and $X_1$ is a prefix of $X_2$.
    Note that then $X_1$ must be a proper prefix of $X_2$, and hence $|X_1| < |X_2|$.
    Thus, we must have $\Pat_1 = X_1$. Combining this with $X_1 \preceq X_2$ (see above), we obtain
    $\Pat_1 \preceq X_2 \preceq \Pat_2$. Consequently, we can write
    $\LexRange{\Pat_1}{\Pat_2}{\Text}$ as a disjoint union:
    \[
      \LexRange{\Pat_1}{\Pat_2}{\Text} = \LexRange{\Pat_1}{X_2}{\Text} \cup \LexRange{X_2}{\Pat_2}{\Text}.
    \]
    We will separately handle each of the two sets, and then combine the result.
    Using \cref{pr:lex-range-minimum-short}\eqref{pr:lex-range-minimum-short-it-1},
    we compute $\min \LexRange{X_1}{X_2}{\Text} \cup \{\infty\}$ (recall that $\Pat_1 = X_1$) in $\bigO(1)$ time.
    Let us now focus on computing $\min \LexRange{X_2}{\Pat_2}{\Text} \cup \{\infty\}$.
    If $|\Pat_2| < 3\tau - 1$, then
    again we compute $\min \LexRange{X_2}{\Pat_2}{\Text} \cup \{\infty\}$ using
    \cref{pr:lex-range-minimum-short}\eqref{pr:lex-range-minimum-short-it-1} in $\bigO(1)$ time.
    Let us now assume that $|\Pat_2| \geq 3\tau - 1$.
    Using \cref{pr:per}, in $\bigO(1)$ time we check if
    $\Pat_2$ is $\tau$-periodic (\cref{def:periodic-pattern}).
    Consider two cases:
    \begin{itemize}
    \item If $\Pat_2$ is $\tau$-periodic, then using \cref{pr:lex-range-minimum-periodic}\eqref{pr:lex-range-minimum-periodic-it-1},
      we compute $\min \LexRange{X_2}{\Pat_2}{\Text} \cup \{\infty\}$ in
      $\bigO(\log \log \Textlen + |\Pat_2| / \log_{\AlphabetSize} \Textlen)$ time. We can use this because
      $X_2 = \Pat_2[1 \dd 3\tau - 1]$.
    \item If $\Pat_2$ is $\tau$-nonperiodic, then using
      \cref{pr:lex-range-minimum-nonperiodic}\eqref{pr:lex-range-minimum-nonperiodic-it-1}, compute
      $\min \LexRange{X_2}{\Pat_2}{\Text} \cup \{\infty\}$ in
      $\bigO(\log \log \Textlen + Q(k,\ell,\AlphabetSize) + |\Pat_2| / \log_{\AlphabetSize} \Textlen)$ time.
    \end{itemize}
    In total, we spend
    $\bigO(\log \log \Textlen + Q(k,\ell,\AlphabetSize) + |\Pat_2| / \log_{\AlphabetSize} \Textlen)$ time.
  \item Let us now assume that $X_1 \neq X_2$ and $X_1$ is not a prefix of $X_2$. Observe
    that, letting $c = \AlphabetSize - 1$, we then have $X_1 c^{\infty} \prec X_2$.
    Combining this with $\Pat_1 \preceq X_1 c^{\infty}$ and $X_2 \preceq \Pat_2$, we can thus write
    $\LexRange{\Pat_1}{\Pat_2}{\Text}$ as a disjoint union:
    \begin{align*}
      \LexRange{\Pat_1}{\Pat_2}{\Text}
        =\,& \LexRange{\Pat_1}{X_1 c^{\infty}}{\Text} \,\cup\\
         \,& \LexRange{X_1 c^{\infty}}{X_2}{\Text} \,\cup\\
         \,& \LexRange{X_2}{\Pat_2}{\Text}.
    \end{align*}
    We will separately compute the minimum of each of the three sets. Based on this, we can determine
    $\min \LexRange{\Pat_1}{\Pat_2}{\Text} \cup \{\infty\}$.
    We begin with the set $\LexRange{\Pat_1}{X_1 c^{\infty}}{\Text}$. If $|\Pat_1| \leq 3\tau - 1$, then
    $\Pat_1 = X_1$, and hence we compute $\min \LexRange{\Pat_1}{X_1 c^{\infty}}{\Text} \cup \{\infty\}$ in
    $\bigO(1)$ time using \cref{pr:lex-range-minimum-short}\eqref{pr:lex-range-minimum-short-it-2}.
    Let us now assume that $|\Pat_1| > 3\tau - 1$. In $\bigO(1)$ time we check if $\Pat_1$ is $\tau$-periodic
    using \cref{pr:per}. We then consider two cases:
    \begin{itemize}
    \item If $\Pat_1$ is $\tau$-periodic, then we compute $\min \LexRange{\Pat_1}{X_1 c^{\infty}}{\Text} \cup \{\infty\}$
      using \cref{pr:lex-range-minimum-periodic}\eqref{pr:lex-range-minimum-periodic-it-2} in
      $\bigO(\log \log \Textlen + |\Pat_1| / \log_{\AlphabetSize} \Textlen)$ time.
    \item Otherwise (i.e., if $\Pat_1$ is $\tau$-nonperiodic), we
      compute $\min \LexRange{\Pat_1}{X_1 c^{\infty}}{\Text} \cup \{\infty\}$
      using \cref{pr:lex-range-minimum-nonperiodic}\eqref{pr:lex-range-minimum-nonperiodic-it-2} in
      $\bigO(\log \log \Textlen + Q(k,\ell,\AlphabetSize) + |\Pat_1| / \log_{\AlphabetSize} \Textlen)$ time.
    \end{itemize}
    Next, we compute $\min \LexRange{X_1 c^{\infty}}{X_2}{\Text} \cup \{\infty\}$ using
    \cref{pr:lex-range-minimum-short}\eqref{pr:lex-range-minimum-short-it-3} in $\bigO(1)$ time.
    Finally, we address the problem of computing $\min \LexRange{X_2}{\Pat_2}{\Text} \cup \{\infty\}$.
    If $|\Pat_2| \leq 3\tau - 1$, then
    $\Pat_2 = X_2$, and hence we have $\LexRange{X_2}{\Pat_2}{\Text} = \emptyset$.
    We thus obtain $\min \LexRange{X_2}{\Pat_2}{\Text} \cup \{\infty\} = \infty$.
    Let us now assume that $|\Pat_2| > 3\tau - 1$. In $\bigO(1)$ time we check if $\Pat_2$ is $\tau$-periodic
    using \cref{pr:per}. We then consider two cases:
    \begin{itemize}
    \item If $\Pat_2$ is $\tau$-nonperiodic, then we compute $\min \LexRange{X_2}{\Pat_2}{\Text} \cup \{\infty\}$
      using \cref{pr:lex-range-minimum-nonperiodic}\eqref{pr:lex-range-minimum-nonperiodic-it-1} in
      $\bigO(\log \log \Textlen + Q(k,\ell,\AlphabetSize) + |\Pat_2| / \log_{\AlphabetSize} \Textlen)$ time.
    \item Otherwise (i.e., if $\Pat_2$ is $\tau$-periodic), we compute $\min \LexRange{X_2}{\Pat_2}{\Text} \cup \{\infty\}$
      using \cref{pr:lex-range-minimum-periodic}\eqref{pr:lex-range-minimum-periodic-it-1} in
      $\bigO(\log \log \Textlen + |\Pat_2| / \log_{\AlphabetSize} \Textlen)$ time.
    \end{itemize}
    We spend
    $\bigO(\log \log \Textlen + Q(k,\ell,\AlphabetSize) +
    |\Pat_1| / \log_{\AlphabetSize} \Textlen +
    |\Pat_2| / \log_{\AlphabetSize} \Textlen)$ time in the above step.
  \end{itemize}
  Over all steps, we spend $\bigO(\log \log \Textlen + Q(k,\ell,\AlphabetSize) +
  |\Pat_1| / \log_{\AlphabetSize} \Textlen +
  |\Pat_2| / \log_{\AlphabetSize} \Textlen)$ time.

  \DSConstruction
  The components of the structure are constructed as follows:
  \begin{enumerate}
  \item In $\bigO(\AlphabetSize^{3\tau} \cdot \tau^2) = \bigO(\Textlen / \log_{\AlphabetSize} \Textlen)$ time
    we apply \cref{pr:per}.
  \item In $\bigO(\Textlen / \log_{\AlphabetSize} \Textlen)$ time we apply \cref{pr:nav-index-short} to text $\Text$.
  \item In the same time as above we apply \cref{pr:lex-range-minimum-short} to text $\Text$.
  \item In $\bigO(\Textlen / \log_{\AlphabetSize} \Textlen + P_t(k,\ell,\AlphabetSize))$ time and using
    $\bigO(\Textlen / \log_{\AlphabetSize} \Textlen + P_s(k,\ell,\AlphabetSize))$ working space we apply
    \cref{pr:lex-range-minimum-nonperiodic} to text $\Text$.
  \item In $\bigO(\Textlen / \log_{\AlphabetSize} \Textlen)$ time we apply
    \cref{pr:lex-range-minimum-periodic} to text $\Text$.
  \end{enumerate}
  In total, the construction takes
  $\bigO(\Textlen / \log_{\AlphabetSize} \Textlen + P_t(k,\ell,\AlphabetSize))$ time uses
  $\bigO(\Textlen / \log_{\AlphabetSize} \Textlen + P_s(k,\ell,\AlphabetSize))$ working space.
\end{proof}

\subsubsection{Alphabet Reduction for Prefix RMQ}\label{sec:prefix-rmq-alphabet-reduction}

\begin{proposition}\label{pr:prefix-rmq-alphabet-reduction}
  Consider a data structure answering prefix RMQ queries
  (see \cref{sec:prefix-rmq-and-lex-range-minimum-problem-def}) that, for any
  array of $m$ integers in the range $[1 \dd m]$,
  and any sequence of $m$ binary strings of length $1 + \lfloor \log m \rfloor$ (where $m \geq 1$),
  achieves the following complexities
  (where both the string at query time as well as input strings
  during construction are given in the packed representation):
  \begin{itemize}
  \item space usage $S(m)$,
  \item preprocessing time $P_t(m)$,
  \item preprocessing space $P_s(m)$,
  \item query time $Q(m)$.
  \end{itemize}
  Let $m' \geq 1$ and let
  $A[1 \dd m']$ be a sequence of $m'$ integers (where for every $i \in [1 \dd m']$, it holds $A[i] \in [1 \dd m']$) and
  $W[1 \dd m']$ be a sequence of $m'$ equal-length strings over alphabet
  $\IntegerAlphabet$ (where $\AlphabetSize \geq 2$) of length
  $\ell \leq (1 + \lfloor \log m' \rfloor) / \lceil \log \AlphabetSize \rceil$.
  Given the arrays $A$ and $W$, with all strings represented in the packed form, we can in
  $\bigO(m' + P_t(m'))$ time and using $\bigO(m' + P_s(m'))$ working space
  construct a data structure of size $\bigO(m' + S(m'))$ that, given the packed
  representation of any $X \in \IntegerAlphabet^{\leq \ell}$
  and any $b,e \in [0 \dd m']$, computes $\PrefixRMQ{A}{W}{b}{e}{X}$ (\cref{def:prefix-rmq})
  in $\bigO(Q(m'))$ time.
\end{proposition}
\begin{proof}
  The above reduction is essentially the same as the one presented in
  \cref{pr:prefix-select-alphabet-reduction}, except we only need
  \cref{lm:prefix-queries-alphabet-reduction}\eqref{lm:prefix-queries-alphabet-reduction-it-1c}.
\end{proof}

\subsubsection{Summary}\label{sec:lex-range-minimum-to-prefix-rmq-summary}

\begin{theorem}\label{th:lex-range-minimum-to-prefix-rmq}
  Consider a data structure answering prefix RMQ
  (see \cref{sec:prefix-rmq-and-lex-range-minimum-problem-def}) that, for any
  sequence of $m$ integers in the range $[1 \dd m]$,
  and any sequence of $m$ binary strings of length $1 + \lfloor \log m \rfloor$ (where $m \geq 1$),
  achieves the following complexities
  (where both the string at query time as well as input strings
  during construction are given in the packed representation):
  \begin{itemize}
  \item space usage $S(m)$,
  \item preprocessing time $P_t(m)$,
  \item preprocessing space $P_s(m)$,
  \item query time $Q(m)$.
  \end{itemize}
  For every $\Text \in \BinaryAlphabet^{\Textlen}$, there exists
  $m = \Theta(\Textlen / \log \Textlen)$ such that, given the packed
  representation of $\Text$, we can in
  $\bigO(\Textlen / \log \Textlen + P_t(m))$ time and using
  $\bigO(\Textlen / \log \Textlen + P_s(m))$ working space construct
  a data structure of size $\bigO(\Textlen / \log \Textlen + S(m))$ that,
  given the packed representation of any $\Pat_1, \Pat_2 \in \BinaryAlphabet^{*}$,
  computes $\min \LexRange{\Pat_1}{\Pat_2}{\Text} \cup \{\infty\}$ (see \cref{def:lex-range})
  in $\bigO(\log \log \Textlen + Q(m) + |\Pat_1| / \log \Textlen + |\Pat_2| / \log \Textlen)$ time.
\end{theorem}
\begin{proof}

  Assume that $\Textlen > 3^{13}-1$ (otherwise, the claim holds trivially).

  We use the following definitions. Let $\AlphabetSize = 3$ and
  let $\Text' \in \IntegerAlphabet^{\Textlen + 1}$ be a string defined so that $\Text'[1 \dd \Textlen] = \Text$ and
  $\Text'[\Textlen + 1] = \two$. Denote $\Textlen' = |\Text'|$. Note that $\Text'[\Textlen']$ does not occur
  in $\Text'[1 \dd \Textlen')$ and it holds $2 \leq \AlphabetSize < (\Textlen')^{1/13}$.
  Let $D$ denote the data structure resulting from applying \cref{pr:prefix-rmq-alphabet-reduction} with alphabet size
  $\AlphabetSize$
  to the structure (answering prefix RMQ on binary strings) form the claim.
  Given any sequence $A[1 \dd m']$ of $m'$ integers in $[1 \dd m']$,
  and any sequence $W[1 \dd m']$ of $m'$ equal-length strings over alphabet $\IntegerAlphabet$ of length
  $\ell' \leq (1 + \lfloor \log m' \rfloor) / \lceil \log \AlphabetSize \rceil$ (where $m' \geq 1$) as input
  (with all strings represented in the packed form),
  the structure $D$ achieves the following complexities (where at query time it is given
  the packed representation of any $X \in \IntegerAlphabet^{\leq \ell'}$ and any
  $b,e \in [0 \dd m']$, and returns $\PrefixRMQ{A}{W}{b}{e}{X}$):
  \begin{itemize}
  \item space usage $S'(m',\ell',\AlphabetSize) = \bigO(m' + S(m'))$,
  \item preprocessing time $P'_t(m',\ell',\AlphabetSize) = \bigO(m' + P_t(m'))$,
  \item preprocessing space $P'_s(m',\ell',\AlphabetSize) = \bigO(m' + P_s(m'))$,
  \item query time $Q'(m',\ell',\AlphabetSize) = \bigO(Q(m'))$,
  \end{itemize}
  where the complexities $S(m')$, $P_t(m')$, $P_s(m')$, and $Q(m')$ refer to the structure from the claim (answering
  prefix RMQ for binary strings).

  \DSComponents
  The data structure answering lex-range minimum queries
  consists of a single component: the data structure from \cref{pr:lex-range-minimum-to-prefix-rmq-nonbinary}
  applied for text $\Text'$ and with the structure $D$ as the structure answering prefix RMQ queries.
  Note that we can use $D$, since it supports prefix RMQ queries for the combination of parameters required
  in \cref{pr:lex-range-minimum-to-prefix-rmq-nonbinary}, i.e., for sequences of strings over alphabet $\IntegerAlphabet$, with the
  length $\ell$ of all strings satisfying
  $\ell \leq (1 + \lfloor \log k \rfloor) / \lceil \log \AlphabetSize \rceil$ (where $k$ is the
  length of the input sequence).
  To bound the space usage of this component, note that by
  \cref{pr:lex-range-minimum-to-prefix-rmq-nonbinary} and the above discussion, there exists
  $m = \Theta(\Textlen' / \log_{\AlphabetSize} \Textlen') = \Theta(\Textlen / \log \Textlen)$ (recall that $\AlphabetSize = 3$)
  such that the structure needs
  $\bigO(\Textlen' / \log_{\AlphabetSize} \Textlen' + S'(m,\ell,\AlphabetSize)) =
  \bigO(\Textlen' / \log_{\AlphabetSize} \Textlen' + m + S(m)) = \bigO(\Textlen / \log \Textlen + S(m))$ space.

  \DSQueries
  The lex-range minimum queries are answered as follows. Let $\Pat_1, \Pat_2 \in \BinaryAlphabet^{*}$.
  Note that by definition of
  $\Text'$, it follows that $\min \LexRange{\Pat_1}{\Pat_2}{\Text} \cup \{\infty\} =
  \min \LexRange{\Pat_1}{\Pat_2}{\Text'} \cup \{\infty\}$.
  By \cref{pr:lex-range-minimum-to-prefix-rmq-nonbinary} and the above
  discussion, computing the value of $\min \LexRange{\Pat_1}{\Pat_2}{\Text'} \cup \{\infty\}$ takes
  $\bigO(\log \log \Textlen' + Q'(m,\ell,\AlphabetSize) + |\Pat_1| / \log_{\AlphabetSize} \Textlen' +
  |\Pat_2| / \log_{\AlphabetSize} \Textlen') =
  \bigO(\log \log \Textlen + Q(m) + |\Pat_1| / \log \Textlen + |\Pat_2| / \log \Textlen)$ time.

  \DSConstruction
  By \cref{pr:lex-range-minimum-to-prefix-rmq-nonbinary}
  and the above discussion, construction of the above data structure
  answering lex-range minimum queries on $\Text'$ (and hence also on $\Text$) takes
  $\bigO(\Textlen' / \log_{\AlphabetSize} \Textlen' + P'_t(m,\ell,\AlphabetSize)) =
  \bigO(\Textlen' / \log_{\AlphabetSize} \Textlen' + m + P_t(m)) =
  \bigO(\Textlen / \log \Textlen + P_t(m))$
  time and uses
  $\bigO(\Textlen' / \log_{\AlphabetSize} \Textlen' + P'_s(m,\ell,\AlphabetSize)) =
  \bigO(\Textlen' / \log_{\AlphabetSize} \Textlen' + m + P_s(m)) =
  \bigO(\Textlen / \log \Textlen + P_s(m))$ working space.
\end{proof}

\section{Equivalence of Pattern Ranking and Pattern SA-Interval Queries}\label{sec:equiv-pattern-ranking-and-pattern-sa-interval}

\subsection{Problem Definitions}\label{sec:equiv-pattern-ranking-and-pattern-sa-interval-problem-def}
\vspace{-1.5ex}

\setlength{\FrameSep}{1.5ex}
\begin{framed}
  \noindent
  \probname{Indexing for Pattern SA-Interval Queries}
  \begin{description}[style=sameline,itemsep=0ex,font={\normalfont\bf}]
  \item[Input:]
    The packed representation of a nonempty string
    $\Text \in \BinaryAlphabet^{\Textlen}$.
  \item[Output:]
    A data structure that, given the packed representation of any
    $\Pat \in \BinaryAlphabet^{*}$, returns the pair
    $(\RangeBegTwo{\Pat}{\Text}, \RangeEndTwo{\Pat}{\Text})$ (\cref{def:occ}).
  \end{description}
  \vspace{-1.2ex}
\end{framed}
\vspace{2ex}

\subsection{Problem Reduction}\label{sec:equiv-pattern-ranking-and-pattern-sa-interval-problem-reduction}

\begin{definition}[Alphabet inversion]\label{def:inv-string}
  Let $\AlphabetSize \geq 2$. For every $S \in \IntegerAlphabet^{*}$,
  by $\InverseString{\AlphabetSize}{S}$ we denote a string
  $S_{\rm inv} \in \IntegerAlphabet^{*}$ of length $|S_{\rm inv}| = |S|$
  such that for every $i \in [1 \dd |S|]$, it holds $S_{\rm inv}[i] = \AlphabetSize - 1 - S[i]$.
\end{definition}

\begin{lemma}\label{lm:inv-string}
  Let $\AlphabetSize \geq 2$ and let $S_1, S_2 \in \IntegerAlphabet^{*}$ be such that $S_1$ is not a prefix of $S_2$.
  Then, $S_1 \succ S_2$ holds if and only if
  $\InverseString{\AlphabetSize}{S_1} \prec \InverseString{\AlphabetSize}{S_2}$ (see \cref{def:inv-string}).
\end{lemma}
\begin{proof}
  Denote $S'_k = \InverseString{\AlphabetSize}{S_k}$, where $k \in \{1,2\}$.

  First, assume that $S_1 \succ S_2$. This implies that for some $\ell \in [0 \dd \min(|S_1|, |S_2|))$, it holds
  $S_1[1 \dd \ell] = S_2[1 \dd \ell]$ and $S_1[\ell + 1] \succ S_2[\ell + 1]$. By \cref{def:inv-string}, we then
  have $\ell < \min(|S'_1|, |S'_2|)$,
  for every $j \in [1 \dd \ell]$, $S'_1[j] = \AlphabetSize - 1 - S_1[j] = \AlphabetSize - 1 - S_2[j] = S'_2[j]$,
  and $S'_1[\ell+1] = \AlphabetSize - 1 - S_1[\ell+1] \prec \AlphabetSize - 1 - S_2[\ell + 1] = S'_2[\ell + 1]$.
  Thus, $S'_1 \prec S'_2$.

  The proof of the opposite implication follows almost identically by noting that
  $\InverseString{\AlphabetSize}{S'_k} = S_k$ holds for $k \in \{1,2\}$.
\end{proof}

\begin{remark}\label{rm:inv-string}
  Note that the assumption that $S_1$ is not a prefix of $S_2$ is necessary in \cref{lm:inv-string},
  since without it, it is possible that $S_1 \prec S_2$ and
  $\InverseString{\AlphabetSize}{S_1} \prec \InverseString{\AlphabetSize}{S_2}$.
\end{remark}

\begin{lemma}\label{lm:range-beg-to-range-end}
  Let $\AlphabetSize \geq 2$, $\Text \in \IntegerAlphabet^{*}$ and $\Pat \in \IntegerAlphabet^{*}$.
  Then, it holds
  \[
    \RangeEndTwo{\Pat}{\Text} = |\Text| - \RangeBegTwo{\Pat_{\rm inv}}{\Text_{\rm inv}},
  \]
  where
  $\Pat_{\rm inv} = \InverseString{\AlphabetSize}{\Pat}$ and
  $\Text_{\rm inv} = \InverseString{\AlphabetSize}{\Text}$ (see \cref{def:inv-string}).
\end{lemma}
\begin{proof}
  Denote $\Textlen = |\Text|$, $\mathcal{S} = \{\Text[j \dd \Textlen] : j \in [1 \dd \Textlen]\}$, and let
  \begin{align*}
    \mathcal{A} &= \{\Text[j \dd \Textlen] : j \in [1 \dd \Textlen]\text{ and }\Text[j \dd \Textlen] \prec \Pat\},\\
    \mathcal{B} &= \{\Text[j \dd \Textlen] : j \in \OccTwo{\Pat}{\Text}\},\\
    \mathcal{C} &= \{\Text[j \dd \Textlen] :
      j \in [1 \dd \Textlen] \setminus \OccTwo{\Pat}{\Text}\text{ and }\Text[j \dd \Textlen] \succ \Pat\}.
  \end{align*}
  By definition, $\RangeBegTwo{\Pat}{\Text} = |\mathcal{A}|$ and
  $\mathcal{S}$ is a disjoint union $\mathcal{S} = \mathcal{A} \cup \mathcal{B} \cup \mathcal{C}$.
  Thus, by $\RangeEndTwo{\Pat}{\Text} = \RangeBegTwo{\Pat}{\Text} + |\OccTwo{\Pat}{\Text}|$, we obtain
  that $\RangeEndTwo{\Pat}{\Text} = |\mathcal{A}| + |\mathcal{B}| = \Textlen - |\mathcal{C}|$.
  Consequently, it suffices to show that $|\mathcal{C}| = \RangeBegTwo{\Pat_{\rm inv}}{\Text_{\rm inv}}$.

  Denote $\mathcal{C}_{\rm inv} = \{\Text_{\rm inv}[j \dd \Textlen] :
  j \in [1 \dd \Textlen]\text{ and }\Text_{\rm inv}[j \dd \Textlen] \prec \Pat_{\rm inv}\}$.
  By definition, $\RangeBegTwo{\Pat_{\rm inv}}{\Text_{\rm inv}} = |\mathcal{C}_{\rm inv}|$.
  Note that for every $j \in [1 \dd \Textlen]$, $\Text_{\rm inv}[j \dd \Textlen] \prec \Pat_{\rm inv}$ implies
  that $j \not\in \OccTwo{\Pat_{\rm inv}}{\Text_{\rm inv}}$.
  Moreover, $\OccTwo{\Pat}{\Text} = \OccTwo{\Pat_{\rm inv}}{\Text_{\rm inv}}$.
  Thus, by \cref{lm:inv-string},
  \begin{align*}
    |\mathcal{C}|
      &= |\{\Text[j \dd \Textlen] : j \in [1 \dd \Textlen]\setminus \OccTwo{\Pat}{\Text}
           \text{ and }\Text[j \dd \Textlen] \succ \Pat\}|\\
      &= |\{\Text[j \dd \Textlen] : j \in [1 \dd \Textlen] \setminus \OccTwo{\Pat}{\Text}
           \text{ and }\InverseString{\AlphabetSize}{\Text[j \dd \Textlen]} \prec \InverseString{\AlphabetSize}{\Pat}\}|\\
      &= |\{\Text[j \dd \Textlen] : j \in [1 \dd \Textlen] \setminus \OccTwo{\Pat}{\Text}
           \text{ and }\Text_{\rm inv}[j \dd \Textlen] \prec \Pat_{\rm inv}\}|\\
      &= |\{\Text_{\rm inv}[j \dd \Textlen] : j \in [1 \dd \Textlen] \setminus \OccTwo{\Pat}{\Text}
           \text{ and }\Text_{\rm inv}[j \dd \Textlen] \prec \Pat_{\rm inv}\}|\\
      &= |\{\Text_{\rm inv}[j \dd \Textlen] : j \in [1 \dd \Textlen] \setminus \OccTwo{\Pat_{\rm inv}}{\Text_{\rm inv}}
           \text{ and }\Text_{\rm inv}[j \dd \Textlen] \prec \Pat_{\rm inv}\}|\\
      &= |\{\Text_{\rm inv}[j \dd \Textlen] : j \in [1 \dd \Textlen] : \Text_{\rm inv}[j \dd \Textlen] \prec \Pat_{\rm inv}\}|\\
      &= |\mathcal{C}_{\rm inv}|
      = \RangeBegTwo{\Pat_{\rm inv}}{\Text_{\rm inv}}.
      \qedhere
  \end{align*}
\end{proof}

\begin{proposition}\label{pr:inv-string}
  For every $u \geq 1$ we can in $\bigO(\sqrt{u} \log u)$ time construct
  a data structure of size $\bigO(\sqrt{u})$ that,
  given the packed representation of any string $S \in \BinaryAlphabet^{*}$,
  returns the packed representation of $\InverseString{2}{S}$ (\cref{def:inv-string})
  in $\bigO(1 + |S| / \log u)$ time.
\end{proposition}
\begin{proof}

  Let $\ell = \lceil (\log (u+1))/2 \rceil$.
  Let $L_{\rm inv}$ be a lookup table such that, for every $X \in \BinaryAlphabet^{\leq \ell}$,
  $L_{\rm inv}$ maps $X$ to the packed representation of $\InverseString{2}{X}$.

  \DSComponents
  The data structure consists of a single component: the lookup table $L_{\rm inv}$. When accessing $L_{\rm inv}$,
  we map $X \in \BinaryAlphabet^{\leq \ell}$ to $\BinStrToInt{X}$ (\cref{def:bin-str-to-int}). The packed
  representation of $X_{\rm inv} = \InverseString{2}{X}$ is also stored as $\BinStrToInt{X_{\rm inv}}$. Thus,
  $L_{\rm inv}$ needs $\bigO(2^{\ell}) = \bigO(\sqrt{u})$ space.

  \DSQueries
  To answer queries, observe that using $L_{\rm inv}$, given the packed representation of any $S \in \BinaryAlphabet^{*}$,
  we can compute $\InverseString{2}{S}$ in $\bigO(1 + |S| / \ell) = \bigO(1 + |S| / \log u)$ time.

  \DSConstruction
  The computation of a single entry of the $L_{\rm inv}$ table takes $\bigO(\ell)$ time. Thus, construction of
  $L_{\rm inv}$ takes $\bigO(2^{\ell} \cdot \ell) = \bigO(\sqrt{u} \log u)$ time in total.
\end{proof}

\subsection{Summary}\label{sec:equiv-pattern-ranking-and-pattern-sa-interval-summary}

\begin{theorem}\label{th:sa-interval-to-pattern-ranking}
  Consider a data structure answering pattern ranking queries
  (see \cref{sec:equiv-pattern-ranking-and-pattern-sa-interval-problem-def})
  that, for any
  text $\Text \in \BinaryAlphabet^{\Textlen}$, achieves the following complexities
  (where in the preprocessing we assume that we are given as input the packed
  representation of $\Text$, and at query time we are given a packed
  representation of $\Pat \in \BinaryAlphabet^{\leq k}$):
  \begin{itemize}
  \item space usage $S(\Textlen)$,
  \item preprocessing time $P_t(\Textlen)$,
  \item preprocessing space $P_s(\Textlen)$,
  \item query time $Q(\Textlen, k)$.
  \end{itemize}
  Then, for every $\Text \in \BinaryAlphabet^{\Textlen}$, we can
  in $\bigO(\Textlen / \log \Textlen + P_t(\Textlen))$ time and
  using $\bigO(\Textlen / \log \Textlen + P_s(\Textlen))$ working space
  construct a data structure of size $\bigO(\Textlen / \log \Textlen + S(\Textlen))$ that, given
  the packed representation of any $\Pat \in \BinaryAlphabet^{\leq k}$,
  returns $(\RangeBegTwo{\Pat}{\Text}, \RangeEndTwo{\Pat}{\Text})$
  in $\bigO(k / \log \Textlen + Q(\Textlen,k))$ time.
\end{theorem}
\begin{proof}

  Denote $\Text_{\rm inv} = \InverseString{2}{\Text}$ (\cref{def:inv-string}).

  \DSComponents
  The data structure consists of the following components:
  \begin{enumerate}
  \item The data structure from \cref{pr:inv-string} for $u = \Textlen$. It needs
    $\bigO(\sqrt{\Textlen})$ space.
  \item The data structure from the claim for $\Text$.
    It needs $\bigO(S(\Textlen))$ space.
  \item The data structure from the claim for $\Text_{\rm inv}$.
    Since $|\Text_{\rm inv}| = |\Text| = \Textlen$,
    it also needs $\bigO(S(\Textlen))$ space.
  \end{enumerate}
  In total, the structure needs
  $\bigO(\sqrt{\Textlen} + S(\Textlen)) =
  \bigO(\Textlen / \log \Textlen + S(\Textlen))$
  space.

  \DSQueries
  The queries are answered as follows. Let $\Pat \in \BinaryAlphabet^{\leq k}$, and assume
  that we are given the packed representation of $\Pat$. To compute
  $(\RangeBegTwo{\Pat}{\Text}, \RangeEndTwo{\Pat}{\Text})$ we proceed as follows:
  \begin{enumerate}
  \item Using the structure from the claim, compute
    $b = \RangeBegTwo{\Pat}{\Text}$ in $\bigO(Q(\Textlen, k))$ time.
  \item Using \cref{pr:inv-string}, in $\bigO(1 + |\Pat|/\log \Textlen) = \bigO(1 + k / \log \Textlen)$ time
    we compute the packed representation of $\Pat_{\rm inv} = \InverseString{2}{\Pat}$.
  \item Using the structure from the claim (built for $\Text_{\rm inv}$),
    compute $e = \Textlen-\RangeBegTwo{\Pat_{\rm inv}}{\Text_{\rm inv}}$ in
    $\bigO(Q(\Textlen, k))$ time. By \cref{lm:range-beg-to-range-end},
    $e = \RangeEndTwo{\Pat}{\Text}$. Thus, we return $(b,e)$ as the answer.
  \end{enumerate}
  In total, the query takes $\bigO(k / \log \Textlen + Q(\Textlen, k))$ time.

  \DSConstruction
  The components of the structure are constructed as follows:
  \begin{enumerate}
  \item In $\bigO(\sqrt{\Textlen} \log \Textlen) = \bigO(\Textlen / \log \Textlen)$ time we apply \cref{pr:inv-string}.
  \item The structure from the claim is built for $\Text$ in $\bigO(P_t(\Textlen))$ time and using $\bigO(P_s(\Textlen))$ space.
  \item In analogous time and space we construct the structure from the claim for $\Text_{\rm inv}$.
  \end{enumerate}
  In total, we spend $\bigO(\Textlen / \log \Textlen + P_t(\Textlen))$ time and use
  $\bigO(\Textlen / \log \Textlen + P_s(\Textlen))$ working space.
\end{proof}

\begin{theorem}\label{th:pattern-ranking-and-pattern-sa-interval-equivalence}
  The problems
  \begin{itemize}
  \item \probname{Indexing for Pattern Ranking Queries} (\cref{sec:prefix-rank-and-pattern-ranking-problem-def}) and
  \item \probname{indexing for Pattern SA-Interval Queries} (\cref{sec:equiv-pattern-ranking-and-pattern-sa-interval-problem-def})
  \end{itemize}
  are equivalent in the following sense.
  If, given the packed representation of a nonempty text $\Text \in \BinaryAlphabet^{\Textlen}$,
  we can in $P_t(\Textlen)$ time and using $P_s(\Textlen)$ working space construct a data structure
  of size $S(\Textlen)$ that, given a packed representation of $\Pat \in \BinaryAlphabet^{\leq k}$, answers
  queries to one of the problems in $Q(\Textlen, k)$ time, then in
  $\bigO(\Textlen / \log \Textlen + P_t(\Textlen))$ time and using
  $\bigO(\Textlen / \log \Textlen + P_s(\Textlen))$ working space we can construct a data structure of size
  $\bigO(\Textlen / \log \Textlen + S(\Textlen))$ that answers queries for the other problem in
  $\bigO(k / \log \Textlen + Q(\Textlen, k))$ time.
\end{theorem}
\begin{proof}
  The reduction in one direction follows by definition. The other direction follows by \cref{th:sa-interval-to-pattern-ranking}.
\end{proof}

\bibliographystyle{alphaurl}
\bibliography{paper}

\newcommand{\etalchar}[1]{$^{#1}$}
\begin{thebibliography}{GMC{\etalchar{+}}14}

\bibitem[ABG{\etalchar{+}}14]{ArroyueloBCMN14}
Diego Arroyuelo, Carolina Bonacic, Veronica Gil{-}Costa, Mauricio Mar{\'{\i}}n,
  and Gonzalo Navarro.
\newblock Distributed text search using suffix arrays.
\newblock {\em Parallel Computing}, 40(9):471--495, 2014.
\newblock \href {https://doi.org/10.1016/J.PARCO.2014.06.007}
  {\path{doi:10.1016/J.PARCO.2014.06.007}}.

\bibitem[ABM08]{bwtbook}
Donald Adjeroh, Tim Bell, and Amar Mukherjee.
\newblock {\em The {B}urrows-{W}heeler Transform: Data Compression, Suffix
  Arrays, and Pattern Matching}.
\newblock Springer, Boston, MA, USA, 2008.
\newblock \href {https://doi.org/10.1007/978-0-387-78909-5}
  {\path{doi:10.1007/978-0-387-78909-5}}.

\bibitem[AKO04]{AbouelhodaKO04}
Mohamed~Ibrahim Abouelhoda, Stefan Kurtz, and Enno Ohlebusch.
\newblock Replacing suffix trees with enhanced suffix arrays.
\newblock {\em Journal of Discrete Algorithms}, 2(1):53--86, 2004.
\newblock \href {https://doi.org/10.1016/S1570-8667(03)00065-0}
  {\path{doi:10.1016/S1570-8667(03)00065-0}}.

\bibitem[BCFR18]{BrisaboaCFR18}
Nieves~R. Brisaboa, Diego Caro, Antonio Fari{\~{n}}a, and M.~Andrea
  Rodr{\'{\i}}guez.
\newblock Using compressed suffix-arrays for a compact representation of
  temporal-graphs.
\newblock {\em Information Sciences}, 465:459--483, 2018.
\newblock \href {https://doi.org/10.1016/J.INS.2018.07.023}
  {\path{doi:10.1016/J.INS.2018.07.023}}.

\bibitem[Bel14]{Belazzougui14}
Djamal Belazzougui.
\newblock Linear time construction of compressed text indices in compact space.
\newblock In David~B. Shmoys, editor, {\em 46th Annual {ACM} Symposium on
  Theory of Computing, {STOC} 2014}, pages 148--193. {ACM}, 2014.
\newblock \href {https://doi.org/10.1145/2591796.2591885}
  {\path{doi:10.1145/2591796.2591885}}.

\bibitem[BGKS15]{WaveletSuffixTree}
Maxim~A. Babenko, Pawe{\l} Gawrychowski, Tomasz Kociumaka, and Tatiana
  Starikovskaya.
\newblock Wavelet trees meet suffix trees.
\newblock In Piotr Indyk, editor, {\em 26th Annual {ACM-SIAM} Symposium on
  Discrete Algorithms, {SODA} 2015}, pages 572--591. {SIAM}, 2015.
\newblock \href {https://doi.org/10.1137/1.9781611973730.39}
  {\path{doi:10.1137/1.9781611973730.39}}.

\bibitem[BHMP14]{BartonHMP14}
Carl Barton, Alice H{\'{e}}liou, Laurent Mouchard, and Solon~P. Pissis.
\newblock Linear-time computation of minimal absent words using suffix array.
\newblock {\em BMC Bioinformatics}, 15:388, 2014.
\newblock \href {https://doi.org/10.1186/S12859-014-0388-9}
  {\path{doi:10.1186/S12859-014-0388-9}}.

\bibitem[BN14]{BelazzouguiN14}
Djamal Belazzougui and Gonzalo Navarro.
\newblock Alphabet-independent compressed text indexing.
\newblock {\em ACM Transactions on Algorithms}, 10(4):23:1--23:19, 2014.
\newblock \href {https://doi.org/10.1145/2635816} {\path{doi:10.1145/2635816}}.

\bibitem[BW94]{bwt}
Michael Burrows and David~J. Wheeler.
\newblock A block-sorting lossless data compression algorithm.
\newblock Technical Report 124, Digital Equipment Corporation, Palo Alto,
  California, 1994.
\newblock URL:
  \url{https://www.hpl.hp.com/techreports/Compaq-DEC/SRC-RR-124.pdf}.

\bibitem[BYRN99]{baeza1999modern}
Ricardo Baeza-Yates and Berthier Ribeiro-Neto.
\newblock {\em Modern information retrieval}.
\newblock ACM Press New York, 1999.

\bibitem[CGN21]{CobasGN21}
Dustin Cobas, Travis Gagie, and Gonzalo Navarro.
\newblock A fast and small subsampled r-index.
\newblock In Pawel Gawrychowski and Tatiana Starikovskaya, editors, {\em 32nd
  Annual Symposium on Combinatorial Pattern Matching, {CPM} 2021}, volume 191
  of {\em LIPIcs}, pages 13:1--13:16. Schloss Dagstuhl--Leibniz-Zentrum
  f{\"{u}}r Informatik, 2021.
\newblock \href {https://doi.org/10.4230/LIPICS.CPM.2021.13}
  {\path{doi:10.4230/LIPICS.CPM.2021.13}}.

\bibitem[CHL07]{AlgorithmsOnStrings}
Maxime Crochemore, Christophe Hancart, and Thierry Lecroq.
\newblock {\em Algorithms on strings}.
\newblock Cambridge University Press, Cambridge, UK, 2007.
\newblock \href {https://doi.org/10.1017/cbo9780511546853}
  {\path{doi:10.1017/cbo9780511546853}}.

\bibitem[CI08]{CrochemoreI08}
Maxime Crochemore and Lucian Ilie.
\newblock Computing longest previous factor in linear time and applications.
\newblock {\em Information Processing Letters}, 106(2):75--80, 2008.
\newblock \href {https://doi.org/10.1016/J.IPL.2007.10.006}
  {\path{doi:10.1016/J.IPL.2007.10.006}}.

\bibitem[CILP20]{Charalampopoulos20}
Panagiotis Charalampopoulos, Costas~S. Iliopoulos, Chang Liu, and Solon~P.
  Pissis.
\newblock Property suffix array with applications in indexing weighted
  sequences.
\newblock {\em ACM Journal of Experimental Algorithmics}, 25:1--16, 2020.
\newblock \href {https://doi.org/10.1145/3385898} {\path{doi:10.1145/3385898}}.

\bibitem[CIS08]{CrochemoreIS08}
Maxime Crochemore, Lucian Ilie, and William~F. Smyth.
\newblock A simple algorithm for computing the {L}empel {Z}iv factorization.
\newblock In {\em 2008 Data Compression Conference, {DCC} 2008}, pages
  482--488. {IEEE} Computer Society, 2008.
\newblock \href {https://doi.org/10.1109/DCC.2008.36}
  {\path{doi:10.1109/DCC.2008.36}}.

\bibitem[CKPR21]{Charalampopoulos21}
Panagiotis Charalampopoulos, Tomasz Kociumaka, Solon~P. Pissis, and Jakub
  Radoszewski.
\newblock Faster algorithms for longest common substring.
\newblock In Petra Mutzel, Rasmus Pagh, and Grzegorz Herman, editors, {\em 29th
  Annual European Symposium on Algorithms, {ESA} 2021}, volume 204 of {\em
  LIPIcs}, pages 30:1--30:17. Schloss Dagstuhl--Leibniz-Zentrum f{\"{u}}r
  Informatik, 2021.
\newblock \href {https://doi.org/10.4230/LIPIcs.ESA.2021.30}
  {\path{doi:10.4230/LIPIcs.ESA.2021.30}}.

\bibitem[Cla98]{Clark98}
David~R. Clark.
\newblock {\em Compact Pat Trees}.
\newblock PhD thesis, University of Waterloo, 1998.

\bibitem[CPS07]{ChenPS07}
Gang Chen, Simon~J. Puglisi, and William~F. Smyth.
\newblock Fast and practical algorithms for computing all the runs in a string.
\newblock In Bin Ma and Kaizhong Zhang, editors, {\em 18th Annual Symposium on
  Combinatorial Pattern Matching, {CPM} 2007}, volume 4580 of {\em LNCS}, pages
  307--315. Springer, 2007.
\newblock \href {https://doi.org/10.1007/978-3-540-73437-6_31}
  {\path{doi:10.1007/978-3-540-73437-6_31}}.

\bibitem[CPZ20]{CaceresPZ20}
Manuel C{\'{a}}ceres, Simon~J. Puglisi, and Bella Zhukova.
\newblock Fast indexes for gapped pattern matching.
\newblock In Alexander Chatzigeorgiou, Riccardo Dondi, Herodotos Herodotou,
  Christos~A. Kapoutsis, Yannis Manolopoulos, George~A. Papadopoulos, and
  Florian Sikora, editors, {\em 46th International Conference on Current Trends
  in Theory and Practice of Informatics, {SOFSEM} 2020}, volume 12011 of {\em
  LNCS}, pages 493--504. Springer, 2020.
\newblock \href {https://doi.org/10.1007/978-3-030-38919-2_40}
  {\path{doi:10.1007/978-3-030-38919-2_40}}.

\bibitem[CT10]{CrochemoreT10}
Maxime Crochemore and German Tischler.
\newblock The gapped suffix array: {A} new index structure for fast approximate
  matching.
\newblock In Edgar Ch{\'{a}}vez and Stefano Lonardi, editors, {\em 17th
  International Symposium on String Processing and Information Retrieval,
  {SPIRE} 2010}, volume 6393 of {\em LNCS}, pages 359--364. Springer, 2010.
\newblock \href {https://doi.org/10.1007/978-3-642-16321-0_37}
  {\path{doi:10.1007/978-3-642-16321-0_37}}.

\bibitem[DKK{\etalchar{+}}04]{DuvalKKLL04}
Jean{-}Pierre Duval, Roman Kolpakov, Gregory Kucherov, Thierry Lecroq, and
  Arnaud Lefebvre.
\newblock Linear-time computation of local periods.
\newblock {\em Theoretical Computer Science}, 326(1-3):229--240, 2004.
\newblock \href {https://doi.org/10.1016/J.TCS.2004.06.024}
  {\path{doi:10.1016/J.TCS.2004.06.024}}.

\bibitem[FA19]{FlickA19}
Patrick Flick and Srinivas Aluru.
\newblock Distributed enhanced suffix arrays: efficient algorithms for
  construction and querying.
\newblock In Michela Taufer, Pavan Balaji, and Antonio~J. Pe{\~{n}}a, editors,
  {\em International Conference for High Performance Computing, Networking,
  Storage and Analysis, {SC} 2019}, pages 72:1--72:17. {ACM}, 2019.
\newblock \href {https://doi.org/10.1145/3295500.3356211}
  {\path{doi:10.1145/3295500.3356211}}.

\bibitem[FGGV06]{FoschiniGGV06}
Luca Foschini, Roberto Grossi, Ankur Gupta, and Jeffrey~Scott Vitter.
\newblock When indexing equals compression: Experiments with compressing suffix
  arrays and applications.
\newblock {\em ACM Transactions on Algorithms}, 2(4):611--639, 2006.
\newblock \href {https://doi.org/10.1145/1198513.1198521}
  {\path{doi:10.1145/1198513.1198521}}.

\bibitem[FGM12]{FerraginaGM12}
Paolo Ferragina, Travis Gagie, and Giovanni Manzini.
\newblock Lightweight data indexing and compression in external memory.
\newblock {\em Algorithmica}, 63(3):707--730, 2012.
\newblock \href {https://doi.org/10.1007/S00453-011-9535-0}
  {\path{doi:10.1007/S00453-011-9535-0}}.

\bibitem[FH11]{FischerH11}
Johannes Fischer and Volker Heun.
\newblock Space-efficient preprocessing schemes for range minimum queries on
  static arrays.
\newblock {\em SIAM Journal on Computing}, 40(2):465--492, 2011.
\newblock \href {https://doi.org/10.1137/090779759}
  {\path{doi:10.1137/090779759}}.

\bibitem[FLMM05]{FerraginaLMM05}
Paolo Ferragina, Fabrizio Luccio, Giovanni Manzini, and S.~Muthukrishnan.
\newblock Structuring labeled trees for optimal succinctness, and beyond.
\newblock In {\em 46th Annual {IEEE} Symposium on Foundations of Computer
  Science, {FOCS} 2005}, pages 184--196. {IEEE} Computer Society, 2005.
\newblock \href {https://doi.org/10.1109/SFCS.2005.69}
  {\path{doi:10.1109/SFCS.2005.69}}.

\bibitem[FM00]{FerraginaM00}
Paolo Ferragina and Giovanni Manzini.
\newblock Opportunistic data structures with applications.
\newblock In {\em 41st Annual Symposium on Foundations of Computer Science,
  {FOCS} 2000}, pages 390--398. {IEEE} Computer Society, 2000.
\newblock \href {https://doi.org/10.1109/SFCS.2000.892127}
  {\path{doi:10.1109/SFCS.2000.892127}}.

\bibitem[FM05]{FerraginaM05}
Paolo Ferragina and Giovanni Manzini.
\newblock Indexing compressed text.
\newblock {\em Journal of the ACM}, 52(4):552--581, 2005.
\newblock \href {https://doi.org/10.1145/1082036.1082039}
  {\path{doi:10.1145/1082036.1082039}}.

\bibitem[FMN09]{FischerMN09}
Johannes Fischer, Veli M{\"{a}}kinen, and Gonzalo Navarro.
\newblock Faster entropy-bounded compressed suffix trees.
\newblock {\em Theoretical Computer Science}, 410(51):5354--5364, 2009.
\newblock \href {https://doi.org/10.1016/J.TCS.2009.09.012}
  {\path{doi:10.1016/J.TCS.2009.09.012}}.

\bibitem[FNI{\etalchar{+}}19]{FujisatoNIBT19}
Noriki Fujisato, Yuto Nakashima, Shunsuke Inenaga, Hideo Bannai, and Masayuki
  Takeda.
\newblock Direct linear time construction of parameterized suffix and {LCP}
  arrays for constant alphabets.
\newblock In Nieves~R. Brisaboa and Simon~J. Puglisi, editors, {\em 26th
  International Symposium on String Processing and Information Retrieval,
  {SPIRE} 2019}, volume 11811 of {\em LNCS}, pages 382--391. Springer, 2019.
\newblock \href {https://doi.org/10.1007/978-3-030-32686-9_27}
  {\path{doi:10.1007/978-3-030-32686-9_27}}.

\bibitem[GB13]{GotoB13}
Keisuke Goto and Hideo Bannai.
\newblock Simpler and faster {L}empel {Z}iv factorization.
\newblock In Ali Bilgin, Michael~W. Marcellin, Joan Serra{-}Sagrist{\`{a}}, and
  James~A. Storer, editors, {\em 2013 Data Compression Conference, {DCC} 2013},
  pages 133--142. {IEEE}, 2013.
\newblock \href {https://doi.org/10.1109/DCC.2013.21}
  {\path{doi:10.1109/DCC.2013.21}}.

\bibitem[GGV03]{wt}
Roberto Grossi, Ankur Gupta, and Jeffrey~Scott Vitter.
\newblock High-order entropy-compressed text indexes.
\newblock In {\em 14th Annual {ACM-SIAM} Symposium on Discrete Algorithms,
  {SODA} 2003}, pages 841--850. {ACM/SIAM}, 2003.
\newblock URL: \url{http://dl.acm.org/citation.cfm?id=644108.644250}.

\bibitem[GHN20]{Gao0N20}
Younan Gao, Meng He, and Yakov Nekrich.
\newblock Fast preprocessing for optimal orthogonal range reporting and range
  successor with applications to text indexing.
\newblock In {\em 28th Annual European Symposium on Algorithms, {ESA} 2020},
  volume 173 of {\em LIPIcs}, pages 54:1--54:18. Schloss Dagstuhl -
  Leibniz-Zentrum f{\"{u}}r Informatik, 2020.
\newblock \href {https://doi.org/10.4230/LIPICS.ESA.2020.54}
  {\path{doi:10.4230/LIPICS.ESA.2020.54}}.

\bibitem[GK12]{GonnellaK12}
Giorgio Gonnella and Stefan Kurtz.
\newblock Readjoiner: a fast and memory efficient string graph-based sequence
  assembler.
\newblock {\em BMC Bioinformatics}, 13:82, 2012.
\newblock \href {https://doi.org/10.1186/1471-2105-13-82}
  {\path{doi:10.1186/1471-2105-13-82}}.

\bibitem[GK17]{GawrychowskiK17}
Paweł Gawrychowski and Tomasz Kociumaka.
\newblock Sparse suffix tree construction in optimal time and space.
\newblock In Philip~N. Klein, editor, {\em 28th Annual {ACM-SIAM} Symposium on
  Discrete Algorithms, {SODA} 2017}, pages 425--439. {SIAM}, 2017.
\newblock \href {https://doi.org/10.1137/1.9781611974782.27}
  {\path{doi:10.1137/1.9781611974782.27}}.

\bibitem[GMC{\etalchar{+}}14]{GogMCTW14}
Simon Gog, Alistair Moffat, J.~Shane Culpepper, Andrew Turpin, and Anthony
  Wirth.
\newblock Large-scale pattern search using reduced-space on-disk suffix arrays.
\newblock {\em IEEE Transactions on Knowledge and Data Engineering},
  26(8):1918--1931, 2014.
\newblock \href {https://doi.org/10.1109/TKDE.2013.129}
  {\path{doi:10.1109/TKDE.2013.129}}.

\bibitem[GNF14]{GonzalezNF14}
Rodrigo Gonz{\'{a}}lez, Gonzalo Navarro, and H{\'{e}}ctor Ferrada.
\newblock Locally compressed suffix arrays.
\newblock {\em ACM Journal of Experimental Algorithmics}, 19(1), 2014.
\newblock \href {https://doi.org/10.1145/2594408} {\path{doi:10.1145/2594408}}.

\bibitem[GNP15]{GogNP15}
Simon Gog, Gonzalo Navarro, and Matthias Petri.
\newblock Improved and extended locating functionality on compressed suffix
  arrays.
\newblock {\em Journal of Discrete Algorithms}, 32:53--63, 2015.
\newblock \href {https://doi.org/10.1016/J.JDA.2015.01.006}
  {\path{doi:10.1016/J.JDA.2015.01.006}}.

\bibitem[GNP20]{Gagie2020}
Travis Gagie, Gonzalo Navarro, and Nicola Prezza.
\newblock Fully functional suffix trees and optimal text searching in
  {BWT}-runs bounded space.
\newblock {\em Journal of the ACM}, 67(1):2:1--2:54, 2020.
\newblock \href {https://doi.org/10.1145/3375890} {\path{doi:10.1145/3375890}}.

\bibitem[Gro11]{Grossi11}
Roberto Grossi.
\newblock A quick tour on suffix arrays and compressed suffix arrays.
\newblock {\em Theoretical Computer Science}, 412(27):2964--2973, 2011.
\newblock \href {https://doi.org/10.1016/J.TCS.2010.12.036}
  {\path{doi:10.1016/J.TCS.2010.12.036}}.

\bibitem[GS04]{gusfield2004linear}
Dan Gusfield and Jens Stoye.
\newblock Linear time algorithms for finding and representing all the tandem
  repeats in a string.
\newblock {\em Journal of Computer and System Sciences}, 69(4):525--546, 2004.
\newblock \href {https://doi.org/10.1016/j.jcss.2004.03.004}
  {\path{doi:10.1016/j.jcss.2004.03.004}}.

\bibitem[Gus97]{gusfield}
Dan Gusfield.
\newblock {\em Algorithms on Strings, Trees, and Sequences: {C}omputer Science
  and Computational Biology}.
\newblock Cambridge University Press, Cambridge, UK, 1997.
\newblock \href {https://doi.org/10.1017/cbo9780511574931}
  {\path{doi:10.1017/cbo9780511574931}}.

\bibitem[GV00]{GrossiV00}
Roberto Grossi and Jeffrey~Scott Vitter.
\newblock Compressed suffix arrays and suffix trees with applications to text
  indexing and string matching (extended abstract).
\newblock In F.~Frances Yao and Eugene~M. Luks, editors, {\em 32nd Annual {ACM}
  Symposium on Theory of Computing, {STOC} 2000}, pages 397--406. {ACM}, 2000.
\newblock \href {https://doi.org/10.1145/335305.335351}
  {\path{doi:10.1145/335305.335351}}.

\bibitem[GV05]{GrossiV05}
Roberto Grossi and Jeffrey~Scott Vitter.
\newblock Compressed suffix arrays and suffix trees with applications to text
  indexing and string matching.
\newblock {\em SIAM Journal on Computing}, 35(2):378--407, 2005.
\newblock \href {https://doi.org/10.1137/S0097539702402354}
  {\path{doi:10.1137/S0097539702402354}}.

\bibitem[Hag98]{Hagerup98}
Torben Hagerup.
\newblock Sorting and searching on the word {RAM}.
\newblock In Michel Morvan, Christoph Meinel, and Daniel Krob, editors, {\em
  15th Annual Symposium on Theoretical Aspects of Computer Science, {STACS}
  1998}, volume 1373 of {\em LNCS}, pages 366--398. Springer, 1998.
\newblock \href {https://doi.org/10.1007/BFb0028575}
  {\path{doi:10.1007/BFb0028575}}.

\bibitem[HL11]{hazelhurst2011kaboom}
Scott Hazelhurst and Zsuzsanna Lipt{\'a}k.
\newblock Kaboom! a new suffix array based algorithm for clustering expression
  data.
\newblock {\em Bioinformatics}, 27(24):3348--3355, 2011.
\newblock \href {https://doi.org/10.1093/bioinformatics/btr560}
  {\path{doi:10.1093/bioinformatics/btr560}}.

\bibitem[HLSS03]{HonLSS03}
Wing{-}Kai Hon, Tak~Wah Lam, Kunihiko Sadakane, and Wing{-}Kin Sung.
\newblock Constructing compressed suffix arrays with large alphabets.
\newblock In Toshihide Ibaraki, Naoki Katoh, and Hirotaka Ono, editors, {\em
  14th International Symposium on Algorithms and Computation, {ISAAC} 2003},
  volume 2906 of {\em LNCS}, pages 240--249. Springer, 2003.
\newblock \href {https://doi.org/10.1007/978-3-540-24587-2_26}
  {\path{doi:10.1007/978-3-540-24587-2_26}}.

\bibitem[HLST11]{HonLST11}
Wing{-}Kai Hon, Chen{-}Hua Lu, Rahul Shah, and Sharma~V. Thankachan.
\newblock Succinct indexes for circular patterns.
\newblock In Takao Asano, Shin{-}Ichi Nakano, Yoshio Okamoto, and Osamu
  Watanabe, editors, {\em 22nd International Symposium on Algorithms and
  Computation, {ISAAC} 2011}, volume 7074 of {\em LNCS}, pages 673--682.
  Springer, 2011.
\newblock \href {https://doi.org/10.1007/978-3-642-25591-5_69}
  {\path{doi:10.1007/978-3-642-25591-5_69}}.

\bibitem[HSS09]{HonSS03}
Wing{-}Kai Hon, Kunihiko Sadakane, and Wing{-}Kin Sung.
\newblock Breaking a time-and-space barrier in constructing full-text indices.
\newblock {\em SIAM Journal on Computing}, 38(6):2162--2178, 2009.
\newblock \href {https://doi.org/10.1137/070685373}
  {\path{doi:10.1137/070685373}}.

\bibitem[IFI11]{ilie2011hitec}
Lucian Ilie, Farideh Fazayeli, and Silvana Ilie.
\newblock Hitec: accurate error correction in high-throughput sequencing data.
\newblock {\em Bioinformatics}, 27(3):295--302, 2011.
\newblock \href {https://doi.org/10.1093/bioinformatics/btq653}
  {\path{doi:10.1093/bioinformatics/btq653}}.

\bibitem[IS11]{IlieS11}
Lucian Ilie and William~F. Smyth.
\newblock Minimum unique substrings and maximum repeats.
\newblock {\em Fundamenta Informaticae}, 110(1-4):183--195, 2011.
\newblock \href {https://doi.org/10.3233/FI-2011-536}
  {\path{doi:10.3233/FI-2011-536}}.

\bibitem[Jac89]{Jac89}
Guy Jacobson.
\newblock Space-efficient static trees and graphs.
\newblock In {\em 30th Annual Symposium on Foundations of Computer Science,
  {FOCS} 1989}, pages 549--554. {IEEE} Computer Society, 1989.
\newblock \href {https://doi.org/10.1109/SFCS.1989.63533}
  {\path{doi:10.1109/SFCS.1989.63533}}.

\bibitem[Kem19]{Kempa19}
Dominik Kempa.
\newblock Optimal construction of compressed indexes for highly repetitive
  texts.
\newblock In Timothy~M. Chan, editor, {\em 30th Annual {ACM-SIAM} Symposium on
  Discrete Algorithms, {SODA} 2019}, pages 1344--1357. {SIAM}, 2019.
\newblock \href {https://doi.org/10.1137/1.9781611975482.82}
  {\path{doi:10.1137/1.9781611975482.82}}.

\bibitem[KK99]{KolpakovK99}
Roman~M. Kolpakov and Gregory Kucherov.
\newblock Finding maximal repetitions in a word in linear time.
\newblock In {\em 40th Annual Symposium on Foundations of Computer Science,
  {FOCS} 1999}, pages 596--604. {IEEE} Computer Society, 1999.
\newblock \href {https://doi.org/10.1109/SFFCS.1999.814634}
  {\path{doi:10.1109/SFFCS.1999.814634}}.

\bibitem[KK03]{KolpakovK03}
Roman~M. Kolpakov and Gregory Kucherov.
\newblock Finding approximate repetitions under {H}amming distance.
\newblock {\em Theoretical Computer Science}, 303(1):135--156, 2003.
\newblock \href {https://doi.org/10.1016/S0304-3975(02)00448-6}
  {\path{doi:10.1016/S0304-3975(02)00448-6}}.

\bibitem[KK19]{sss}
Dominik Kempa and Tomasz Kociumaka.
\newblock String synchronizing sets: Sublinear-time {BWT} construction and
  optimal {LCE} data structure.
\newblock In Moses Charikar and Edith Cohen, editors, {\em 51st Annual {ACM}
  {SIGACT} Symposium on Theory of Computing, {STOC} 2019}, pages 756--767.
  {ACM}, 2019.
\newblock \href {https://doi.org/10.1145/3313276.3316368}
  {\path{doi:10.1145/3313276.3316368}}.

\bibitem[KK22]{dynsa}
Dominik Kempa and Tomasz Kociumaka.
\newblock Dynamic suffix array with polylogarithmic queries and updates.
\newblock In Stefano Leonardi and Anupam Gupta, editors, {\em 54th Annual {ACM}
  {SIGACT} Symposium on Theory of Computing, STOC 2022}, pages 1657--1670.
  {ACM}, 2022.
\newblock \href {https://doi.org/10.1145/3519935.3520061}
  {\path{doi:10.1145/3519935.3520061}}.

\bibitem[KK23a]{breaking}
Dominik Kempa and Tomasz Kociumaka.
\newblock Breaking the ${O(n)}$-barrier in the construction of compressed
  suffix arrays and suffix trees.
\newblock In Nikhil Bansal and Viswanath Nagarajan, editors, {\em 34th Annual
  {ACM-SIAM} Symposium on Discrete Algorithms, SODA 2023}, pages 5122--5202.
  {SIAM}, 2023.
\newblock \href {https://doi.org/10.1137/1.9781611977554.ch187}
  {\path{doi:10.1137/1.9781611977554.ch187}}.

\bibitem[KK23b]{collapsing}
Dominik Kempa and Tomasz Kociumaka.
\newblock Collapsing the hierarchy of compressed data structures: Suffix arrays
  in optimal compressed space.
\newblock In {\em 64th {IEEE} Annual Symposium on Foundations of Computer
  Science, {FOCS} 2023}, pages 1877--1886. {IEEE}, 2023.
\newblock \href {https://doi.org/10.1109/FOCS57990.2023.00114}
  {\path{doi:10.1109/FOCS57990.2023.00114}}.

\bibitem[KK24]{sublinearlz}
Dominik Kempa and Tomasz Kociumaka.
\newblock {L}empel-{Z}iv {(LZ77)} factorization in sublinear time.
\newblock In {\em 65th {IEEE} Annual Symposium on Foundations of Computer
  Science, {FOCS} 2024}, pages 2045--2055. {IEEE}, 2024.
\newblock Full version:
  \href{https://arxiv.org/pdf/2409.12146}{arXiv:2409.12146}.
\newblock \href {https://doi.org/10.1109/FOCS61266.2024.00122}
  {\path{doi:10.1109/FOCS61266.2024.00122}}.

\bibitem[KK25]{hierarchy}
Dominik Kempa and Tomasz Kociumaka.
\newblock On the hardness hierarchy for the ${O}(n\sqrt{\log n})$ complexity in
  the word {RAM}.
\newblock In {\em Proceedings of the 57th Annual {ACM} Symposium on Theory of
  Computing, {STOC} 2025}, pages 290--300. {ACM}, 2025.
\newblock \href {https://doi.org/10.1145/3717823.3718291}
  {\path{doi:10.1145/3717823.3718291}}.

\bibitem[KK26]{dichotomy}
Dominik Kempa and Tomasz Kociumaka.
\newblock Tight lower bounds for central string queries in compressed space.
\newblock In {\em Proceedings of the 2026 {ACM-SIAM} Symposium on Discrete
  Algorithms, {SODA} 2026}, 2026.
\newblock To appear.

\bibitem[KKP03]{KimKP03}
Dong~Kyue Kim, Yoo~Ah Kim, and Kunsoo Park.
\newblock Generalizations of suffix arrays to multi-dimensional matrices.
\newblock {\em Theoretical Computer Science}, 302(1-3):223--238, 2003.
\newblock \href {https://doi.org/10.1016/S0304-3975(02)00828-9}
  {\path{doi:10.1016/S0304-3975(02)00828-9}}.

\bibitem[KKP14]{srm}
Juha K{\"{a}}rkk{\"{a}}inen, Dominik Kempa, and Simon~J. Puglisi.
\newblock String range matching.
\newblock In Alexander~S. Kulikov, Sergei~O. Kuznetsov, and Pavel~A. Pevzner,
  editors, {\em 25th Annual Symposium on Combinatorial Pattern Matching, {CPM}
  2014}, volume 8486 of {\em LNCS}, pages 232--241. Springer, 2014.
\newblock \href {https://doi.org/10.1007/978-3-319-07566-2_24}
  {\path{doi:10.1007/978-3-319-07566-2_24}}.

\bibitem[KKP15]{pSAscan}
Juha K{\"{a}}rkk{\"{a}}inen, Dominik Kempa, and Simon~J. Puglisi.
\newblock Parallel external memory suffix sorting.
\newblock In Ferdinando Cicalese, Ely Porat, and Ugo Vaccaro, editors, {\em
  26th Annual Symposium on Combinatorial Pattern Matching, {CPM} 2015}, volume
  9133 of {\em Lecture Notes in Computer Science}, pages 329--342. Springer,
  2015.
\newblock \href {https://doi.org/10.1007/978-3-319-19929-0\_28}
  {\path{doi:10.1007/978-3-319-19929-0\_28}}.

\bibitem[KKR{\etalchar{+}}20]{KociumakaKRRW20}
Tomasz Kociumaka, Marcin Kubica, Jakub Radoszewski, Wojciech Rytter, and Tomasz
  Waleń.
\newblock A linear-time algorithm for seeds computation.
\newblock {\em ACM Transactions on Algorithms}, 16(2):27:1--27:23, 2020.
\newblock \href {https://doi.org/10.1145/3386369} {\path{doi:10.1145/3386369}}.

\bibitem[KNSW08]{kurtz2008new}
Stefan Kurtz, Apurva Narechania, Joshua~C Stein, and Doreen Ware.
\newblock A new method to compute $k$-mer frequencies and its application to
  annotate large repetitive plant genomes.
\newblock {\em BMC Genomics}, 9(1):1--18, 2008.
\newblock \href {https://doi.org/10.1186/1471-2164-9-517}
  {\path{doi:10.1186/1471-2164-9-517}}.

\bibitem[KP13]{KempaP13}
Dominik Kempa and Simon~J. Puglisi.
\newblock {L}empel-{Z}iv factorization: Simple, fast, practical.
\newblock In Peter Sanders and Norbert Zeh, editors, {\em 15th Meeting on
  Algorithm Engineering and Experiments, {ALENEX} 2013}, pages 103--112.
  {SIAM}, 2013.
\newblock \href {https://doi.org/10.1137/1.9781611972931.9}
  {\path{doi:10.1137/1.9781611972931.9}}.

\bibitem[KPD{\etalchar{+}}04]{kurtz2004versatile}
Stefan Kurtz, Adam Phillippy, Arthur~L. Delcher, Michael Smoot, Martin Shumway,
  Corina Antonescu, and Steven~L. Salzberg.
\newblock Versatile and open software for comparing large genomes.
\newblock {\em Genome Biology}, 5:1--9, 2004.
\newblock \href {https://doi.org/10.1186/gb-2004-5-2-r12}
  {\path{doi:10.1186/gb-2004-5-2-r12}}.

\bibitem[LD09]{bwa}
Heng Li and Richard Durbin.
\newblock Fast and accurate short read alignment with {B}urrows-{W}heeler
  transform.
\newblock {\em Bioinformatics}, 25(14):1754--1760, 2009.
\newblock \href {https://doi.org/10.1093/bioinformatics/btp324}
  {\path{doi:10.1093/bioinformatics/btp324}}.

\bibitem[LS12]{bowtie2}
Ben Langmead and Steven~L. Salzberg.
\newblock Fast gapped-read alignment with {B}owtie 2.
\newblock {\em Nature methods}, 9(4):357, 2012.
\newblock \href {https://doi.org/10.1038/nmeth.1923}
  {\path{doi:10.1038/nmeth.1923}}.

\bibitem[LSSY02]{LamSSY02}
Tak~Wah Lam, Kunihiko Sadakane, Wing{-}Kin Sung, and Siu{-}Ming Yiu.
\newblock A space and time efficient algorithm for constructing compressed
  suffix arrays.
\newblock In Oscar~H. Ibarra and Louxin Zhang, editors, {\em 8th Annual
  International Conference on Computing and Combinatorics, {COCOON} 2002},
  volume 2387 of {\em LNCS}, pages 401--410. Springer, 2002.
\newblock \href {https://doi.org/10.1007/3-540-45655-4_43}
  {\path{doi:10.1007/3-540-45655-4_43}}.

\bibitem[LV88]{LandauV88}
Gad~M. Landau and Uzi Vishkin.
\newblock Fast string matching with $k$ differences.
\newblock {\em Journal of Computer and System Sciences}, 37(1):63--78, 1988.
\newblock \href {https://doi.org/10.1016/0022-0000(88)90045-1}
  {\path{doi:10.1016/0022-0000(88)90045-1}}.

\bibitem[Mai89]{main1989detecting}
Michael~G Main.
\newblock Detecting leftmost maximal periodicities.
\newblock {\em Discrete Applied Mathematics}, 25(1-2):145--153, 1989.
\newblock \href {https://doi.org/10.1016/0166-218X(89)90051-6}
  {\path{doi:10.1016/0166-218X(89)90051-6}}.

\bibitem[M{\"{a}}k03]{Makinen03}
Veli M{\"{a}}kinen.
\newblock Compact suffix array: {A} space-efficient full-text index.
\newblock {\em Fundamenta Informaticae}, 56(1-2):191--210, 2003.

\bibitem[MB91]{ManberB91}
Udi Manber and Ricardo~A. Baeza{-}Yates.
\newblock An algorithm for string matching with a sequence of don't cares.
\newblock {\em Information Processing Letters}, 37(3):133--136, 1991.
\newblock \href {https://doi.org/10.1016/0020-0190(91)90032-D}
  {\path{doi:10.1016/0020-0190(91)90032-D}}.

\bibitem[MM90]{ManberM90}
Udi Manber and Gene Myers.
\newblock Suffix arrays: {A} new method for on-line string searches.
\newblock In David~S. Johnson, editor, {\em 1st Annual {ACM-SIAM} Symposium on
  Discrete Algorithms, {SODA} 1990}, pages 319--327. {SIAM}, 1990.
\newblock URL: \url{http://dl.acm.org/citation.cfm?id=320176.320218}.

\bibitem[MM93]{sa}
Udi Manber and Eugene~W. Myers.
\newblock Suffix arrays: {A} new method for on-line string searches.
\newblock {\em SIAM Journal on Computing}, 22(5):935--948, 1993.
\newblock \href {https://doi.org/10.1137/0222058} {\path{doi:10.1137/0222058}}.

\bibitem[MN05]{MakinenN05}
Veli M{\"{a}}kinen and Gonzalo Navarro.
\newblock Succinct suffix arrays based on run-length encoding.
\newblock {\em Nordic Journal of Computing}, 12(1):40--66, 2005.
\newblock \href {https://doi.org/10.1007/11496656_5}
  {\path{doi:10.1007/11496656_5}}.

\bibitem[MNN17]{MunroNN17}
J.~Ian Munro, Gonzalo Navarro, and Yakov Nekrich.
\newblock Space-efficient construction of compressed indexes in deterministic
  linear time.
\newblock In Philip~N. Klein, editor, {\em 28th Annual {ACM-SIAM} Symposium on
  Discrete Algorithms, {SODA} 2017}, pages 408--424. {SIAM}, 2017.
\newblock \href {https://doi.org/10.1137/1.9781611974782.26}
  {\path{doi:10.1137/1.9781611974782.26}}.

\bibitem[MNN20]{MunroNN20}
J.~Ian Munro, Gonzalo Navarro, and Yakov Nekrich.
\newblock Text indexing and searching in sublinear time.
\newblock In Inge~Li G{\o}rtz and Oren Weimann, editors, {\em 31st Annual
  Symposium on Combinatorial Pattern Matching, {CPM} 2020}, volume 161 of {\em
  LIPIcs}, pages 24:1--24:15. Schloss Dagstuhl--Leibniz-Zentrum f{\"{u}}r
  Informatik, 2020.
\newblock \href {https://doi.org/10.4230/LIPICS.CPM.2020.24}
  {\path{doi:10.4230/LIPICS.CPM.2020.24}}.

\bibitem[MNV16]{MunroNV16}
J.~Ian Munro, Yakov Nekrich, and Jeffrey~Scott Vitter.
\newblock Fast construction of wavelet trees.
\newblock {\em Theoretical Computer Science}, 638:91--97, 2016.
\newblock \href {https://doi.org/10.1016/j.tcs.2015.11.011}
  {\path{doi:10.1016/j.tcs.2015.11.011}}.

\bibitem[MSST20]{MatsudaSST20}
Kotaro Matsuda, Kunihiko Sadakane, Tatiana Starikovskaya, and Masakazu
  Tateshita.
\newblock Compressed orthogonal search on suffix arrays with applications to
  range {LCP}.
\newblock In Inge~Li G{\o}rtz and Oren Weimann, editors, {\em 31st Annual
  Symposium on Combinatorial Pattern Matching, {CPM} 2020}, volume 161 of {\em
  LIPIcs}, pages 23:1--23:13. Schloss Dagstuhl--Leibniz-Zentrum f{\"{u}}r
  Informatik, 2020.
\newblock \href {https://doi.org/10.4230/LIPICS.CPM.2020.23}
  {\path{doi:10.4230/LIPICS.CPM.2020.23}}.

\bibitem[Nav01]{Navarro01}
Gonzalo Navarro.
\newblock A guided tour to approximate string matching.
\newblock {\em {ACM} Computing Surveys}, 33(1):31--88, 2001.
\newblock \href {https://doi.org/10.1145/375360.375365}
  {\path{doi:10.1145/375360.375365}}.

\bibitem[NM07]{NavarroM07}
Gonzalo Navarro and Veli M{\"{a}}kinen.
\newblock Compressed full-text indexes.
\newblock {\em ACM Computing Surveys}, 39(1):2:1--2:61, 2007.
\newblock \href {https://doi.org/10.1145/1216370.1216372}
  {\path{doi:10.1145/1216370.1216372}}.

\bibitem[NPL{\etalchar{+}}13]{NaPLHLMP13}
Joong~Chae Na, Heejin Park, Sunho Lee, Minsung Hong, Thierry Lecroq, Laurent
  Mouchard, and Kunsoo Park.
\newblock Suffix array of alignment: {A} practical index for similar data.
\newblock In Oren Kurland, Moshe Lewenstein, and Ely Porat, editors, {\em 20th
  International Symposium on String Processing and Information Retrieval,
  {SPIRE} 2013}, volume 8214 of {\em LNCS}, pages 243--254. Springer, 2013.
\newblock \href {https://doi.org/10.1007/978-3-319-02432-5_27}
  {\path{doi:10.1007/978-3-319-02432-5_27}}.

\bibitem[OG10]{OhlebuschG10}
Enno Ohlebusch and Simon Gog.
\newblock Efficient algorithms for the all-pairs suffix-prefix problem and the
  all-pairs substring-prefix problem.
\newblock {\em Information Processing Letters}, 110(3):123--128, 2010.
\newblock \href {https://doi.org/10.1016/J.IPL.2009.10.015}
  {\path{doi:10.1016/J.IPL.2009.10.015}}.

\bibitem[OG11]{OhlebuschG11}
Enno Ohlebusch and Simon Gog.
\newblock {L}empel-{Z}iv factorization revisited.
\newblock In Raffaele Giancarlo and Giovanni Manzini, editors, {\em 22nd Annual
  Symposium on Combinatorial Pattern Matching, {CPM} 2011}, volume 6661 of {\em
  LNCS}, pages 15--26. Springer, 2011.
\newblock \href {https://doi.org/10.1007/978-3-642-21458-5_4}
  {\path{doi:10.1007/978-3-642-21458-5_4}}.

\bibitem[Ohl13]{ennobook}
Enno Ohlebusch.
\newblock {\em Bioinformatics algorithms: Sequence analysis, genome
  rearrangements, and phylogenetic reconstruction}.
\newblock Oldenbusch Verlag, Ulm, Germany, 2013.

\bibitem[OS09]{OkanoharaS09}
Daisuke Okanohara and Kunihiko Sadakane.
\newblock A linear-time {B}urrows--{W}heeler transform using induced sorting.
\newblock In Jussi Karlgren, Jorma Tarhio, and Heikki Hyyr{\"{o}}, editors,
  {\em 16th International Symposium on String Processing and Information
  Retrieval, {SPIRE} 2009}, volume 5721 of {\em LNCS}, pages 90--101. Springer,
  2009.
\newblock \href {https://doi.org/10.1007/978-3-642-03784-9_9}
  {\path{doi:10.1007/978-3-642-03784-9_9}}.

\bibitem[Pre18]{Prezza18}
Nicola Prezza.
\newblock In-place sparse suffix sorting.
\newblock In Artur Czumaj, editor, {\em 29th Annual {ACM-SIAM} Symposium on
  Discrete Algorithms, {SODA} 2018}, pages 1496--1508. {SIAM}, 2018.
\newblock \href {https://doi.org/10.1137/1.9781611975031.98}
  {\path{doi:10.1137/1.9781611975031.98}}.

\bibitem[RNO11]{RussoNO11}
Lu{\'{\i}}s M.~S. Russo, Gonzalo Navarro, and Arlindo~L. Oliveira.
\newblock Fully compressed suffix trees.
\newblock {\em ACM Transactions on Algorithms}, 7(4):53:1--53:34, 2011.
\newblock \href {https://doi.org/10.1145/2000807.2000821}
  {\path{doi:10.1145/2000807.2000821}}.

\bibitem[Sad02]{cst}
Kunihiko Sadakane.
\newblock Succinct representations of {LCP} information and improvements in the
  compressed suffix arrays.
\newblock In {\em 13th Annual {ACM-SIAM} Symposium on Discrete Algorithms, SODA
  2002}, pages 225--232. {ACM/SIAM}, 2002.
\newblock URL: \url{http://dl.acm.org/citation.cfm?id=545381.545410}.

\bibitem[Shi96]{shi1996suffix}
Fei Shi.
\newblock Suffix arrays for multiple strings: a method for on-line multiple
  string searches.
\newblock In Joxan Jaffar and Roland H.~C. Yap, editors, {\em 2nd Asian
  Conference on Computing Science, {ASIAN 1996}}, volume 1179 of {\em LNCS},
  pages 11--22. Springer, 1996.
\newblock \href {https://doi.org/10.1007/bfb0027775}
  {\path{doi:10.1007/bfb0027775}}.

\bibitem[Wei73]{Weiner73}
Peter Weiner.
\newblock Linear pattern matching algorithms.
\newblock In {\em 14th Annual Symposium on Switching and Automata Theory, {SWAT
  (FOCS)} 1973}, pages 1--11. {IEEE} Computer Society, 1973.
\newblock \href {https://doi.org/10.1109/SWAT.1973.13}
  {\path{doi:10.1109/SWAT.1973.13}}.

\end{thebibliography}

\end{document}